\documentclass[aps,nofootinbib,endfloat,showkeys]{revtex4}

\usepackage{graphicx}
\usepackage{epsfig}
\usepackage{subeqn}

\newcommand{\aln}[1]{\alpha_{0}(#1,\omega)}
\newcommand{\al}[1]{\alpha(#1,\omega)}
\newcommand{\eps}[1]{\varepsilon_{#1}(\omega)}
\newcommand{\ooc}[1]{\frac{\omega^{#1}}{c^{#1}}}
\newcommand{\qint}[1]{\int^{\infty}_{-\infty}\!\frac{d#1}{2\pi}}
\renewcommand{\r}[1]{$(\!\!~\ref{#1})$}  
\newcommand{\nn}{\nonumber}

\newcommand{\citen}[1]{\cite{#1}}
\newcommand{\subsubsubsection}[1]{\vspace{\baselineskip}\begin{center}{\noindent \em #1}\end{center}\vspace{\baselineskip}}

\newcommand{\myfigpath}{./}

\begin{document}

\title{A random walk through surface scattering phenomena:\\ 
Theory and phenomenology}

\author{Ingve Simonsen}
\email{Ingve.Simonsen@phys.ntnu.no}

\affiliation{Department of Physics, 
  Norwegian University of Science and Technology, 
  NO-7491 Trondheim, Norway}

\begin{abstract}
  No surface is perfectly planar at all scales. The notion of flatness
  of a surface therefore depends on the size of the probe used to
  observe it. As a consequence rough interfaces are abundant in
  nature.  Here the old, but still active field of rough surface
  scattering of electromagnetic waves is addressed. This topic has
  implications and practical applications in fields as diverse as
  observational astronomy and the electronics industry. This article
  reviews the theoretical and computational foundation and methods
  used in the study of rough surface scattering. Furthermore, it
  presents and explains the physical origin of a series of multiple
  scattering surface phenomena. In particular what is discussed are:
  the enhanced backscattering and satellite peak phenomena, coherent
  effects in angular intensity correlation functions and second
  harmonic generated light (a non-linear effect).
\end{abstract}

\date{\today} 
\keywords{Wave scattering; random systems; multiple scattering; coherent effects}

\maketitle
\tableofcontents


\section{Introduction}
\label{Chap:Introduction}

We are surrounded by waves, and they effect our daily life in a way
that many of us are not aware.  Sound and light are our main tools for
observing our immediate surroundings. Light is for example responsible
for you being able to read these lines, and more important, to get
access to the vast majority of the knowledge accumulated in writings
by man throughout centuries of intellectual activities. X-ray and
ultra sound techniques have given tremendous contribution to the
success of modern medicine.  Radio- and micro-waves are invaluable in
modern communication technology including cellular phones and radio
and TV broadcastings.  Understanding of quantum waves, and their
behavior, constitutes the foundation of electronics and semiconductor
technologies --- an essential ingredient in the past and future
progress of computer hardware. The above list is not at all, or
intended to be, complete.  It could in fact easily been made much
longer.  However, the bottom line that we want to make here is that
with the ubiquitous presence of wave phenomena in various
applications, it is not surprising to find that wave phenomena have
had, and still have, a prominent position in our studies of the physical
world, and even today such phenomena are of out-most importance in
science, medicine and technology.

If you take an average introductionary text on wave phenomena, you
will find discussions of how plane waves of constant frequency
propagates in a homogeneous, isotropic medium. Thereafter, the authors
typically discuss the scattering and transmission of such waves at a
{\em planar} interface separating two semi-infinite media of different
dielectric properties~\cite{Book:Kong} --- the Fresnel formulae. These
formulae serve to accurately describe the scattering of light from for
example a mirror.  However, from our everyday experience, we know that
most surfaces are not mirror like, and naturally occurring objects are
more complicated then two semi-infinite media.  Most naturally
occurring surfaces are actually not smooth at all. They are, however,
rough in some sense.  In fact, all objects, man-made or not, {\em
  have} to be rough at atomic scales, but such small length scales are
normally not resolved by our probes.

It should be kept in mind that the characterization of a surface as
rough or smooth is noticeably not unique, and it is not a intrinsic
property of the surface. Instead, however, it depends on the
wavelength used to ``observe'' the surface.  If the typical roughness
is on a scale much smaller then the wavelength of the probe, this
surface is considered as smooth.  However, by reducing the wavelength
of the light, the same surface might also be characterized as being
rough. It is, among other factors, the surface topography and the
wavelength of the probe, as we will see below, that together go into
the characterization of a surface as being rough\footnote{When
  discussing the Rayleigh criterion later in this section
  we will see that also the angle of incidence of the light will play
  an important role.}.

\begin{figure}[t!]
  \begin{center}
    \leavevmode
    \includegraphics[width=3cm,height=13cm,angle=-90]{\myfigpath/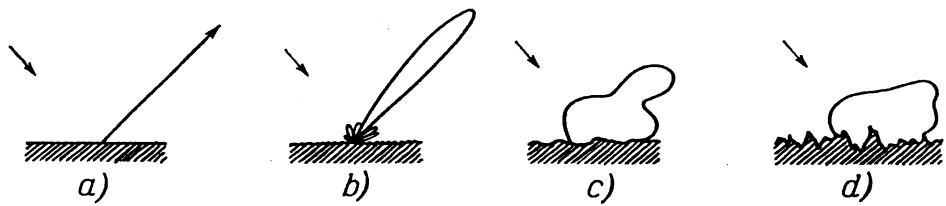}
    \caption{An illustration showing the transition from
      specular~(Fig.~\protect\ref {Fig:planar_to_rough}a) to diffuse
      scattering~(Fig.~\protect\ref {Fig:planar_to_rough}d) of light
      from a surface of increasing roughness. The arrows indicate the
      direction of the incident light (After Ref.~\citen{Book:Beckmann}).}
    \label{Fig:planar_to_rough}
  \end{center}
\end{figure}

Let us from now on assume an electromagnetic probe, {\it i.e.} light.
If the surface can be considered as smooth, light is scattered
(coherently) into the specular direction. As the roughness of the
surface is increased so that the surface becomes weakly rough, a small
fraction of the incident light will be scattered into other directions
than the specular one. This non-specular scattering is called {\em
  diffuse scattering} or by some authors {\em incoherent scattering}.
As the roughness is increased even further, the diffuse (incoherent)
component of the scattered light is increased on the expense of the
specular component.  When the surface roughness is so that the
specular component can be more-or-less neglected as compared to the
diffuse component, the surface is said to be strongly rough. This
transmission from as smooth to a strongly rough surface is depict in
Figs.~\ref{Fig:planar_to_rough}.


\vspace{\baselineskip}

Due to the practical applications of waves, and the number of
naturally occurring surfaces being rough, it is rather remarkable that
it took several hundreds years from the birth of optics as a
scientific discipline to someone started to considered wave
scattering from rough surfaces. As far as we know today, the first
such theoretical study was made at the end of the 19th century
(probably in the year of 1877) by one of the greatest scientists of
its time, the British physicist Lord
Rayleigh~\cite{Rayleigh1,Rayleigh2}. He considered the scattering of
light incident normally onto a sinusoidal surface.

In 1913 Mandel'shtam studied how light was scattered from liquid
surfaces~\cite{Mandelshtam}. By doing so, he became the first to
consider scattering from {\em randomly} rough surfaces. This, as it
turned out, should define the beginning of an active research area ---
{\em wave scattering from randomly rough surfaces} --- which still
today is an active field. However, it was first after the last world
war that the research effort put into the field stated to
accelerate~\cite{Rice}.  Since that time, a massive body of research
literature has been generated in the
field~\cite{Book:Beckmann,Book:Ogilvy,Book:Voro,Book:Nieto}.

Up to the mid 1980's most of the theories used in this field were single
scattering theories~\cite{Book:Beckmann,Book:Ogilvy,Book:Voro}. However,
from then on the main focus of the research has been on multiple scattering
theories. In addition, advances in experimental techniques has lately
enabled experimentalists to fabricate surfaces under well controlled
conditions by using a holographic grating technique~\cite{fabrication}.
This has opened up a unique possibility for direct comparison of theory and
experiments in a way not possible a few decades ago.

\vspace{\baselineskip}

Inspired by the works of Lord Rayleigh~\cite{Rayleigh1,Rayleigh2}
researchers developed a criterion --- the Rayleigh criterion --- that
could be used to determine when a given surface was to be considered
as rough. Here both the wavelength of the incident light as well as
its angle of incidence are incorporated~\cite{Book:Ogilvy}. 

To illustrate how this comes about, let us consider a rough surface
defined by $x_3=\zeta(x_1)$. On this surface we pick two arbitrary
points $(\xi,\zeta(\xi))$ and $(\xi',\zeta(\xi'))$. It could now be
asked: What is the phase difference between two waves being scattered
from these two points? For simplicity we will here only consider the
specular direction. Under this assumption it is straight forward to
show that the phase difference is given by the following expression
\begin{eqnarray}
  \label{Eq:Intro:Phase}
  \Delta \phi &=& 2 \left| {\mathbf k} \right|
             \left| \zeta(\xi)-\zeta(\xi')  \right|
             \cos \theta_0,
\end{eqnarray}
where $|{\mathbf k}| =2\pi/\lambda$ is the modulus of the wave vector of
the incident light of wavelength $\lambda$, and $\theta_0$ is the
angle of incidence of the light as measured from the normal to the mean
surface. From Eq.~\r{Eq:Intro:Phase} we immediately observe that if
the surface is planar, so that $\zeta(\xi)=\zeta(\xi')$, the phase
difference (in the specular direction) is always zero independent of
the angle of incidence. However, if the surface is rough,
$\Delta\phi\neq0$ in general. If $\Delta\phi\ll\pi$, the two waves
will be in, or almost in, phase and they will thus interfere
constructively. On the other hand, if $\Delta\phi\simeq\pi$, they will be
(more-or-less) completely out off phase and as a result interfere
destructively, and no, or almost no, energy will be scattered into the
specular direction. In terms of the phase, a smooth surface would
correspond to $\Delta\phi\ll\pi$, and a rough one to
$\Delta\phi\simeq\pi$. Thus, $\Delta\phi=\pi/2$ might be considered as
the borderline between a smooth and a rough surface; if $\Delta\phi<
\pi/2$ the surface is smooth, and otherwise
($\pi/2<\Delta\phi\leq\pi$) it is rough. The criterion
$\Delta\phi<\pi/2$ is the famous {\rm Rayleigh criterion} for a smooth
surface.

If the surface is randomly rough, it is practical to replace the
height difference $\zeta(\xi)-\zeta(\xi')$ by a typical height fluctuation
as provided, for example, by the {\sc rms}-height, $\delta$, of the surface.
Hence, the Rayleigh criterion can be expressed as
\begin{eqnarray}
  \label{Rayleigh-parameter}
  R_a &=& \left|{\mathbf k}\right|\delta\cos\theta_0 
          \;\;<\;\; \frac{\pi}{4},
\end{eqnarray}
where $R_a$ is the so-called Rayleigh parameter.  From the Rayleigh
criterion, $R_a<\pi/4$, it should be observed that in addition to the
surface topography itself and the wavelength of the light, also its
angle of incidence goes into determining if a surface is rough or not.
This is probably the most important lesson to be learned today from
the Rayleigh criterion.

\vspace{\baselineskip}

The present review consists of basically two main parts --- one focus
theoretical methods whilst the other one is devoted to rough surface
scattering phenomenology. In the first part we try to present an
overview of some of the main theories and methods used in the study of
wave scattering from randomly rough surfaces. We start in
Sect.~\ref{Chapter:Elmag} by recapitulating the basic results of
electromagnetic theory including Maxwell's equations.  This section
serves among other things to define our notation. Then we continue by
describing how to characterize randomly rough surfaces
(Sect.~\ref{Chap:Characterization}). Sect.~\ref{Chap:Theory} is
devoted to the quantities and main techniques used in the field of
electromagnetic wave scattering from randomly rough surfaces. We here
review classical theories like small amplitude perturbation theory,
many-body perturbation theory as well as numerical simulation
approaches. Finally in Sect.~\ref{Chapt:Phenomena} we discuss some of
the phenomena that may occur when light is scattered from rough
surfaces.  Such effects include the backscattering and satellite peaks
phenomena (weak localization), Anderson localization, angular
intensity correlation effects and nonlinear effects (second harmonic
generation).


\section{Elements of Electromagnetic theory}
\label{Chapter:Elmag}

The present review mainly concern itself with rough surfaces and the
scattering of electromagnetic wave from such. In this section we
therefor review some of the basic results of electromagnetic theory,
including surface polaritons.  The present section also serves to
define our notation that we will use extensively in the following
sections.  The style of this review is kept quite brief, since all the
material should be well known. A more thorough treatments
can be found for example in the classical text on electrodynamics by
J.\ D.\ Jackson~\cite{Jackson}.

\subsection{Maxwell's Equations and the Constitutive Equations} 
\label{Sect:Maxwell-equations-const-rel}


\begin{table}[tbhp]
  \begin{center}
    \begin{tabular}{c|ll}
      \hline
      Quantity & SI-unit & Name \\
      \hline\hline
       ${\mathbf E}$ & $V/m$            & Electric field  \\
       ${\mathbf H}$ & $A/m$            & Magnetic field  \\
       ${\mathbf D}$ & $C/m^2$          & Electric displacement  \\
       ${\mathbf B}$ & $Wb/m^2$         & Magnetic induction  \\
       $\rho$        & $C/m^3$          & Charge density \\
       ${\mathbf J}$ & $A/m^2$          & Current density  \\
      \hline
    \end{tabular}
    \label{Table:Maxwell}  
    \caption{Summary of the quantities contained in Maxwell's
      equations, as well as their SI-units.}
  \end{center}
\end{table}


\subsubsection{Maxwell's Equations}

The Maxwell's equations, which unify in one magnificent theory all
the phenomena of electricity and magnetism, were put forward by the
Scottish physicist James Clerk Maxwell~(1831--1879).  These equations
are the fundamental equations of electromagnetism, in the same way
that Newton's law is to classical mechanics. In fact, the Maxwell's
equations are in a way even more fundamental since they are consistent
with the theory of special relativity that Einstein develop years
later. Because all of electromagnetism is contained within this set of
equations, they are definitely among one of the greatest triumphs of
the human mind.

Strictly speaking the equations put forward by Maxwell only applies to point
charges in vacuum. A dielectric, for example, is a collection of a very huge
number of point charges. To deal with them all individually is an impossible
task. It is therefore practical to introduce effective fields, ${\mathbf D}$
and ${\mathbf H}$, to represent their collective behavior. This dielectric
approach to electromagnetism represents great simplifications for many
(near-to-natural) systems. It is based on the following two
assumptions~\cite{Jackson}: ({\em i}) the response of the background medium
is dipole like as well as linear in the applied fields, and ({\em ii}) the
medium is homogeneous (or close to) throughout a given region. The first
assumption obviously breaks down if the fields becomes to strong while the
latter breaks down on short length scales. Hence the resulting effective
field theory, or effective Maxwell theory as we might call it, should be
treated as a long-wavelength approximation to electromagnetism for weak
fields. In most practical situations the above approximations are
fortunately well satisfied, and in particular they are valid for the type of
scattering system that we will be considering.

In the SI-system, the (effective) Maxwell's equations take on the
following form:
\begin{subequations}
   \label{Elmag:MaxwellEq}  
\begin{eqnarray}
  {\mathbf \nabla} \cdot {\mathbf D} &=& \rho,   
          \label{Elmag:MaxwellEq1}  \\
  {\mathbf \nabla} \cdot {\mathbf B} &=& 0,
          \label{Elamag:MaxwellEq2} \\
  {\mathbf \nabla} \times {\mathbf E} &=& 
        -\frac{\partial{\mathbf B}}{\partial t},      
          \label{Elmag:MaxwellEq3}  \\
  {\mathbf \nabla} \times {\mathbf H} &=& 
     \frac{\partial {\mathbf D}}{\partial t} + {\mathbf J}.
          \label{Elmag:MaxwellEq4}
\end{eqnarray}
\end{subequations}
Here ${\mathbf E}$ and ${\mathbf H}$ denote the electric and magnetic
field vectors respectively. These field vectors make together up what
is known as the electromagnetic field. The field quantities ${\mathbf
  D}$ and ${\mathbf B}$, known as the electrical displacement and the
magnetic induction respectively, are included in order to describe the
effect of the electromagnetic field on matter. Finally,
$\rho$ and ${\mathbf J}$ denote the charge density and the current
density respectively. Those two latter quantities act like sources for
the electromagnetic field, ${\mathbf E}$ and ${\mathbf H}$, and they
fulfill the continuity relation
\begin{eqnarray}
  \frac{\partial \rho}{\partial t} + {\mathbf \nabla}\cdot {\mathbf J}
  &=& 0.
\end{eqnarray}
The various quantities appearing in the Maxwell's equations, and
related formulae, are summarized in Table~\ref{Table:Maxwell} where
also their SI-units are given.

\subsubsection{Constitutive Equations}
\label{Subsect:Constitutive Equations}
   
The Maxwell's equations (\ref{Elmag:MaxwellEq}) consist of eight scalar
equations. However, on the other hand the field vectors, ${\mathbf
  E}$, ${\mathbf H}$, ${\mathbf D}$, and ${\mathbf B}$, represent in
total 12 (scalar) variables, 3 for each of the 4 vectors.  Thus,
obviously, the Maxwell's equations alone do not uniquely specify a
solution. Therefore, in order to obtain a unique solution to the
Maxwell's equations, those are supplemented by so-called constitutive
relations also known as material equations.  These relations read
\begin{subequations}
  \label{Elmag:constitutive_1}
\begin{eqnarray}
       {\mathbf D} &=& \varepsilon  {\mathbf E}, \\ 
       {\mathbf B} &=& \mu          {\mathbf H}.
\end{eqnarray}
\end{subequations}   
Here $\varepsilon$ and $\mu$ are the constitutive parameters which are
tensors of 2nd order and known as the {\em permittivity}\footnote{This
  quantity is also known as the dielectric function.}, and the {\em
  permeability} tensor respectively.  In general these tensors are
rather complicated functions of the spatial variable ${\mathbf x}$ and
the field vectors ${\mathbf E}$ and ${\mathbf H}$. However, for an
isotropic and homogeneous medium, these tensors reduce to scalars, and
if the field-strengths are not too large, they may be considered as
independent of the field vectors. In this latter case we are dealing
with linear electromagnetic theory.  The fascinating, but complicated
nonlinear electromagnetic theory~\cite{Book:Keller90} where per
definition $\varepsilon$ and $\mu$ depend on ${\mathbf E}$ and
${\mathbf H}$, will not be discussed here in any depth.

Eqs.~\r{Elmag:constitutive_1} can within linear electromagnetic
theory be cast  into the equivalent form 
\begin{subequations}
  \label{Elmag:constitutive}
\begin{eqnarray}
       {\mathbf D} &=& \varepsilon_0  {\mathbf E} + {\mathbf P},  \\
       {\mathbf B} &=& \mu_0  {\mathbf H} + {\mathbf M},
\end{eqnarray}
\end{subequations}   
where ${\mathbf P}$ and ${\mathbf M}$ are the electric and magnetic
polarizations respectively.  The constants $\varepsilon_0$ and $\mu_0$
are the permittivity and permeability of vacuum respectively.  In the
SI-system they have the following values
\begin{subequations}
\begin{eqnarray}
  \varepsilon_0 &=& 8.854 \times 10^{-12} \; \mbox{F/m}, \\
  \mu_0         &=& 4\pi  \times 10^{-7}  \; \mbox{H/m}.
\end{eqnarray}
\end{subequations}

\subsection{The Electromagnetic Wave Equations}
 
Probably the two most important consequences of the Maxwell's
equations are the {\em wave equations} and the
existence of solutions to these which are known as {\em
  electromagnetic waves} due to the wave-like nature of such
solutions. In this section we derive the wave equations in a material
medium. For simplicity, and since it is the most relevant case for
this review, we will limit ourselves to a region of space which is
source free and isotropic.

The derivation of the wave equations for the ${\mathbf E}$-field in a
source-free region ({\it i.e.} $\rho=0$ and ${\mathbf J}={\mathbf
  0}$), is achieved by eliminating the ${\mathbf H}$-field from the
Maxwell's equations. This is done by taking the curl of
Eq.~\r{Elmag:MaxwellEq3}, substituting Eq.~\r{Elmag:MaxwellEq4}, and
taking advantage of the constitutive
relations~\r{Elmag:constitutive_1}. The result is
\begin{eqnarray}
  \label{Elmag:temp}
  {\mathbf \nabla} \times ({\mathbf \nabla} \times {\mathbf E}) 
  + \varepsilon \mu \frac{\partial^2 {\mathbf E}}{\partial t^2}
  &=& 0.
\end{eqnarray}
By applying the vector identity $\nabla \times(\nabla\times {\mathbf
  A})= \nabla(\nabla \cdot {\mathbf A}) -\nabla^2{\mathbf A}$ to
Eq.~\r{Elmag:temp} and taking advantage of Eq.~\r{Elmag:MaxwellEq1}
we obtain the well-known standard (space-time) wave
equation for the electrical field in a source-free, homogeneous and
isotropic medium
\begin{eqnarray}
  {\mathbf \nabla}^2{\mathbf E}
       -\mu\varepsilon\frac{\partial^2{\mathbf E}}{\partial t^2} 
       &=& 0.
\label{Elmag:WEq}  
\end{eqnarray} 
In a similar way one can obtain a wave equation for the magnetic field 
by eliminating the electric field from
the Maxwell's equations.

It should be notice that not every solution to the wave equation is
also a solution to the Maxwell's equations. For it to be, it must in
addition satisfy Gauss's law, ${\mathbf \nabla} \cdot{\mathbf
  E}={\mathbf k}\cdot{\mathbf E}=0$ in order to also be a solution of
Maxwell's equations\footnote{One explicate example of this is provided
  by ${\mathbf E}= \hat{\mathbf e}_3\, E_0 \cos(kx_3-\omega t)$ that
  satisfies the wave-equation, but not ${\mathbf \nabla} \cdot{\mathbf
    E}=0$. It must therefore be discarded as a solution of the
  Maxwell's equations.}.  As the reader readily may check the wave
equation has a solution ${\mathbf E} = \exp(i{\mathbf k} \cdot{\mathbf
  r}-i\omega t)$ if $\omega=ck$. This solution is the plane wave
solution.


\begin{figure}[t!]
  \begin{center}
    \leavevmode
    \includegraphics[width=12cm,height=3.5cm]{\myfigpath/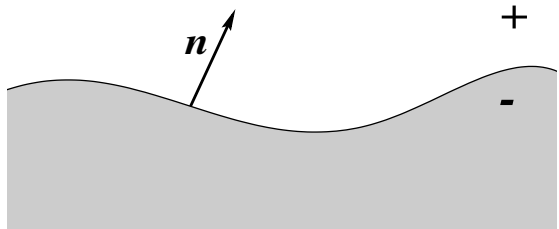}    
    \caption{A sketch of a general interface separating two dielectric
      media.}
    \label{Fig:Elmag:genral-geomegry}
  \end{center}
\end{figure}


\subsection{Boundary Conditions}
\label{Sec:BoundaryConditions}

In Sect.~\ref{Sect:Maxwell-equations-const-rel} we introduced the
Maxwell's equations and the constitutive relations. These equations
can be solved for the field vectors in a region of space containing no
boundaries. However, no real media are infinite, {\em i.e.} without
boundaries. For practical applications of the electromagnetic theory it
is therefore important to know how to treat the boundaries between two
media of different electromagnetic properties. It is this question
that we will address in this section.

Let us consider the geometry of
Figure~\ref{Fig:Elmag:genral-geomegry}. It shows an arbitrary
interface separating the otherwise homogeneous, isotropic and linear
media labeled $\pm$.  We have also introduced a normal vector for the
interface, ${\mathbf n}$, which is directed into medium~$+$.  The
question we now address is: How are the electromagnetic field vectors
for the two media in the immediate vicinity of the interface related
to each other?  The answer to this question should be well-known and
can be found in nearly any book on electromagnetic theory, {\it e.g.}
in Refs.~\citen{Book:Kong} and \citen{Jackson}.  The results, for
which the derivation will not be repeated here, are
\begin{subequations}
  \label{Elmag:BC}
\begin{eqnarray}
  {\mathbf n  \cdot } \left( {\mathbf B}_- - {\mathbf B}_+\right) &=& 0, 
        \label{Elmag:BC1}\\
  {\mathbf n  \cdot } \left( {\mathbf D}_- - {\mathbf D}_+\right) &=& \rho_s, 
        \label{Elmag:BC2}\\
  {\mathbf n \times }   \left( {\mathbf E}_- - {\mathbf E}_+\right)
     &=& 0, 
        \label{Elmag:BC3}\\
  {\mathbf n \times }   \left( {\mathbf H}_- - {\mathbf H}_+\right) 
     &=& {\mathbf J}_s, 
        \label{Elmag:BC4}
\end{eqnarray}
\end{subequations}
where the vector subscripts, $\pm$, are referring to the media where
the field vectors are evaluated.  In Eqs.~\r{Elmag:BC} $\rho_s$ and
${\mathbf J}_s$ denote the surface charge density and the surface
current density respectively, while  the other quantities have been
defined earlier. In many areas of optics one deals with situations
where the surface charge density and the surface current density are
zero. Under such circumstances the normal component of ${\mathbf B}$
and ${\mathbf D}$ are continuous, while the vectors ${\mathbf E}$
and ${\mathbf H}$ have continuous tangential components.

It should be stressed that in arriving at the results~\r{Elmag:BC}, it
has been assumed that the electromagnetic properties take on their
bulk values all the way to the surface.  This is obviously not true,
but is a good approximation  whenever the mean field theory applies.

\subsubsection{Boundary Condition at a General One-Dimensional Surface}

Most of this review will concern itself with randomly rough surfaces
that are effectively one-dimensional, {\it i.e.} the surface profile
function $\zeta$ has a non-trivial dependence only on $x_1$, say, and
does not depend explicitly on $x_2$. In this case the boundary
conditions~\r{Elmag:BC} simplifies somewhat. This is what we plan to
outline in this section.
 
Let us start by assuming, without loss of generality, that the plane of
incidence is the $x_1x_3$-plane and that the incident light is either
$p$- or $s$-polarized. In such case, there is only one non-trivial
field component needed in order to fully describe the electromagnetic
filed. For $p$-polarization this component is $H_2$, while for
$s$-polarization it is $E_2$. Thus the primary field for a
one-dimensional interface problem can be written as 
\begin{eqnarray}
    \label{Elmag:Primary-fields}
    \Phi_\nu(x_1,x_3 | \omega ) 
      &=& \left\{
               \begin{array}{ll}
                  H_2(x_1,x_3|\omega), & \quad \nu=p, \\
                  E_2(x_1,x_3|\omega), & \quad \nu=s,
               \end{array} 
           \right. 
\end{eqnarray}
where a harmonic time-dependence, $\exp(-i\omega t)$, has been
assumed, but suppressed.  This form for the primary field will be used
frequently throughout this review.  Notice the fact that the primary
field can be fully described by a single vector component. This
represents a dramatic simplification of the problem since it
is reduced form a vector problem down to a scalar one.

When $\Phi_\nu(x_1,x_3 | \omega )$ is known, the remaining component
of the electromagnetic field can be calculated from it alone. These
components are given for $p$-polarization by
\begin{subequations}
  \label{Elmag:BS:p-pol}
\begin{eqnarray}
  E_1(x_1,x_3|\omega) 
        &=&  - \frac{i}{\omega\eps{}}
               \frac{\partial}{\partial x_3}  
                H_2(x_1,x_3|\omega),\\
  E_3(x_1,x_3|\omega) 
        &=&    \frac{i}{\omega\eps{}}
               \frac{\partial}{\partial x_1}  
                H_2(x_1,x_3|\omega),
\end{eqnarray}
\end{subequations}
and for $s$-polarization   
\begin{subequations}
  \label{Elmag:BS:s-pol}
\begin{eqnarray}
  H_1(x_1,x_3|\omega) 
        &=&   \frac{i}{\omega\mu(\omega)}
               \frac{\partial}{\partial x_3}  E_2(x_1,x_3|\omega),\\
  H_3(x_1,x_3|\omega) 
        &=&   - \frac{i}{\omega\mu(\omega)}
                \frac{\partial}{\partial x_1}  E_2(x_1,x_3|\omega).
\end{eqnarray}
\end{subequations}
In the above equations $\eps{}$ and $\mu(\omega)$ denoted the
dielectric function and the magnetic permeability respectively of the
medium where the fields are being evaluated.  The relations
\r{Elmag:BS:p-pol} and \r{Elmag:BS:s-pol} are easily derived by using
the two curl-equations contained in the Maxwell's equations,
Eqs.~\r{Elmag:MaxwellEq3} and \r{Elmag:MaxwellEq4}, together with the
constitutive relations, Eqs.~\r{Elmag:constitutive}.

Let us now try to focus on the boundary conditions that the primary
field $\Phi_\nu(x_1,x_3| \omega)$ will be subjected to. By
construction  $\Phi_\nu(x_1,x_3| \omega)$ is a tangential field
independent of polarization. Therefore it follows automatically from 
Eqs.~\r{Elmag:BC1} and \r{Elmag:BC4} ($\rho_s={\mathbf J}_s=0$) that
\begin{subequations}
\label{Elmag:BCS}
\begin{eqnarray}
   \left. \Phi_\nu^+(x_1,x_3| \omega) \right|_{x_3=\zeta(x_1)}
    &=& \left. \Phi_\nu^-(x_1,x_3| \omega) \right|_{x_3=\zeta(x_1)},
   \label{Elmag:BCSa}
\end{eqnarray} 
where $\zeta(x_1)$ denotes the interface separating the two materials
of different dielectric properties. 

In order to satisfy the remaining boundary conditions expressed in
Eqs.~\r{Elmag:BC}, we notice that for respectively $p$- and
$s$-polarization we have
\begin{eqnarray}
  {\mathbf n \times \mathbf E} &=&  
           \hat{\mathbf e}_2 \; \frac{i}{\omega \eps{}} 
                  \partial_n  \Phi_p , \nn\\
  {\mathbf n \times \mathbf H} &=&  \hat{\mathbf e}_2 \; 
       \frac{i}{\omega\mu(\omega)}  \partial_n  \Phi_s, \nn
\end{eqnarray}
where $\partial_n$ denotes the normal derivative to the surface.
If the one-dimensional interface can be represented as
$x_3=\zeta(x_1)$, where $\zeta(x_1)$ is a single-valued function of
$x_1$ the normal derivative becomes
\begin{eqnarray}
  \label{Elmag:normal-derivative}
  \partial_n &=&  {\mathbf n \cdot \nabla} \; =\; 
       \frac{-\zeta'(x_1) \partial_{x_1}
                      + \partial_{x_3} }{
         \sqrt{1+(\zeta'(x_1))^2}  }, \nn
\end{eqnarray}
where $\partial_{x_i}= \partial/\partial x_i$ and 
\begin{eqnarray}
  \label{Elmag:normal-vector}
  {\mathbf n} &=& \frac{\zeta'(x_1) \hat{{\mathbf e}}_1 +\hat{{\mathbf
        e}}_3}{ \sqrt{1+(\zeta'(x_1))^2} }.
\end{eqnarray}
Here $\hat{{\mathbf e}}_i$ are the standard unit vectors.  Hence the
remaining boundary conditions can be expressed as
\begin{eqnarray}
   \frac{1}{\kappa^+_\nu(\omega)}
   \left. \partial_n \Phi_\nu^+(x_1,x_3| \omega)
    \right|_{x_3=\zeta(x_1)} 
   =
   \frac{1}{\kappa^-_\nu(\omega)}
   \left. \partial_n \Phi_\nu^-(x_1,x_3| \omega)
    \right|_{x_3=\zeta(x_1)}, \quad
    \label{Elmag:BCSb}
\end{eqnarray}
where $\kappa^\pm_\nu(\omega)$ are defined as 
\begin{eqnarray}
  \label{Elmag:BCSc}
  \kappa_\nu^\pm(\omega) &=& 
    \left\{
    \begin{array}{ll}
       \eps{\pm}, & \quad \nu=p \\
         \mu_\pm(\omega),  & \quad \nu=s
    \end{array}.
   \right.
\end{eqnarray}
\end{subequations}
Eqs.~\r{Elmag:BCS} are the final result for the boundary conditions to
be satisfied by the primary field $\Phi_\nu(x_1,x_3| \omega)$ on a
one-dimensional interface $x_3=\zeta(x_1)$.

\subsection{Surface Plasmon Polaritons}

%

In subsequent sections, we will see that so-called surface plasmon
polaritons, or for short just SPPs, will play an important roll for the
rough surface scattering problem. We will therefore in this section
define and discuss some of the distinguishing properties of such modes.

Before starting our discussion, we have to know what a polariton is:
According to its classical definition a polariton is defined to be an
elementary electromagnetic wave, and therefore a solution of the
Maxwell's equations, that may couple to one of several possible
excitations possible in a condensed medium.  Examples of such
excitations are  plasmons, phonons, mognons etc., and in
such cases one talks of plasmon polaritons, phonon polaritons and
magnon polaritons.  With the notion of polariton established, one
might definition an SPP as follows: {\em A surface plasmon polariton is a
plasmon polariton where the associated electromagnetic field is
confined to the surface separating two dielectric medium.}

\subsubsection{SPPs on a plan surface geometry}

To see under which condition SPPs might exist, and to discuss some of
their properties, we will consider a planar interface separating two
isotropic and homogeneous media. For simplicity, the coordinate system
will be chosen so that the interface is located at $x_3=0$. The
materials above~($x_3>0$) and below~($x_3<0$) this surface will be
characterized by frequency dependent dielectric functions
$\varepsilon_+(\omega)$ and $\varepsilon_-(\omega)$ respectively.  For
simplistic reasons, which are not essential for the present
discussion, we will assume that the imaginary part of the dielectric
functions can be neglected.  The conclusion that we arrive at herein
will, however, be independent of this assumption.  Furthermore, we
will assume either pure p- or s-polarization of the incident
light. Hence the scalar wave equation might be used. A more complete
discussion using vector fields can be found in
Refs.~\citen{Book:Surface-Polaritons1} and
~\citen{Book:Surface-Polaritons2}.

According to the definition of SPP, we are interested in solutions to
the Maxwell equations, equivalent in our case to the scalar wave
equation, that are wave-like parallel to the surface $x_3=0$ and that
decays exponentially with increasing distance from the surface into
each of the two media. Such a solution can be represented as
\begin{eqnarray}
    \label{Elmag:SPP-fields}
    \Phi^\pm_\nu(x_1,x_3|\omega)
      &=& {\cal A}^\pm_\nu \:e^{\mp\beta_\pm(\omega) x_3} \,e^{ikx_1},
      \qquad \nu=p,s,
\end{eqnarray}
where ${\cal A}^\pm_\nu$ represents the amplitudes (to be determined).
The decay constants $\beta_\pm(\omega)$ are defined as
\begin{eqnarray}
  \label{Elmag:beta-def}
  \beta_\pm(\omega) &=& \sqrt{ k^2-\varepsilon_\pm(\omega) \ooc{2} },
\end{eqnarray}
and they must be real and positive for Eq.~\r{Elmag:SPP-fields} to
describe an electromagnetic wave localized to the surface\footnote{If
  we had allowed the dielectric functions of the problem to be complex
  with $Im\, \varepsilon_\pm(\omega)>0$, we would have to require that
  $Re\,\beta_\pm(\omega) >0$.}.  To investigate if
Eq.~\r{Elmag:SPP-fields} is an acceptable solution for our scattering
system, we have to impose the boundary conditions, given in
Eqs.~\r{Elmag:BCS}, for the two polarizations.
By utilizing the continuity of the fields on the flat surface,
Eq.~\r{Elmag:BCSa}, one finds that
\begin{eqnarray}
   {\cal A}^+_\nu &=&  {\cal A}^-_\nu \;\equiv\; {\cal A}_\nu,
 \nn
\end{eqnarray}
for all locations along the surface. Moreover, the normal derivative
condition, Eq.~\r{Elmag:BCSb}, gives the following condition for the
existence for surface plasmon polaritons on a flat surface
($\partial_n=\partial_{x_3}$)
\begin{eqnarray}
  \label{Elmag:SPP-condition}
  \left( \frac{\beta_+(\omega)}{\kappa^+_\nu(\omega)} +
    \frac{\beta_-(\omega)}{\kappa^-_\nu(\omega)} \right) {\cal A_\nu}
    &=& 0,
\end{eqnarray}
where we recall the definitions of $\kappa^\pm_\nu(\omega)$ from
Eq.~\r{Elmag:BCSc}.  The most immediate consequence of this relation
is the following: Since for $s$-polarization $\kappa^\pm_s(\omega) =
\mu_0$ (for non-magnetic materials) and $\beta_\pm(\omega)$ are
assumed to be real and {\em positive}, the only solution to
Eq.~\r{Elmag:SPP-condition} is ${\cal A}_s\equiv 0$. Thus, a
$s$-polarized surface plasmon polariton (surface wave) {\em cannot}
exist for the scattering system that we are considering.

However, for $p$-polarization, where $\kappa^\pm_p(\omega) =
\varepsilon_\pm(\omega)$ a non-trivial solutions might exist. It is
given by (assuming that ${\cal A}_p \neq 0$)
\begin{eqnarray}
  \label{Elmag:SPP-dipersion-relation-1}
  \frac{\varepsilon_+(\omega)}{\varepsilon_-(\omega)} &=&
  - \frac{\beta_+(\omega)}{\beta_-(\omega)},
\end{eqnarray}
which is the dispersion relation for surface plasmon polaritons on a
flat interface. For this relation to be satisfied, since
$\beta_\pm(\omega)$ are both assumed to be positive, the two
dielectric functions of the scattering system have to have different
signs due to the presence of the negative sign on the right-hand-side
of Eq.~\r{Elmag:SPP-dipersion-relation-1}. Only such combination of
materials will support surface plasmon polaritons.  An important
example of such a system at optical frequencies, is a metal with a
planar interface to vacuum.

By squaring both sides of Eq.~\r{Elmag:SPP-dipersion-relation-1} as
well as taking advantage of Eq.~\r{Elmag:beta-def}, the dispersion
relation can be expressed as 
\begin{eqnarray}
  \label{Elmag:dispersion-relation}
  k_{sp}(\omega) &=& \pm \sqrt{
 \frac{\varepsilon_+(\omega)\varepsilon_-(\omega)}{
            \varepsilon_+(\omega)+\varepsilon_-(\omega)}} \ooc{}.
\end{eqnarray}
This equation gives an explicit expression for the wave vector of the
surface plasmon polariton. However, this formulae should be used with
some care since it may, from the way it is derived from
Eq.~\r{Elmag:SPP-dipersion-relation-1}, introduce some spurious
solutions. The additional and sufficient requirement that have to be
satisfied is that $\beta_\pm(\omega)$ are positive while the two
dielectric functions, $\varepsilon_\pm(\omega)$, have different sign.

\subsubsection{SPPs at a planar free electron metal surface}

Let us for illustrative purposes consider a free electron metal with a
planar interface to vacuum. For such a metal the dielectric function
is known to be~\cite{Book:Kitel}
\begin{eqnarray}
  \label{Elmag:eps-free-electron-metal}
  \varepsilon_-(\omega) 
     &=& \eps{\infty}\left(1-\frac{\omega_p^2}{\omega^2} \right),
\end{eqnarray}
while the one for vacuum is $\varepsilon_+(\omega)=1$. 
In the above equation $\eps{\infty}$ is the background dielectric constant
of the material while $\omega_p$ is the electronic plasma
frequency. With Eq.~\r{Elmag:eps-free-electron-metal} the frequency of
the SPP can be shown to be 
\begin{eqnarray}
  \label{Elmag:SPP-free-metal}
  \omega_{sp}(k) &=& 
  \left[
      \frac{1}{2}\left(\frac{k^2c^2}{\eps{\infty}}
       \left(1+\eps{\infty}\right)
            +\omega^2_p \right) 
      -  \frac{1}{2}
         \sqrt{ 
         \left(\frac{k^2c^2}{\eps{\infty}}
                      \left(1+\eps{\infty}\right) +\omega^2_p\right)^2
            -4k^2c^2\omega_p^2  } \;\; 
    \right]^\frac{1}{2}. \qquad
\end{eqnarray}
It should be noticed from this equation that
\begin{eqnarray}
  \label{Eq:Elmag:assymp}
   \omega_{sp}(k) &=&      
   \left\{
     \begin{array}{cl}
       \displaystyle
       kc, & \quad k \rightarrow 0, \\
       \displaystyle
       \sqrt{\frac{\eps{\infty}}{\eps{\infty}+1}} \,\omega_p, & \quad k \rightarrow \infty, 
     \end{array}
   \right.  .
\end{eqnarray}

\begin{figure}[t!]
  \begin{center}
    \leavevmode
    \includegraphics[width=10cm,height=8cm]{\myfigpath/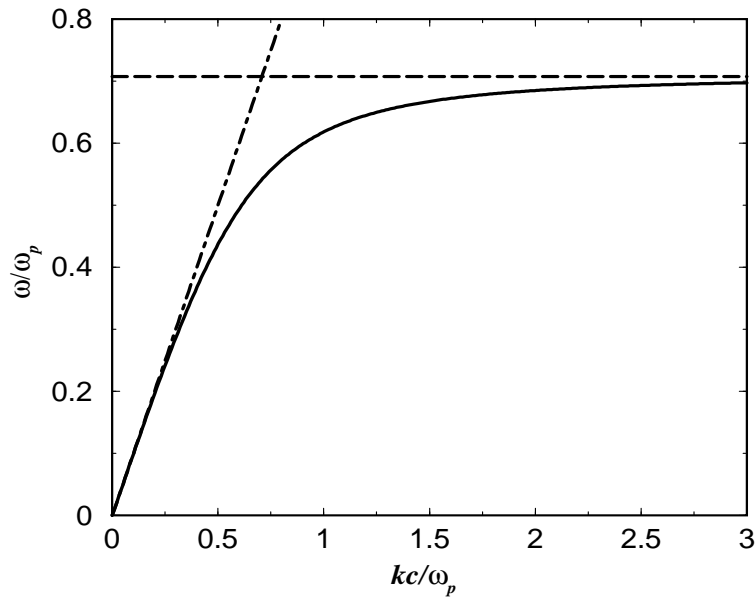}
    \caption{Dispersion relation curve,
      Eq.~\protect\r{Elmag:SPP-free-metal}, for a surface plasmon
      polariton (solid line) at a flat interface between a simple
      metal and vacuum (assuming $\eps{\infty}=1$). The dash-dotted
      line represents the light line $\omega=kc$, while the dashed line
      is the large momentum asymptotic limit
      $\omega_{sp}=\omega_p/\sqrt{2}$ ~(see
      Eq.~\protect\r{Eq:Elmag:assymp}).}
    \label{Fig:Elmag:Dispersion-relation}
  \end{center}
\end{figure}


This means that in the small wave vector limit the surface plasmon
polariton is photon-like, while it in the large wave vector limit it
is plasmon-like.  In Fig.~\ref{Fig:Elmag:Dispersion-relation} the
dispersion relation, Eq.~\r{Elmag:SPP-free-metal}, for a free electron
metal is plotted. In this figure we have also included the
light-line~(dash-dotted line) $\omega=kc$, as well as the large wave
vector limit~(dashed line) of $ \omega_{sp}(k)$.

From Fig.~\ref{Fig:Elmag:Dispersion-relation} we see that the
dispersion curve for the SPP lies entirely to the right of the
light-line, $\omega=kc$. The physical consequence of this is that
there is no coupling between the surface plasmon polariton and light
in vacuum for a flat vacuum-metal interface. Or put another way, light
incident onto a planar vacuum-metal interface cannot excite surface
plasmon polaritons. Later, however, we will see that if the surface is
rough such coupling is possible.  This will give rise to many new and
interesting multiple-scattering effects, as we discuss in some detail
in Sect.~\ref{Chapt:Phenomena}.


\subsection{Characterization of Random Rough Surfaces}
\label{Chap:Characterization}
  
Almost everyone grows up with some kind of intuitive ``feeling'' of
what is meant by a rough surface. A fractured stone, say, is normally
looked upon as being rough, while a piece of paper as being smooth.
However, on the micro-scale, where the human eye is not very
sensitive, also the paper has some kind of structure.  So in a strict
sense, both the paper and the stone surface are rough. Paper is made
out of fibers which is quite different from the crystals that are seen
on the micro scale of the surface of the fractured stone. So the
question is: How shall we quantify the difference in roughness between
say the paper and the stone surface? One possibility is to measure by
some suitable technique the surface topography. Such measurements will
of course produce different results for the paper and stone surface.
However, if we move to another area of the fractured stone and measure
the surface topography here, we will obviously not get the same result
as obtained in the previous measurement taken from another area of
the same surface.  So, how shall we be able to characterize the rough
surfaces at hand, so that we are able to distinguish them from each
other?  In this section we intend to discuss in some detail how to
characterize randomly rough surfaces in a quantitative way.

However, before we do so, let us take a look at what kind of rough
surfaces we have. Depending on how the surface height fluctuates
around some reference surface, we may categorize them as being
deterministic or randomly rough.  For random surfaces, one may in
addition group them as correlated or uncorrelated surfaces, and they
might occur as fractal or non-fractal surfaces depending on under
which conditions they were formed.  Rough surfaces that are found in
nature are normally randomly rough, correlated surfaces.  We will
therefore proceed by discussing how to characterize such surfaces.

\subsection{A Statistical description of Randomly Rough Surfaces}
\label{Sect:Stat-surf}

Two randomly rough surfaces are never identical. Thus the knowledge of the
surface topography alone is therefore not enough to be able to say if two
rough surfaces were generated by the same underlying process, and therefore
have to be looked upon as being identical. However, if we assume that the
randomly rough surface can be considered as a continuous random
process~\cite{Book:Ross,Book:Feller-1,Book:Feller-2,Book:Parzen62,Book:Papoulis84,Book:Papoulis90},
then a statistical description might be relevant and useful. We will now
introduce this method of characterization.

Under experimental conditions, the surface topography is measured
relative to some reference surface. In our case we will assume that
this reference surface is a planer surface. Other choices might be
practical in some cases, but this will not be discussed here.
Furthermore, it is convenient to choose our coordinate system so that
this planar surface is located at $x_3=0$. In this case the randomly
rough surface is just the roughness that perturb the plane $x_3=0$.
For simplicity, we limit our discussion to one-dimensional surfaces.
The extension to (isotropic) two-dimensional surfaces is trivial. For
the purpose of this introduction it will be assumed that the surface
does not possess any overhangs\footnote{Such surfaces are also known
  as reentrant surfaces.}, that is to say that the surface profile
function, that we will denote by $\zeta(x_1)$, is a single-valued
function of the lateral coordinate $x_1$. For characterization of
surfaces where the surface profile function does not fulfill this
property the reader is invited to consult Ref.~\citen{Mendaza-Suarez}.

In order for the surface profile function $\zeta(x_1)$ to be planar on
average, it must, with our choice for the coordinate system, have a
vanishing mean, {\it i.e.} we must require that 
\begin{eqnarray}
  \label{CHAR:surface-average}
  \left< \zeta(x_1)\right> &=& 0.
\end{eqnarray}
Here the angle brackets are used to denote a spatial average over a
large spatial region. If, however, the surface is {\em
  ergodic}~\cite{Book:Feller-1,Book:Feller-2,Book:Parzen62,Book:Papoulis84,Book:Papoulis90},
as we will assume here, this spatial average is equal to an
average over an ensemble of realizations of
$\zeta(x_1)$.  It
is therefore, under the assumption of ergodicity, more convenient to
think of $\left<\cdot\right>$ as an ensemble average.
 
Another notion  that is important when  characterizing rough surfaces,
and  studying  light  scattering  from   such,  is  the  one  of  {\em
  stationarity}~\cite{Book:Ogilvy}.    A  surface   is   said  to   be
stationary,  or translation invariant,  if its  statistical properties
are  independent of which  portion of  the surface  was used  in their
determination.  That  the surface  roughness  possess stationarity  is
crucial for  the applicability of many  of the theories  used to study
rough surface scattering. Rigorous numerical 
simulations~(Sect.~\ref{Sect:Theory:Sect:NumSim}), however, can
still handle non-stationary surfaces.

\subsubsection{Gaussian Random Surfaces}

In theoretical studies of light scattering from rough surfaces, the
random surfaces have in the overall majority of the studies been
assumed to possess Gaussian height statistics. Such a statistics is
rather appealing from a theoretical point of view since all moments
can be related to the two first moments. such moments either
vanish~(odd moments), or they are related to the second moment~(even
moments)~\cite{Book:Ogilvy}.

The zero-mean property, Eq.~\r{CHAR:surface-average}, does not specify
how the different heights are located relative to one another along
the surface. Such information is provided by the height-height
correlation function.  Under the assumption of $\zeta(x_1)$ being
stationary we can write
\begin{eqnarray}
  \label{CHAR:surface-correlation}
  \left< \zeta(x_1)\zeta(x_1')\right> &=& \delta^2 W(|x_1-x_1'|),
\end{eqnarray}
where $\delta$ is the {\sc rms}-height of the surface profile
function, and $W(|x_1|)$ is the height auto-correlation function
normalized so that $W(0)=1$. In cases where $W(|x_1|)=1$
($W(|x_1|)=-1$) one speaks of perfect correlation (anti-correlation).
Furthermore it can be shown that $-1\leq W(|x_1|)\leq 1$.  Notice,
that since the heights-distribution is Gaussian,
Eqs.~\r{CHAR:surface-average} and \r{CHAR:surface-correlation}
together determines uniquely the statistical properties of the
surface since all higher order moments can (for a Gaussian surface) be
related to the first two.

In many of the perturbation theories developed for rough surface
scattering, the power spectrum of the surface randomness is a quantity
that appear more-or-less naturally. It is defined as the 
Fourier transform of the (normalized) correlation function
\begin{eqnarray}
  \label{CHAR:def-power-spectrum}
  g(|k|) &=& \int^{\infty}_{-\infty} dx_1 \; W(|x_1|)\; e^{-ikx_1}.
\end{eqnarray}

In order to get an intuitive picture of how the surface height varies
along the surface, it is often useful to supply the mean slope, $s$,
and the mean distance between consecutive peaks and valleys,
$\left<D\right>$, as measured along the (lateral) $x_1$-direction.
For a stationary zero-mean, Gaussian random process, the {\sc
  rms}-slope, $s$, is related to the power spectrum
by~\cite{Michel1990}
\begin{eqnarray}
   \label{rms-slope}
   s &=& \left<(\zeta'(x_1))^2\right>^{1/2}
      =  \delta \sqrt{\int_{-\infty}^{\infty} 
              \frac{dk}{2\pi} \; k^2  g(|k|)},
\end{eqnarray}
and a good estimator for $\left<D\right>$ has been shown to be~\cite{Michel1990}
\begin{eqnarray}
   \label{peaks-valleys}
   \left<D\right> &\simeq&
   \pi
   \sqrt{
          \frac{\int_{-\infty}^{\infty} dk \; k^2 g(|k|)}{
                \int_{-\infty}^{\infty} dk\; k^4 g(|k|)}
        }.
\end{eqnarray}

In the literature many different forms for the correlation function
$W(|x_1|)$ has been considered~(see {\it e.g.}
Ref.~\citen{Book:Ogilvy} and references therein). However, here we
will only be dealing with two such forms. They are the Gaussian
form given by 
\begin{subequations}
  \label{Eq:Char:W-Gaussian}
  \begin{eqnarray}
    W(|x_1|) &=& \exp \left( - \frac{x^2_1}{a^2}\right),\\
    g(|k|) &=& \sqrt{\pi} a\, \exp \left(-\frac{a^2 k^2}{4} \right),     
  \end{eqnarray}
\end{subequations}
where $a$ is the transverse correlation length, and the so-called
West-O'Donnell (or rectangular) form  
\begin{subequations}
  \label{Eq:Char:W-WOD}
  \begin{eqnarray}
    W(|x_1|) &=& \frac{\sin k_{+}x_1-\sin k_{-} x_1}{(k_{+}-k_{-})x_1} \\
    g(|k|) &=& \frac{\pi}{k_{+}-k_{-}} [\theta(k_{+}-k)\theta 
    (k-k_{-}) + \theta (k_{-}+k)\theta (-k-k_{-})]. \quad
    \qquad
  \end{eqnarray}
\end{subequations}
where $\theta(\cdot)$ is the Heaviside unit step function. In
Eqs.~\r{Eq:Char:W-WOD} the quantities $k_\pm$, with $0< k_- < k_+$,
denote the lower and upper momentum cut-off for the spectrum, and they
will be given a more precise definition in later sections.  The latter
power spectrum was recently used by West and O'Donnell~\cite{west95}
in an experimental study of the enhanced backscattering phenomenon
from weakly rough surfaces.


\begin{figure}[t!]
  \begin{center}
    \leavevmode
    \includegraphics[height=8cm,width=14cm]{\myfigpath/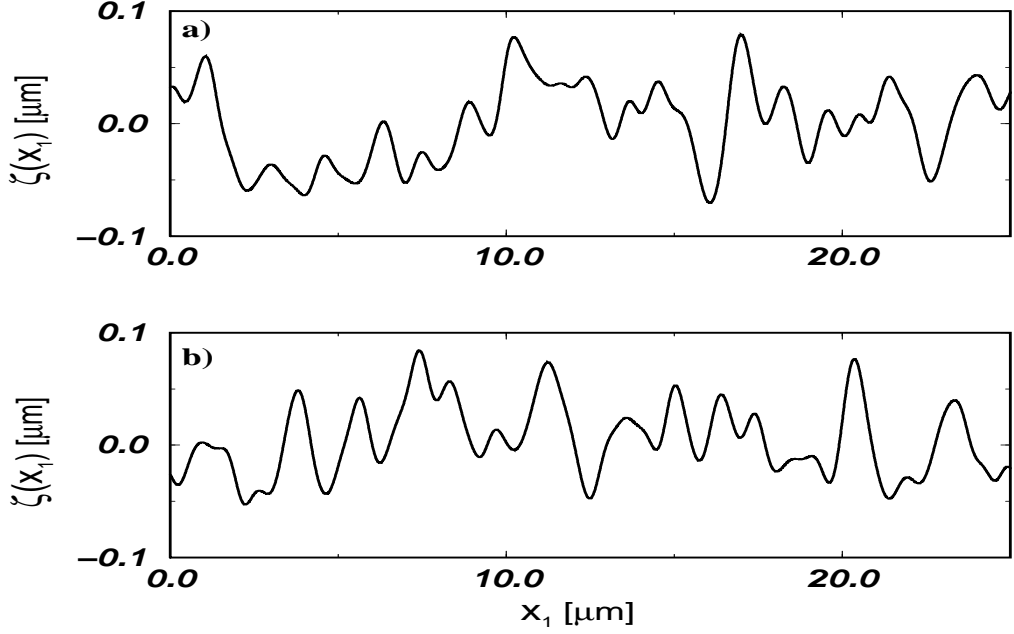}  
    \caption{Examples of two rough profiles both with Gaussian height
      distributions and with an {\sc rms}-value $\delta=30 {\rm nm}$.
      The power spectrum is of the (a) Gaussian type, with $a=100 {\rm
        nm}$, and the (b) West-O'Donnell type, with
      $k_{-}=0.82(\omega/c)$ and $k_{+}=1.97(\omega/c)$.  Here the
      wavelength is $\lambda = 632.8 {\rm nm}$.  With these parameters
      the {\sc rms}-slope and distance between consecutive peaks and
      valleys are respectively $s=0.424$ and $\left<D\right>=128.3
      {\rm nm}$ for the Gaussian power spectrum, and $s=0.427$ and
      $\left<D\right>=201.1 {\rm nm}$ in case of the West-O'Donnell
      power spectrum.  Note that there are different scales on the
      first and second axes, with the result that the profiles appear
      much rougher than they are in reality. The two surface profiles
      where generated from the same underlying uncorrelated random
      numbers.}
    \label{Fig:CHAR:Surface-Profiles-Ex}
  \end{center}
\end{figure}


For the two above power spectra the mean slope, $s$, and the distance
between consecutive peaks and valleys, $\left<D\right>$, then
become~\cite{Michel1990}
\begin{eqnarray}
     s   &=& \left\{
         \begin{array}{ll}
             \sqrt{2}\frac{\delta }{a}, & \qquad \mbox{Gaussian} \\
             \;\frac{\delta }{\sqrt{3}}\sqrt{k_+^2+k_+k_-+k_-^2},
             & \qquad \mbox{West-O'Donnell}
         \end{array}
       \right. ,
\end{eqnarray}
and~\cite{Michel1990}
\begin{eqnarray}
   \left<D\right> &=&  \left\{
         \begin{array}{ll}
             \frac{\pi}{\sqrt{6}}a, & \qquad \mbox{Gaussian} \\
             \pi \sqrt{ \frac{5}{3}
                         \frac{k_+^3-k_-^3}{k_+^5-k_-^5}},
             & \qquad \mbox{West-O'Donnell}
         \end{array}
       \right. .
\end{eqnarray}
Two surface profiles with the same (Gaussian) height distribution, but
with a Gaussian and a West-O'Donnell power spectrum possessing nearly
the same value of the {\sc rms}-slope, $s$, are plotted in
Figs.~\ref{Fig:CHAR:Surface-Profiles-Ex}.

It will later in explicate calculations prove useful to also have the
Fourier representation of the surface profile function (and its
inverse) at our disposal. They are  defined as  
\begin{subequations}
  \begin{eqnarray}
    \zeta(x_1) &=& \int^{\infty}_{-\infty} \frac{dk}{2\pi} \;\tilde{\zeta}(k )e^{ikx_1}, \\
    \tilde{\zeta}(k) &=& \int^{\infty}_{-\infty}  dx_1 \;\zeta(x_1) e^{-ikx_1}.
\end{eqnarray}
\end{subequations}
The Fourier transform, $\tilde{\zeta}(k)$, of the surface profile
function also constitutes a zero-mean Gaussian random process with
statistical properties
\begin{subequations}
  \begin{eqnarray}
    \left< \zeta(k) \right> &=& 0, \\
    \left< \tilde{\zeta}(k) \tilde{\zeta} (k')\right> 
      &=& 2\pi\delta (k+k')\delta^2g(|k|),
  \end{eqnarray}
\end{subequations}
where $\delta(\cdot)$ denotes the Dirac delta  function.

\subsubsection{Non-Gaussian Random Surfaces}

Naturally occurring surfaces have often more complicated height
distributions then the Gaussian~\cite{Book:Ogilvy}. To fully
characterize such surfaces are quite difficult and probably explains
why they have gotten less attention in the literature then they
probably deserve.  The main problem is that in order to characterize
them statistically, moments of in principle infinite order has to be
known.  These moments are not, as for Gaussian surfaces, related to
moments of lower order in a trivial way since the characteristic
function is in general not known for non-Gaussian surfaces. We do not
intend in this introduction to discuss non-Gaussian random surfaces in
any detail, since we will not focus on them later. However, we would
like to mention that as long as this kind of surfaces can be generated
numerically, the scattering problem for non-Gaussian surfaces are not
hard to handle by numerical simulations~\cite{AnnPhys, PhysRep}. On
the other hand, small amplitude perturbation theory, say, can not be
utilized in its standard form to non-Gaussian surfaces.

\subsection{Self-affine surfaces}

It has been known for quite some time that self-affine surfaces are
abundant in nature. They can be found in various areas of natural
science such as surface
growth~\cite{Meakin,Book:Barabasi,Book:Meakin}, fractured
surfaces~\cite{Bouchaud}, geological
structures~\cite{Schmittbuhl,Nes}, metallurgy~\cite{laminage}, and
biological systems~\cite{Vicsek} to mention a few.

A surface, $\zeta(x_1)$, is self-affine, according to its definition,
between the scales $\xi_{-}$ and $\xi_{+}$, if it remains (either
exactly or statistically) invariant in this region under
transformations of the form
\begin{subequations}
  \label{Scaling-relation-self-affine}
  \begin{eqnarray}
     x_1         &\to&   \lambda x_1, \\
     \zeta       &\to&   \lambda^{H} \zeta,
   \end{eqnarray}
\end{subequations}
for all positive real numbers $\lambda$. Here $H$ is the {\em
  roughness exponent}, also known as the {\em Hurst exponent}, and it
characterizes this invariance. It is usually found in the
range from zero to one. When $H=1/2$ the surface is an example of the
famous random (Brownian) walk where the surface is uncorrected.
However, if $H\neq 1/2$ the profile shows correlations; for
$H>1/2$ it is said to be {\em persistent}~(correlated), and for
$H<1/2$ it is {\em anti-persistent}~(anti-correlated). The reason for
this naming is that if the self-affine ``walker'' when moving from the previous to
the present space step went up, say, it is more likely that it
will go up (down) in the next one if $H>1/2$ ($H<1/2$).  

\begin{figure}[t!]
  \begin{center}
    \leavevmode
    \includegraphics[height=14cm,width=14cm]{\myfigpath/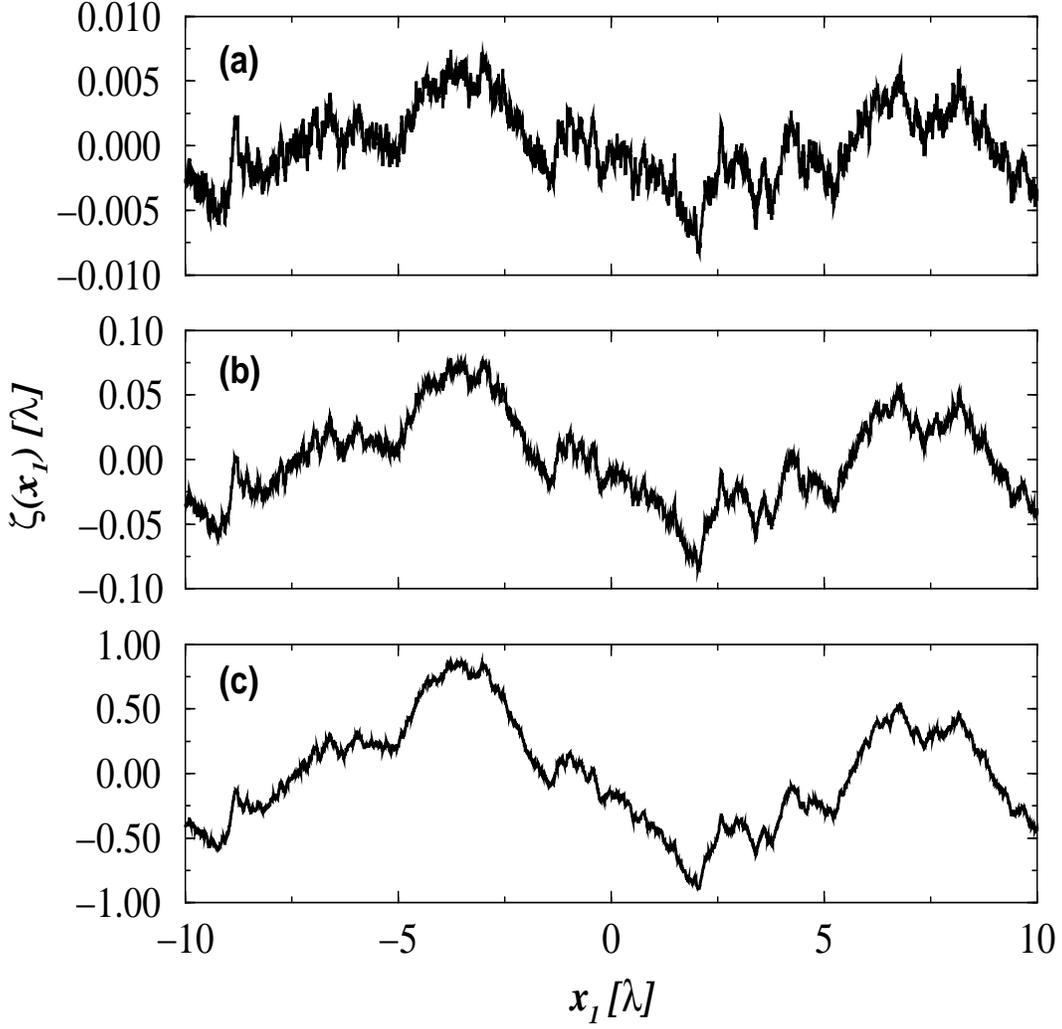}  
    \caption{Examples of self-affine profiles. The Hurst exponents
      were (a) $H=0.3$, (b) $H=0.5$, and (c) $H=0.7$, and for all
      cases the topothesy where $\ell=10^-3\lambda$ where $\lambda$ is
      an arbitrary length scale. The surfaces were generated by the
      Fourier filtering method from the same uncorrelated Gaussian
      distributed number. Notice how the {\sc rms}-height of the
      surface as measured over its total length increases as we
      increase the Hurst exponent. This is in agreement with
      Eq.~\protect\r{Eq:sigma}.  }
    \label{Fig:CHAR:Self-affine-profiles}
  \end{center}
\end{figure}

The scaling relation \r{Scaling-relation-self-affine} is often put in
the more compact, but equivalent form
\begin{eqnarray}
  \label{Eq:CHAR:Scaling-rel}
  \zeta(x_1)   &\simeq&  \lambda^{-H}\zeta(\lambda x_1),
\end{eqnarray}
where $\simeq$ is used to indicate statistically equality. This
relation says that if we take the original profile $\zeta(x_1)$,
enlarge (or contract) the lateral direction by rescaling $x_1$ into
$\lambda x_1$, and {\em simultaneously} scaling $\zeta$ to
$\lambda^{-H}\zeta$, the profile $\zeta(x_1)$ and its rescaled version
$\lambda^{-H}\zeta(\lambda x_1)$ should be indistinguishable. Of
course, this holds true in an exact sense only for deterministic
surfaces. In the statistical case, however, it is the statistical
properties of the profile and its rescaled version that are
indistinguishable. In Figs.~\ref{Fig:CHAR:Self-affine-profiles} we
presents some examples of self-affine surfaces of Hurst exponent
$H=0.3$~(Fig.~\ref{Fig:CHAR:Self-affine-profiles}a),
$H=0.5$~(Fig.~\ref{Fig:CHAR:Self-affine-profiles}b), and
$H=0.7$~(Fig.~\ref{Fig:CHAR:Self-affine-profiles}c).  As can be seen
from these figures the landscapes become more ``calm'' the larger the
Hurst exponent becomes. 

The scaling relation Eq.~\r{Eq:CHAR:Scaling-rel} does not fully
specify the self-affine surface. In particular no information is
contained in Eq.~\r{Eq:CHAR:Scaling-rel} about the amplitude of the
surface. Such information is provided by the length scale, $\ell$,
known as the {\em topothesy}. This length scale is define as the
length, $\ell$, measured along the $x_1$-direction, for over which the
root-mean-square of the height-difference between two points separated
by $\ell$ is just $\ell$. To make this even more clear, let us
introduce 
\begin{eqnarray}
  \label{Eq:CHAR:rms-top}
 \sigma(\Delta x_1) &=& \left<\left\{
                             \zeta(x_1+\Delta x_1)-\zeta(x_1) 
                        \right\}^2 \right>^\frac{1}{2}_{x_1},  
\end{eqnarray}
as the {\sc rms}-value of the height-difference measured over a window
of size $\Delta x_1$. With this definition the topothesy is defined
as the length scale for which
\begin{eqnarray}
  \label{Eq:CHAR:Topothesy}
  \sigma(\ell) &=& \ell.
\end{eqnarray}
From Eq.~\r{Eq:CHAR:rms-top} it follows immediately that
$\sigma(x_1)\sim x_1^H$, so that with Eq.~\r{Eq:CHAR:Topothesy} we get
\begin{eqnarray}
  \label{Eq:sigma}
   \sigma \left( \Delta x \right) = \ell^{1-H} \Delta x_1^{H}.
\end{eqnarray} 
Notice that Eq.~\r{Eq:CHAR:Topothesy} allows for a geometrical
interpretation of the topothesy as the length scale over which the
profile has a mean slope of 45 degrees.  The smaller $\ell$, the
flatter the profile appears on a macroscopic scale.  It should be
stressed that in spite of the geometrical interpretation of $\ell$,
there is nothing a priori that restricts the topothesy to length
scales where the self-affinity can be found.  However, for the
surfaces usually considered in scattering problems, we rather expect
that $\ell \ll \xi_{-}$.  When $\xi_{-} < \ell < \xi_{+}$, the
topothesy makes the transition between the scales, below $\ell$, for
which a fractal dimension $D=2-H$ can be measured using {\em e.g.} the
box counting
method~\cite{Book:Feder,Book:Vicsek,Book:Meakin,Mandelbrot} and the
scales, above $\ell$, for which this dimension is just unity. For
length scales $\xi_-<\Delta x_1<\ell$ the fractal dimension is
therefore nontrivial (read different from one) and we have an example
of a self-affine
fractal~\cite{Book:Feder,Book:Meakin,Book:Vicsek,Mandelbrot}.  It
should be noticed that the fractal property of the self-affine surface
crucially depends on which length scale the surface is being observed.
This essential point seems often to be overlooked in the literature
where one too often treat self-affine surfaces as they were
fractals~\cite{Book:Feder,Mandelbrot} at any length scale.

Even if the self-affine correlations of the profile is fully
characterized by its Hurst exponent $H$, its topothesy parameter
$\ell$ and the bounds of the self-affine regime $\xi_{-}$ and
$\xi_{+}$, nothing is said about its height-distribution. It is
therefore not uncommon to talk about for example a Gaussian
self-affine surfaces meaning that the surface correlation is of the
self-affine type, while the distribution of heights is Gaussian. Thus
by specifying the self-affine parameters, {\it i.e.} $H$, $\ell$, and
$\xi_\pm$, in addition to the parameters needed in order to
characterize the height-distribution, the surface is completely
specified. Under the assumption that the surface has Gaussian height
distribution it can be shown that the probability, $p(\zeta;x_1)$ for
finding height $\zeta$ at position $x_1$ given that $\zeta(0)=0$, can
be written as~\cite{Book:Meakin}
\begin{eqnarray}
  \label{dist}
  p(\zeta; x_1) &=& \frac{1}{\sqrt{2\pi} \ell^{1-H} x^{H}_1}
  \exp \left[ -\frac{1}{2} \left( 
      \frac{\zeta} {\ell^{1-H} x^{H}_1}
    \right)^2 \right].
\end{eqnarray}
However, independent of the height-distribute being Gaussian or not,
$p(\zeta; x_1)$ should satisfy the following scaling
relation~\cite{Book:Feder,Book:Meakin}
\begin{eqnarray}
   p(\zeta; x_1) &=& \lambda^{H}  p(\lambda^{H} \zeta; \lambda x_1),
\end{eqnarray}
which can be derived from the scaling relation
Eq.~\r{Scaling-relation-self-affine}.

In fact, the scaling relation~\r{Scaling-relation-self-affine}, or the
equivalent form given in Eq.~\r{Eq:CHAR:Scaling-rel}, is extremely
powerful and can be used to derive most, if not all, of the properties
of a self-affine surface. To show an explicit example of this, we
would like, before closing this section, to derive the scaling
relation of the power spectrum of the surface. This scaling relation
is the most popular one to use for both generating self-affine
surfaces as well as to measure the Hurst exponent. For a surface,
$\zeta(x_1)$, of length, $L_1$, the power spectrum is defined as
\begin{eqnarray}
  \label{power-spec-self-affine-def}
  g(|k|) &=& \frac{1}{L_1}\int^{\frac{L_1}{2}}_{-\frac{L_1}{2}} dx_1
                \;\; e^{ikx_1} \!
                \left< \zeta(y_1+x_1)\zeta(y_1)\right>_{y_1}.
\end{eqnarray}
where, as we recall, $\left< \zeta(y_1+x_1)\zeta(y_1)\right>_{y_1}$ is
the (two-point) correlation function.  By now taking advantage of the
scaling relations~\r{Scaling-relation-self-affine} and
\r{Eq:CHAR:Scaling-rel}, one finds
\begin{eqnarray}
  \label{G-scale}
  g\left(\left| \frac{k}{\lambda} \right|\right)
  &\simeq&
     \frac{1}{\lambda L_1}\int^{\frac{\lambda L_1}{2}}_{-\frac{\lambda
     L_1}{2}} d(\lambda x_1)
               \;\; e^{ikx_1} \!
                \left< \lambda^{H} \zeta(y_1+x_1)
                       \; \lambda^{H}\zeta( y_1)\right>_{y_1}. 
                  \qquad
\end{eqnarray}
Hence, one obtains from Eq.~\r{G-scale} that
\begin{eqnarray}
  g\left(\left| \frac{k}{\lambda} \right|\right)
   &\simeq& \lambda^{2H+1} g\left(\left|k\right|\right),
\end{eqnarray}
so that the power spectrum itself has to scale like
\begin{eqnarray}
  g\left(\left| k\right|\right) &\sim&  k^{-2H-1}.
\end{eqnarray}

For more details about self-affine surfaces and their properties the
reader is referred to the literature~\cite{Book:Meakin,Book:Feder,Mandelbrot}.

\subsection{Numerical Generation of Randomly Rough Surfaces} 

Earlier in the sections we have discussed how to statistically
characterize randomly rough surfaces. In analytical work, this is all
what we need. However in a numerical Monte Carlo simulation approaches
to be presented in a later section, individual surface, called
realizations, have to be generated so that they possess the right
statistical properties. The question therefore is: How can we do this?
We do not intend to give a detailed discussion here, but will instead
sketch how it can be done.

As long as the power spectrum of the surface is known, an efficient way of
generating the surface is by using the so-called Fourier filtering
method~\cite{Book:Feder,Book:Meakin}. This method basically consists of to
main steps. First, {\em uncorrelated} random numbers of the type wanted for
the height distribution of the surface are generated in Fourier space.
Second, these numbers are filtered by the square root of the power spectrum
$g(|k|)$, and the result transform by an inverse Fourier transform to real
space. It was in this way that the surfaces shown in
Figs.~\ref{Fig:CHAR:Surface-Profiles-Ex} and
\ref{Fig:CHAR:Self-affine-profiles} were generated. For more details the
reader is advised to consult Refs.~\citen{AnnPhys,PhysRep,Book:Meakin} and
\citen{Book:Feder}.

\section{Quantities and Techniques used in 
         Rough Surface Scattering Studies }
\label{Chap:Theory}

The intention of the present section is to introduce some of the main
quantities and techniques, both analytical and numerical, used in the
field of wave scattering from randomly rough surfaces. The idea is to
cover in some detail the most central techniques at the sacrifice of
a wide coverage. The first part of this section is devoted to the
discussion of some general properties of the scattering problem. This
discussion is independent of the techniques used for its solution. In
the second part, however, some of the central theoretical approaches
towards the solution of the scattering problem are presented. Here the
outlined theories will only sparsely be applied to a concrete problem.
However, this is done in the next section where phenomena in rough
surface scattering are discussed.

\subsection{The Scattering Geometry}
\label{Theory:Sect:scattering geometry}

\begin{figure}[t!]
  \begin{center}
    \leavevmode    
    \includegraphics[width=12cm,height=6cm]{\myfigpath/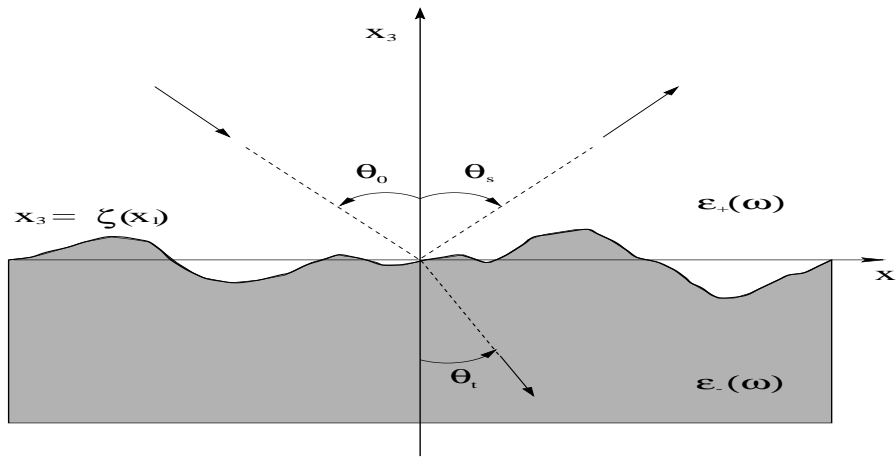}
    \caption{The main scattering geometry used throughout this
      section for the wave scattering from a rough surface defined by
      $x_3=\zeta(x_1)$. The region above the surface,
      $x_3>\zeta(x_1)$, is assumed to be vacuum
      ($\varepsilon_+(\omega)=1)$, while the one below is metal or a
      dielectric characterized by a frequency-dependent dielectric
      function $\varepsilon_-(\omega)=\eps{}$.  Notice for which
      direction the incident ($\theta_0$), scattering ($\theta_s$),
      and transmission ($\theta_t$) angles are defined positive. An
      angle of transmission is only well-defined if the lower medium
      is transparent, {\it i.e.} if $Re\, \varepsilon(\omega)>0$.}
    \label{Fig:Theory:Geometry}
  \end{center}
\end{figure}

The scattering geometry that we will mainly concern ourselves with in
this section is depicted in Fig.~\ref{Fig:Theory:Geometry}. It consists
of vacuum ($\eps{+}=\eps{0}=1$) in the region $x_3 > \zeta(x_1)$, and
a metal or dielectric characterized by an isotropic,
frequency-dependent, dielectric function $\eps{-}=\eps{}$, in the
region $x_3 < \zeta(x_1)$. Here $\zeta(x_1)$ denotes the surface
profile function and it is assumed to be a single-valued function of
$x_1$ that is differentiable as many times as is necessary.
Furthermore, it constitutes a zero mean, stationary, Gaussian random
process which we from Sect.~\ref{Sect:Stat-surf} recall is defined by
\begin{subequations}
\begin{eqnarray}
  \left< \zeta(x_1)\right> &=& 0, \\ 
  \left< \zeta(x_1)\zeta (x'_1)\right> &=& \delta^2 W(|x_1-x'_1|).
\end{eqnarray}
\end{subequations}
Here $W(|x_1|)$ denotes the auto-correlation function and it will be
specified later.

The incident wave is assumed to be either $p$- or $s$-polarized, as
indicated by the index $\nu$, and the plane of incidence will be the
$x_1x_3$-plane.  Furthermore, the angle of incidence, reflection, and
transmission, $\theta_0$, $\theta_s$, and $\theta_t$ respectively, are
measured positive according to the convention indicated in
Fig.~\ref{Fig:Theory:Geometry}.

\subsection{The Scattered Field}
\label{Sect:Theory:Scattered-fields}

From Sect.~\ref{Chapter:Elmag} we recall that in order to solve the
scattering problem, we have to solve the Helmholtz equation and
satisfy the boundary conditions, Eqs.~\r{Elmag:BCS}, at the rough
interface $\zeta(x_1)$ as well as the boundary conditions at infinity.
In the present section we will give the form of the far fields that
automatically satisfy the Helmholtz equation and the boundary
conditions at infinity. We will discuss separately the case where the
incident field is a plane wave and where it is a wave of finite
width.

However, before we do so, we recall that for the scattering geometry
depicted in Fig.~\ref{Fig:Theory:Geometry}, the Maxwell's equations
are equivalent to the scalar Helmholtz equation for the field
$\Phi_\nu(x_1,x_3|\omega)$ defined by Eq.~\r{Elmag:Primary-fields},
{\it i.e.}
\begin{eqnarray}
    \Phi_\nu(x_1,x_3 | \omega ) 
      &=& \left\{
               \begin{array}{ll}
                  H_2(x_1,x_3|\omega), & \quad \nu=p, \\
                  E_2(x_1,x_3|,\omega), & \quad \nu=s,
               \end{array} 
           \right. .
\end{eqnarray}
It is the asymptotic, far-field behavior of $\Phi_\nu(x_1,x_3|\omega)$
that we are trying to determine.

\subsubsection{Plane Incident Wave}

Let us first consider the case where the incident field is (a either
$p$- or $s$-polarized) plane wave of the form
\begin{eqnarray}
  \label{Eq:Theory:plane-inc-wave}
  \Phi^{inc}_{\nu}(x_1,x_3|\omega) &=&  e^{ikx_1-i\aln{k}x_3},
\end{eqnarray}
where\footnote{We will use the notation $\alpha_0(q,\omega)$ instead
  of $\alpha_+(q,\omega)$ in order follow the notation frequently used
  in the literature.}  
\begin{eqnarray}
    \label{Eq:Theory:alpha0}
    \alpha_0(q,\omega) &\equiv&
     \alpha_+(q,\omega) \;=\;
      \left\{
      \begin{array}{ll}
          \sqrt{\frac{\omega^2}{c^2}-q^2}, & \quad |q|<\frac{\omega}{c},
          \\
          i \sqrt{q^2-\frac{\omega^2}{c^2}}, & \quad |q|>\frac{\omega}{c}.
      \end{array}
      \right. .\qquad
\end{eqnarray} 
Then the form of the field in region $x_3> \max \zeta(x_1)$ that
satisfied both the Helmholtz equation as well as the boundary
conditions at infinity ($x_3=\infty$) can be written as
\begin{subequations}   
  \label{Eq:Theory:Asymp-fields}
  \begin{eqnarray}
    \label{Eq:Theory:plane-tot-wave}
    \Phi_\nu^+(x_1,x_3|\omega) &=& 
    \Phi^{inc}_{\nu}(x_1,x_3|\omega) 
    + \qint{q}\; R_\nu(q|k) e^{iqx_1+i\aln{q}x_3}. \qquad \quad  
  \end{eqnarray}
  Similarly, a solution to the Helmholtz equation in the region $x_3<
  \min \zeta(x_1)$ that satisfy the boundary condition at
  $x_3=-\infty$ is
  \begin{eqnarray}
    \label{Eq:Theory:plane-tot-wave-below}
    \Phi_\nu^-(x_1,x_3|\omega) &=& 
    \qint{q}\; T_\nu(q|k) e^{iqx_1-i\al{q}x_3}, \qquad   
  \end{eqnarray}
\end{subequations}
where ($\eps{-}\equiv \eps{}$)
\begin{eqnarray}
  \label{Eq:Theory:alpha}
  \alpha(q,\omega) &\equiv&
  \alpha_-(q,\omega) \;=\;
       \sqrt{\varepsilon(\omega )\frac{\omega^2}{c^2}-q^2},
        \qquad Re\,\alpha, Im\,\alpha > 0.
\end{eqnarray}   
In these equations $R_\nu(q|k)$ and $T_\nu(q|k)$ denote the scattering
and transmission amplitudes respectively.  Notice that
these asymptotic expressions does note say anything about how the
fields look like in the surface region $\min \zeta(x_1) < x_3 < \max
\zeta(x_1)$. This and its consequence, will be discussed in more
detail in Sect.~\ref{Theory:Sect:RRE} when we derive the so-called
reduced Rayleigh equation.

\subsubsection{Finite Width Incident Wave}
\label{Subsect:Finite}

If the incident field is not a plane wave, but instead has a finite
width, then the above expressions will have to be changed
somewhat. In this case the incident field can be written as 
\begin{subequations}
  \begin{eqnarray}
    \Phi^{inc}_{\nu}(x_1,x_3|\omega) 
    &=&  \int^{\frac{\omega}{c}}_{-\frac{\omega}{c}} \,
    \frac{dk}{2\pi}\,F(k) e^{ikx_1-i\aln{k}x_3},
  \end{eqnarray}
  {\it i.e.} as a weighted sum of plane waves.
  Here $F(k)$ is in principle an arbitrary function for which the
  integral exits. Due to the linearity of the Maxwell's equations, the
  scattered field becomes
  \begin{eqnarray}
    \Phi^{sc}_\nu(x_1,x_3|\omega) &=&
    \qint{q} R_\nu(q,\omega) e^{iqx_1+i\aln{q}x_3},
  \end{eqnarray}
  where 
  \begin{eqnarray}
    \label{Eq:Theory:R-finite}
    R_\nu(q,\omega)  &=& \int^{\frac{\omega}{c}}_{-\frac{\omega}{c}} 
    \frac{dk}{2\pi} \, R_\nu(q|k) F(k).
  \end{eqnarray}
\end{subequations}
The total field in the region $x_3> \max\zeta(x_1)$ therefore is
$\Phi_\nu^+(x_1,x_3|\omega)= \Phi_\nu^{inc}(x_1,x_3|\omega)+
\Phi_\nu^{sc}(x_1,x_3|\omega)$.

In a similar way the field in the region $x_3<\min\zeta(x_1)$ can be
written as
\begin{eqnarray}
  \Phi_\nu^-(x_1,x_3|\omega) &=& 
    \qint{q}\; T_\nu(q,\omega) e^{iqx_1-i\al{q}x_3}, \qquad 
\end{eqnarray}
where $T_\nu(q,\omega)$ is given by an expression similar to
Eq.~\r{Eq:Theory:R-finite}.

In order to fully define the asymptotic forms of the field, the
envelope $F(k)$ has to be given.  Here we will only consider so-called
Gaussian finite beams.  Such beams are obtained if $F(k)$ has the
Gaussian form. If the half-width of the incident beam is denoted by
$w$ the Gaussian envelope $F(k)$ can be written as~\cite{AnnPhys}
\begin{eqnarray} 
  F(k) &=& \frac{w\omega}{2\sqrt{\pi} c} \frac{1}{\aln{k}} 
            \exp\left[-\frac{w^2\omega^2}{4c^2}
            \left(\arcsin\frac{kc}{\omega}-\theta_0\right)^2 \right]. 
\end{eqnarray}

\subsection{The mean differential reflection coefficient}
\label{Theory:Sect:mean-DRC}

In the previous section, we obtained the asymptotic forms of the
scattered and transmitted fields. These fields are known whenever the
scattering and transmission amplitudes $R_\nu(q|k)$ and $T_\nu(q|k)$
are known. We will later in this section describe methods for how to
determine these amplitudes.

However, these two amplitudes are not accessible in experiments.
Since, of course, our ultimate goal is to compare the theoretical
predictions to those of experimental measurements, one has to relate
these amplitudes to measurable quantities. Such quantities are
provided by the so-called mean differential reflection and
transmission coefficients.  These are not the only experimentally
accessible quantities possible.  However, other such quantities must
necessarily be related to the reflection or transmission
amplitudes, since they fully specify the scattering and transmission
problem.

The mean differential reflection coefficient\footnote{If the surface
  is two-dimensional, which we however will not discuss in great
  detail here, one has to consider scattering into solid angle
  $d\Omega_s$ around the scattering direction $(\theta_s, \phi_s)$
  instead of into the angular interval $d\theta_s$ around the
  scattering angle $\theta_s$ as is the case if the surface is
  one-dimensional.  Furthermore, one also has to take into account
  that depolarizations may occur in scattering from two-dimensional
  surfaces. Hence the mean differential reflection and transmission
  coefficients in the $2D$-case have polarization indices referring to
  the polarization of the incident and scattered light respectively.}
is defined as the fraction of the total incident power scattered, by
the randomly rough surface, into an angular interval of width
$d\theta_s$ about the scattering angle $\theta_s$.  Thus, in order to
obtain an expression for this quantity one has to find an expression
for the power incident onto the rough surface and the power scattered
from it.  We recall that the total power contained in an
electromagnetic wave of electric and magnetic field vectors ${\mathbf
  E}$ and ${\mathbf H}$ respectively is given by the real part of the
complex Poynting vector ${\mathbf S}={\mathbf E}\times {\mathbf H}^*$,
where the asterisk denotes complex conjugate. More useful to us is in
fact the time-averaged of this (complex) quantity. It is given
by~\cite{Jackson,Book:Kong,Book:Born}
\begin{eqnarray}
\left<{\mathbf S}\right>_t=\frac{1}{2}\, {\mathbf E}\times {\mathbf H}^*, 
\end{eqnarray}
where $\left< \cdot\right>_t$ indicates time average. Hence the
time-averaged power incident onto the rough surface, and scattered
from it, are given by the real part of the $3$-component of
$\left<{\mathbf S}\right>_t$, evaluated for the fields involved.  The
corresponding time-averaged total energy flux therefore
becomes\footnote{Recall that the coordinate system is chosen so that
  the $x_1x_2$-plane coincide with the average (planar) surface, and
  thet the incident plane is the $x_1x_3$-plane.}
\begin{eqnarray}
  P  \; = \; \int dx_1dx_2 \;Re \left<{\mathbf S}\right>_t
     \; = \;  L_2 \int dx_1 \;Re \left<{\mathbf S}\right>_t.
\end{eqnarray}
In writing the above equation we have taken advantage of the fact that
for a one-dimensional rough surface with its generator along the
$x_1$-direction (as we consider here), the $x_2$-integration becomes
trivial and only contributes with a factor $L_2$, the length of the
surface in the $x_2$-direction.

\subsubsection{Plane incident wave}
\label{Theory:Subsect:mean DRC-PlaneWave}

We recall that if the incident wave is a plane wave of the form given
in Eq.~\r{Eq:Theory:plane-inc-wave}, then the scattered field is given
by the second term of Eq.~\r{Eq:Theory:plane-tot-wave}.
The incident and scattered energy fluxes thus becomes
\begin{subequations}
  \begin{eqnarray}
    P_{inc} &=&  \frac{L_1 L_2}{2}\frac{c^2}{\omega} \aln{k},
  \end{eqnarray}
  and 
  \begin{eqnarray}
    P_{sc} &=& 
    \frac{L_2}{2}\frac{c^2}{\omega} 
    \int^{\ooc{}}_{-\ooc{}}
    \frac{dq}{2\pi}\;\aln{q} \left|R_\nu(q|k)\right|^2,
      \nonumber \\
    &=& \int^{\frac{\pi}{2}}_{-\frac{\pi}{2}} d\theta_s \;
         p_{sc}(\theta_s),
  \end{eqnarray}
  where 
  \begin{eqnarray}
    p_{sc}(\theta_s) &=&
    \frac{L_2}{4\pi} \omega \cos^2\theta_s \left|R_\nu(q|k)\right|^2.
  \end{eqnarray}
\end{subequations}
Hence, the differential reflection coefficient, according to its
definition, is given by the following expression
\begin{eqnarray}
  \frac{\partial R_\nu}{\partial \theta_s}
      \;=\; \frac{p_{sc}(\theta_s)}{P_{inc}} \;=\;
      \frac{1}{L_1} 
        \frac{\omega}{2\pi c}\frac{\cos^2\theta_s}{\cos\theta_0}
            \left| R_\nu(q|k)\right|. \nonumber 
\end{eqnarray}
In the above expression it is understood that the momenta $k$ and $q$
are related to the angles $\theta_0$ and $\theta_s$ according to 
\begin{subequations}
  \label{Eq:mom-angles}
  \begin{eqnarray}
    \label{Eq:Theory:k-angle}
    k &=&  \ooc{} \;\sin\theta_0, \\
    \label{Eq:Theory:q-angle}
    q &=& \ooc{} \;\sin\theta_s.
  \end{eqnarray}
\end{subequations}
Notice that $\partial R_\nu/\partial \theta_s$ includes the
contribution from only one single realization of the rough surface.
However, we are more interested in the mean of this quantity obtained
by making an average over an ensemble of realizations of the rough
surface profile. In consequence we obtain the following expression for
the {\em mean differential reflection coefficient}~(DRC)
\begin{eqnarray}
  \label{Theory:meanDRC-total}
  \left< \frac{\partial R_\nu}{\partial \theta_s}\right>
      &=&
      \frac{1}{L_1} 
          \frac{\omega}{2\pi c}\frac{\cos^2\theta_s}{\cos\theta_0}
            \left<\left| R_\nu(q|k)\right|^2\right>.
\end{eqnarray}
When light is scattered from a randomly rough surface both
coherent~(specular) and incoherent~(diffuse) scattering processes will
normally occur. The scattered power due to both these processes are
contained in Eq.~\r{Theory:meanDRC-total}.  In theoretical studies of
wave scattering from rough surfaces it has proven useful to separate
these two contributions even though such a separation is not possible
under experimental conditions. The separation is done by noticing that
$\left<\left| R_\nu(q|k)\right|^2\right>$ can trivially be written as
\begin{eqnarray}
  \label{Theory:R-subdiv}
   \left<\left| R_\nu(q|k)\right|^2\right> &=& 
     \left<\left| R_\nu(q|k)\right|^2\right> 
     -  \left|\left<R_\nu(q|k)\right>\right|^2
     +  \left|\left<R_\nu(q|k)\right>\right|^2.
\end{eqnarray}
Here the last term (on the right hand side) corresponds to the
coherently scattered light, while the first two terms are related to
the light scattered incoherently. By using this result we find that
the mean DRC can be subdivided into a coherent and an incoherent part,
and they are respectively given by
\begin{subequations}
\label{Theory:meanDRC}
\begin{eqnarray}
  \label{Theory:meanDRC-cor}
  \left< \frac{\partial R_\nu}{\partial \theta_s}\right>_{\mbox{\small
      coh}}
      &=&
      \frac{1}{L_1}
          \frac{\omega}{2\pi c}\frac{\cos^2\theta_s}{\cos\theta_0}
            \left|\left< R_\nu(q|k)\right>\right|^2, 
\end{eqnarray}
and 
\begin{eqnarray}
  \label{Theory:meanDRC-incohr}
  \left< \frac{\partial R_\nu}{\partial \theta_s}\right>_{\mbox{\small
      incoh}}
      &=&
      \frac{1}{L_1} 
      \frac{\omega}{2\pi c}\frac{\cos^2\theta_s}{\cos\theta_0}
            \left[ 
                   \left<\left| R_\nu(q|k)\right|^2\right>
               -   \left|\left< R_\nu(q|k)\right>\right|^2
            \right]. \qquad
\end{eqnarray}
\end{subequations}

Expressions for the {\em mean differential transmission coefficient},
$\left< \partial T_\nu/\partial \theta_s
\right>=\left<p_{tr}(\theta_t)/P_{inc}\right>$, can be obtained in an
analogous way by calculating $p_{tr}(\theta_t)$, the equivalent of
$p_{sc}(\theta_s)$ in transmission. The results of such a calculation
is that the expressions for $\left<\partial T_\nu/\partial
  \theta_s\right>$ can be obtained from those of $\left<\partial
  R_\nu/\partial \theta_s\right>$ by substituting the
transmission amplitude $T_\nu(q|k)$ for the reflection amplitude,
$R_\nu(q|k)$ and multiplying the final expression with a factor of 
$\sqrt{\eps{}}$.

\subsubsection{Finite width incident beam}

In the previous subsection we considered a plane incident wave. Such
waves can never be achieved under experimental conditions, and it
is therefore desirable in some cases to work with a incident beam of
finite width. 

Such a beam has already been defined in
Subsection.~\ref{Subsect:Finite} and with these expressions one gets
for the incident and scattered energy fluxes
\begin{subequations}
  \begin{eqnarray}
    P_{inc} &=&  L_2 \frac{w c}{2\sqrt{2\pi}}
        \left[
            \mbox{erf}\left(
             \frac{w}{\sqrt{2}}\ooc{}
                  \left(
                     \frac{\pi}{2}-\theta_0\ 
                  \right)
             \right)
         +
           \mbox{erf}\left(
             \frac{w}{\sqrt{2}}\ooc{}
                  \left(
                     \frac{\pi}{2}+\theta_0\ 
                  \right)
             \right)
        \right], \nonumber \\
  \end{eqnarray}
  where $\mbox{erf}(x)$ is the error-function~\cite{Stegun}, and 
  \begin{eqnarray}
    p_{sc}(\theta_s) &=&  
       L_2    \frac{\omega}{2\pi^2}\cos^2\theta_s 
              \left|R_\nu(q|k)\right|^2.
  \end{eqnarray}
\end{subequations}
Hence, the differential reflection coefficient, according to its
definition, is given by the expression
\begin{eqnarray}
  \label{Theory:DRC-total-beam}
  \left< \frac{\partial R_\nu}{\partial \theta_s}\right>
      &=&
      \frac{2}{(2\pi)^{\frac{3}{2}}}\frac{\omega}{cw}\cos^2\theta_s
         \frac{\left<\left|R_\nu(q,\omega)\right|^2\right>}{\frac{1}{2}
         \left[
        \mbox{erf}\left(\frac{w\omega}{\sqrt{2}c}\left(\frac{\pi}{2}+\theta_0\right)\right) 
      + \mbox{erf}\left(\frac{w\omega}{\sqrt{2}c}\left(\frac{\pi}{2}-\theta_0\right)\right) 
 \right]}.\nn \\
\end{eqnarray}
Also here $q$ is understood to be related to the scattering angle by
Eq.~\r{Eq:Theory:q-angle}. The coherent and incoherent part of the
mean differential reflection coefficient are obtained in the same way
as for the case of a plane incident wave. These expressions will not
be explicitly included here since they follow from
Eq.~\r{Theory:DRC-total-beam} by a simple substitution for of
Eq.~\r{Theory:R-subdiv}.

\subsection{General Properties of the Scattering Problem}
\label{Theory:Sect:GenScatProp}

In order to solve the scattering problem, we have to calculate the
scattering and transmission amplitudes, $R_\nu(q|k)$ and $T_\nu(q|k)$.
However, before we start discussing various methods for obtaining this
goal, we will introduce some general features that the scattering
problem should fulfill.  These properties are, among others, {\em
  reciprocity} and {\em unitarity}.

\subsubsection{Reciprocity}
\label{Theory:Sect:Reciprocity}

A general property of the scattering problem is the one of
reciprocity. It involves the scattering matrix, $S_\nu(q|k)$, defined
via the scattering amplitude according to
\begin{eqnarray}
  \label{Theory:Eq:S-matrix-def}
  S_\nu(q|k) &=& \frac{\sqrt{\aln{q}}}{\sqrt{\aln{k}}} 
    R_\nu(q|k) . 
\end{eqnarray}
The reciprocity theorem states that this scattering matrix, or just
S-matrix for short, should satisfy the following relation
\begin{eqnarray}
  \label{Theory:Eq:Reciprocity}
  S_\nu(q|k) &=&  S_\nu(-k|-q). 
\end{eqnarray}
This relation for the surface scattering problem can under rather
general assumptions be derived rigorously from Lorentz's reciprocity
theorem~\cite{Book:Nieto}.  However, we will not present such an
interesting, but lengthy derivation here. We would, however, like to
point out that such a derivation does not assume anything about the
dielectric functions involved.  Furthermore, there is no restriction
on how strongly rough the surface is, neither how it is correlated.
Hence, the reciprocity theorem is generally valid. It should also be
pointed out that there seems to be no equivalent theorem to
Eq.~\r{Theory:Eq:Reciprocity} that involves the transmission amplitude
$T_\nu(q|k)$. Reciprocity is therefore a property of the scattering
amplitude.

\subsubsection{Unitarity}
\label{Theory:Sect:Unitarity}

In cases where the scattering medium is a perfect reflector, {\it
  i.e.} if ${\rm Re}\,\eps{}<0$ and ${\rm Im}\, \eps{}=0$, the
scattering matrix, $S_\nu(q|k)$, possess an additional property call
{\em unitarity}. Since there is no absorption (${\rm Im}\, \eps{}=0$)
and no transmission in the system the energy incident on the rough,
perfectly reflecting surface must be conserved. Without going into
details, this has the consequence that the following relation has to
be satisfied~\cite{Book:Nieto}
\begin{eqnarray}
  \label{Theory:Eq:Unitarity}
  \int^{\frac{\omega}{c}}_{-\frac{\omega}{c}} \!\frac{dq}{2\pi} \, 
      S_\nu(q|k) S^*_\nu(q|k') &=& 2\pi \delta(k-k'),
         \qquad |k|,|k'|< \frac{\omega}{c}.
\end{eqnarray}
It can be derived by calculating the total energy flux scattered
from the surface that, due to energy conservation, should equal the
incident energy flux. Eq.~\r{Theory:Eq:Unitarity} expresses the
unitarity of the scattering matrix, and it is a consequence of the
conservation of energy in the scattering process.

Even if Eq.~\r{Theory:Eq:Unitarity} is derived under the assumption
that energy conservation is satisfied, let us for a moment show how
this indeed follows from the unitarity condition. Let us assume that
the rough surface has length $L_1$. Then we know from the sampling
theorem~\cite{NR} that the smallest momentum variable that we can
resolve is $2\pi/L_1$. By multiplying each side of
Eq.~\r{Theory:Eq:Unitarity} by $dk'/(2\pi)$ and integrating the
resulting expression over an interval of length $2\pi/L_1$ that
contains $k'=k$ one finds
\begin{eqnarray}
  \frac{1}{L_1} \int^{\frac{\omega}{c}}_{-\frac{\omega}{c}} \!\frac{dq}{2\pi} \, 
      \left| S_\nu(q |k)\right|^2   &=& 1.
\end{eqnarray}
By now using the definition of the scattering matrix,
Eq.~\r{Theory:Eq:S-matrix-def}, together with
Eq.~\r{Eq:Theory:q-angle}, one arrives at
\begin{eqnarray}
  \label{Theory:Eq:meanDRC-conservation}
  \int^{\pi/2}_{-\pi/2} 
     \left< \frac{\partial R_\nu}{\partial \theta_s}\right> \;
     d\theta_s &=& 1,
\end{eqnarray}
where we have taken advantage of Eq.~\r{Theory:meanDRC-total}.  From
the definition of the mean differential reflection coefficient given
in Sect.~\ref{Theory:Subsect:mean DRC-PlaneWave}, we understand that
Eq.~\r{Theory:Eq:meanDRC-conservation} is just the conservation of
energy for the scattering system considered.

\subsubsection{Energy Conservation}
\label{Theory:Sect:energy_conservation}

If, however, the lower medium is {\em not} a perfect conductor, but
still is a  non-absorbing medium (${\rm Im}\, \eps{}=0$) the unitarity
condition Eq.~\r{Theory:Eq:Unitarity} will no longer hold true.
However, we should still have conservation of energy.  This means that
all energy incident onto the rough surface should be either scattered
from it or transmitted trough it. This fact is expressed by the
following equation
\begin{eqnarray}
  \label{Theory:Eq:Energy-Conservation}
   {\cal U}^{\rm sc}_\nu(\theta_0,\omega) + 
   {\cal U}^{\rm tr}_\nu(\theta_0,\omega)
       &=& 1,
\end{eqnarray}
where $\theta_0$ is the angle of incidence of the light, and 
\begin{subequations}
  \begin{eqnarray}
    {\cal U}^{\rm sc}_\nu(\theta_0,\omega) &=& \int^{\pi/2}_{-\pi/2} 
     \left< \frac{\partial R_\nu}{\partial \theta_s}\right> \;
     d\theta_s, \\
    {\cal U}^{\rm tr}_\nu(\theta_0,\omega) &=& \int^{\pi/2}_{-\pi/2} 
     \left< \frac{\partial T_\nu}{\partial \theta_t}\right> \;
     d\theta_t.
   \end{eqnarray}
\end{subequations}
Physically ${\cal U}^{\rm sc}_\nu(\theta_0,\omega)$ expresses the
fraction of the incident energy scattered from the surface, while in a
similar way ${\cal U}^{\rm tr}_\nu(\theta_0,\omega)$ expresses the
energy fraction transmitted through the system.  Notice that the
energy conservation condition should hold true for all incident angles
and polarizations as well as being independent of the width of the
incident beam. The only restriction being than there should be no
medium that absorbs energy.  However, if absorption is present,
Eq.~\r{Theory:Eq:Energy-Conservation} might be modified by adding an
absorption term to the right hand side.  Unfortunately, this
absorption term is hard to calculate in a rigorous way.
  
For practical purposes, the
conditions Eqs.~\r{Theory:Eq:meanDRC-conservation} and 
\r{Theory:Eq:Energy-Conservation} are most frequently used as a
test of the quality of numerical simulations~(see
Sect.~\ref{Sect:Theory:Sect:NumSim}).  In such an approach these conditions
are necessary, but not sufficient conditions for the correctness of
the simulations.

\subsection{Derivation of the Reduced Rayleigh Equation}
\label{Theory:Sect:RRE}

The reduced Rayleigh equation~(RRE), under which name we know it
today, was first derived by Toigo, Marvin and Celli~\cite{Toigo} in
the last half of the 1970's.  This equation is the single integral
equation satisfied by the reflection or transmission amplitudes. This
equation, even if its precise region of validity is hard to quantify
in detail, has served as the starting point for many, if not all, of
the perturbative techniques developed in the field of wave scattering
from rough surface.  We would, however, already at this early stage
like to stress that the reduced Rayleigh equation is not restricted to
the same limitations as perturbation theory, and that its validity
goes beyond that of such theories. It can in fact also be used in
numerical simulations to obtain non-perturbative results.

We will below give the detailed derivation of the RRE for reflection.
The scattering geometry that we consider is the one presented in
Fig.~\ref{Fig:Theory:Geometry}. This geometry is illuminated from
above by a plane incident wave of either $p$- or $s$-polarization, and
the incident plane is the $x_1x_3$-plane.

\subsubsection{The Rayleigh Hypothesis}
\label{Theory:The Rayleigh hypothesis}

It should be clear that for the region above the maximum point of the
surface the total field takes the form given by
Eq.~\r{Eq:Theory:plane-tot-wave} and similarly the total field below
the minimum point of the surface can be expressed according to
Eq.~\r{Eq:Theory:plane-tot-wave-below}.  However, in order to solve
the scattering problem, one has to take into account the boundary
conditions to be satisfied at the randomly rough surface $\zeta(x_1)$
separating the two media above and below it. The problem is that we
don't know the form of the total field close to the surface, or to be
more precise, in the region $\min \zeta(x_1) < \zeta(x_1) < \max
\zeta(x_1)$.  It should be obvious from a ray optical point of view,
that at least for rather rough surfaces, expansions of the form
\r{Eq:Theory:Asymp-fields} are not adequate to describe the total
field in this region due to the lack of not allowing downward
propagating scattered modes.  However, as the surface becomes smother
and smother, the asymptotic expansions of the field given earlier
should represent a better and better approximation for the total
field. This lead Lord Rayleigh~\cite{Rayleigh1,Rayleigh2}, when
studying scattering from sinusoidal surfaces, to assume that the
asymptotic expansions for the total field was not only valid in the
region far away from the rough surface, but could also be used all the
way down to the rough surface.  Under this assumption, known today as
the {\em Rayleigh hypothesis}, he could satisfy the boundary
conditions on the rough surface and thereby derive equation which lead
to the solution of his scattering problem.

\subsubsubsection{The Validity of the Rayleigh Hypothesis} 

We will in the next subsection demonstrate this procedure when applied
to the wave scattering from a randomly rough surface. However, before
we do so, we will dwell a little upon the validity of the Rayleigh
hypothesis.  Theories based on this approximation do not properly
include, as mentioned above, downward propagating scattered or upward
propagating transmitted waves. From a naive geometrical optics
argument, we realize that a scattering process producing incoming
scattered or transmitted waves has to be a multiple scattering
process.  So, scattering geometries where the Rayleigh hypothesis is
not valid, has thus to be dominated by multiple scattering, and has
therefore to be rather rough. It should, however, be stressed that
this do not imply that for any scattering geometry dominated by
multiple scattering, the Rayleigh hypothesis is doomed to break down.
One might very well have processes dominated by multiple scattering
without receiving essential contributions from downward propagating
scattered waves. Good examples of this are provided by the ability of
perturbative and numerical studies based on the Rayleigh hypothesis to
show multiple scattering phenomena like the enhanced backscattering
and satellite peaks~\cite{AnnPhys,Madrazo,MP13}.

The Rayleigh hypothesis is hence a good approximation if the surface
is not too rough. H owever, at what level of roughness must
we say that this approximation no longer is valid? There has been many
papers devoted to the study of  the validity of the Rayleigh
approximation. There seems today to be consensus on the
criterion~\cite{Rayleigh-citerion}
\begin{eqnarray}
  \frac{\delta}{a} &\ll& 1,
\end{eqnarray}
where $\delta$ and $a$ are the {\sc rms}-height and correlation length
of the surface respectively.  The reader is encouraged to consult the
literature for more details~\cite{Rayleigh-citerion}.

\subsubsection{The Rayleigh Equations}

In this subsection we will derive a set of two inhomogeneous coupled
integral equation for the scattering and transmission amplitudes,
$R_\nu(q|k)$ and $T_\nu(q|k)$. These equations are referred to as the
Rayleigh equations, and we will now demonstrate how they are obtained.

From Sect.~\ref{Sec:BoundaryConditions} we recall that the boundary
conditions to be satisfied by the field on the surface are the
continuity of the field and its normal derivative, {\it i.e.}
\begin{subequations}
\label{Theory:BCS}
\begin{eqnarray}
   \left. \Phi_\nu^+(x_1,x_3| \omega) \right|_{x_3=\zeta(x_1)}
    &=& \left. \Phi_\nu^-(x_1,x_3| \omega) \right|_{x_3=\zeta(x_1)},
   \label{Theory:BCSa}\\
   \left. 
           \partial_n \Phi_\nu^+(x_1,x_3| \omega) 
    \right|_{x_3=\zeta(x_1)} 
   &=&
   \left. \frac{\partial_n \Phi_\nu^-(x_1,x_3| \omega)}{\kappa_\nu(\omega)}
    \right|_{x_3=\zeta(x_1)}, \qquad
    \label{Theory:BCSb}
\end{eqnarray}
\end{subequations}
where the normal derivative, $\partial_n$, and the symbol
$\kappa_\nu(\omega)$ have been defined earlier in
Eqs.~\r{Elmag:normal-derivative} and \r{Elmag:BCSc}.

If now the Rayleigh hypothesis holds true, the asymptotic field
expansions of Sect.~\ref{Sect:Theory:Scattered-fields}, can be used in
order to fulfill those boundary conditions. By substituting the
asymptotic expansions, Eqs.~\r{Eq:Theory:Asymp-fields}, into the
boundary condition for the field, Eq.~\r{Theory:BCSa}, one is lead to
the following integral equation
\begin{eqnarray}
 e^{ikx_1-i\aln{k}\zeta(x_1)}
       + \int\!\frac{dq}{2\pi}\;
       R_\nu(q|k)e^{iqx_1+i\aln{q}\zeta(x_1)} 
  &=&  \int\!\frac{dq}{2\pi}\; T_\nu(q|k)e^{iqx_1-i\al{q}\zeta(x_1)}.
\nonumber
\end{eqnarray} 
If we now rewrite this equation, which will prove useful later on, by
using the properties of the Dirac $\delta$-function we get
\begin{subequations}
\label{Theory:RE}
\begin{eqnarray}
\label{Theory:RE1}
 \int\!\frac{dq}{2\pi}\; 
  e^{iqx_1} \left[ 2\pi\delta(q-k) e^{-i\aln{q}\zeta(x_1)}
                   +R_\nu(q|k)e^{i\aln{q}\zeta(x_1)}
            \right] 
    &=& \int\!\frac{dq}{2\pi}\;  e^{iqx_1} \,   
             T_\nu(q|k)\,e^{-i\al{q}\zeta(x_1)}.   
\end{eqnarray}   
By doing the same for the boundary condition for the normal
derivative, {\it i.e.} Eq.\r{Theory:BCSb}, one arrives at
\begin{eqnarray}
\label{Theory:RE2}
 \int\!\frac{dq}{2\pi}\; 
  e^{iqx_1} \left[ -2\pi\delta(q-k)
                     \left\{\zeta'(x_1)q+\aln{q}\right\}  
                     e^{-i\aln{q}\zeta(x_1)}
     +\; R_\nu(q|k)       
                     \left\{-\zeta'(x_1)q+\aln{q}\right\}  
                     e^{i\aln{q}\zeta(x_1)}
            \right]         
      \\
  && \hspace*{-14cm} \nn
  = 
     -\frac{1}{\kappa_\nu(\omega)}
    \int\!\frac{dq}{2\pi}\;  e^{iqx_1} \,   T_\nu(q|k)\,
            \left\{\zeta'(x_1)q+\al{q}\right\}  
                 e^{-i\al{q}\zeta(x_1)}.
\end{eqnarray}
\end{subequations}   
Together Eqs.~\r{Theory:RE} constitute a set of coupled inhomogeneous
integral equations announced earlier --- the Rayleigh equations.

\subsubsection{The Reduced Rayleigh Equations}
           
We will now continue to derive the so-called reduced Rayleigh
equation~(RRE) for reflection~\cite{Toigo} and
transmission~\cite{Tamara-unp}. The RRE is a single integral equation
satisfied by the reflection or transmission amplitude. They are
derived from the Rayleigh equations by eliminating respectively the
transmission and reflection amplitudes.

\subsubsubsection{The Reduced Rayleigh Equation for Reflection}

In order to obtain the reduced Rayleigh equation for reflection, we
have to eliminate the transmission amplitude, $ T_\nu(q|k)$ from the
(coupled) Rayleigh equations given in the previous subsection. By
multiply Eq.\r{Theory:RE1} by
\begin{eqnarray} 
 e^{-ipx_1-i\al{p}\zeta(x_1)}\left[-\zeta'(x_1)p+\al{p}\right],
\end{eqnarray}
and Eq.\r{Theory:RE2} by
\begin{eqnarray} 
     \kappa_\nu(\omega) e^{-ipx_1-i\al{p}\zeta(x_1)},
\end{eqnarray}
adding the two resulting equations, and integrating the final
result over $x_1$, one finds that the terms containing the
transmission amplitude vanishes identically. 
In detail what one gets for terms proportional to $T_\nu(q|k)$ are
\begin{eqnarray} 
    \label{Eq:Theory:T-vanish}
     \int\!dx_1\,\frac{dq}{2\pi}\; T_\nu(q|k)\, 
      \left[ -\zeta'(x_1)(p+q)+\al{p}-\al{q}\right]\,    
       e^{-i(p-q)x_1}e^{i(-\al{p}-\al{q})\zeta(x_1)}. \quad
\end{eqnarray}
This expression is simplified by introducing an integral defined
according to\footnote{Be aware that various sign conventions seems
  to appear in the literature for this quantity.}  
\begin{eqnarray}
  \label{Theory:zeta-int}
  I(\gamma | q)  &=&     \int\!dx_1 \; e^{-i\gamma\zeta(x_1)} e^{-iqx_1}.
\end{eqnarray}
From this definition it follows that
\begin{eqnarray}
  \label{Theory:zeta-int-a}
  \frac{qI(\gamma | q)}{\gamma}  &=&     
       \int\!dx_1 \;\zeta'(x_1) e^{-i\gamma\zeta(x_1)} e^{-iqx_1}.   
\end{eqnarray}
With Eqs.~\r{Theory:zeta-int} and \r{Theory:zeta-int-a},
Eq.~\r{Eq:Theory:T-vanish} can be written in the form
\begin{eqnarray} 
     \int\!\frac{dq}{2\pi}\; T_\nu(q|k)\, 
      \left[ \frac{(p-q)(p+q)}{\al{p}+\al{q}} +\al{p}-\al{q}\right]\,    
           I(\al{p}+\al{q} | p-q).  \nn 
\end{eqnarray} 
By some simple algebra it can readily be shown that the expression in
the square brackets is identically zero, and thus the transmission
amplitude $T_\nu(q|k)$ has been eliminated from the Rayleigh equations.

The reduced Rayleigh equation for reflection now follows from the
remaining non-vanishing parts of the equation. It reads 
\begin{subequations}
  \label{Theory:RRE}
\begin{eqnarray} 
   \int\!\frac{dq}{2\pi}\; M^+_\nu(p|q)\, R_\nu(q|k) 
     &=&   M^-_\nu(p|q)
\end{eqnarray}
where
\begin{eqnarray}
  \label{Theory:RRE-matrix-elements}
  M^\pm_\nu(p|q) &=& \pm 
    \left[ \frac{(p+\kappa_\nu(\omega)q)(p-q)}{\al{p}\mp\aln{q}}
           +\al{p} \pm \kappa_\nu(\omega)\aln{q}
    \right] \, 
     I\left( \al{p}\mp \aln{q}|p-q \right).
\end{eqnarray}
\end{subequations}

\subsubsubsection{P-polarization}

If we restrict ourselves to $p$-polarization, the reduced Rayleigh
equation as presented in Eqs.~\r{Theory:RRE} takes on a simpler form. By a
straight forward calculation one finds\footnote{Here the matrix
  elements $M^\pm_p(p|q)$ and $N^\pm_p(p|q)$ are related by
  $M^\pm_p(p|q)=(\eps{ }-1)N^\pm_p(p|q)$.}
\begin{subequations}
  \label{Theory:RRE-p}
\begin{eqnarray} 
  \label{Theory:RRE-p-a}
   \int\!\frac{dq}{2\pi}\; N^+_p(p|q)\, R_p(q|k) 
     &=&
   N^-_p(p|q),
\end{eqnarray}
where
\begin{eqnarray}
  \label{Theory:RRE-p-b}
 N^\pm_p(p|q) &=& 
   \pm \frac{pq\pm \al{p}\aln{q}}{\al{p}\mp\aln{q}}
    I\left( \al{p}\mp \aln{q}|p-q \right). \qquad \quad
\end{eqnarray}
\end{subequations}

\subsubsubsection{S-polarization}

Similarly for $s$-polarization one finds\footnote{Here is 
  $M^\pm_s(p|q)= (\eps{}-1) (\omega/c)^2 N^\pm_s(p|q)$.}
\begin{subequations}
  \label{Theory:RRE-s}
\begin{eqnarray} 
  \label{Theory:RRE-s-a}
   \int\!\frac{dq}{2\pi}\; N^+_s(p|q)\, R_s(q|k) 
     &=&
   N^-_s(p|q),
\end{eqnarray}
where
\begin{eqnarray}
  \label{Theory:RRE-s-b}
 N^\pm_s(p|q) &=& 
   \pm \frac{1}{\al{p}\mp\aln{q}}
    I\left( \al{p}\mp \aln{q}|p-q \right). \qquad \quad
\end{eqnarray}
\end{subequations}

\subsubsection{The Reduced Rayleigh Equation for Transmission}

In the previous subsection it was shown that by eliminating the terms
containing the transmission amplitude from the two coupled Rayleigh
equations, the reduced Rayleigh equation for reflection was obtained.
In a similar way, the scattering amplitude might be eliminated from
the same equations, resulting in the reduced Rayleigh equation for
transmission.

We will not go into details about how this is done, since the
derivation mimics the one given in the previous subsection. Here we
will only give the results that are~\cite{Tamara-unp}
\begin{subequations}
  \begin{eqnarray}
    \qint{q} \frac{I(\al{q}-\aln{p}|p-q)}{\al{q}-\aln{p}} \;
    \left[pq+\aln{p}\al{q}\right] T_p(q|k)  
      &=& \; 2\pi \delta(p-k) 
            \frac{2\eps{}\aln{k}}{\eps{}-1}. 
  \end{eqnarray}
  for $p$-polarization, and~\cite{Tamara-unp} 
  \begin{eqnarray}
    \qint{q} \frac{I(\al{q}-\aln{p}|p-q)}{\al{q}-\aln{p}} \;
    T_s(q|k) 
      &=&  2\pi \delta(p-k) 
            \frac{2\aln{k}}{\ooc{2}\left(\eps{}-1\right)}. 
            \nn \\
  \end{eqnarray}
  for $s$-polarization. 
\end{subequations}

This concludes the section on the reduced Rayleigh equation.

\subsection{Small Amplitude Perturbation Theory}
\label{Sect:Theory:SAPT}

Among the oldest theories addressing rough surface scattering we find
the small amplitude perturbation theory~\cite{Rice}.  The starting
point for this perturbation theory, like most of the perturbation
theories developed for handling wave scattering from rough surface, is
the reduced Rayleigh equation~\r{Theory:RRE}.  If the rough surface is
weakly rough, most of the light incident upon it is scattered into the
specular direction. However, due to the surface roughness, a small
fraction of the incident power is scattered away from the specular
direction.  Theoretically, this non-specular scattering is taken into
account by assuming an expansion for the scattering amplitude in
powers of the surface profile function of the form\footnote{The
  prefactors in this expansion is included for later convenience.}
\begin{eqnarray}
  \label{Theory:R-expansion}
  R_\nu(q|k) &=& \sum^{\infty}_{n=0} \frac{R_\nu^{(n)}(q|k)}{n!}.
\end{eqnarray}
Here $R_\nu^{(n)}(q|k)$ is assumed to be of order ${\cal O}(\zeta^n)$
in the surface profile function $\zeta(x_1)$.  In order to solve the
scattering problem in this way, we therefore have to determine the
expansion coefficients $\{ R_\nu^{(n)}(q|k)\}$.  However, to determine
all these coefficients is obviously not practically possible if
$\zeta(x_1)$ is a rough surface.  Therefore, the expansion
\r{Theory:R-expansion} is terminated at some upper value $N$,
resulting in an $N$'th order perturbation theory. In practical
application one usually has $N\leq 3$--$5$. If the surface is weakly
rough the sum of these $N$ terms will provide a good approximation to
the total scattering amplitude $R_\nu(q|k)$. However, as the surface
roughness becomes stronger and stronger, a higher number of terms has
to be included in the expansion, and the method becomes cumbersome and
not very practical since $R_\nu^{(n)}(q|k)$ for big values of $n$
easily becomes complicated.  Thus, the small amplitude perturbation
theory is only of interest for weakly rough surfaces.  Hence, it
should therefore not represent any restriction to assume that the
Rayleigh hypothesis is valid and therefore that $R_\nu(q|k)$ should
satisfy the reduced Rayleigh equation, Eq.~\r{Theory:RRE}.

Under this assumption, the various terms in the expansion for the
scattering amplitude can in principle be obtained by substituting its
expansion, Eq.~\r{Theory:R-expansion}, into the reduced Rayleigh
equation, Eq.~\r{Theory:RRE}, and satisfy the resulting equation
order-by-order in the surface profile function $\zeta(x_1)$. However,
before this can be done, $N^\pm_\nu(p|q)$, or equivalently
$M^\pm_\nu(p|q)$, that enter via the reduced Rayleigh equation, also
have to be expanded in powers of the surface profile function. Since
the matrix-elements, $N^\pm_\nu(p|q)$, only depend on $\zeta(x_1)$
through the integrals $I(\gamma|q)$, defined in
Eq.~\r{Theory:zeta-int}, one makes the following expansion
\begin{subequations}
  \label{Theory:int-expansion}
\begin{eqnarray}
  I(\gamma|q) \;\equiv\; \int\!dx_1 \; e^{-i\gamma\zeta(x_1)}
   e^{-iqx_1}
   \;=\; \sum^{\infty}_{n=0} \frac{(-i\gamma)^n}{n!}\,{\tilde \zeta}^{(n)}(q),
\end{eqnarray}
where
\begin{eqnarray}
  {\tilde \zeta}^{(n)}(q) &=& \int^{\infty}_{-\infty} dx_1\; 
                              \zeta^n(x_1) e^{-iqx_1},
\end{eqnarray}
\end{subequations}
is the (inverse) Fourier transform of the $n$th power of the surface
profile function.  From the above equations it should be apparent why
we made the choice we did for the prefactors in the expansion for
$R_\nu(q|k)$.

When the expansion~\r{Theory:int-expansion} is substituted into the
expressions for $N_\nu^\pm(p|q)$,
Eqs.~\r{Theory:RRE-p-b} and \r{Theory:RRE-s-b}, one obtains
\begin{subequations}
  \label{Eq:Theorey:SMPT:N-expansion}
  \begin{eqnarray}
    N_\nu^\pm(p|q) &=& \sum_{n=0}^{\infty}
         \frac{{\cal N}^{\pm\,(n)}_{\nu}(p|q)}{n!}\tilde{\zeta}^{(n)}(p-q),
  \end{eqnarray}
  where for $p$-polarization 
  \begin{eqnarray}
    {\cal N}^{\pm\,(0)}_{p}(p|p) &=& 
        \frac{\eps{}\aln{p}\pm\al{p}}{\eps{}-1},
  \end{eqnarray}
  and 
  \begin{eqnarray}
    {\cal N}^{\pm\,(n)}_{p}(p|q) &=& 
     (-i)^n\left[\pm pq+\al{p}\aln{q}\right]
           \left[ \al{p}\mp \aln{q}\right]^{n-1}, \qquad \quad
  \end{eqnarray}
  when $n\geq1$, while for $s$-polarization we have
  \begin{eqnarray}
    {\cal N}^{\pm\,(0)}_{s}(p|p) &=& 
        \frac{\aln{p}\pm\al{p}}{\ooc{2}\left(\eps{}-1\right)},
  \end{eqnarray}
  and 
  \begin{eqnarray}
    {\cal N}^{\pm\,(n)}_{s}(p|q) &=& 
     \pm(-i)^n \left[ \al{p}\mp \aln{q}\right]^{n-1},
  \end{eqnarray}
  when $n\geq1$.
\end{subequations} 

With these relations available, a recurrence relations for
$\{R_\nu^{(n)}(p|q)\}$, is readily obtained by substituting
Eqs.~\r{Theory:R-expansion} and \r{Eq:Theorey:SMPT:N-expansion} into
the reduced Rayleigh equation, Eqs.~\r{Theory:RRE-p-a} and
\r{Theory:RRE-s-a}, and equating terms of
the same order in $\tilde{\zeta}(q)$. The recurrence relation reads
\begin{eqnarray}
  \label{Eq:Theory:SAPT:recurrence-rel}
  \sum_{m=0}^{\infty}
   \left(
     \begin{array}{c}
       n \\ m
     \end{array}
  \right)
  \qint{p} \; {\cal N}^{+\,(n-m)}_\nu(q|p) 
         \tilde{\zeta}^{(n-m)}(q-p) R^{(m)}_\nu(p|k)
  &=& \;\;  {\cal N}^{-\,(n)}_\nu(q|k) \tilde{\zeta}^{(n)}(q-k). 
\end{eqnarray}

Now the expansion coefficients for the scattering amplitude,
$R_\nu^{(n)}(p|q)$, should be rather straight forward to obtain (at
least in principle).  The lowest order term, $n=0$, is given by
\begin{subequations}
  \begin{eqnarray}
    \label{Theory:R-zero}
    R^{(0)}_\nu (q|k) \;=\;  2\pi\, \delta(p-k) R^{(0)}_\nu(k,\omega)
  \end{eqnarray}
  where $R^{(0)}_\nu(k,\omega)$ given by~($\varepsilon=1$)
  \begin{eqnarray}
    \label{Theory:Fresnel}
    R^{(0)}_\nu(k,\omega) 
    &=& \left\{
      \begin{array}{ll}
        \frac{\displaystyle \eps{}\aln{k}-\eps{0}\al{k}}{
               \displaystyle \eps{}\aln{k}+\eps{0}\al{k}}, & \quad 
        \nu=p, \\
        \frac{\displaystyle \aln{k}-\al{k}}{
               \displaystyle \aln{k}+\al{k}}, & \quad
        \nu=s.
      \end{array}
    \right.
  \end{eqnarray}
\end{subequations}
In these expressions, the momentum variables $k$ and $q$ are
understood to be related to the angles of incidence and scattering,
$\theta_0$ and $\theta_s$ respectively, through
\begin{subequations}
\begin{eqnarray}
  k  &=&   \ooc{} \sin\theta_0, \\
  q  &=&   \ooc{} \sin\theta_s.
\end{eqnarray}
\end{subequations}
The above results are, as expected, the result that we would get for
the scattering from a planar surface, and we note that
$R^{(0)}_\nu(k,\omega)$ is nothing else then the Fresnel reflection
coefficients~\cite{Book:Born}. Notice that the $\delta$-function in
Eq.~\r{Theory:R-zero}, coming from ${\tilde \zeta}^{(0)}$, guarantees
that the scattering is only into the specular direction.

The results for the higher order terms ($n\geq1$) in the
expansion of $R_\nu(q|k)$, describe the light scattered by the
roughness into directions other than the specular. These terms can be
calculated recursively from Eq.~\r{Eq:Theory:SAPT:recurrence-rel},
but, unfortunately, such expressions easily become rater cumbersome
for higher the order terms. However, the first few terms are
manageable, and they are given by the following
expressions~\cite{Shchegrov-PhD-thesis}
\begin{subequations}
  \label{Eq:theory:SAPT:R-first-terms}
  \begin{eqnarray}
     R^{(1)}_\nu (q|k) &=& \chi^{(1)}_\nu(q|k) \, \tilde{\zeta}(q-k), \\ 
     R^{(2)}_\nu (q|k) &=& \frac{1}{2!} \qint{p} 
            \chi^{(2)}_\nu(q|p|k) \,
             \tilde{\zeta}(q-p)\tilde{\zeta}(p-k), \\
     R^{(3)}_\nu (q|k) &=& \frac{1}{3!} \qint{p_1}  \qint{p_2}  
          \chi^{(3)}_\nu(q|p_1|p_2|k)\,
    \tilde{\zeta}(q-p_1)\tilde{\zeta}(p_1-p_2)\tilde{\zeta}(p_2-k), \qquad 
  \end{eqnarray}
\end{subequations}
where, the functions $\chi^{(1)}_\nu(q|k)$, $\chi^{(2)}_\nu(q|p|k)$,
$\ldots$ are somewhat lengthy functions of their arguments, and are
therefore given separately in Appendix~\ref{APP:SAPT}.

As discussed earlier, the first few terms of the
expansion~\r{Theory:R-expansion} with
Eqs.~\r{Eq:theory:SAPT:R-first-terms} substituted, should hence for a
weakly rough surfaces represent a good ($3$rd order) approximation to
$R_\nu(q|k)$.  Experimental accessible quantities could therefore be
calculated based on this approximation.  For instance, the mean
differential reflection coefficient from the incoherent component of
the scattered field is then, according to
Eq.~\r{Theory:meanDRC-incohr}, through fourth order in the surface
profile function, given by
\begin{subequations}
  \label{Eq:SAPT-expansion-total}
\begin{eqnarray}
  \label{Eq:DRC-small-amp-pert-theory}
  \left< \frac{\partial R_\nu}{\partial \theta_s}\right>_{\rm incoh}  
    &=& 
    \frac{1}{L_1} \frac{\omega}{2\pi c} 
       \frac{\cos^2\theta_{s}}{\cos\theta_{0} } 
       \left[
         \left<\left| R^{(1)}_\nu(q|k)  \right|^2\right> 
        +\frac{1}{4}
           \left\{ 
              \left<\left| R^{(2)}_\nu(q|k)\right|^2\right>
              -\left|\left< R^{(2)}_\nu(q|k)\right>\right|^2
           \right\}
           \right. 
           \nn \\ && \left. \hspace*{2.5cm}
         +\frac{1}{3}
           Re\left<R^{(3)}_\nu(q|k)R^{(1)\,*}_\nu(q|k)\right>
       \right] + {\cal O}(\delta^6). \qquad \qquad  
\end{eqnarray}
In arriving at Eq.~\r{Eq:DRC-small-amp-pert-theory} it has been
assumed that the surface profile function, $\zeta(x_1)$, constitutes a
zero-mean, stationary, Gaussian random process. Due to the Gaussian
character of the surface only terms that contain an even number of
surface profile function survives the averaging process.  The
different averaged contained in Eq.~\r{Eq:DRC-small-amp-pert-theory}
can all be related to the set of functions $\{\chi^{(n)}_\nu(q|k)\}$
according to
\begin{eqnarray}
  \label{Eq:DRC-small-amp-pert-theory:1}
  \left<\left| R^{(1)}_\nu(q|k)  \right|^2\right> 
     &=&  L_1 \delta^2 g(|q-k|) \left|\chi^{(1)}_\nu(q|k)\right|^2, \\
  \label{Eq:DRC-small-amp-pert-theory:2}
 \left<\left| R^{(2)}_\nu(q|k)\right|^2\right>
              -\left|\left< R^{(2)}_\nu(q|k)\right>\right|^2
     &=&  L_1 \delta^4  \qint{p}   g(|q-p|)  g(|p-k|) 
        \nn \\ && \mbox{}  \hspace*{1.5cm} \times
           \left\{  \left|\chi^{(2)}_\nu(q|p|k)\right|^2
                    +
               \chi^{(2)}_\nu(q|p|k)\chi^{(2)\,*}_\nu(q|q+k-p|k) 
          \right\}, 
       \nn \\
\end{eqnarray}
and 
\begin{eqnarray}
  \label{Eq:DRC-small-amp-pert-theory:3}
 \left<R^{(3)}_\nu(q|k)R^{(1)\,*}_\nu(q|k)\right>
     &=&   L_1 \delta^4  g(|q-k|) \chi^{(1)\,*}_\nu(q|k)
      \nn  \\ && \hspace*{-3cm} \times
           \qint{p} 
         \left\{  \chi^{(3)}_\nu(q|p|q|k) g(|p-q|)
                 + \chi^{(3)}_\nu(q|k|p|k) g(|p-k|)
                + \chi^{(3)}_\nu(q|p|p+k-q|k) g(|p-q|)
             \right\}.  
\end{eqnarray}
\end{subequations}
Notice that the term~\r{Eq:DRC-small-amp-pert-theory:1} represents
single scattering, while the terms in
Eq.~\r{Eq:DRC-small-amp-pert-theory:2} give the contribution due to
double scattering. Eq.~\r{Eq:DRC-small-amp-pert-theory:3} represents a
``mixed'' contribution.

\subsubsubsection{Accuracy of the Small Amplitude Perturbation Theory} 

So what accuracy can we expect to achieve by using the small amplitude
perturbation theory, and for what range of surface parameters is it
valid? There have been many studies in the past, both theoretical,
numerical, and experimental, regarding this issue.  Probably the most
relevant for this introduction is the one by E.\ I.\ Thorsos and D.\ 
R.\ Jackson~\cite{Thorsos1989} who for an acoustic scattering problem
studied the validity of small amplitude perturbation theory for
Gaussian surfaces by comparing its prediction to the one obtained by
rigorous numerical simulations~(see
Sect.~\ref{Sect:Theory:Sect:NumSim}).  They found that the small
amplitude perturbation theory is good if $k\delta\ll 1$ and $ka\leq
1$, where $\delta$ and $a$ respectively are the {\sc rms}-height and
the correlation length of the surface.  For an electromagnetic
scattering problem this translate into
\begin{subequations}
  \label{Small-amp-pert}
  \begin{eqnarray}
    \label{Small-amp-pert-a}
    \sqrt{|\eps{}|} \ooc{} \delta &\ll& 1,
  \end{eqnarray}
  and 
  \begin{eqnarray}
    \label{Small-amp-pert-b}
    \ooc{}a &\leq& 1,
  \end{eqnarray}
\end{subequations}
The first condition~\r{Small-amp-pert-a} comes from the fact that
quantities containing the surface profile function should be
expandable in a Taylor series about their mean surface. The second
criterion stating that the correlation length of the surface can not
be too big originates from the fact that if the correlation length of
the surface is too large the Gaussian height distribution becomes
close to a $\delta$-function with the result that the second order
term in the perturbative expansion can be of the same order as the
first order term even if Eq.~\r{Small-amp-pert-a} is satisfied.

If the surface is non-Gaussian it is found that the above
criteria  can be relaxed somewhat~\cite{Chen1987}

\subsection{Unitary and Reciprocal Expansions}
\label{Theory:Conserving Approximations}

In the previous section we presented small-amplitude perturbation
theory, which is based on the expansion of the scattering amplitude
$R_\nu(q|k)$ in powers of the surface profile function $\zeta(x_1)$.
In Sect.~\ref{Theory:Sect:Reciprocity} we claimed that a valid theory
for rough surface scattering should satisfy reciprocity, {\it i.e.}
the theory should satisfy the relation $S_\nu(q|k)=S_\nu(-k|-q)$,
where $S_\nu(q|k)$ is the scattering matrix as defined in
Eq.~\r{Theory:Eq:S-matrix-def}. By inspecting the formulae obtained in
the previous section for $R_\nu(q|k)$, it is at least not obvious that
reciprocity is satisfied.  Does this mean that small amplitude
perturbation theory does not respect the principle of reciprocity, and
therefore is an incorrect theory? The answer to this question is no,
as you might have guest, and the small amplitude perturbation theory
does in fact respect reciprocity. However, to see this is not straight
forward since an extensive rewriting of the expressions are need for.
Theories where reciprocity is not apparent at first glance is normally
referred to by saying that the theory is not manifestly reciprocal.

Due to the lack of manifest reciprocity in the small amplitude
perturbation theory as well as the desire to map the classical
scattering problem onto the formalism of a quantum mechanical
scattering problem~\cite{Book:Newton}, Brown {\it et
  al}.~\cite{Brown83,Brown84}, in the fist half of the 1980's,
constructed a theory which was manifestly reciprocal.  This theory
goes today under the name of many-body perturbation theory, but is
also known as self-energy perturbation theory.  It is this kind of
perturbation theory that we will concern ourselves with in this
section.

\subsubsection{The Transition Matrix}

The starting point for the many-body perturbation theory is to make
the postulation that the scattering amplitude $R_\nu(q|k)$ should
satisfy the relation~\cite{Brown84}
\begin{eqnarray}
    \label{Theory:postulat-mbpt}
    R_\nu(q|k) &=& 2\pi \delta(q-k) R^{(0)}_{\nu}(k,\omega)
                - 2i G^{(0)}_\nu(q,\omega){\cal T}_\nu(q|k)
                  G^{(0)}_\nu(k,\omega)\alpha_0(k,\omega).
\end{eqnarray}
Here $R^{(0)}_{\nu}(k,\omega)$ is the Fresnel reflection coefficient
and defined by Eq.~\r{Theory:Fresnel}. The second term of this
equation, containing the transition matrix ${\cal T}_\nu(q|k)$, also
known as the T-matrix, represents the field scattered away from the
specular direction.  Furthermore, $G^{(0)}_\nu(k,\omega)$ is the
surface plasmon polariton Green's function for the planar vacuum-metal
interface.  This Green's function can be defined from the relation
\begin{subequations}
\begin{eqnarray}
    \label{Theory:def-greens-func}
    2i\alpha_0(k,\omega)G^{(0)}_\nu(k,\omega)+R^{(0)}_{\nu}(k,\omega) &=& -1,
\end{eqnarray}
That leads to the following expressions 
\begin{eqnarray}
  G^{(0)}_\nu(k,\omega) &=& 
  \left\{
    \begin{array}{ll}
      \frac{\displaystyle i\eps{}}{\displaystyle \eps{}\aln{k}+\al{k}}, & \quad \nu=p, \\
      \frac{\displaystyle i}{\displaystyle \aln{k}+\al{k}}, & \quad \nu=s.
    \end{array}
    \right.
\end{eqnarray}
\end{subequations}

\subsubsection{The Scattering Potential}

The transition matrix is postulated to satisfy the following 
equation~\cite{Brown84}
\begin{subequations}
  \label{Theory:transition-matrix}
  \begin{eqnarray}
    \label{Theory:transition-matrix_A}
    {\cal T}_\nu(p|k) &=& 
        V_\nu(p|k) + 
        \qint{q}\, V_\nu(p|q)G_\nu^{(0)}(q,\omega){\cal T}_\nu(q|k), \\
    &=&
       \label{Theory:transition-matrix_B}
        V_\nu(p|k) + 
        \qint{q}\, {\cal T}_\nu(p|q)G_\nu^{(0)} (q,\omega)V_\nu(q|k),  
  \end{eqnarray}
\end{subequations}
where $V(q|k)$ is known as the {\em scattering potential}. It is
supposed to be a non-resonant function of its arguments, {\it i.e.}
not containing the Green's function $G^{(0)}_\nu$. In arriving at
Eq.~\r{Theory:transition-matrix_B} we have used explicitly
that both $V_\nu(p|k)$ and ${\cal T}_\nu(p|k)$ are reciprocal, {\it
  i.e.} that $V_\nu(p|k)=V_\nu(-k|-p)$ with a similar expression for
the transition amplitude\footnote{That this is indeed the case might
  be confirmed from the expressions to be derived later for these
  quantities.}.

We now seek an (integral) equation satisfied by the T-matrix.  This is
done by substituting Eq.~\r{Theory:postulat-mbpt} into the reduced
Rayleigh equation~\r{Theory:RRE}, and thereby obtain
\begin{eqnarray}
    \label{Theory:T-matrix-eq}
    \int\!\frac{dq}{2\pi} \; N^{+}_\nu(p|q)G^{0}_\nu(q,\omega)
          {\cal T}_\nu(q|k) 
    &=&
    \frac{R^{(0)}_\nu(k,\omega)N^{+}_\nu(q|k)-N^{-}_\nu(p|k)}{
            2i\alpha_0(k,\omega)G^{(0)}_\nu(k,\omega)}. \qquad
\end{eqnarray}
Even though, the many-body perturbation theory could have proceeded
from this equation, it has proven useful from a purely algebraic point
of view, to instead of the T-matrix work in terms of the scattering
potential $V_\nu(q|k)$. Thus we aim to obtain an integral equation for
this quantity which our perturbation theory will be based directly
upon. By substituting the right hand side of
Eq.~(\ref{Theory:transition-matrix_A}) into
Eq.~(\ref{Theory:T-matrix-eq}), making a change of variable and using
Eq.~(\ref{Theory:transition-matrix_A}) once more, one obtains the
desired integral equation for the scattering potential. It reads
\begin{subequations}
  \label{Theory:int-eq-scattering-pot:total}
  \begin{eqnarray}
    \label{Theory:int-eq-scattering-pot}
    \int\!\frac{dq}{2\pi}\; A_\nu(p|q) V_\nu(q|k) &=& B_\nu(p|k),
  \end{eqnarray}
  where the matrix elements are given by
  \begin{eqnarray}
    \label{Theory:a-matrix}
    A_\nu(p|q) &=& 
      \frac{
          \left[ 
             2i\alpha_0(q,\omega)G^{(0)}_\nu(q,\omega)+R^{(0)}_\nu(q,\omega)
          \right]
             N^{+}_\nu(p|q)-N^{-}_\nu(p|q)
       }{
         2i\alpha_0(k,\omega)
       }
        \;=\;
      - \frac{ N^{+}_\nu(p|q)+N^{-}_\nu(p|q) }{
         2i\alpha_0(q,\omega)
       },
   \end{eqnarray}
   and
   \begin{eqnarray}
    \label{Theory:B-matrix}
    B_\nu(p|k) &=& 
      \frac{ 
             R^{(0)}_\nu(p,\omega)N^{+}_\nu(p|k)-N^{-}_\nu(p|k)
       }{
         2i\alpha_0(k,\omega)G^{(0)}_\nu(k,\omega)
       }.
   \end{eqnarray}
\end{subequations}
To obtain $A_\nu(p|q)$ we have explicitly taken advantage of
Eq.~\r{Theory:def-greens-func}. With the expression presented for the
matrix elements for the reduced Rayleigh equation, $N^\pm_\nu(q|k)$,
it is now straightforward to obtain closed form expressions for
$A_\nu(p|q)$ and $B_\nu(p|k)$, but we will not present such
expressions here.

The integral equation~\r{Theory:int-eq-scattering-pot:total} will be
the starting point for our manifestly reciprocal many-body perturbation
theory.  The essence of the theory is to expanding the scattering
potential in powers of the surface roughness $\zeta(x_1)$ according to
\begin{eqnarray}
  \label{Theory:expansion}
  V_\nu(q|k) &=& \sum_{n=0}^{\infty} \frac{(-i)^n}{n!} V^{(n)}_\nu(q|k).
\end{eqnarray}
with similar expansions for $A_\nu(p|q)$ and $B_\nu(p|q)$.  In these
expressions the superscripts denotes, as earlier, the order of the
corresponding terms in the surface profile function. 

One of the advantages of this theory is that even a lower order
approximation to the scattering potential corresponds to a resummation
of an infinite number of terms in an expansion in powers of the surface
profile function\footnote{This is also the case for the so-called
  self-energy perturbation theory to be presented in the next section.}

We will not go into details here, but it can be shown that 
the results for the few first terms in the expansion of the scattering
potential are~\cite{Maradudin93}
\begin{subequations}
  \label{Eq:Scattering-pot}
  \begin{eqnarray}
    V^{(1)}_\nu(q|k) &=&  
    \left\{ 
      \begin{array}{ll}
        i\frac{\eps{}-1}{\varepsilon^2(\omega)}
          \left[ \eps{} qk-\al{q}\al{k}\right]
          \tilde{\zeta}^{(1)}(q-k), \quad & \nu=p \\
        i\frac{\omega^2}{c^2}  (\eps{}-1) \tilde{\zeta}^{(1)}(q-k),
        \quad &\nu=s, 
        \end{array}
      \right. \nn \\
  \end{eqnarray}
  for the $1$st oder term, and 
  \begin{eqnarray}
   V^{(2)}_p(q|k) &=&
         i\frac{\eps{}-1}{\varepsilon^2(\omega)}
         \left[ \al{q}+\al{k}\right] 
          \left[ qk -\al{q}\al{k}\right] 
          \tilde{\zeta}^{(2)}(q-k)   \nn \\ \qquad
     && + 2i \frac{(\eps{}-1)^2}{\varepsilon^3(\omega)}
          \al{q}  
          \qint{p}\tilde{\zeta}^{(1)}(q-p)
                \al{p}\tilde{\zeta}^{(1)}(p-k)\al{k},
  \end{eqnarray}
  and
  \begin{eqnarray}
    V^{(2)}_s(q|k) &=& i \frac{\omega^2}{c^2}
    (\eps{}-1) \left[ \al{q}+\al{k}\right] \tilde{\zeta}^{(2)}(q-k),\qquad
  \end{eqnarray}
\end{subequations}
for the second order terms. Higher order terms can be found in {\it
  e.g.}  Ref.~\citen{PRB57-13209-1998}. It should be mentioned that by
definition the lowest non-vanishing order of the scattering potential
is 1st order in the surface profile function. In other words
$V^{(0)}_\nu(p|q)=0$ always. 

As the reader easily may check, the above expressions are manifest
reciprocal, {\it i.e.}
\begin{eqnarray}
V^{(n)}_\nu(q|k)=V^{(n)}_\nu(-k|-q),  \nn  
\end{eqnarray}
and this property should hold
true to all orders in the surface profile function~\cite{Brown84}.
However, it should be emphasized that expressions of this form are not
obtained directly from the solution of
Eq.~\r{Theory:int-eq-scattering-pot:total}, but that some rewritings
instead are needed for~\cite{Maradudin93}.

If the transmission matrix is expanded in the same way as in
 Eqs.~\r{Theory:expansion}, {\it i.e.} 
 \begin{eqnarray}
    {\cal T}_\nu(q|k) &=& \sum_{n=0}^{\infty} \frac{(-i)^n}{n!} 
    {\cal T}^{(n)}_\nu(q|k),
\end{eqnarray}
a recurrence relation for $\{ {\cal T}^{(n)}_\nu(p|q)\}$
in terms of $V^{(m)}_\nu(p|q)$ can be derived and thus $T_\nu(q|k)$ to
some order  can be calculated. 

Hence, through the calculation of the scattering potential the
T-matrix is now known. The contribution to the mean differential
reflection coefficient from the incoherent component of the scattered
light is
\begin{eqnarray}
  \left< \frac{\partial R_\nu}{\partial \theta_s}\right>_{{\rm incoh}}  
  &=& \frac{1}{L_1} \frac{2}{\pi} \left(\ooc{}\right)^3 
  \cos^2\theta_s\cos\theta_0  
        \left| G_\nu^{(0)}(q,\omega)\right|^2\,
        \left[
          \left<\left| {\cal T}_\nu(q|k) \right|^2 \right>
          - \left|\left< {\cal T}_\nu(q|k) \right> \right|^2
        \right]
        \left| G_\nu^{(0)}(k,\omega)\right|^2.\quad \qquad 
\end{eqnarray}
This expression is obtained by substituting
Eq.~\r{Theory:postulat-mbpt} into the defining expression for the
incoherent component of the mean DRC
(Eq.~\r{Theory:meanDRC-incohr}).  The expression in the square
rackets for the


\subsection{Many-Body Perturbation Theory}

The Green's function $G_0(q,\omega)$, we recall, is the surface
plasmon polariton Green's function at a planar interface.  In addition
to this Green's function it is also useful to define a rough surface
Green's function, $G_\nu(k,\omega)$, or more formally the Green's
function of a $\nu$-polarized electromagnetic field at the randomly
rough interface.  Some authors refer to this function as the {\em
  renormalized} Green's function. It is defined as the solution of the
following equation~\cite{Brown84}
\begin{subequations}
  \begin{eqnarray}
    \label{Theory:Rough-green-a}
    G_\nu(q|k) &=& 2\pi \delta(q-k) G^{(0)}_\nu(k)
    + G^{(0)}_\nu(q) \qint{p}\, V_\nu(q|p)G_\nu(p|k). \qquad
  \end{eqnarray}
  This equation is often in the literature referred to as the
  Lippmann-Schwinger equation for the renormalized Green's function
  $G_\nu(q|k)$.  Notice that from Eqs.~\r{Theory:postulat-mbpt} and
  \r{Theory:Rough-green-a} it follows that
  $R(q|k)=-2\pi\delta(q-k)-2iG(q|k)\alpha_0(k)$.  An alternative way of
  expressing the above equation is obtained by iterating on
  $G_\nu(q|k)$.  The result can be written as
  \begin{eqnarray}
    G_\nu(q|k) &=&    2\pi \delta(q-k) G^{(0)}_\nu(k)
    + G^{(0)}_\nu(q) {\cal T}_\nu(q|p) G^{(0)}_\nu(k), \qquad
  \end{eqnarray}
\end{subequations}
where we have used a Born series~\cite{Book:Newton} expansion for the
of the T-matrix in Eq.~\r{Theory:transition-matrix}. This equation is
often for simplicity expressied in operator form like $G=G_0+G_0 T G_0$.

In terms of the renormalized Green's function the mean DRC takes on
the form
\begin{eqnarray}
  \left< \frac{\partial R_\nu}{\partial \theta_s}\right>_{\mbox{incohr}}
  &=&
  \frac{1}{L_1} \frac{2}{\pi} 
  \ooc{3} \cos^2\theta_s \cos\theta_0
  \left[  \left<\left| G_\nu(q|k)\right|^2\right>
    -\left|\left< G_\nu(q|k)\right>\right|^2   
  \right],\qquad
\end{eqnarray}
where the method of smoothing~\cite{Brown85} has been applied to
Eq.~\r{Theory:postulat-mbpt} as well as the reduced Rayleigh equation. In
these expressions the mean of the renormalized Greens function satisfies the
{\em Dyson equation}
\begin{subequations}
  \begin{eqnarray}
    \label{Eq:Dyson-equation}
    \left<G_\nu(q|k)\right> 
     &=& 2\pi \delta(q-k) G^{(0)}_\nu(k)
    + G^{(0)}_\nu(q) \qint{p}\, \left<M_\nu(q|p)\right>
          \left<G_\nu(p|k)\right>, \qquad
  \end{eqnarray}
  where the unaveraged {\em proper self-energy} $M_\nu(q|k)$ is a
  solution of the equation
  \begin{eqnarray}
    M_\nu(q|k) &=& V_\nu(q|k)  
    + \qint{p}\, M_\nu(q|p)G^{(0)}_\nu(p)  
       \left[V_\nu(p|k)-\left<V_\nu(p|k)\right> \right]. \qquad
  \end{eqnarray}
\end{subequations}
Since the surface profile function is stationary, both the
renormalized Green's function and the proper self-energy are diagonal
in the momentum variables $q$ and $k$, {\it i.e.}  
\begin{eqnarray}
\left<G_\nu(q|k)\right> &=& 2\pi\delta(q-k)G_\nu(k),   
\end{eqnarray}
with a similar expression for the proper self-energy. Under this
assumption the renormalized Green's function can formally be written
as
\begin{eqnarray}
  G_\nu(k) &=& \frac{1}{(G_\nu^{(0)}(k))^{-1}-M_\nu(k)}.
\end{eqnarray}
Hence the surface polariton poles in
$G_\nu(k)$ are shifted as compared to those of $G^{(0)}_\nu$ due to the
presence $M_\nu(k)$. The self-energy can be calculated perturbatively
as an expansion in powers of the surface profile function. The
resulting perturbation theory is known as self-energy perturbation
theory~\cite{Sanchez-Gil1995}.

The two-particle average Green's satisfies a {\em Bethe-Salpeter}
equation~\cite{Book:Newton} of the form\footnote{In arriving at this
  equation also here the method of smoothing~\cite{Brown85} has been
  applied.}
\begin{eqnarray}
  \label{Eq:two-partical}
 \left< \left| G_\nu(q|k) \right|^2\right>  
      &=& L_1 2\pi \delta(q-k)
      \left|G_\nu(q)\right|^2
      +\left|G_\nu(q)\right|^2  \qint{p}\, U_\nu(q|p)
       \left< \left|G_\nu(p|k) \right|^2\right>,  
\end{eqnarray}
where $U_\nu(q|p)$ is the so-called irreducible vertex function.
Formally, one may write the solution to Eq.~\r{Eq:two-partical} as
\begin{eqnarray}
  \left< \left| G_\nu(q|k) \right|^2\right>  &=&
   L_1 2\pi \delta(q-k) \left|G_\nu(q)\right|^2
 + L_1 \left|G_\nu(q)\right|^2 X_\nu(q|k)
        \left|G_\nu(k)\right|^2 \nn \\ 
\end{eqnarray}
where $X_\nu(q|k)$ is the reducible vertex function.
With this equation the mean DRC takes on the form
\begin{eqnarray}
  \label{Eq:vortex-DRC}
 \left< \frac{\partial R_\nu}{\partial \theta_s}\right>_{\mbox{incohr}}
      &=&
      \frac{2}{\pi} 
      \ooc{3} \cos^2\theta_s \cos\theta_0
      \left| G_\nu(q)\right|^2  X_\nu(q|k) 
      \left|G_\nu(k)\right|^2.   \qquad
\end{eqnarray}
The reducible vertex function can be shown to be related to
$U_\nu(Q|k)$ though the equation
\begin{eqnarray}
  X_\nu(q|k) &=& U_\nu(q|k) 
    +  \qint{p}\, U_\nu(q|p)\left| G_\nu(p)\right|^2 X_\nu (p|k). 
\end{eqnarray}
Unfortunately we do not know, in general, how to solve the Bethe-salpeter
equation~\r{Eq:two-partical}.  Hence some approximative methods have to be
employed. The most frequently used methods are the Freilikher
factorization~\cite{Freilikher-factorization}, or a diagrammatic
method~\cite{McGurn1985,Book:Newton,Book:Sheng1995}.  In this latter
approach $X_\nu(q|k)$ is approximated by a subset of (an infinite number of)
diagrams. Those usually are the ladder diagrams and the maximally-crossed
diagrams, where the former describes wave diffusion in the random media,
while the latter is related to wave localization.

With this approach it can be shown that the reducible vertex function
for $p$-polarized light incident on a rough metal surface can be
written as~\cite{McGurn1985}
\begin{eqnarray}
  \label{Eq:vertex-function}
  X_p(q|k) &=&
  \left[ 
    U^{(0)}_p(q|k)
    + \frac{A(q|k)}{4\Delta^2}
    + \frac{A\left( \frac{q-k}{2}|\frac{k-q}{2}\right)
      }{(q+k)^2+4\Delta^2} 
  \right],   
\end{eqnarray}
which when substituted into Eq.~\r{Eq:vortex-DRC} defines the mean DRC
in this approximation. Here, $A(q|k)$ is a smooth function of its
arguments and $\Delta=\Delta_\varepsilon+\Delta_{sp}$ is the (total)
decay rate of surface plasmon polaritons due to Ohmic
losses~($\Delta_\varepsilon$) in the metal and conversion into other
surface plasmon polaritons~($\Delta_{sp}\simeq Im(M(k_{sp}))$). Their
mathematical expressions, as well as the other quantities appearing in
Eq.~\r{Eq:vertex-function}, can be found in Ref.~\citen{McGurn1985}.
The first term in Eq.~\r{Eq:vertex-function} is due to
single-scattering, the second arises from the ladder diagrams, while
the last one is the contribution from the maximally-crossed diagrams.
This last term is the one that is responsible for the enhanced
backscattering phenomenon that we will discuss in the next section.

It should be mentioned that in arriving at Eq.~\r{Eq:vertex-function}
the {\em pole-approximation} for the renormalized Green's function has
been utilized. This approximation amount to writing 
\begin{eqnarray}
  \label{Eq:Pole-appr}
  G_p(k) \simeq 
      \frac{C(\omega)}{k-k_{sp}-i\Delta}
     - \frac{C(\omega)}{k+k_{sp}+i\Delta},            
\end{eqnarray}
where $k_{sp}$ is the wave vector of the surface plasmon
polariton~(See. Eq.~\r{Elmag:dispersion-relation}), and $C(\omega)$ is a constant. 
The Green's function for $s$-polarization does not have poles, and the
pole-approximation is thus not relevant in this case.


\subsection{Numerical Simulation Approach}
\label{Sect:Theory:Sect:NumSim}

In the previous sections various perturbation theories were discussed.
Such theories catch the main physics of the scattering problem if the
surface is not too rough. However, for interfaces that are strongly
rough, none of the perturbative approaches can be used because too
many terms in the expansion have to be included in order for the
approach to be practical.

At the present time, there does not exist any analytic
non-perturbative theory that is valid for an arbitrary roughness.  The
reason for the lack of such general analytic theory is that for
strongly rough surfaces higher order scattering processes become
important. In consequence the boundary conditions to be satisfied on
the random interface become dominated by non-local effects. This
means that the total field on the surface at some point depends on the
total field in other locations on the surface. These non-local
boundary conditions hamper the development of analytic theories for
strongly rough surfaces.

The best one can do at the present for these strongly rough surfaces
is to resort to a numerical simulation approach. This approach, as we
will see below, is based on deriving a set of coupled integral
equation for the source functions, the field and its normal derivative
evaluated on the surface. With the knowledge of these sources, the
total field, and therefore the solution to the scattering problem, can
be obtained from the extinction theorem in any point above the
surface.  We will now outline how all this comes about.

\subsubsection{The Extinction Theorem}
\label{Theory:Sect:The-Extinction-Theorem}

We will now derive the so-called {\em Ewald-Oseen extiction theorem}
first formulated by P.\ P.\ Ewald and C.\ W.\ Oseen in the beginning
of this century. The numerical simulation approach to be presented
later in this section will be based directly upon this theorem.

Let us start by recall from Sect.~\ref{Sect:Theory:Scattered-fields}
that the primary field, $\Phi_\nu({\mathbf r}|\omega)$, satisfies the
wave equation
\begin{eqnarray}
  \label{Eq:Theory:recall-wave-eq}
  \left( \partial^2_{{\mathbf r}} +\eps{}\ooc{} \right)
  \Phi_\nu({\mathbf r}|\omega) &=& -J^{ext}_\nu({\mathbf r},\omega)
\end{eqnarray}
where $\eps{}$ is the dielectric function of the medium where the
field is evaluated, $\partial_{\mathbf r}=\nabla$ is the
nabla-operator in the number of spatial dimensions considered, while
$J^{ext}_\nu({\mathbf r})$ is an external source term for the field.
In order to solve the scattering problem in question, one may solve
this equation in the regions of constant dielectric function and match
these solutions by the boundary conditions that the field, and its
normal derivative, should satisfy on any interface~(See
Sect.~\ref{Sec:BoundaryConditions}).  However, it is often more
convenient to take advantage of certain integral theorems that is a
consequence of the wave equation~\r{Eq:Theory:recall-wave-eq}.  The
extinction theorem provides such an example.

The wave equation~\r{Eq:Theory:recall-wave-eq} is accompanied by
the following equation for its Green's function~\cite{Book:Morse}
\begin{eqnarray}
  \label{Eq:Theory:Greens-eq}
  \left( \partial^2_{\mathbf r} +\eps{}\ooc{} \right)
  G({\mathbf r} | {\mathbf r}' ; \omega) &=& 
    -4\pi \delta({\mathbf r}-{\mathbf r}').
\end{eqnarray}
Furthermore, we are only interested in out-going solutions to this
equation that fulfill the Sommerfeld's radiation condition (at
infinity)~\cite{Sommerfeld}
\begin{eqnarray}
  \lim_{r\rightarrow\infty} r
     \left( \partial_{r}G-ikG\right) = 0,
\end{eqnarray}
where $r=|{\mathbf r}|$.   

In two-dimensions, that we consider in the present introduction, an
explicit representation of the out-going, free space Green's function
is provided by~\cite{Book:Morse,Jackson}
\begin{eqnarray}
  \label{Eq:Theory:Greens-func}
  G({\mathbf r} | {\mathbf r}' ; \omega) &=& 
   i \pi H_0^{(1)} \;
      \left( 
        \eps{}\ooc{}\left|{\mathbf r}-{\mathbf r}'\right|
      \right),
\end{eqnarray}
where, $H_0^{(1)}(z)$, is the Hankel-function of the first kind and
zeroth-order~\cite{Book:Morse,Stegun} and ${\mathbf
  r}=(x_1,x_3)$.

\begin{figure}[t!]
  \begin{center}
    \leavevmode
    \includegraphics[width=6cm,height=3cm]{\myfigpath/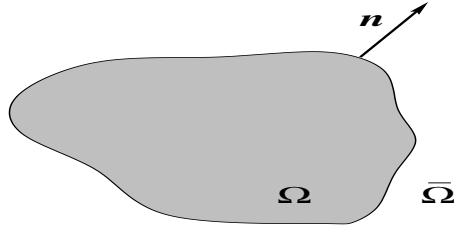}
     \caption{The geometry considered in the extinction theorem.}
     \label{Fig:Extiction-theorem}
  \end{center}
\end{figure}

Let us start by considering a spatial region $\Omega$ containing a
homogeneous, isotropic dielectric medium. This region has a boundary
$\partial \Omega$~(See Fig.~\ref{Fig:Extiction-theorem}).  The
exterior of the region $\Omega$ will be denoted by $\bar{\Omega}$
where its boundary is $\partial\bar{\Omega}$. Notice that
$\partial\bar{\Omega}$ includes $\partial{\Omega}$ in addition to the
surface at infinity.  We assume that an external source is present
somewhere in the external region $\bar{\Omega}$ and that no sources
are present within $\Omega$.


If we multiply Eqs.~\r{Eq:Theory:recall-wave-eq} and
\r{Eq:Theory:Greens-eq} by respectively $G({\mathbf r} | {\mathbf
  r}' ; \omega)$ and $-\Phi_\nu({\mathbf r}|\omega)$, add the
resulting equations, and finally integrate the result over the
exterior region $\bar{\Omega}$ we are left with\footnote{We have here
  interchanged ${\mathbf r}$ and ${\mathbf r}'$ for later
  convenience.}  (${\mathbf r}'\in \bar{\Omega}$)
\begin{eqnarray}
  \label{Eq:Theory:Extinction-on-the-way}
 - \frac{1}{4\pi}\,
  \int_{\bar{\Omega}} \!d{\mathbf r}' 
   \left[   
     \Phi_\nu({\mathbf r}'|\omega)
     \partial_{{\mathbf r}'}^2 G({\mathbf r}' | {\mathbf r} ; \omega)  
     -
     \partial_{{\mathbf r}'}^2  \Phi_\nu({\mathbf r}'|\omega)
        G({\mathbf r}' | {\mathbf r} ; \omega) 
   \right] \qquad \qquad   \nonumber \\
   =
    - \frac{1}{4\pi}  
      \int_{\bar{\Omega}} \!d{\mathbf r}'  
          J^{ext}_\nu({\mathbf r}',\omega)  
           G({\mathbf r}' | {\mathbf r} ; \omega)  
   +
   \left\{
     \begin{array}{ll}
       \Phi_\nu({\mathbf r}|\omega), & {\mathbf r}\in\bar{\Omega}\\
       0,                            & {\mathbf r}\not\in\bar{\Omega}
     \end{array}
     \right. .
\end{eqnarray}
Since $G({\mathbf r}' | {\mathbf r} ; \omega)$ is the out-going free
space Green's function the first term of the right hand side is just
the incident field due to the source, {\it i.e.}
\begin{eqnarray}
  \label{Eq:Theory:incident-field-identity}
   \frac{1}{4\pi}  
      \int_{\bar{\Omega}} \!d{\mathbf r}'  
          J^{ext}_\nu({\mathbf r}',\omega)  G({\mathbf r}' | {\mathbf r} ;
      \omega)  
     &=& \Phi^{inc}_\nu({\mathbf r}|\omega).
\end{eqnarray}
This relation holds true independent of ${\mathbf r}$ is located in the
exterior~($\bar\Omega$) or interior~($\Omega$) region.

Furthermore, by taking advantage of Green's second integral identity
that for two well-behaved\footnote{By well-behaved we here mean
  functions that at least are differential two times.}  functions
$u({\mathbf r})$ and $v({\mathbf r})$ defined on a region $V$,
reads~\cite{Book:Morse,Jackson}
\begin{eqnarray}
  \int_V d{\mathbf r} 
  \left[ u({\mathbf r})\partial_{{\mathbf r}}^2 v({\mathbf r}) 
          - v({\mathbf r}) \partial_{{\mathbf r}}^2 u({\mathbf r})
  \right]
    &=&   \int_{\partial V} dS
    \left[
      u({\mathbf r})\partial_{n} v({\mathbf r}) 
         - v({\mathbf r}) \partial_{n} u({\mathbf r})
    \right],  \nonumber \\
\end{eqnarray}
where $\partial_{n}$ denotes the {\em outward} normal derivative to
$\partial V$, Eq.~\r{Eq:Theory:Extinction-on-the-way} can be written
as
\begin{eqnarray}
  \label{Eq:Theory:Extinction-theorem}
  \Phi^{inc}_\nu({\mathbf r}|\omega) 
  +
  \frac{1}{4\pi}\,
  \int_{\partial{\Omega}} \!dS' 
   \left[   
     \Phi_\nu({\mathbf r}'|\omega)
     \partial_{n'} G({\mathbf r} | {\mathbf r}' ; \omega)  
     -
     \partial_{n'}  \Phi_\nu({\mathbf r}'|\omega)
        G({\mathbf r} | {\mathbf r}' ; \omega) 
   \right]   
     &=&
   \left\{
     \begin{array}{ll}
       \Phi_\nu({\mathbf r}|\omega), & {\mathbf r} \in\bar\Omega\\
       0,                            & {\mathbf r} \not\in\bar\Omega
     \end{array}
     \right.   \quad 
\end{eqnarray}
where $dS'$ is a surface element.  In writing this equation we have
explicitly used the fact that the portion of the surface
integral over $\partial\bar{\Omega}$, that corresponds to the surface
at infinity vanishes due to Sommerfeld's radiation condition
satisfied by $G({\mathbf r} | {\mathbf r}' ; \omega)$. Hence the only
surface left in the surface integral is $\partial\Omega$ as indicated
in the above equation. In addition we have also utilized the relation
$\partial_n=-\partial_{\bar{n}}$ for the {\em outward} normal
derivative to the region $\Omega$, while $\partial_{\bar{n}}$ is the
outward normal derivative for the same surface, but for region
$\bar\Omega$.  Notice that the incident field term is present due to
the face that the volume $\bar\Omega$ contains a source. If this
region is source-less this term is missing.

Eq.~\r{Eq:Theory:Extinction-theorem} with the right-hand-side set to
zero is the extinction theorem. It is so named because the incident
field is extinguished in region $\Omega$ by the induced field as
represented by the second term of the left-hand-side of this equation.
Furthermore, Eq.~\r{Eq:Theory:Extinction-theorem} with ${\mathbf r}
\in \Omega$ expresses the fact that the field at any point outside
$\Omega$ can be found by performing a surface integral over $\partial
\Omega$. In order to do so, however, the total field and its normal
derivative on the surface $\partial\Omega$ has to be known. Hence the
scattering problem is equivalent to finding the field and the normal
derivative on the surface.

\subsubsubsection{The Scattered and Transmitted fields}

From the above discussion, we learned that the essential quantities to
look for is the field and its normal derivative evaluated on the
surface.  We will now see how these two quantities can be calculated
by taking advantage of the extinction theorem. This is done by
applying Eq.~\r{Eq:Theory:Extinction-theorem} in turn to the different
regions naturally defined by the scattering geometry as the regions of
constant dielectric properties. For the scattering system depict in
Fig.~\ref{Fig:Theory:Geometry} this means to apply
Eq.~\r{Eq:Theory:Extinction-theorem} separately to the regions
$x_3>\zeta_1(x_1)$ and $x_3<\zeta_2(x_1)$.  The result is\footnote{In
  these equations, and some to come, we have suppressed an explicit
  reference to the frequency of the incident light in order to make
  the formulae more compact.}
\begin{subequations}
  \label{Eq:Theory:Int_eq}
\begin{eqnarray}
    \label{Eq:Theory:Int_eq_1}
    \theta(x_3-\zeta(x_1)) \Phi^+_\nu({\mathbf r}) 
      &=&     \Phi^{inc}_\nu({\mathbf r})  
  + \frac{1}{4\pi}
    \int \! dx_1' \; \gamma(x_1')\!
      \left. \left[ 
          \Phi^+_\nu({\mathbf r}') 
            \partial_{n'} G_+({\mathbf r}|{\mathbf r}')
       -   \partial_{n'} \Phi^+_\nu({\mathbf r}') 
            G_+({\mathbf r}|{\mathbf r}')
      \right] \right|_{x_3'=\zeta(x_1')},  
    \\
    \label{Eq:Theory:Int_eq_2}
    \theta(\zeta(x_1)-x_3) \Phi^-_\nu({\mathbf r})  
      &=&   
     - \frac{1}{4\pi}
    \int \! dx_1' \; \gamma(x_1')\!
      \left. \left[ 
          \Phi^-_\nu({\mathbf r}') 
            \partial_{n'} G_-({\mathbf r}|{\mathbf r}')
       -   \partial_{n'} \Phi^-_\nu({\mathbf r}') 
            G_-({\mathbf r}|{\mathbf r}')
      \right] \right|_{x_3'=\zeta(x_1')}.
\end{eqnarray}
\end{subequations}
where the superscripts $\pm$ indicate solutions to the wave
equation~\r{Eq:Theory:recall-wave-eq} in regions of dielectric
function $\varepsilon_\pm(\omega)$.  Furthermore, we have defined
\begin{subequations}
\begin{eqnarray}
  \partial_n  &=& \frac{\partial_{x_3}-\zeta'(x_1)\partial_{x_1}
                        }{ \gamma(x_1)},
\end{eqnarray}
  where
  \begin{eqnarray}
    \label{Eq:Theory:gamma-def}
    \gamma(x_1) &=& \sqrt{1+\zeta'(x_1)^2}.
  \end{eqnarray}
\end{subequations}
In writing Eqs.~\r{Eq:Theory:Int_eq} we have taken advantage of the
assumption made earlier that the surface, $\zeta(x_1)$, is a
single-valued function of $x_1$ so that its surface element becomes
\begin{eqnarray}
  dS &=& \gamma(x_1) dx_1.
\end{eqnarray}
If this assumption does not hold true, the discussion becomes
considerably more difficult. A treatment of such a case can be found
in {\it e.g.} Ref.~\citen{Mendaza-Suarez}. However, we will not here
considered this possibility any further.

Notice that the integral equations~\r{Eq:Theory:Int_eq} are uncoupled.
However, by taking into account the boundary conditions to be
satisfied on the rough surface $x_3=\zeta(x_1)$, {\it i.e.}
\begin{subequations}
\begin{eqnarray}
  \left. \Phi^+_\nu(x_1,x_3;\omega)\right|_{x_3=\zeta(x_1)}
     &=& \left. \Phi^-_\nu(x_1,x_3;\omega)\right|_{x_3=\zeta(x_1)},  \\
 \left. \frac{\partial_n \Phi^+_\nu(x_1,x_3;\omega)}{\kappa^+_\nu(\omega)}
       \right|_{x_3=\zeta(x_1)}
   &=& 
    \left. \frac{\partial_n\Phi^-_\nu(x_1,x_3;\omega)}{\kappa^-_\nu(\omega)} 
    \right|_{x_3=\zeta(x_1)},
      \qquad \quad 
\end{eqnarray}
\end{subequations}
the two integral equations will be coupled, and Eqs.~\r{Eq:Theory:Int_eq}
take on the form
\begin{subequations}
  \label{Eq:Theory:int_eq}
\begin{eqnarray}
    \label{Eq:Theory:int_eq_1}
    \theta(x_3-\zeta(x_1)) \Phi^+_\nu({\mathbf r}|\omega) 
      &=&     \Phi^{inc}_\nu({\mathbf r}|\omega)  
    \int \! dx_1' 
       \left[ 
           A_+({\mathbf r}|x_1';\omega)
           {\cal F}_\nu(x_1'|\omega)
         -
              B_+({\mathbf r}|x_1';\omega) {\cal N}_\nu(x_1'|\omega) 
      \right],
 \nonumber \\
    \label{Eq:Theory:int_eq_2}
    \theta(\zeta(x_1)-x_3) \Phi^-_\nu({\mathbf r}|\omega)  
      &=&    
     - \int \! dx_1' 
      \left[ 
           A_-({\mathbf r}|x_1';\omega) {\cal F}_\nu(x_1'|\omega)
         -  
           \frac{\kappa^-_\nu(\omega)}{\kappa^+_\nu(\omega)} 
            B_-({\mathbf r}|x_1';\omega) {\cal N}_\nu(x_1'|\omega)
     \right] ,  \nonumber
\end{eqnarray}
\end{subequations}
where the symbols ${\kappa^\pm_\nu(\omega)}$ have been defined earlier
in Eq.~\r{Elmag:BCSc}.  Here we have introduced the source
functions\footnote{Notice that the operator $\gamma(x_1)\partial_n$
  appearing in ${\cal N}_\nu(x_1|\omega)$ is nothing else then the
  {\em unnormalized} normal derivative.}
\begin{subequations}
  \label{Eq:Theory:Source-func}
  \begin{eqnarray}
    \label{Eq:Theory:Source-func-field}
    {\cal F}_\nu(x_1|\omega) &=&
          \left.\Phi^+_\nu(x_1,x_3|\omega)\right|_{x_3=\zeta(x_1)}, \\
    {\cal N}_\nu(x_1|\omega) &=&
          \left. \gamma(x_1) \partial_n
          \Phi^+_\nu(x_1,x_3|\omega)\right|_{x_3=\zeta(x_1)},
  \end{eqnarray}
\end{subequations}
as well as the kernels
\begin{subequations}
\label{Eq:Theory:Kernels}
  \begin{eqnarray}
    A_\pm({\mathbf r}|x_1';\omega) &=& 
            \left. 
              \frac{1}{4\pi}\; \gamma(x_1')\;
                \partial_{n'} G_\pm(x_1,x_3|x_1',x_3')
            \right|_{x_3'=\zeta(x_1')}, \quad \\
    B_\pm({\mathbf r}|x_1';\omega) &=&
    \left.
      \frac{1}{4\pi}\;G_\pm(x_1,x_3|x_1',x_3')
            \right|_{x_3'=\zeta(x_1')}.
  \end{eqnarray}
\end{subequations}

Notice that the second term on the right-hand-side of
Eq.~\r{Eq:Theory:int_eq_1} represents the field scattered from the
rough surface, $\Phi_\nu^{sc}({\mathbf r}|\omega)$. By substituting
the following Fourier representation for the Green's
function~\cite{Stegun}
\begin{eqnarray}
  \label{Eq:Theory:Green-fuc-exp}
  G_+({\mathbf r}|{\mathbf r}'; \omega) &=&
     \qint{q} \frac{2\pi i}{\alpha_+(q,\omega)}
     \; e^{ iq(x_1-x_1')+i\alpha_+(q,\omega)|x_3-x_3'|},
\end{eqnarray}
into Eqs.~\r{Eq:Theory:Kernels}, and the resulting expression into
Eq.~\r{Eq:Theory:int_eq_1}, we find that the scattered field far above
the surface, $x_3\gg\zeta(x_1)$, can be written as
\begin{subequations}
\begin{eqnarray}
  \Phi_\nu^{sc}({\mathbf r}|\omega) &=& \qint{q} \;
       R_\nu(q,\omega) e^{iqx_1+i\alpha_+(q,\omega)x_3},
\end{eqnarray}
where the scattering amplitude is given by the following expression
\begin{eqnarray}
  \label{Eq:Theory:R-num}
  R_\nu(q,\omega) &=& \frac{i}{2\alpha_+(q,\omega) } 
        \int^{\infty}_{-\infty} dx_1\; 
         e^{-iqx_1-i\alpha_+(q,\omega)\zeta(x_1)}  \, 
         \left[ i\left\{ q\zeta'(x_1)- \alpha_+(q,\omega) \right\} 
             {\cal F}_\nu(x_1)
         - {\cal N}_\nu(x_1) \right] . \quad
       \nonumber
\end{eqnarray}
\end{subequations} 
In these expressions $\alpha_+(q,\omega)$, and later to be used
$\alpha_-(q,\omega)$, are defined as in Eqs.~\r{Eq:Theory:alpha0} and 
\r{Eq:Theory:alpha}.

If the medium occupying the region $x_3<\zeta(x_1)$ is transparent, a
transmitted field will also exist. It is given by the right-hand-side
of Eq.~\r{Eq:Theory:int_eq_2}. Under this assumption, a 
Fourier representation for $G_-({\mathbf r}|{\mathbf r}';
\omega)$, equivalent the one given in Eq.~\r{Eq:Theory:Green-fuc-exp},
will give a transmitted field in the region $x_3\ll\zeta(x_1)$ of the
form
\begin{subequations}
\begin{eqnarray}
  \Phi_\nu^{tr}({\mathbf r}|\omega) &=& \qint{q} \;
       T_\nu(q,\omega) e^{iqx_1-i\alpha_-(q,\omega)x_3},
\end{eqnarray}
where the transmission amplitude is defined as
\begin{eqnarray}
  \label{Eq:Theory:T-num}
  T_\nu(q,\omega) &=&
      -\frac{i}{2\alpha_-(q,\omega) } 
        \int^{\infty}_{-\infty} dx_1\; 
         e^{-iqx_1+i\alpha_-(q,\omega)\zeta(x_1)} \,
         \left[ i\left\{ q\zeta'(x_1)+ \alpha_-(q,\omega) \right\} 
             {\cal F}_\nu(x_1)
         - \frac{\kappa^-_\nu(\omega)}{\kappa^+_\nu(\omega)}  
        {\cal N}_\nu(x_1)  \right] . \quad
       \nonumber
\end{eqnarray}
\end{subequations}

\subsubsubsection{The Equations for the Source Functions}

In order to solve the scattering problem we see from
Eq.~\r{Eq:Theory:R-num} that we need to know the source functions
${\cal F}_\nu(x_1|\omega)$ and ${\cal N}_\nu(x_1|\omega)$.  The
question therefore is: How to calculate these source functions?  A
coupled set of equations for these sources are most easily obtained by
setting $x_3=\zeta(x_1)+\eta$, with $\eta\rightarrow 0^+$, in
Eqs.~\r{Eq:Theory:int_eq}. Doing so results in the following set of
inhomogeneous, coupled integral equations for the sources
\begin{subequations}
  \label{Eq:Theory:source-eq}
\begin{eqnarray}
  \label{Eq:Theory:source-eq_1}
     {\cal F}_\nu(x_1)
       &=&     {\cal F}_\nu^{inc}(x_1)    \int \! dx_1' 
       \left[ 
           {\cal A}_+(x_1|x_1')
           {\cal F}_\nu(x_1')
         -
           {\cal B}_+(x_1|x_1') {\cal N}_\nu(x_1') 
      \right], \qquad \qquad \\ 
    \label{Eq:Theory:source-eq_2} 
    0  &=&   \int \! dx_1' 
      \left[ 
           {\cal A}_-(x_1|x_1') {\cal F}_\nu(x_1')
         -  
           \frac{\kappa^-_\nu}{\kappa^+_\nu}
            {\cal B}_-(x_1|x_1') {\cal N}_\nu(x_1')
     \right] ,  
   \end{eqnarray}
\end{subequations}
where the kernels are defined as
\begin{subequations}
  \label{Eq:theory:kernel} 
  \begin{eqnarray}
    \label{Eq:theory:A-kernel} 
    {\cal A}_\pm(x_1|x_1') &=& 
         \lim_{\eta\rightarrow 0^+} 
          \left. A_\pm({\mathbf r}|x_1')\right|_{x_3=\zeta(x_3)+\eta},  
          \\
    \label{Eq:theory:B-kernel} 
    {\cal B}_\pm(x_1|x_1') &=& 
         \lim_{\eta\rightarrow 0^+} 
          \left. B_\pm({\mathbf r}|x_1')\right|_{x_3=\zeta(x_3)+\eta}.  
\end{eqnarray}
\end{subequations}
In order to solve Eqs.~\r{Eq:Theory:source-eq}, the integral equations
are converted into matrix equations by discretizing the spatial
variables $x_1$ and $x_1'$ and using some kind of quadrature scheme
for approximating the integrals that they contain. First of all, the
infinitely long surface is restricted to a finite length $L_1$, so
that the spatial integration range from $-L_1/2$ to $L_1/2$.  Second,
a grid defined according to
\begin{eqnarray}
  \xi_n &=& [x_1]_n \;\;=\;\; -\frac{L_1}{2}+\left(n-\frac{1}{2}
  \right) \Delta\xi,  \qquad n=1,2,3,\ldots, N,\quad
\end{eqnarray}
with $\Delta\xi=L_1/N$ is introduced for $x_1'$. If we assume that the
source functions are slowly varying functions over a grid cell (of
size $\Delta\xi$), they can be considered as constant over this
distance and therefore put outside the integral. The integral
equations~\r{Eq:Theory:source-eq} are thus converted into the
following coupled matrix equations by putting $x_1=\xi_m$
\begin{subequations}
  \label{Eq:Theory:matrix-eq}
  \begin{eqnarray}
    \label{Eq:Theory:matrix-eq_1}
     {\cal F}_\nu(\xi_m)
       &=&     {\cal F}_\nu^{inc}(\xi_m) +   \sum_{n=1}^N 
       \left[ 
           {\cal A}^+_{mn}
           {\cal F}_\nu(\xi_n')
         -
              {\cal B}^+_{mn} {\cal N}_\nu(\xi_n') 
      \right], \qquad  \\ 
    \label{Eq:Theory:matrix-eq_2} 
    0  &=&   \sum_{n=1}^N 
      \left[ 
           {\cal A}^-_{mn} {\cal F}_\nu(\xi_n')
         -  
           \frac{\kappa^-_\nu}{\kappa^-_\nu }
            {\cal B}^-_{mn} {\cal N}_\nu(\xi_n')
     \right] ,
   \end{eqnarray}
\end{subequations}
where ${\cal F}_\nu^{inc}(\xi_m)$ is defined from
Eq.~\r{Eq:Theory:Source-func-field} by using
$\Phi^{inc}_\nu(x_1,x_3|\omega)$ for the field
$\Phi^+_\nu(x_1,x_3|\omega)$.  Moreover, the matrix elements $ {\cal
  A}^\pm_{mn}$ and $ {\cal B}^\pm_{mn}$ are defined as
\begin{subequations}
  \label{Eq:Theory:Matrix-elements}
  \begin{eqnarray}
    {\cal A}^\pm_{mn} &=&
        \int^{\xi_n+\Delta\xi/2}_{\xi_n-\Delta\xi/2} dx_1'\;
                    {\cal A}_\pm(\xi_m|x_1'), \\
    {\cal B}^\pm_{mn} &=&
        \int^{\xi_n+\Delta\xi/2}_{\xi_n-\Delta\xi/2} dx_1'\;
                    {\cal B}_\pm(\xi_m|x_1'). 
  \end{eqnarray}
\end{subequations}

It should be kept in mind that these matrix elements are related to
the Hankel function, $H^{(1)}_0(z)$, and its derivative, through the
(two-dimensional) Green's function that enters via
Eqs.~\r{Eq:Theory:Kernels} and \r{Eq:theory:kernel}.  Care has to be
taken when evaluating these matrix elements since the Hankel functions
are singular when their arguments vanish.  Hence the kernels, $ {\cal
  A}_\pm(x_1|x_1')$ and $ {\cal B}_\pm(x_1|x_1')$, are also singular
when $x_1=x_1'$. Fortunately these singularities are integrable so
that the matrix elements, ${\cal A}^\pm_{mn}$ and ${\cal B}^\pm_{mn}$,
in contrast to the kernels, are well define everywhere. The somewhat
technical procedure for showing this is presented in
Appendix~\ref{App:Matrix-element-expansion}, from where we obtain that
(see Eqs.~\r{App:A-matrix-elemnets} and \r{App:B-matrix-elemnets})
\begin{subequations}
  \label{Eq:Theory:Matrix-elements-final}
\begin{eqnarray}
    {\cal A}^\pm_{mn} 
       &=&
       \left\{
     \begin{array}{ll}
      \Delta \xi \, A_\pm(\xi_m|\xi_n),   & \quad m\neq n, \\
            \frac{1}{2} + \Delta \xi\;
            \frac{\zeta''(x_m)}{4\pi \gamma^2(\xi_m)},
              & \quad  m=n, \\
   \end{array}
       \right. \qquad
    \label{A-matrix-main} 
\end{eqnarray}
and 
\begin{eqnarray}
    {\cal B}^{\pm}_{mn} 
       &=&
       \left\{
     \begin{array}{ll}
      \Delta \xi \, B_\pm(\xi_m|\xi_n),   & \quad m \neq n \\
      -\frac{i}{4} \Delta\xi \, H^{(1)}_0\left(
          \sqrt{\varepsilon_\pm} \ooc{} \frac{\gamma(\xi_m)\Delta\xi}{2e} \right),
     & \quad  m=n. \\
   \end{array}
       \right. \qquad
    \label{B-matrix-main} 
\end{eqnarray}
\end{subequations}

The matrix equations~\r{Eq:Theory:matrix-eq}, together with the
expressions for the matrix elements
Eqs.~\r{Eq:Theory:Matrix-elements-final}, can readily be put onto the
computer and solved by standard techniques from linear
algebra~\cite{NR,Matrix} in order to obtain the source
functions. With these source functions available, the scattering
amplitude, and, if defined, the transmission amplitude, can be
obtained from respectively Eqs.~\r{Eq:Theory:R-num} and
\r{Eq:Theory:T-num}. These amplitudes are again related to physical
observable quantities, like the mean differential reflection or
transmission coefficients, as discussed earlier.  Hence the scattering
problem is in principle solved!

It should also be mentioned that the approach presented here can be
generalized to more complicated scattering geometries like film
systems etc.~\cite{Lu1991,PhysRep,MP15}. However, in such cases the
higher demand is put on computational resources.

\subsubsection{A Remark on the Accuracy of the Numerical Simulation Approach}

The numerical approach described above is formally exact since {\em
  no} approximations have been introduced. It is therefore in
principle applicable to scattering from surfaces of any roughness.  It
has proven useful in many situations, and serve today as a standard,
and invaluable, tool for rough surface scattering studies.  This is in
particular true for scattering from strongly rough surfaces where it
represents the only rigorous available method at present.

Even if this approach is formally exact, it has some practical
limitations.  Imagine a weakly rough metal surface that is illuminated 
by $p$-polarized light. In this case the incident light can
excite surface plasmon polaritons that will travel along the rough
surface. The mean free path of these surface plasmon polaritons will
in the present case be quite large.

In computer simulations we are, of course, not able to represent
infinitely long surfaces. Instead we are limited to surface of finite
length.  To avoid essential contributions to the simulation results
from artificial scattering processes {\it e.g.} where surface plasmon
polaritons are being scattered from the edges of our (finite length)
surface, its length needs to be long. In order not to compromise the
spatial resolution used in the simulations, big demands on computer
memory and {\sc cpu}-time is a consequence. This sets a practical
limit for the use of rigorous numerical simulations for weakly rough
surfaces.  However, for such kind of roughness, perturbation theory,
where we by construction are using surfaces of infinite length, are
adequate and accurate as discussed earlier.  The present limitation of
the rigorous numerical simulation approach should therefore not
represent a too severe restriction from a practical point of view.

We should also mention that there exists another numerical technique
that for the same amount of memory used in the rigorous approach can
handle much longer surfaces (and therefore reduce edge effects). This
technique is based on a numerical solution of the reduced Rayleigh
equation, that was introduced in Eq.~\r{Theory:RRE} of
Sect.~\ref{Theory:Sect:RRE}, and is the single integral equation
satisfied by the scattering amplitude.  The interested reader should
consult Refs.~\citen{Madrazo} and \citen{MP13} for details regarding
this numerical technique. This approach is obviously restricted to
surfaces for which the Rayleigh hypothesis is valid. It can therefore
not be applied to surfaces of arbitrary roughness, but it is valid for
surfaces that practically can not be treated within perturbation
theory. A direct numerical solution of the reduced Rayleigh (integral)
equation can therefore be looked upon as a bridge between 
perturbation theory and the rigorous numerical simulation approach.







\section{Physical Phenomena in Electromagnetic Rough Surface Scattering}
\label{Chapt:Phenomena}

Wave scattering from randomly rough surfaces has a long history in
science~\cite{Rayleigh1,Rayleigh2,Mandelshtam,Rice}. In the overall
majority of theoretical studies conducted up to the early 1980's,
single-scattering approaches were
used~\cite{Book:Beckmann,Book:Ogilvy,Book:Voro,Book:Nieto}. However,
around this time people started getting interested in the effects and
consequences of incorporating multiple-scattering events into the
theories. It created a lot of excitement in the field when new and
interesting multiple scattering phenomena were either predicted
theoretically and/or observed in experiments.  During the period of
time that has passed since the early 1980's, multiple scattering
effects from randomly rough surfaces have attracted much attention
from theorists and experimentalists alike, and today the research in
this field is concentrated around different kinds of multiple
scattering effects~\cite{PhysRep}.

It ought to be mentioned that many of the effects to be discussed here
are not exclusive to surface scattering. Quite a few of them have in
fact their analogies in light scattering from volume disordered
systems. For discussion of light scattering from volume disordered
systems the interested reader is referred to the
literature~\cite{Rossum1999,Multiple}.

In this section we aim at discussing some of the new multiple
scattering effects that might take place when electromagnetic waves
are scattered from a randomly rough surface.  The technical details on
which the present section rely were mainly presented in the previous
section.  We have therefore tried to keep the discussion at a
phenomenological, and hopefully pedagogical, level. Unnecessary
technical details have been avoided whenever possible. As a service to
the more technical oriented reader, an extensive reference to the
original literature has been made.


\subsection{Coherent  Effects in Multiple-Scattered Fields: Weak
  Localization of Light on a Randomly Rough Surface}
\label{Sect:Phen:Coherence Effects}
\sectionmark{Coherent  Effects in Multiple-Scattered Field}

In 1985 McGurn, Maradudin, and Celli~\cite{McGurn1985} predicted
theoretically the existence of what later has been known as the {\em
  enhanced backscattering phenomenon} in surface scattering.  This
phenomenon express itself as a well-defined peak in the
retroreflection direction of the angular dependence of the light
scattered incoherently from a rough surface.  The work in
Ref.~\citen{McGurn1985} was the first to report on an effect that
was shown to be caused by multiple scattering processes taking place
at the rough surface.  The enhanced backscattering phenomenon is an
example of what is known as a {\em coherent effect} in the multiple
scattered field.  Later on, other coherent phenomena, like the
enhanced transmission~\cite{McGurn89}, satellite
peaks~\cite{Freilikher94,Freilikher94a} and enhancements due to the
excitations of magnetoplasmons~\cite{Lu1991,McGurn1991} were
predicted.

\subsubsection{Enhanced backscattering}
\label{Sect:Phen:Enhanced backscattering}

In this subsection the backscattering enhancement phenomenon will be
discussed. Since the mechanisms that give rise to it are different for
weakly and strongly rough surfaces, they will be treated separately.
The scattering system that will be considered is depicted in
Fig.~\ref{Fig:Theory:Geometry} and consists of a single rough
vacuum-metal surface.


\begin{figure}[bt!]
  \begin{center}
    \leavevmode
    \begin{tabular}{@{}c@{\hspace{1.0cm}}c@{}}
      \includegraphics[height=6.cm,width=7cm,]{\myfigpath/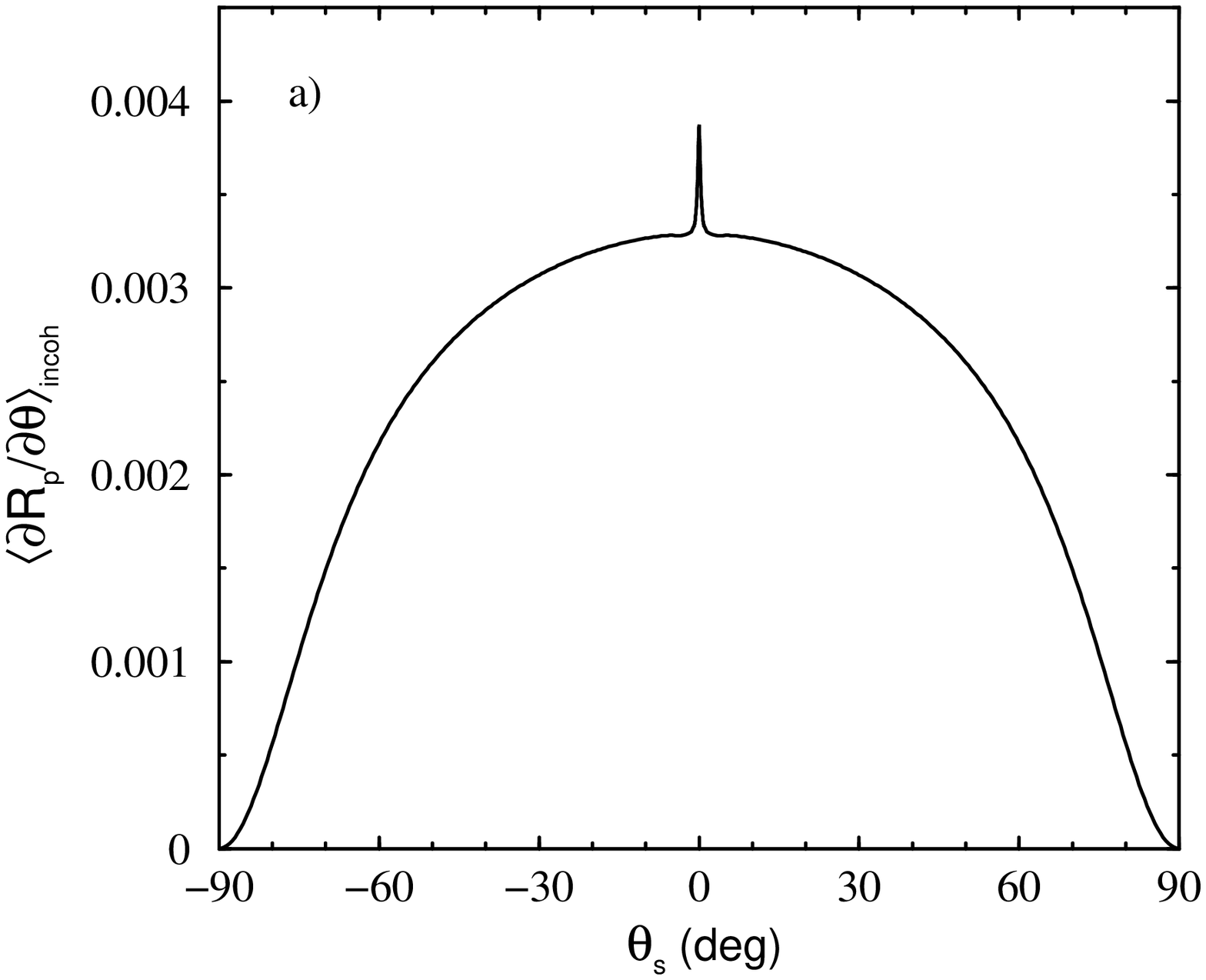} 
      & 
      \includegraphics[height=6.cm,width=7.cm]{\myfigpath/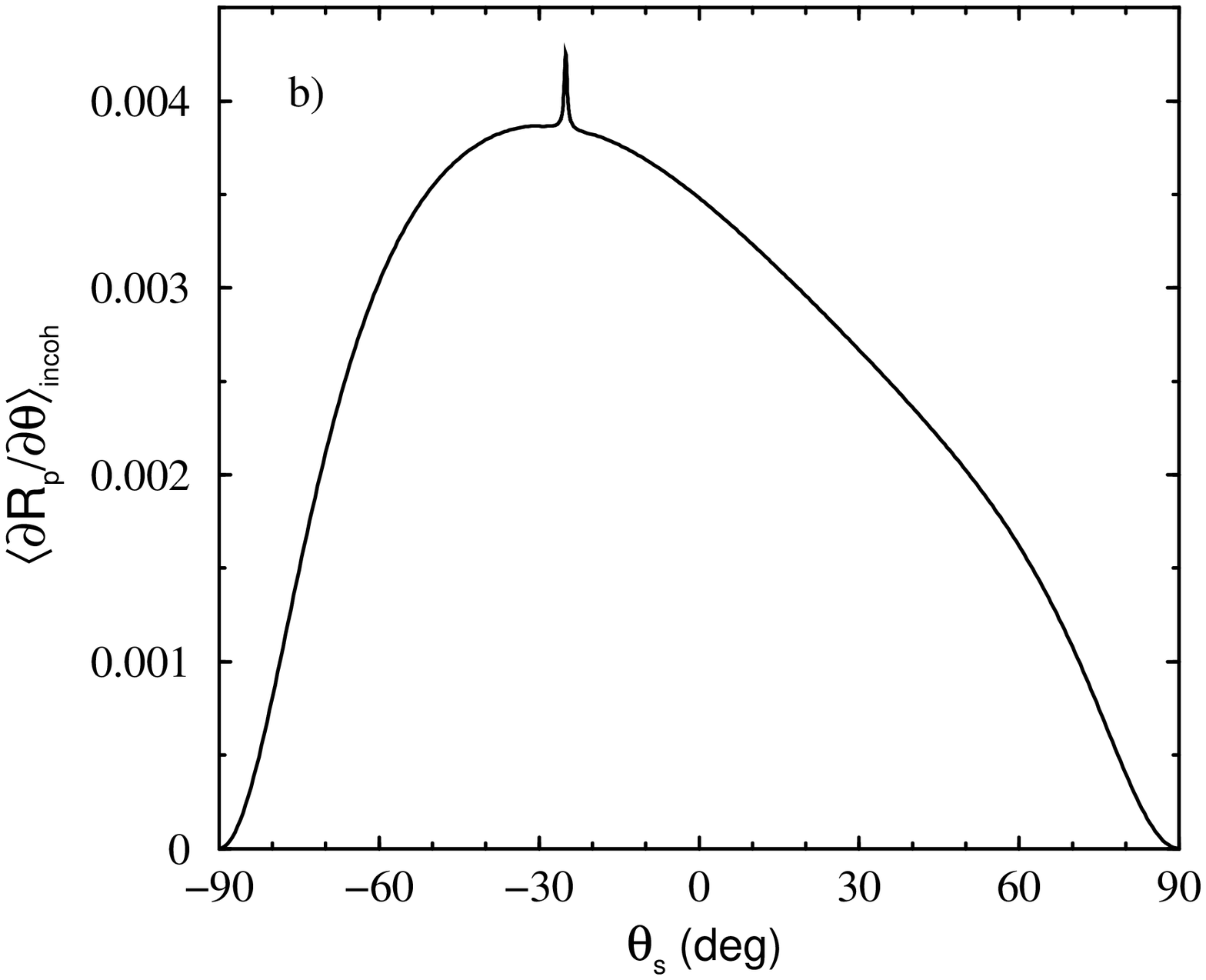}
  \end{tabular}
    \caption{Perturbative calculations for the  mean differential 
      reflection coefficient for the incoherent component of the light
      scattered from a randomly rough silver surface. The incident
      angles of the light of wavelength $\lambda=632.8 {\rm nm}$ were
      (a)~$\theta_0=0^\circ$ and (b)~$\theta_0=25^\circ$.  The
      dielectric constant of silver at this wavelength is
      $\eps{}=-7.5+i0.24$.  The surface was characterized by a
      Gaussian height distribution of {\sc rms}-height $\delta=5 {\rm nm}$
      and a Gaussian height-height correlation function of correlation
      length $a=100 {\rm nm}$.}
    \label{fig:Backscattering:fig-1}
  \end{center}
\end{figure}  


\subsubsubsection{Weakly Rough Surfaces}

It is familiar from every day life that if the surface is not too
rough, the waves incident on it will mostly be scattered into the
specular direction. That is, if the angle of incidence is $\theta_0$,
then most of the energy will be scattered into the direction
$\theta_s=\theta_0$, which defines the specular direction.  For a
weakly rough surfaces the intensity, or equivalently the mean
differential reflection coefficient~(DRC), will have a maximum --- a
specular peak --- for the scattering angle $\theta_s=\theta_0$.
Normally the specular peak is not of any interest to us, and it is
therefore in theoretical studies usually subtracted of, leaving only
the intensity that results from light scattered incoherently by the
rough surface.

In 1985 McGurn, Maradudin, and Celli~\cite{McGurn1985} predicted based
on a perturbation theoretical study, that also in the
retroreflection~(anti-specular) direction of the angular dependence of
the mean DRC there might be an enhancement.  This effect, known today
as enhanced backscattering, manifest itself as a well-defined peak in
the retroreflection direction of the angular dependence of the
intensity of the light that has been scattered incoherently from the
random surface.


\begin{figure}[tb!]
  \begin{center}
    \leavevmode
    \begin{tabular}{@{}c@{\hspace{1.0cm}}c@{}}
      \includegraphics[height=4cm,width=6cm]{\myfigpath/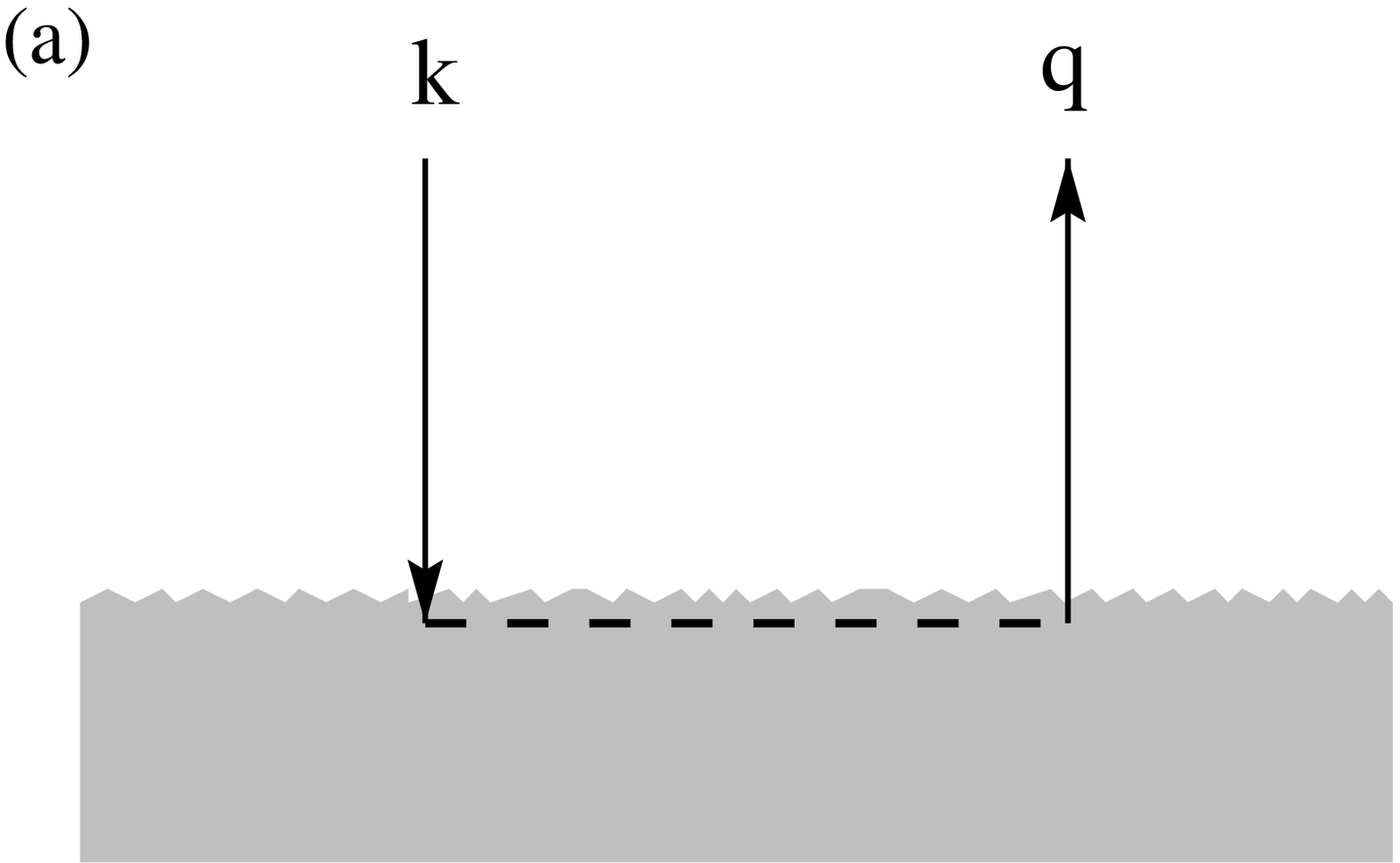} 
      & 
      \includegraphics[height=4cm,width=6cm]{\myfigpath/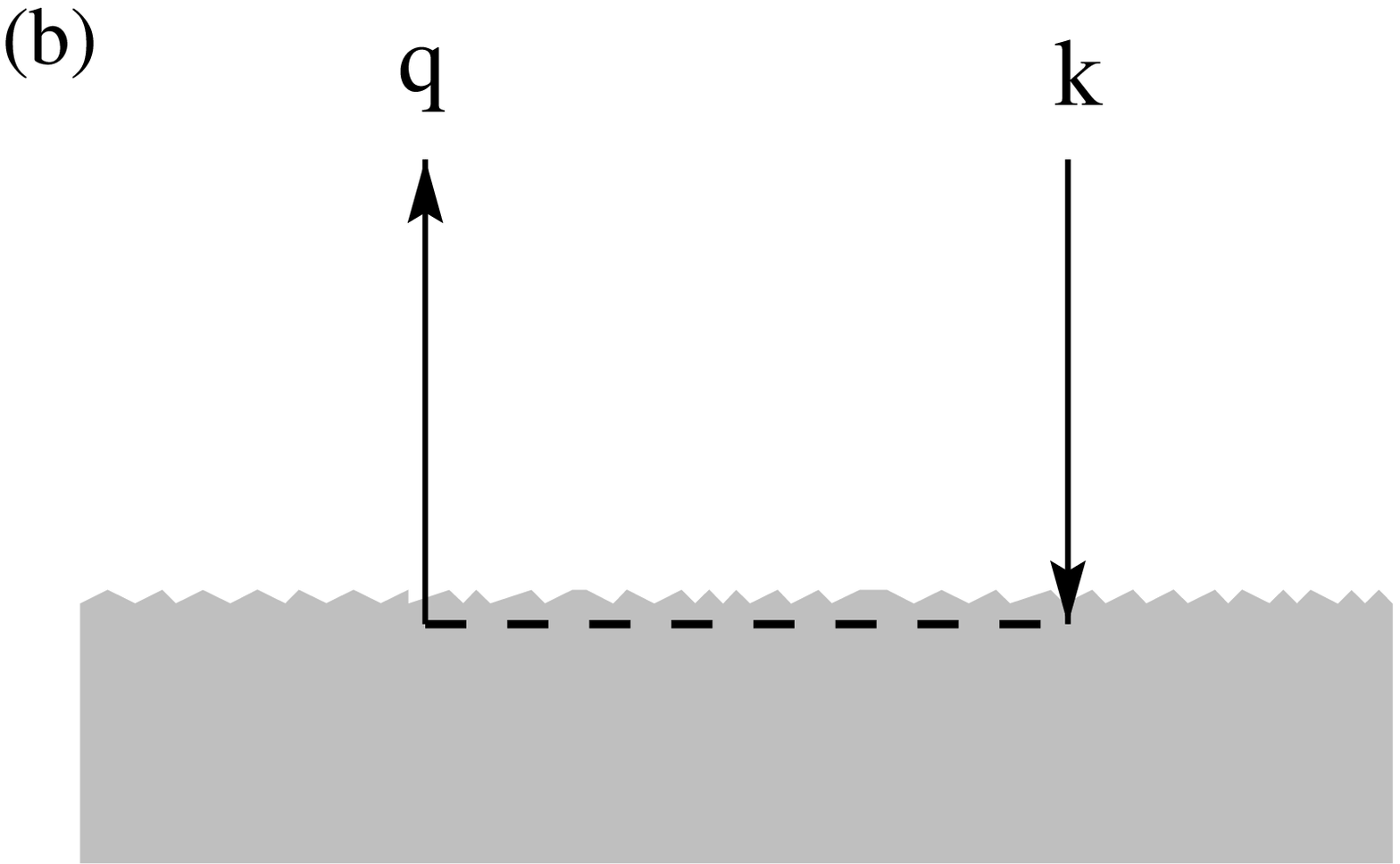} 
    \end{tabular}
    \caption{Diagrams showing two of the scattering events that
      through interference give rise to the enhanced backscattering
      peak for weakly rough surfaces.}
    \label{Fig:Phen:BS-origin}
  \end{center}
\end{figure}



\begin{figure}[b!]
  \begin{center}
    \leavevmode
      \includegraphics[height=7cm,width=9cm]{\myfigpath/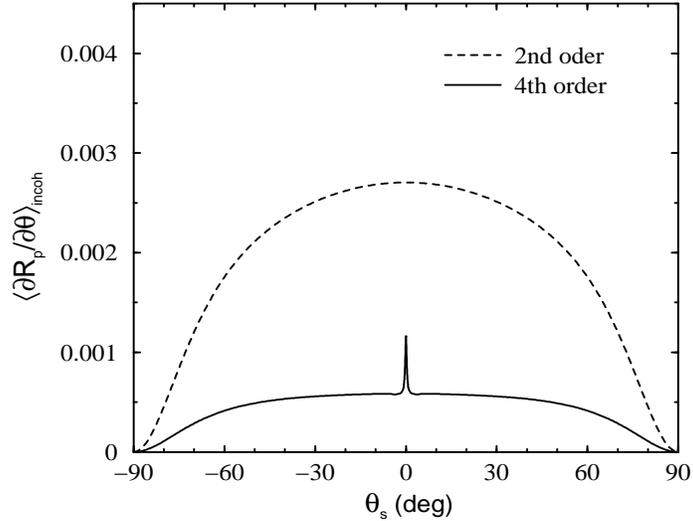}     
    \caption{The same as Fig.~\protect\ref{fig:Backscattering:fig-1}a,
      but now showing separately the 2nd and 4th order contribution in
      the surface profile function to the mean differential reflection
      coefficient. The sum of these terms gives the curve in
      Fig.~\protect\ref{fig:Backscattering:fig-1}a. Notice that the
      enhanced backscattering peak comes from the 4th order
      contribution, {\it i.e.} form the double scattering
      contribution.}
    \label{Fig:Phen:BC-diff-order}
  \end{center}
\end{figure}


In the original paper by McGurn {\it et\ al.}~\cite{McGurn1985}, the
calculation of the enhanced backscattering peak was carried out for
$p$-polarized light scattered from a weakly rough silver surface.
Their calculations, based on a many-body perturbation theory, took
into account multiple scattering events in the calculation of the
intensity scattered incoherently by the surface.  In
Figs.~\ref{fig:Backscattering:fig-1} we show the results of a small
amplitude perturbative calculation, like the one given in
Sect.~\ref{Sect:Theory:SAPT}, for the incoherent contribution to the
mean differential reflection coefficient for $p$-polarized light
incident at angles
$\theta_0=0^\circ$~(Fig.~\ref{fig:Backscattering:fig-1}a) and
$\theta_0=25^\circ$~(Fig.~\ref{fig:Backscattering:fig-1}b) on the
rough silver surface. Terms to 4'th order in the surface profile
function were included which is enough to include all double
scattering processes.  The wavelength of the incident light was
$\lambda=632.8 {\rm nm}$, and the dielectric constant of silver at
this wavelength is $\eps{}=-7.5+i0.24$.  The surface was assumed to be
characterized by a Gaussian height distribution and the height-height
correlation function was also of the Gaussian type. The
root-mean-square~({\sc rms}) height of the surface was $\delta=5 {\rm
  nm}$ while the correlation length was $a=100 {\rm nm}$. From
Fig.~\ref{fig:Backscattering:fig-1} we see well-pronounced peaks for
the retroreflection directions.  It should be stressed that it is the
{\em incoherent} component of the mean DRC that is plotted in these
figures, so that the peak seen at $\theta_s=0^\circ$ in
Fig.~\ref{fig:Backscattering:fig-1}a is no specular
effect\footnote{Recall that for normal incidence the specular and
  anti-specular directions coincide.} since all contributions from
specular scattering have been subtracted off. That this is the case
should be obvious from the position of the enhanced backscattering
peak seen in Fig.~\ref{fig:Backscattering:fig-1}b corresponding to the
incident angle $\theta_0=25^\circ$.


\begin{figure}[t!]
  \begin{center}
    \leavevmode
      \includegraphics[width=15cm,height=12cm]{\myfigpath/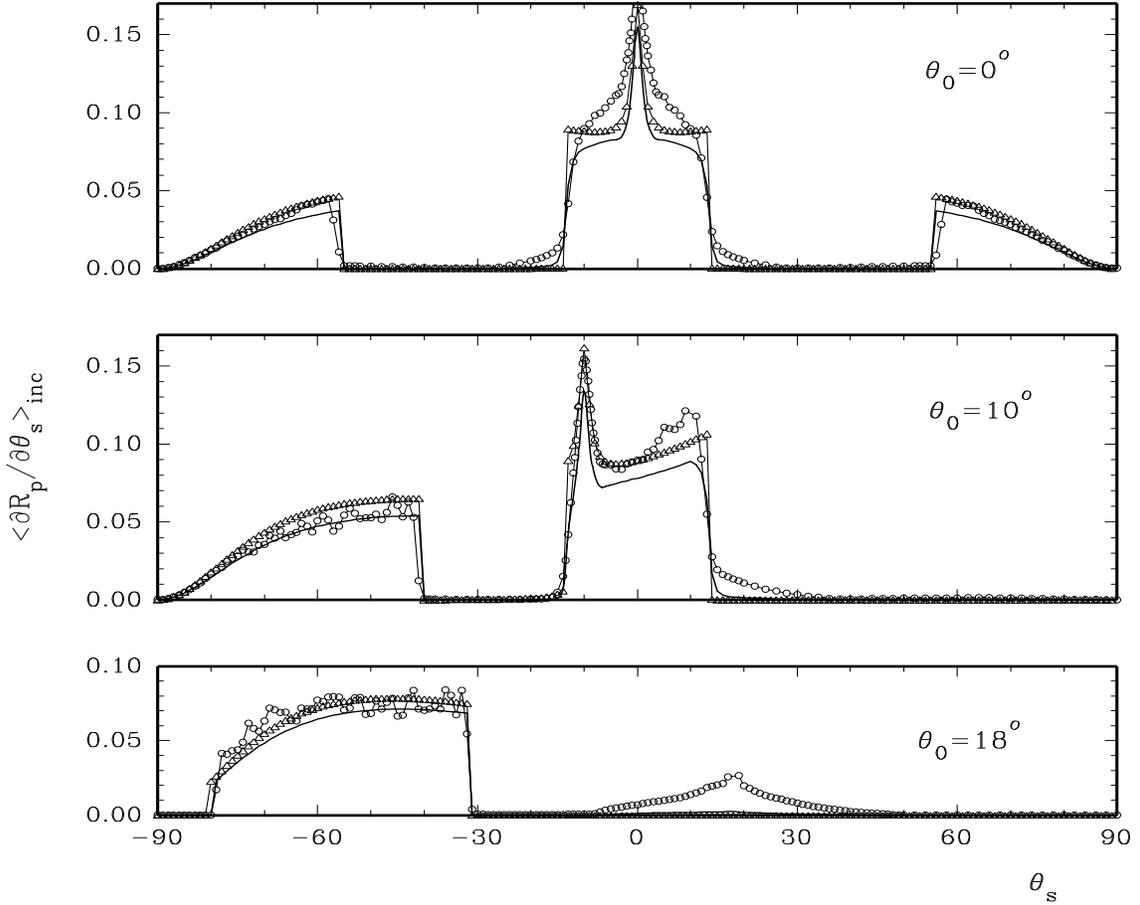}     
    \caption{Experimental results~(open circles) for 
      $\left< \partial R_p/\partial \theta_s\right>_{incoh.}$ as a
      function of the scattering angle $\theta_s$ for three different
      incident angles when $p$-polarized light of wavelength
      $\lambda=612.7 {\rm nm}$ is incident on a one-dimensional random
      gold surface ($\eps{}=-9.00+i1.29$). For the random surface a
      West-O'Donnell power spectrum with $k_-=0.82(\omega/c)$ and
      $k_+=1.29 (\omega/c)$ was used. The {\sc rms}-height of the
      surface was $\delta=10.9 {\rm nm}$. The solid lines and the open
      triangles are perturbation theoretical results based on
      respectively small amplitude and many-body perturbation theory.
      (After Ref.~\protect\citen{west95}). }
    \label{Fig:Phen:Exp-results}
  \end{center}
\end{figure}


The natural question is now: What is the origin of the enhanced
backscattering peak? It was realized that it had to be caused by
multiple scattering since it had not been seen earlier when using
single scattering theories. It turned out that the origin lies in the
interference between a multiple scattered path with its reciprocal
partner~\cite{McGurn1985,west95}.  To illustrate this, let us consider
the double scattering path shown in Fig.~\ref{Fig:Phen:BS-origin}a.
Here an incident wave excites through the breakdown of infinitesimal
translation invariance of the system, a surface plasmon polariton that
propagates along the surface. At the next scattering event this
surface polariton is converted back into a volume electromagnetic wave
that propagating away from the surface.  This path has a reciprocal
partner (Fig.~\ref{Fig:Phen:BS-origin}b) where the scattering takes
place from the same scattering centers at the rough surface, but now
in the opposite order. For the backscattering direction these two
paths will have exactly the same amplitude and phase, {\it i.e.} they
will be coherent, and hence they will interfere {\em constructively}.
However, as we move away from the backscattering direction, the two
paths fast become incoherent so that their intensities just add. Thus,
due to the interference nature of the enhanced backscattering peak the
amplitudes at the position of the peak would in the absence of single
scattering be twice that of its background due to the cross-terms
originating from the square modulus of the amplitudes needed in order
to calculate the intensity. However, notice that it is not uncommon
that single scattering gives considerable contribution to the mean
differential reflection coefficient of the light scattered
incoherently from the surface. In such cases, the height of the peak
is not twice of its background.


\begin{figure}[t!]
  \begin{center}
    \leavevmode
    \begin{tabular}{@{}c@{\hspace{1.0cm}}c@{}}
      \includegraphics[height=6cm,width=7cm]{\myfigpath/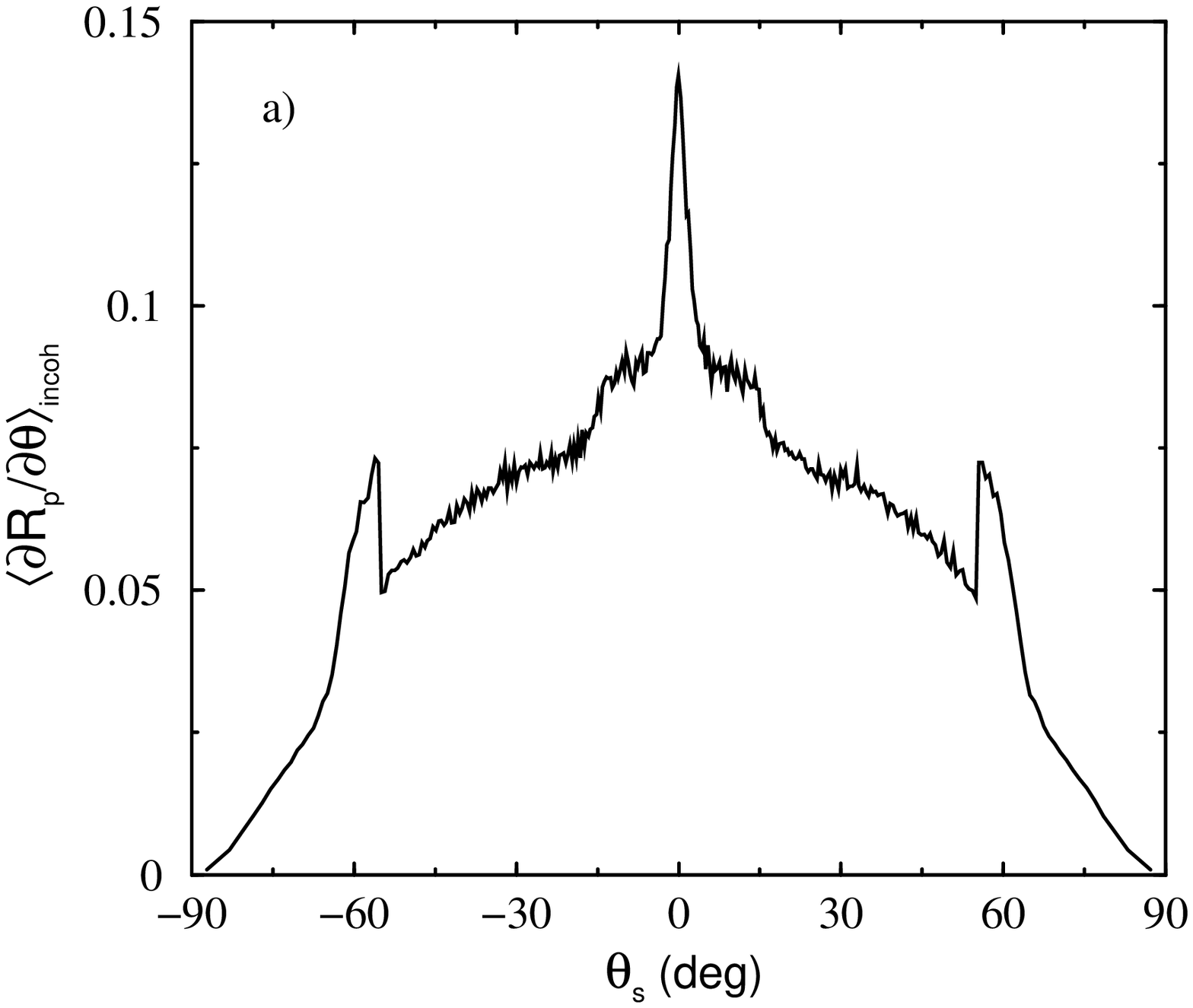} 
      & 
      \includegraphics[height=6cm,width=7cm]{\myfigpath/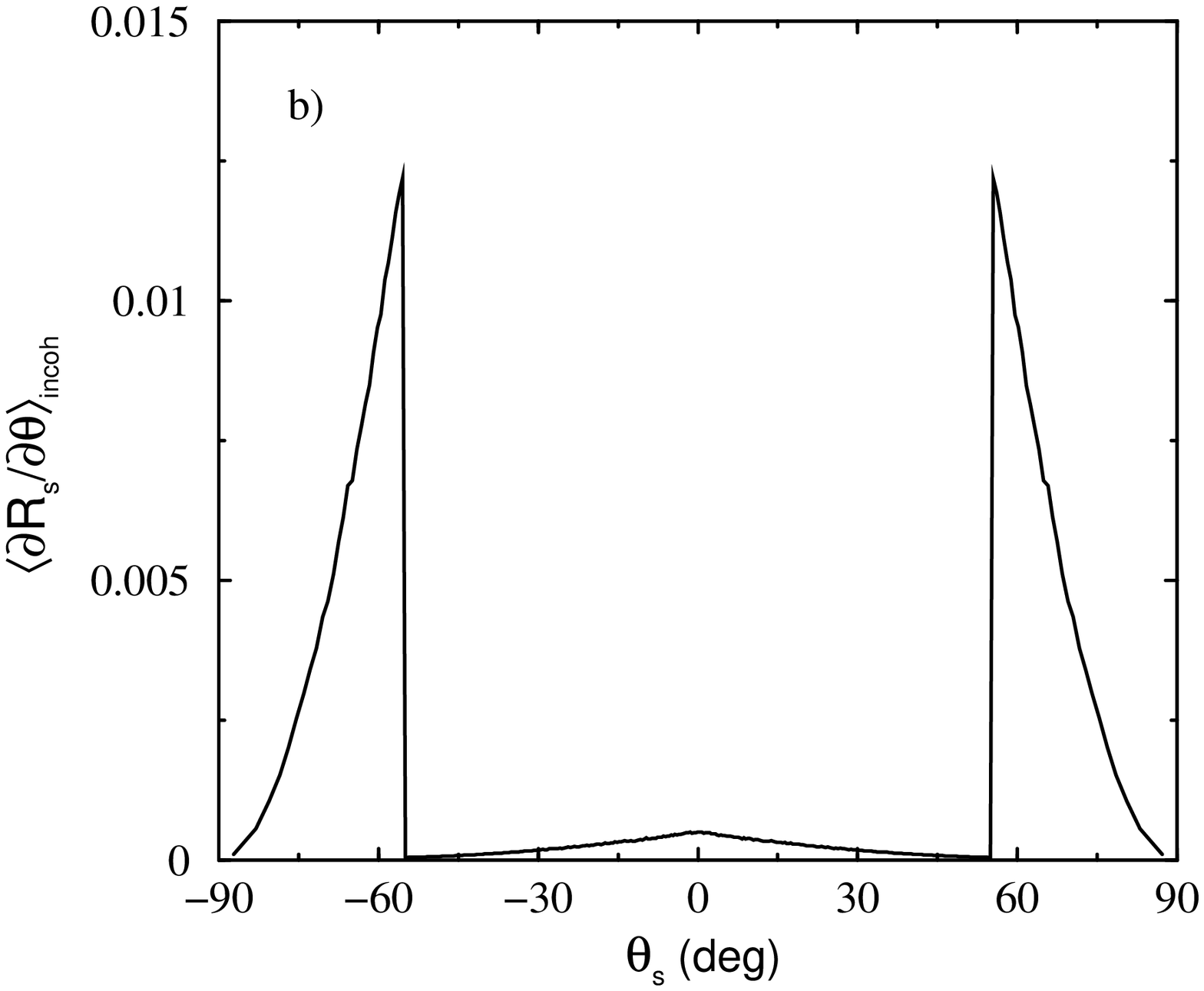}
    \end{tabular}
    \caption{Numerical Monte Carlo simulation results for the 
      mean differential reflection coefficient, $\left< \partial
        R_\nu/\partial \theta_s\right>_{incoh}$, for (a)~$p$- and
      (b)~$s$-polarized light scattered from a rough silver surface of
      {\sc rms}-height $\delta=15 {\rm nm}$. The angle of incidence
      was $\theta_0=0^\circ$ and the wavelength of the incident light
      was $\lambda=632.8 {\rm nm}$. At this wavelength the dielectric
      constant of silver is $\eps{}=-7.5+i0.24$. Furthermore, the
      surface was of the West-O'Donnell type characterized by the
      parameters $k_-=0.82(\omega/c)$ and $k_+=1.92(\omega/c)$. The
      simulations were performed by numerically solving the reduced
      Rayleigh equation that the scattering amplitude
      satisfies~\protect\cite{Madrazo,MP13}. The number of samples
      used in obtain these results was $N_\zeta=3000$.  Notice that in
      the case of $s$-polarization there is no enhanced backscattering
      peak in contrast to what is the case for $p$-polarization. This
      difference is caused by the fact that $s$-polarized incident
      light can not excite surface plasmon polaritons at a rough
      vacuum-metal interface.}
    \label{Fig:Phen:BC-s-pol}
  \end{center}
\end{figure}


To show that multiple scattering indeed is the origin of the enhanced
backscattering phenomenon, we show in Fig.~\ref{Fig:Phen:BC-diff-order} 
the different contributions to the incoherent component of the mean
DRC obtained from Eq.~\r{Eq:DRC-small-amp-pert-theory}. We recall that
the first term of this equation is the single scattering contribution,
{\it i.e.} it is of 2nd order in the surface profile function
$\zeta(x_1)$.  The next two terms are both double scattering
contributions or, equivalently, 4th order contributions in the surface
profile function. From Fig.~\ref{Fig:Phen:BC-diff-order} it is seen
that the single scattering contribution (2nd order in $\zeta(x_1)$) is
a smooth function of the scattering angle. Furthermore, it is seen
that the peak stems from the double scattering contribution, {\it
  i.e.} it comes from the second and third terms of
Eq.~\r{Eq:DRC-small-amp-pert-theory}. In a diagrammatic language this
term comes from the maximally crossed diagrams. The
interested reader should consult Ref.~\citen{McGurn1985} for
additional details.

It should be noted that even if we earlier only included fully the lowest
order multiple scattering process~(double scattering), higher order
processes will not change the statement that the enhanced backscattering
phenomenon is caused by multiple scattering through the constructive
interference between a scattering path with its reciprocal partner.

The enhanced backscattering effect from weakly rough vacuum-metal
surfaces was observed in experiments by West and
O'Donnell~\cite{west95} in 1995 in the scattering of $p$-polarization
light from a rough gold surface of {\sc rms}-height $\delta=10.9 {\rm
  nm}$.  The power-spectrum used in these experiments was of the
rectangular type also known as the West-O'Donnell power-spectrum. The
remaining parameters used are defined in the caption of
Figs.~\ref{Fig:Phen:Exp-results}.  We have in
Figs.~\ref{Fig:Phen:Exp-results} reproduced their experimental
results~(open circles) together with some perturbation theoretically
results (solid lines and open triangles). At least for the two
smallest angles of incidence well-defined peaks around the
retroreflection direction in the experimental results are seen.

For weakly rough surfaces we just argued that the origin of the enhanced
backscattering effect involves surface plasmon polaritons. In
$s$-polarization, a rough (one-dimensional) vacuum-metal interface does {\em
not} support such surface waves. Hence, one does not expect to see any
backscattering peak for this polarization for weakly rough surfaces. This is
indeed seen from Figs.~\ref{Fig:Phen:BC-s-pol} showing numerical simulation
results for $p$- and $s$-polarized incident light based on the solution of
the reduced Rayleigh equation that the scattering amplitude satisfies. The
power spectrum used for the surface was again of the West-O'Donnell type,
and it was defined by the parameters $k_-=0.82(\omega/c)$ and
$k_+=1.92(\omega/c)$. It is seen from Figs.~\ref{Fig:Phen:BC-s-pol} that
only in $p$-polarization do we see an enhanced backscattering peak.  From
Eq.~\r{Eq:SAPT-expansion-total} we see that the single scattering
contribution is proportional to the power spectrum, $g(|k|)$, of the surface
roughness. Hence, if $k_->0$ the incoherent component to the mean DRC,
$\left<\partial R_\nu/\partial\theta_s\right>_{incoh}$ should not contain
any contribution from single scattering events in the angular range
$-\theta_-< \theta_s < \theta_-$, where $\theta_-=\arcsin\left( (k_-
c)/\omega\right)$. With the parameters used in obtaining
Figs.~\ref{Fig:Phen:BC-s-pol} this gives $\theta_-=55.1^\circ$. For
scattering angels $|\theta_s|>\theta_-=55.1^\circ$ single scattering is
allowed.  This can be seen as a jump in Figs.~\ref{Fig:Phen:BC-s-pol} at
this angle. Furthermore, around the backscattering peak, single scattering
should be absent and indeed the enhanced backscattering peak is twice that
of its background as predicted above.  Notice also that the overall fraction
of the light scatterer incoherently from the surface is at least one order
of magnitude lower for $s$- then $p$-polarization.


\begin{figure}[t!]
  \begin{center}
    \leavevmode
    \begin{tabular}{@{}c@{\hspace{1.0cm}}c@{}}
      \includegraphics[height=6cm,width=7cm]{\myfigpath/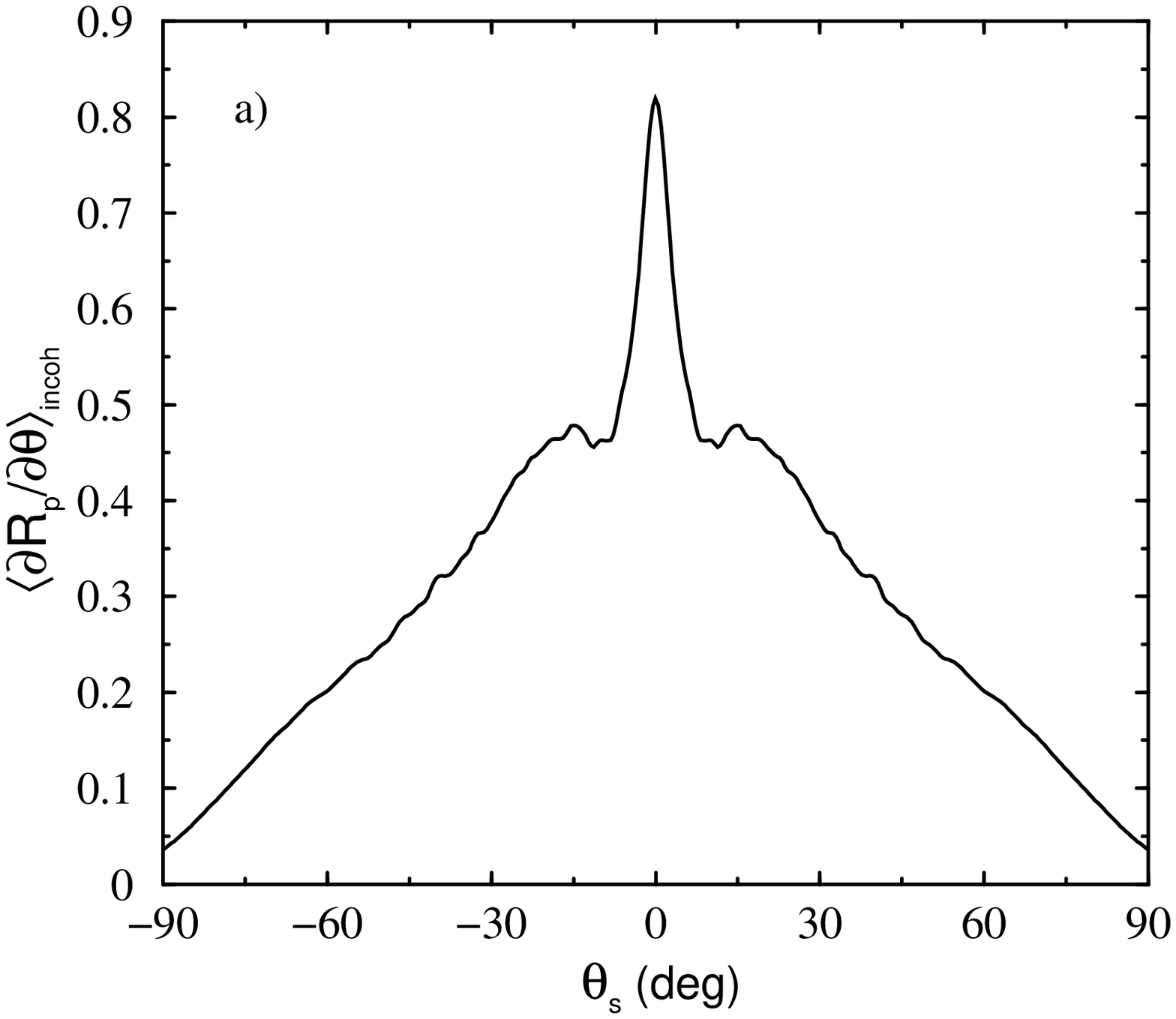} 
      & 
      \includegraphics[height=6cm,width=7cm]{\myfigpath/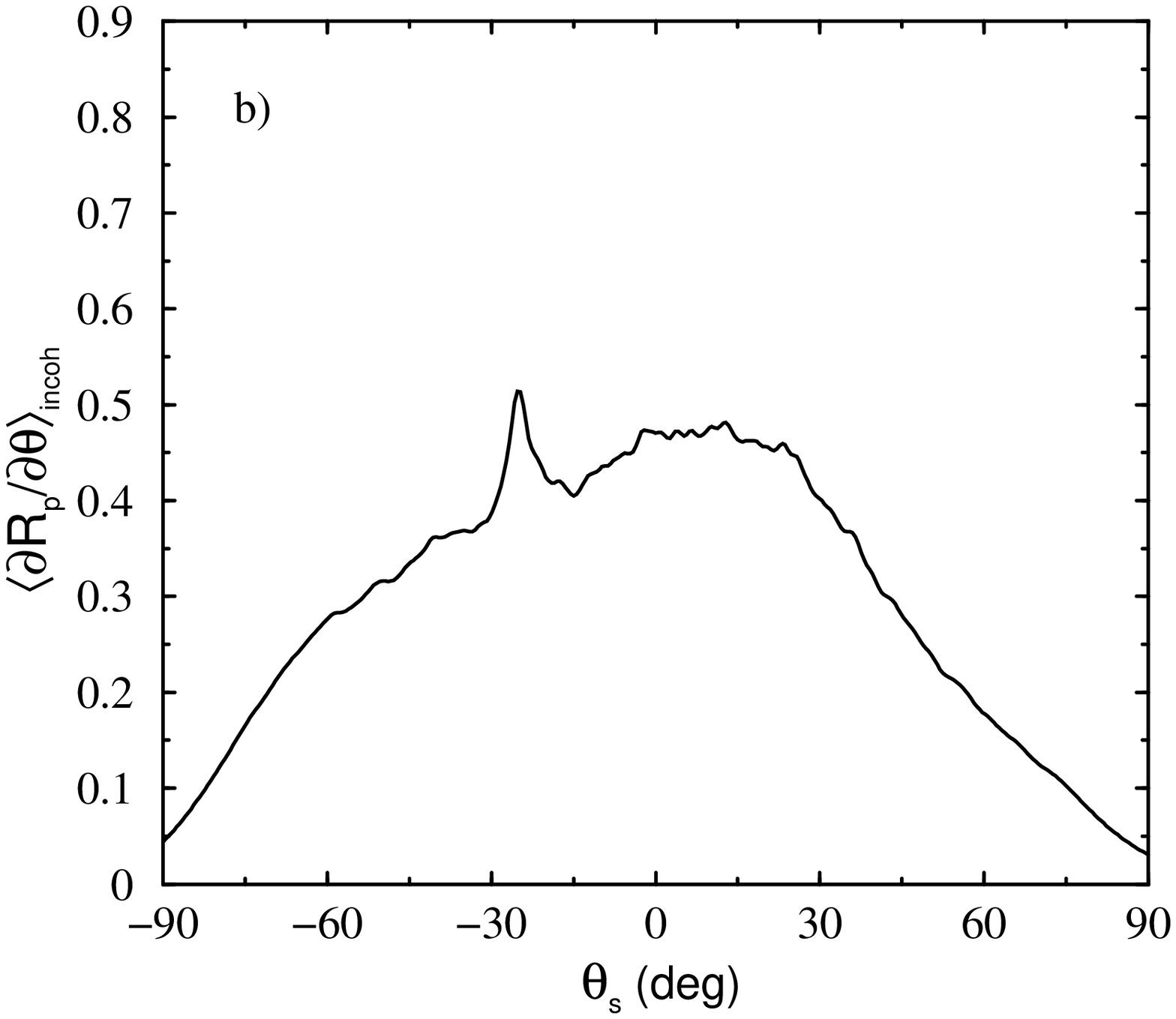} 
      \\
      \includegraphics[height=6cm,width=7cm]{\myfigpath/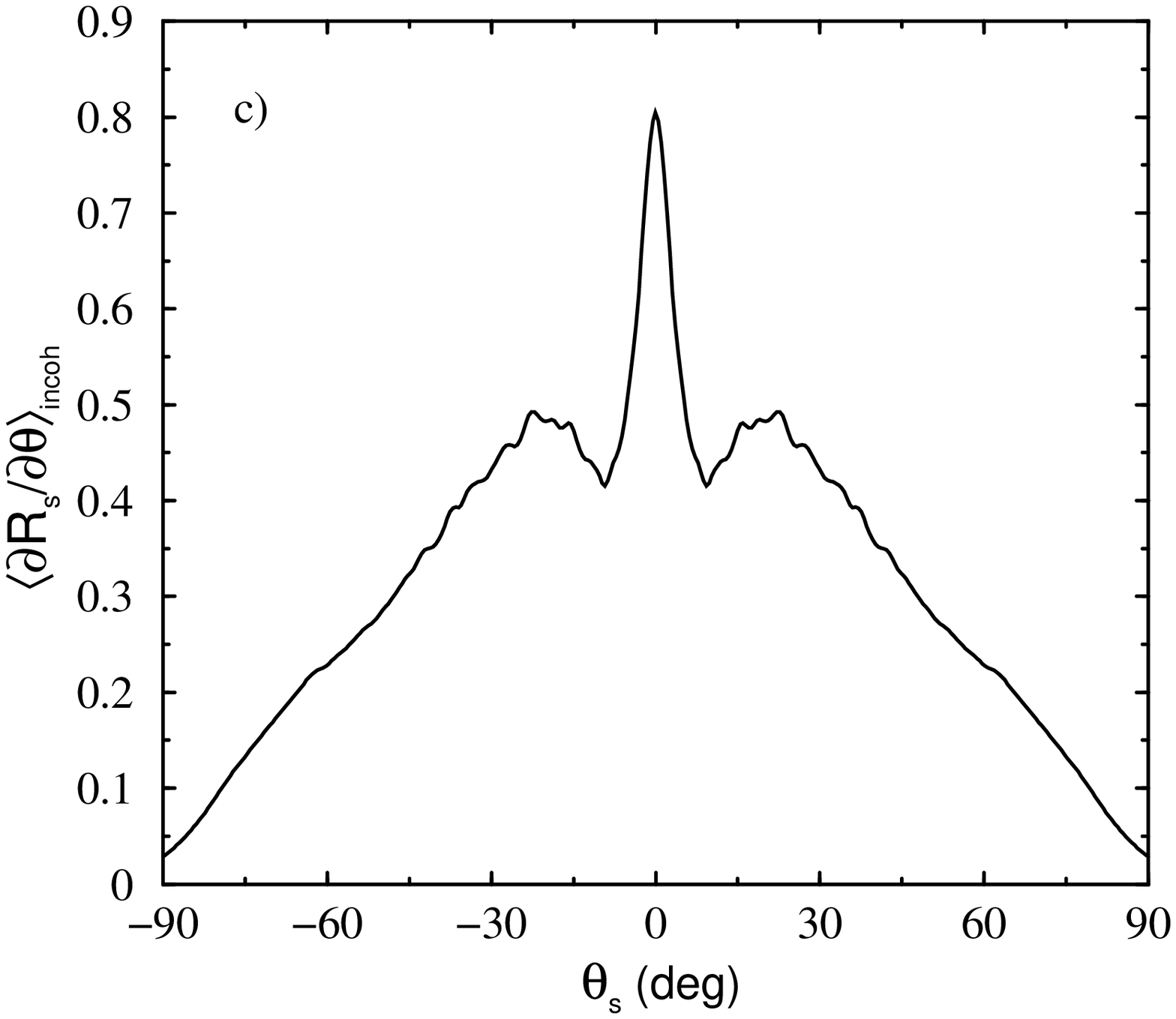} 
      & 
      \includegraphics[height=6cm,width=7cm]{\myfigpath/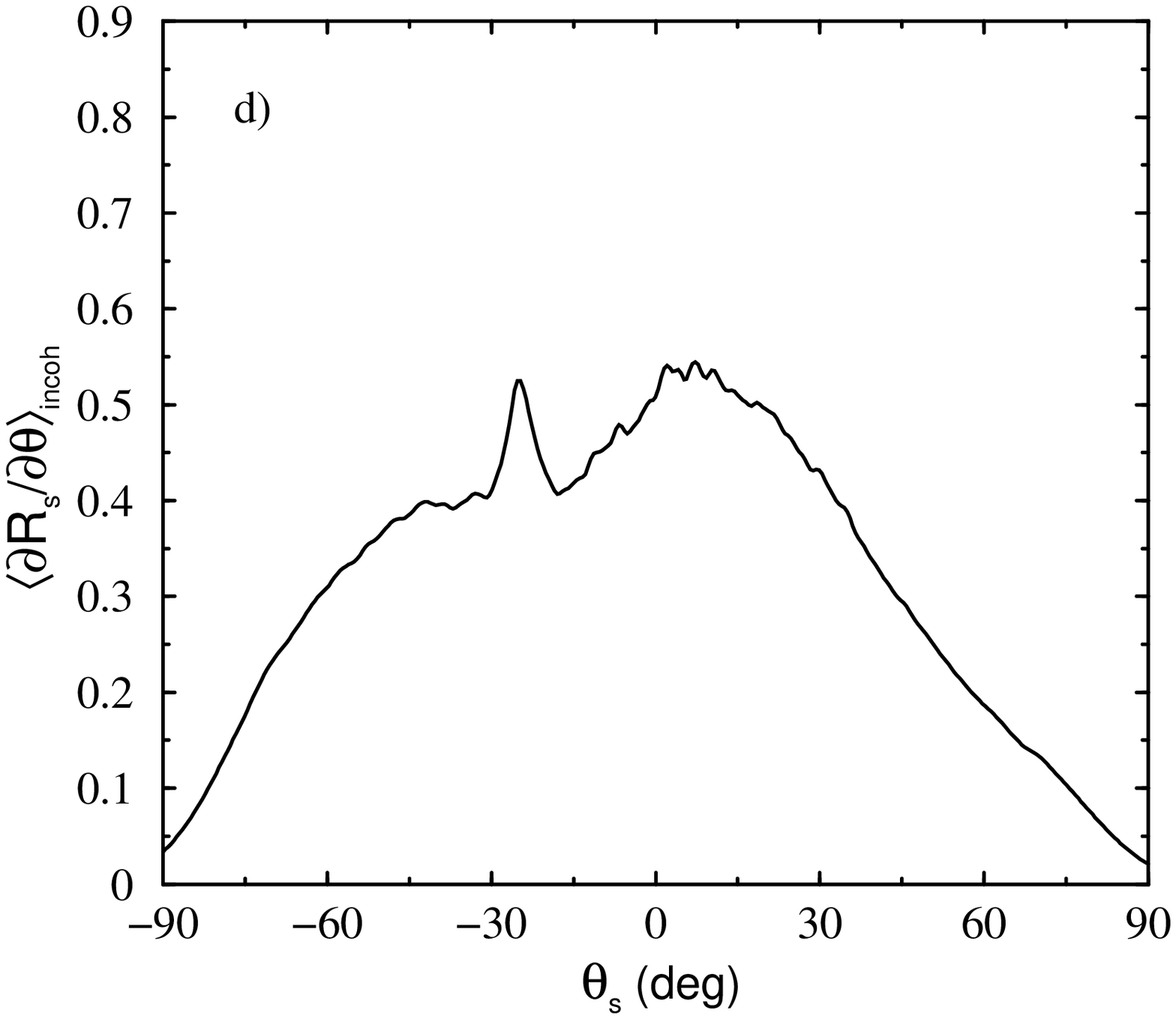} 
  \end{tabular}
    \caption{Rigorous Monte Carlo simulation results for the
      angular dependence of the incoherent component of the mean
      differential reflection coefficient for the scattering of (a and
      b) $p$- and (c and d) $s$-polarized incident light from a
      strongly rough silver surface. The wavelength of the incident
      light was $\lambda= 612.7 {\rm nm}$ for which the dielectric
      constant of silver is $\eps{}= -17.2+i0.498$. The incident
      angles of the light were (a and c) $\theta_0=0^\circ$ and (b and
      d) $\theta_0=25^\circ$.  The strongly rough surface was
      characterized by Gaussian height distribution of {\sc
        rms}-height $\delta=1.2 {\rm \mu m}$ and the transverse
      correlation length for the Gaussian correlation surface was
      $a=2 {\rm \mu m}$.  The length of the surface was $L=25.6 {\rm
        \mu m}$ and a finite sized beam of width $g=6.4 {\rm \mu m}$
      was used in the simulations. The number of discretization points
      was $N=500$. The numerical results were all based on an ensemble
      average over $N_\zeta=3000$ realizations of the randomly rough
      surface.}
    \label{Fig:Phen:BC-num-calc}
  \end{center}
\end{figure}



\begin{figure}[t!]
  \begin{center}
    \leavevmode
    \begin{tabular}{@{}c@{\hspace{1.0cm}}c@{}}
      \includegraphics[height=2.in,width=2.5in,]{\myfigpath/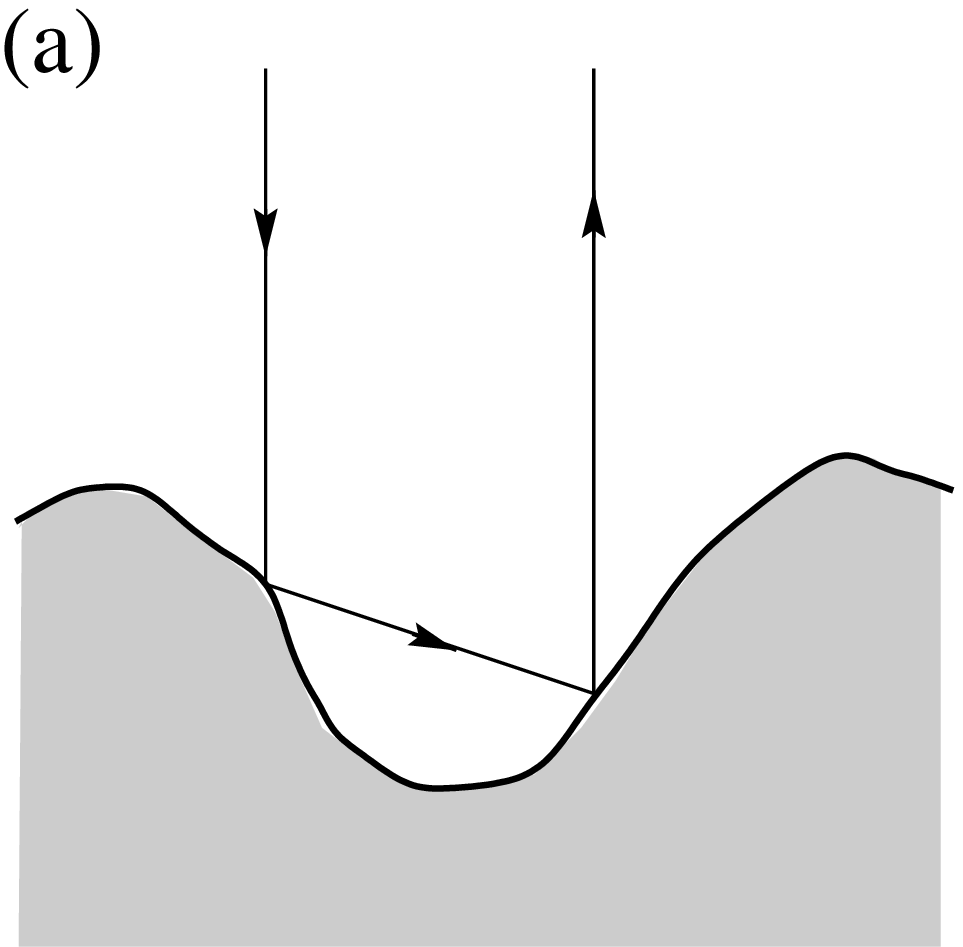} 
      & 
      \includegraphics[height=2.in,width=2.5in]{\myfigpath/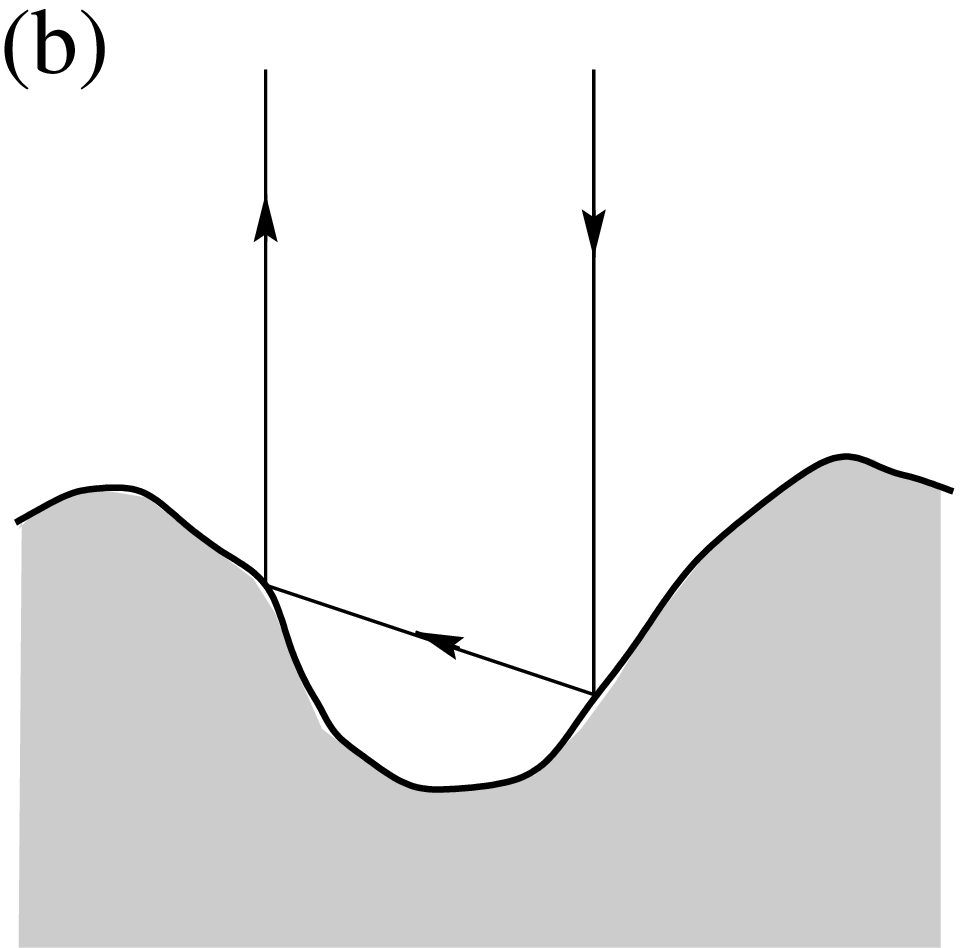} 
    \end{tabular}
    \caption{Diagrams showing two double scattering paths that for 
      strongly rough surfaces through interference represent the main
      contributes to the enhanced backscattering peak phenomenon.}
    \label{Fig:Phen:BS-origin-strongly-rough}
  \end{center}
\end{figure}


\subsubsubsection{Strongly rough surfaces}

We will now consider strongly rough surfaces. In order to study the
backscattering phenomenon for such surfaces we have to resort to
numerical simulations like {\em e.g.} the approach outlined in
Sect.~\ref{Sect:Theory:Sect:NumSim}. In
Figs.~\ref{Fig:Phen:BC-num-calc} we present the results of such
simulations for the angular dependence of the incoherent component of
the mean DRC for (a and b) $p$- and (c and d) $s$-polarized light
incident on a rough vacuum-metal surface of {\sc rms}-height
$\delta=1.2 {\rm \mu m}$. The correlation length for the Gaussian
correlated surface was $a=2 {\rm \mu m}$. The main difference between
these results and those for the weakly rough surfaces presented
earlier~(Fig.~\ref{Fig:Phen:BC-s-pol}) is that we now also observe an
enhanced backscattering peak in the case of $s$-polarization.  So what
is the reason for this difference between weakly and strongly rough
surfaces when it comes to the backscattering phenomenon?  The
explanation lies in the mechanism causing the backscattering peak for
strongly rough surfaces~\cite{mendez87,odonnell87,nieto87,AnnPhys}.
Since the excitation of surface plasmon polaritons is weak for
strongly rough surface, it is unlikely that the reason for the
backscattering peak is caused by this type of surface waves.  Such a
mechanism could not in any case explain the presence of the
backscattering peak observed for $s$-polarization.  Instead the
backscattering peak for strongly rough surfaces arises due to the
constructive interference between multiple scattered volume paths
like {\it e.g.} those shown in
Figs.~\ref{Fig:Phen:BS-origin-strongly-rough}. In this case no surface
waves are excited, but instead the multiple scattering takes place
within the valleys of the now strongly rough surface. The incident
wave, that after its first encounter with the rough surface, is
scattered at least one more time before leaving the surface for good.
Also in this case for the backscattering direction this path has a
reciprocal partner that is phase coherent with the first one and with
which the latter path can interfere constructively.  Since this
mechanism does not involve any surface plasmon polaritons, there is no
reason why the backscattering phenomenon should not show up also in
$s$-polarization from strongly rough surfaces. In fact as can be seen
from Fig.~\ref{Fig:Phen:BC-num-calc}c and d, the backscattering peak
in $s$-polarization is as pronounced as for $p$-polarization. Observe
also that the energy scattered incoherently, which for strongly rough
surfaces is close to the total scattered energy, is of roughly the
same order for both polarizations. This is in contrast to the
situation found for weakly rough surfaces.

The enhanced backscattering phenomenon from strongly rough surfaces
was experimental confirmed as early as 1987 by M\'endez and
O'Donnell~\cite{mendez87}. This was just two years after its
theoretically prediction by McGurn~{\it et al.}~\cite{McGurn1985} for
weakly rough surfaces. In fact these experiments provided the first
experimental evidence what-so-ever for the enhanced backscattering
phenomenon from rough surfaces.

\subsubsection{Satellite Peaks}

The backscattering phenomenon discussed in the previous subsection is
not the only coherent effect that might exist when light is scattered
from a randomly rough surface. Another such effect is the existence of
so-called {\em satellite peaks} predicted by Freilikher, Pustilnik,
and Yurkevich~\cite{Freilikher94} in 1994.  Satellite peaks are
enhancements in the angular distribution of light scattered
incoherently from scattering systems that supports more the one
surface~\cite{Freilikher94,Freilikher94a,McGurn1991,Lu1991} or guided
waves~\cite{McGurn87,sanchez95b,sanchez94,sanchez95,Madrazo,MP13}.  As
will be shown in detail in
Subsect.~\ref{Subsect:Phen:Fundamentral-BS-SP}, they are not caused by
interference between reciprocal paths as was the case for the
backscattering phenomenon, but instead by interference of {\em
  nonreciprocal} paths.  These enhancements should occur for
scattering angles that are located symmetrically with respect to the
position of the enhanced backscattering peaks that the scattering
system also gives rise to.


\begin{figure}[t!]
  \begin{center}
    \leavevmode
    \includegraphics[height=2.5in,width=4.5in]{\myfigpath/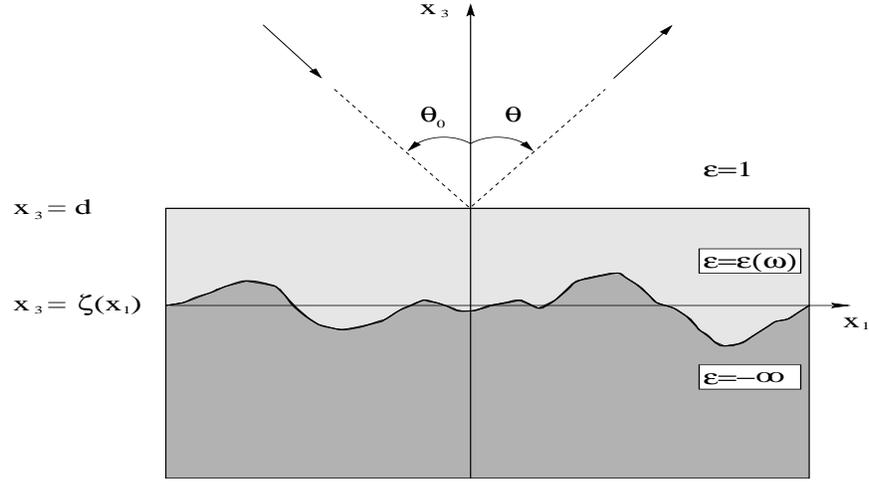} 
    \caption{A sketch of a film scattering geometry that supports guided
    waves and that may give rise to satellite peaks in the angular
    dependence of the scattered light.}
    \label{Fig:Phen:SP-fig} 
  \end{center}
\end{figure}



\begin{figure}[b!]
  \begin{center}
    \leavevmode
    \begin{tabular}{@{}c@{\hspace{1.0cm}}c@{}}
      \includegraphics[height=6cm,width=7cm]{\myfigpath/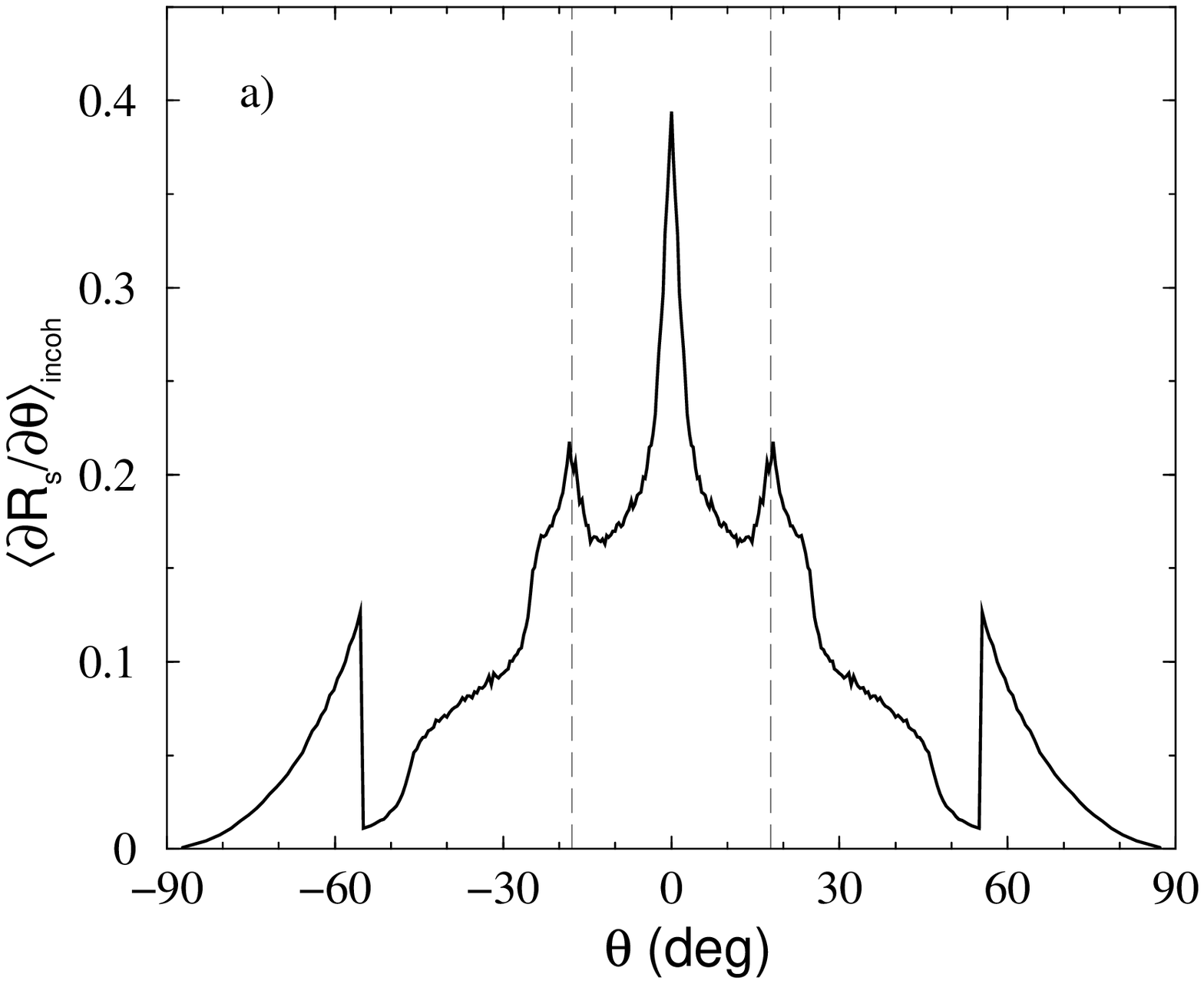} 
      & 
      \includegraphics[height=6cm,width=7cm]{\myfigpath/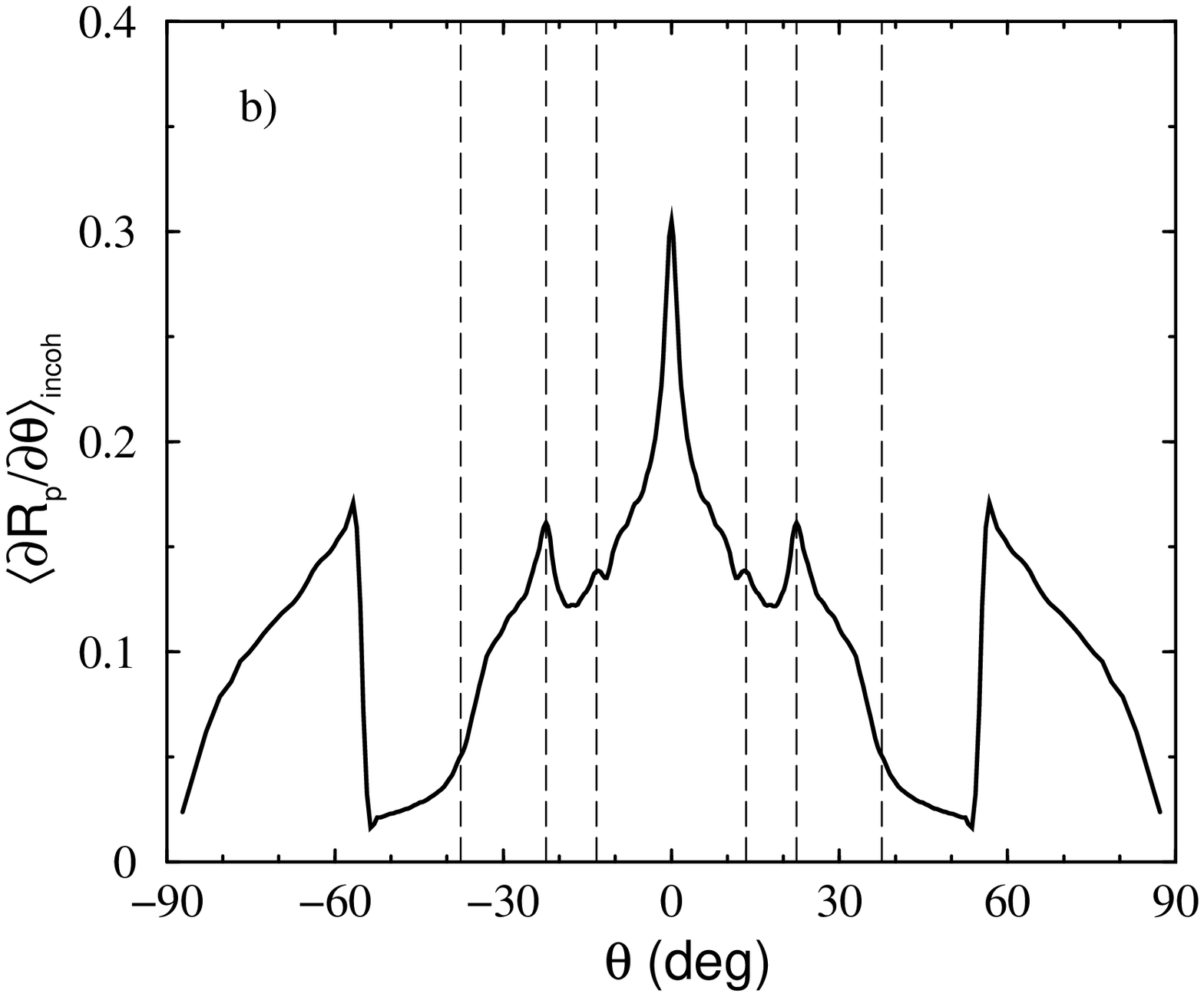}
  \end{tabular}
  \caption{The  contribution to the mean differential
    reflection coefficient from the incoherent component of the
    scattered light $\left< \partial R_\nu / \partial \theta
    \right>_{\mathrm incoh}$ as a function of the scattering angle
    $\theta_s$ when an $s$-~(Fig.~\protect\ref{Fig:Phen:SP}a) or
    $p$-polarized~(Fig.~\protect\ref{Fig:Phen:SP}b) plane wave of
    wavelength $\lambda =633 {\rm nm}$ is incident normally
    ($\theta_0=0^\circ$) on the film scattering geometry shown in
    Fig.~\protect\ref{Fig:Phen:SP-fig}. The dielectric constant of the
    film at the wavelength of the incident light is
    $\varepsilon(\omega) =2.6896+i0.01$, and the films mean thickness
    was $d=500 {\rm nm}$. The semi-infinite medium which the film is
    ruled on was a perfect conductor. The surface profile function
    $\zeta(x_1)$ of the film-conductor interface was characterized by
    a Gaussian surface height distribution of {\sc rms}-height $\delta
    = 30 {\rm nm}$ and a West-O'Donnell power spectrum defined by the
    parameters $k_-=0.82 (\omega/c)$ and $k_+=1.97 (\omega/c)$. The
    length of the surface used in the simulations was $L=160\lambda$.
    The dashed vertical lines indicates the estimated positions of the
    satellite peaks~(see Ref.~\citen{MP13} for details). The results
    were obtained by numerical simulations based on the reduced
    Rayleigh equation. The data in Fig.~\protect\ref{Fig:Phen:SP}b
    have bee smoothed to make the positions of the satellite peaks
    more apparent. (After Ref.~\citen{MP13}.)}
    \label{Fig:Phen:SP}
  \end{center}
\end{figure}


To illustrate this, let us study the film scattering system shown in
Fig.~\ref{Fig:Phen:SP-fig}. Here the lower interface is rough and the
upper one is planar. Furthermore, the lower semi-infinite medium is
assumed to be a perfect conductor, while the incident medium is
assumed to be vacuum.  In Figs.~\ref{Fig:Phen:SP} we show the
results of numerical simulations for the mean differential reflection
coefficient in the case of $s$- (Fig.~\ref{Fig:Phen:SP}a) and
$p$-polarized (Fig.~\ref{Fig:Phen:SP}b) incident light. The remaining
surface parameters are given in the caption of
Figs.~\ref{Fig:Phen:SP}.  In the case of $s$-polarization, the
scattering system of  mean thickness $d=500 {\rm nm}$ supports
two satellite peaks, while in the case of $p$-polarization it can at
most support six such peaks~\cite{MP13}.  The positions of these peaks
are indicated by dashed vertical lines in Figs.~\ref{Fig:Phen:SP}.
From Fig.~\ref{Fig:Phen:SP}a the two satellite peaks that the
scattering system supports in this case are easily distinguished from
the background. However, from Fig.~\ref{Fig:Phen:SP}b one sees that
only four out of the six possible peaks can be observed. There are
two reasons why some of these satellite peaks may not be observable:
First, some of them may lie in the non-radiative part of the
spectrum, and are therefore not even in principle observable.  Second,
their strength might be to low to be observable~\cite{Madrazo,MP13},
{\it i.e.}  one (or both) of the channels involved in the interference
process that gives rise to the satellite peaks might have too low
intensity~\cite{Madrazo,MP13} (see
Subsect.~\ref{Subsect:Phen:Fundamentral-BS-SP}).

It can be shown (result not shown) that by reducing the thickness of
the film, and thus reducing the number of guided waves that the system
supports say to one, all the satellite peaks vanish while the enhanced
backscattering peak is still present~\cite{sanchez95}. In an analogous
way, if the film thickness is increased, more then two satellite peaks
might be seen~\cite{sanchez95}.

\subsubsection{A Formal Approach to Enhanced Backscattering and Satellite Peaks}
\label{Subsect:Phen:Fundamentral-BS-SP}

In the previous two subsections the enhanced backscattering and
satellite peaks phenomena were discussed. In the present subsection a
more detailed analysis and formal approach towards these two phenomena
will be presented. In particular we will determine at which positions
the satellite peaks are to be expected.

Let us consider a general film scattering system, where at least one
of the interfaces are rough. Fig.~\ref{Fig:Phen:SP-fig} provides one
example of such a system.  Depending on the thickness of the film, the
scattering system supports $N > 0$ guided waves at the frequency
$\omega$ of the incident light. The wavenumbers of these modes, or
``channels'' as some authors prefer to call them, will be denote by
$q_n(\omega)$ where $n=1,\ldots N$.  Through the surface roughness
the incident light may couple to these guided waves.


\begin{figure}[t!]
  \begin{center}
    \leavevmode
      \includegraphics[height=5cm,width=10cm]{\myfigpath/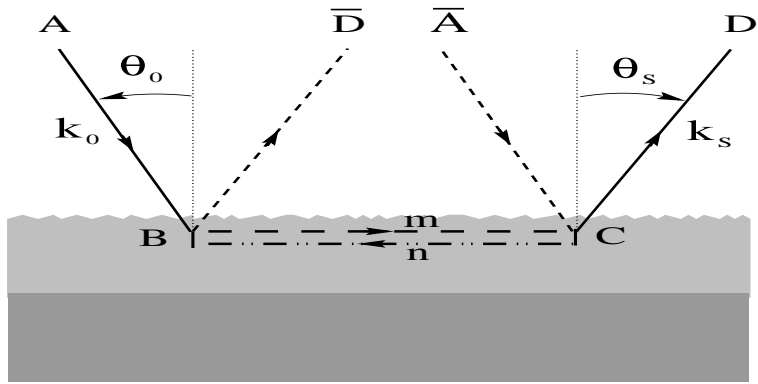}
    \caption{Illustration of two double   scattering sequences occurring
      in the scattering of electromagnetic waves from a bounded
      systems that supports more the one guided (or surface) wave. }
    \label{Fig:Phen:scat-path} 
  \end{center}
\end{figure}


In Fig.~\ref{Fig:Phen:scat-path} we show two general double scattering
paths\footnote{We here consider double scattering for simplicity. Higher
order scattering processes can be treated the same way, but doing so will
not change the main conclusions.} where the scattering takes place at the
same scattering centers, but in the reverse order. Such paths will in
general be phase incoherent due to the randomness of the rough surface.
However, we will now   looking into if there are particular angles of
incidence and scattering for which these two paths are phase coherent.  Let
us start by assume that path $ABCD$ goes through channel $m$ and path
$\bar{A}CB\bar{D}$ through the $n$-channel. The phase difference between
these two paths can then be expressed as
\begin{eqnarray}
  \label{Eq:Phen:Phase-diff}
  \Delta\phi_{nm} &=&
  {\mathbf r}_{BC} \cdot 
         \left({\mathbf k}_0 + {\mathbf k}_{s}\right)
  + \left| {\mathbf r}_{BC} \right| 
          \left[ q_n(\omega) - q_m(\omega) \right].
\end{eqnarray}
Here ${\mathbf k}_0$ and ${\mathbf k}_s$ are the wave vectors of the
incident and scattered waves, respectively, while ${\mathbf r}_{ BC}$
is the distance (vector) from point $B$ to $C$. According to its
definition, we will have phase coherence when this phase-difference is
zero, {\it i.e.}  when $\Delta\phi_{nm}=0$. Now let us consider
separately two cases: ({\it i}) $n=m$ and ({\it ii}) $n\neq m$. In the
first case the last term in Eq.~\r{Eq:Phen:Phase-diff} is zero with
the consequence that one has phase coherence if ${\mathbf k}_{s}=-
{\mathbf k}_{0}$. This coherence is obviously what gives rise to the
enhanced backscattering phenomenon. In the second case when $m\neq n$,
the last part of Eq.~\r{Eq:Phen:Phase-diff} does not vanish. The
condition for phase coherence then becomes
\begin{eqnarray}
  \label{Eq:Phen:Angels}
  \sin\theta_s &=&
   -\sin\theta_0 
    \pm \frac{1}{\sqrt{\eps{0}}} \frac{c}{\omega} 
     \left| q_n(\omega) - q_m(\omega) \right|.
\end{eqnarray}
In this equation we have also allowed for the case $m=n$ since it
naturally includes the position of the backscattering peaks. Hence,
Eq.~\r{Eq:Phen:Angels} defines the angles for which peaks due to
coherent effects are expected in the angular dependence of the light
scattered incoherently from the randomly rough surface. The angle
obtained for $m=n$ is the position of the backscattering peak, while
the angles obtained for $m\neq n$ correspond to satellite peaks.  The
reader should check that the angles obtained from
Eq.~\r{Eq:Phen:Angels} fit the position of the satellite peaks shown
in Fig.~\ref{Fig:Phen:SP}. The values for $q_n(\omega)$ can 
be found in Refs.~\citen{Madrazo} and \citen{MP13}.

This concludes our discussion of coherent effects in the scattered
field.  Even though we have focused on the reflected light, it should
be pointed out that there also exist similar effects in the
transmitted light~\cite{McGurn89}. For a discussion of this case the
reader is referred to the literature for details~\cite{PhysRep,McGurn89}.


\subsection{Localization}

The notion of localization was introduced into physics by
P.W.~Anderson in his famous 1958 paper~\cite{Anderson1958}, a work
that he was awarded the Nobel Prize for. Anderson studied the
transport properties of electrons in materials with bulk disorder.
This study led him to what today is known as the Anderson localization
phenomenon, sometimes also called strong localization.  The phenomenon
expresses itself by a disorder-induced phase transition in the
transport behavior of the electrons. As the disorder is increased in a
three dimensional system, the scattering evolves from a diffusion
regime, for which the well-known Ohm's law holds, to a localized
regime in which the material behaves as an insulator and all states
are localized in space.  These two phases are separated by the
mobility edge. Anderson suggested~\cite{Anderson1958} that the
localization regime was caused by strong interference that resulted in
an exponential decay of the wave function of the electrons in all
directions and a subsequent vanishment of the diffusion constant.
Hence localization is a multiple scattering phenomenon.  In
contrast to what is the case for three dimensional systems, all states
are expected to be localized for systems that are one- and
two-dimensional~\cite{Book:Sheng1995}. However, in this latter case it
might happen that the localization length is large, and even exceeds the
sample size.  For a more detailed introduction to localization the
reader is directed to Refs.~\citen{Book:Sheng1995} and
\citen{Book:Sheng1990}.

It was realized shortly after Anderson published his pioneering
work~\cite{Anderson1958}, that a similar phenomenon should also be
expected for multiple-scattering of electromagnetic waves. At room
temperature, photons can be treaded as non-interacting. They are
therefore not hampered by the troublesome self-interaction that
electrons have and that are known to represent another, but different
mechanism towards an insulator regime~\cite{Mott1990}.  This fact
makes photon disordered systems ideal for studying Anderson
localization~\cite{Lagendijk1996}.  However, it should still take
several decades before localization of electromagnetic waves was
confirmed experimentally.  Finally in 1997 Wiersma, Bartolini,
Lagendijk, and Righini were able to obtain direct experimental
evidence that confirmed the localization hypothesis for
electromagnetic waves in disordered
media~\cite{Localization_of_light}. These experiments were performed
on a system containing very strongly scattering semiconductor powders.
Thus, the scattering system involved was of the bulk disordered type.

It is still, however, an open question if Anderson localization of
electromagnetic waves can be observed experimentally for systems
containing only surface disorder, even though it should exist in
principle due to the system being two-dimensional (and in some cases
effectively one-dimensional). Since there is only disorder on the
surface, localization can only exist for electromagnetic waves that
happen to``live'' on or close to the surface. Such waves are called
surface waves, and we will here focus on surface plasmon
polaritons~(SPP) that might exist on {\it e.g.} a vacuum-metal
interface.  SPP localization should be characterized by the
exponential decay of the transmitted intensity as a function of
distance traveled by the SPP along the rough surface.

However, the problem of observing SPP localization for surface
disordered systems is that Anderson localization might be masked by
more dominating effects giving rise to the same type of signature as
localization itself --- the decay
$\exp\left(-L/\ell_T(\omega)\right)$ with system size $L$ of the
transmittance where $\ell_T(\omega)$ is a decay length.  The competing
effects are in addition to the Anderson localization: ({\it i})
ohmic losses in the metal due to $Im\,\varepsilon(\omega) \neq 0$, and
({\it ii}) roughness-induced conversion of surface plasmon polaritons
into volume waves above the surface --- so-called {\em leakage}. Hence the
decay length, $\ell_T(\omega)$, of the transmission coefficient (that
we will define below), should be related to the decay length due to
ohmic losses, $\ell_\epsilon(\omega)$, the one due to leakage,
$\ell_{rad}(\omega)$, and the Anderson localization length
$\ell(\omega)$, according to the formulae
\begin{eqnarray}
  \label{Eq:Loc:Total-length}
  \frac{1}{\ell_T(\omega )} 
  &=&
    \frac{1}{\ell (\omega )} 
  + \frac{1}{\ell_{rad}(\omega )} 
  + \frac{1}{\ell_{\varepsilon}(\omega )} .
\end{eqnarray}   
In order to determine the Anderson localization length one has to sort out
the contribution from each of these competing effects. In other words,
we have to identify the lengths $\ell_{\varepsilon}(\omega)$ and
$\ell_{rad}(\omega )$ in order to be able to estimate $\ell(\omega)$.


\begin{figure}[t!]
  \begin{center}
    \leavevmode
    \includegraphics[width=9cm]{\myfigpath/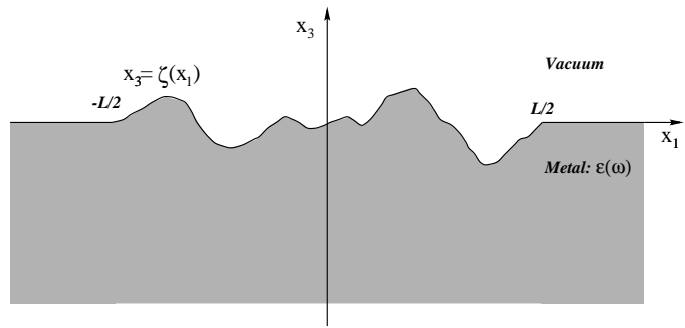}
    \caption{The scattering system considered in the study of Anderson
      localization of surface plasmon polaritons on a rough surface.}
    \label{Fig:SPP-localization-geometry}
  \end{center}
\end{figure}


The decay rate due to ohmic losses is easily determined and doesn't
represent any serious problem~(see discussion below). However, a much
more severe problem is how to separate the contribution from leakage
and localization. Leakage is expected to be a rather strong effect
with the consequence that $\ell_{rad}(\omega )$ is small compared to
the other lengths appearing on the right-hand-side of
Eq.~\r{Eq:Loc:Total-length}. If this is the case, a measurement of the
transmission coefficient for the SPP will result in
$\ell_T(\omega)\simeq \ell_{rad}(\omega).$ It is therefore important,
if we are trying to estimate $\ell (\omega)$, to be able to separate
the localization length from the one of leakage, or to be able to
suppress leakage.  The approach we will follow here is the latter one
--- the suppression of leakage. This can be done by specially designing
surfaces that suppress leakage.  How this can be done will
be presented briefly below~(see Ref.~\citen{MP2} for
more details).

\subsubsection{The Scattering System}

The scattering system that will be considered in this section is
depicted in Fig.~\ref{Fig:SPP-localization-geometry}.  We study the
scattering of a p-polarized surface plasmon polariton of frequency
$\omega$ propagating in the positive $x_1$-direction and is incident
onto a segment of a one-dimensional randomly rough surface defined by
the equation $x_3 = \zeta(x_1)$.  The surface profile function
$\zeta(x_1)$ is assumed to be a single-valued function of $x_1$ that
is nonzero only in the interval $-L/2 < x_1 < L/2$. The region $x_3 >
\zeta(x_1)$ is vacuum; the region $x_3 < \zeta(x_1)$ is a metal
characterized by an isotropic, frequency-dependent, complex dielectric
function $\eps{} = \eps{1}+i\eps{2}$.  We are interested in the
frequency range in which $\eps{1} < -1, \eps{2}> 0$, within which
surface plasmon polaritons exist.  Furthermore, the rough portion of
the surface is assumed to constitute a Gaussian random process and
with the other ``standard'' properties described in
Sect.~\ref{Sect:Stat-surf}.


\begin{figure}[b!]
  \begin{center}
    \leavevmode
    \includegraphics[width=12cm]{\myfigpath/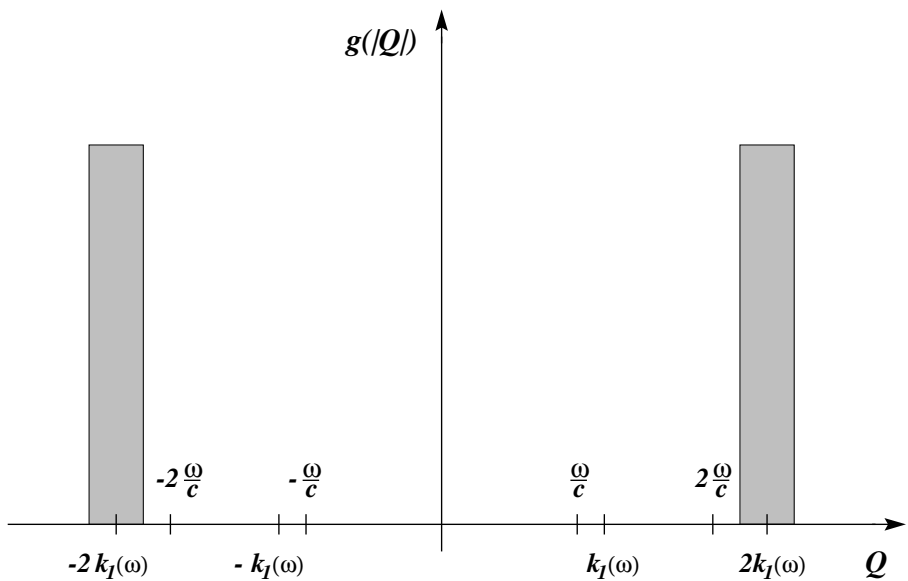}
    \caption{A sketch of the power-spectrum used in order to suppress
      leakage. See text for details}
    \label{Fig:SPP-R_power-spectrum}
  \end{center}
\end{figure}


\subsubsection{Surfaces that Suppress Leakage}

The first step towards the estimation of the Anderson localization
length for this scattering system is to construct surfaces that
suppress leakage. We recall that the incident surface plasmon
polariton has a wave vector given by
\begin{eqnarray}
  k_{sp}(\omega) &=& 
      \ooc{} \left[ \frac{\eps{}}{\eps{} + 1}\right]^{\frac{1}{2}}
           = k_1(\omega ) + ik_2(\omega ),
\end{eqnarray}
where $k_1(\omega )$ and $k_2(\omega )$ are the real and imaginary
part of the (complex) SPP wave vector and defined explicitly
in Ref.~\citen{MP2}.

By interaction with the surface roughness, the incident SPP picks up
momenta available in the power-spectrum of the roughness, and due to
scattering, changes its wave vector into $q$. If this momenta
satisfies $|q|\leq \omega/c$, leakage has occurred. To prevent this, or
at least reduce this effect, we might use an intelligently choice for
the power-spectrum.  Such a choice is provided by a
(rectangular)~West-O'Donnell power-spectrum of width $2\Delta k$
located around $\pm k_1(\omega)$~(see
Fig.~\ref{Fig:SPP-R_power-spectrum}), {\it i.e.} a power-spectrum of
the form
\begin{eqnarray}
  \label{Eq:Loc:power}
  g(|Q|)  = \frac{\pi}{2\Delta k} [\theta (Q-k_{-})\theta (k_{+}-Q) 
  + \theta (-Q-k_{-})\theta (k_{+}+Q)] ,
\end{eqnarray}
where $k_\pm = 2k_1(\omega ) \pm \Delta k$ and $\Delta k$ must satisfy
the inequality $\Delta k < k_1(\omega ) - \ooc{}$.

That a surface characterized by the power spectrum~\r{Eq:Loc:power}
suppresses leakage can be seen from the following argument: The
incident surface plasmon polariton has a wave number whose real part
is $k_1(\omega )$.  After its first interaction with the surface
roughness the real part of its wave number will lie in the two
intervals $(3k_1(\omega ) - \Delta k, 3k_1(\omega ) + \Delta k)$ and
$(-k_1(\omega )-\Delta k,-k_1(\omega )+\Delta k)$.  For the same
reason, after its second interaction with the surface roughness the
real part of the wave number of the surface plasmon polariton will lie
in the three intervals $(5k_1(\omega ) - 2\Delta k, 5k_1(\omega )
+2\Delta k)$, $(k_1(\omega ) - 2\Delta k, k_1(\omega ) + 2\Delta k)$,
and $(-3k_1(\omega )-2\Delta k, -3k_1(\omega ) + 2\Delta k)$.  After
three interactions with the surface roughness the real part of its
wave number will lie in the four intervals $(7k_1(\omega)-3\Delta k,
7k_1(\omega ) + 3\Delta k)$, $(3k_1(\omega )-3\Delta k, 3k_1(\omega
)+3\Delta k)$, $(-k_1(\omega ) - 3\Delta k, -k_1(\omega ) + 3\Delta
k)$, and $(-5k_1(\omega ) - 3\Delta k,-5k_1(\omega )+3\Delta k)$, and
so on.  Thus, for example, if $-k_1(\omega )+3\Delta k < -(\omega
/c)$, so that $\Delta k < \frac{1}{3} (k_1(\omega ) - (\omega /c))$,
after three scattering processes the surface plasmon polariton will
not have been converted into volume electromagnetic waves.  In
general, if we wish the surface plasmon polariton to be scattered $n$
times from the surface roughness without being converted into volume
electromagnetic waves, we must require that
\begin{eqnarray}
  \Delta k < \frac{1}{n} (k_1(\omega ) - (\omega /c)).
\end{eqnarray}


\begin{figure}[t!]
  \begin{center}
    \leavevmode
    \includegraphics[height=7cm,width=11cm]{\myfigpath/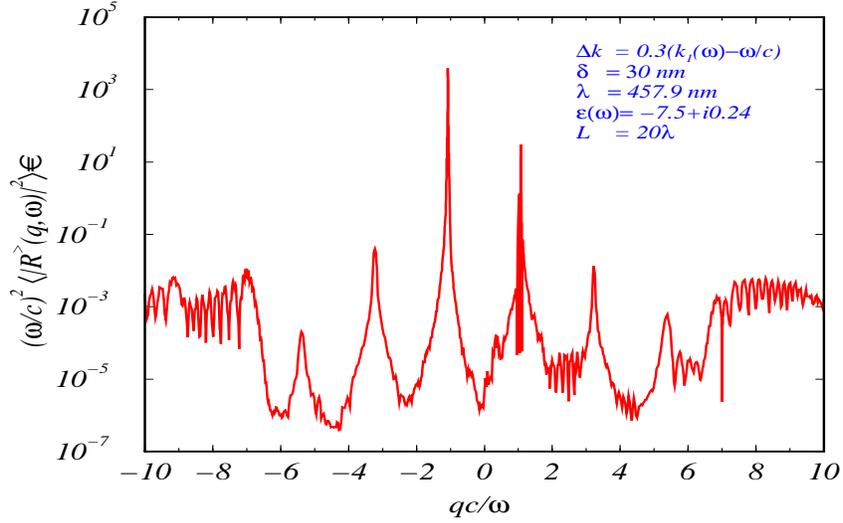}
    \caption{Numerical calculations for
      $(\omega /c)^2\left<|R(q|k)|^2\right>$ as a function of
      $cq/\omega$ for a rough silver surface characterized by the
      parameters $\Delta k = 0.3 (k_1(\omega )-(\omega/c))$ and
      $\delta = 30 {\rm nm} $.  The rough portion of the surface had
      length $L=20\lambda$.  The frequency of the surface plasmon
      polariton, $k(\omega)=k_1(\omega)+ik_2(\omega)=(1.0741+i
      0.0026)\omega/c$, corresponds to a vacuum wavelength of $\lambda
      = 457.9 {\rm nm}$, and the dielectric constant of silver at this
      frequency was $\eps{} = -7.5 + i0.24$.  The results for 50
      realizations of the surface profile function were averaged to
      obtain the results plotted in this figure. }
    \label{Fig:SPP-R_power_2}
  \end{center}
\end{figure}


To illustrate that the above procedure really works, we present in
Fig.~\ref{Fig:SPP-R_power_2} numerical simulation results for the scattering
amplitude above the surface, $(\omega /c)^2\left< |R^>(q,\omega )|^2\right>$
as a function of $cq/\omega$ for a silver surface characterized by the
parameters $\Delta k = 0.3 (k_1(\omega )-(\omega/c))$ and $\delta = 30 {\rm
nm} $. This surface should thus suppress leakage up to and including third
order scattering processes. We see from Fig.~\ref{Fig:SPP-R_power_2} that
$\left< |R^>(q,\omega )|^2\right>$ indeed becomes heavily suppressed in the
range $-(\omega/c) < q < (\omega/c)$.  Notice that the six peaks seen in
Fig.~\ref{Fig:SPP-R_power_2} correspond to the real parts of the wavenumbers
of the scattered surface plasmon polaritons resulting from the scattering of
an incident surface plasmon polariton of wave vector
$k(\omega)=k_1(\omega)+ik_2(\omega)=(1.0741+i 0.0026)\omega/c$.


\begin{figure}[t!]
  \begin{center}
    \leavevmode
    \includegraphics[width=11cm]{\myfigpath/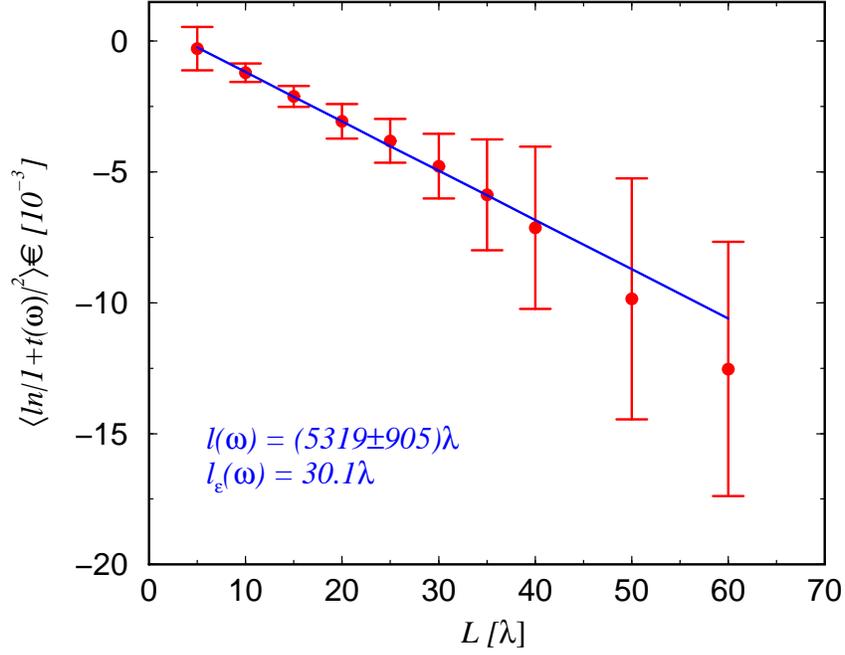}
    \caption{Numerical simulation results for $\left<\ln
        \left|1+t(\omega)\right|^2\right>$ vs. the length of the rough
      portion of the surface $L$. The remaining surface parameters
      were the same used in Fig.~\protect\ref{Fig:SPP-R_power_2}.  The
      number of realizations used for the rough surface in calculating
      the ensemble average was for each system size of the order of
      $10^2$. The error-bars indicate the spread in
      $\left|1+t(\omega)\right|^2$ that was part of this average.  The
      solid line, which slope is related to the Anderson localization
      length, $\ell(\omega)$, is a $\chi^2$-fit to the numerical data
      corresponding to a localization length of $\ell(\omega) =
      (5319\pm905) \,\lambda$. The length due to ohmic losses in the
      metal is $\ell_\varepsilon(\omega)=30.1\lambda$.}
    \label{Fig:SPP-lnT}
  \end{center}
\end{figure}


\subsubsection{The Anderson Localization Length for Surface Plasmon
  Polaritons Localized on a Randomly Rough Surface}

The transmission coefficient for surface plasmon polariton, $T(L)$, is
defined as the fraction of the energy flux entering the random segment
of the metal surface at $x_1=-L/2$ and that leaves it at $x_1=L/2$,
{\it i.e.}
\begin{eqnarray}
  \label{Eq:Loc:def-transmission}
  T(L) &=& \frac{P_{\rm tr}\left(\frac{L}{2}\right)}{
    P_{\rm inc}\left(-\frac{L}{2}\right)},
\end{eqnarray}
where $P_{\rm inc}(x_1)$ and $P_{\rm tr}(x_1)$ denote the incident and
transmitted flux at position $x_1$. 

Above the surface the field can be written as~\cite{MP2}
\begin{subequations}  
  \begin{eqnarray}
    H^>_2(x_1,x_3|\omega) &=& \int^{\infty}_{-\infty}
                               \frac{dq}{2\pi}\, G_0(q,\omega) T(q,\omega)
                               e^{iqx_1+i\alpha_0(q,\omega)x_3}
    \nn \\
          &\sim&  
                t(\omega) e^{ik_{sp}(\omega)x_1-\beta_0(\omega)x_3}, \quad 
                 x_1 \gg L/2,                    
  \end{eqnarray}      
  with $\beta_0(\omega) =i\alpha_0(k_{sp}(\omega),\omega)$ and
  \begin{eqnarray}
    \label{Eq:log:t}
    t(\omega) &=& i \frac{(-\varepsilon(\omega))^{3/2}}{\varepsilon^2(\omega)-1}
                  T(k_{sp}(\omega),\omega).   
  \end{eqnarray}
\end{subequations}  
Thus the transmission coefficient can alternatively be written as~\cite{MP2}  
\begin{eqnarray}
  \label{Eq:Loc:def-transmission-rewritten}
  T(L) &=& \left|1+t(\omega)\right|^2
   \exp\left(-\frac{L}{\ell_\varepsilon(\omega)}\right)
\end{eqnarray}
where $\ell_\varepsilon(\omega)=1/(2k_2(\omega))$ is the SPP mean free
path due to ohmic losses. Notice that
Eq.~\r{Eq:Loc:def-transmission-rewritten} separates the contribution
due to ohmic losses from the one of Anderson localization (and
leakage).

The quantity that we will be interested in when studying the
localization phenomenon is not the transmission coefficient itself, but
instead $\ln T(L)$ or more precisely  its (ensemble) average given by
\begin{eqnarray}
  \label{Eq:Phen:loc-length}
  \left<\ln T(L)\right>  &=&
  \left<\ln \left|1+t(\omega)\right|^2\right> 
           -\frac{L}{\ell_\epsilon(\omega)}.
\end{eqnarray}
If the effect of leakage can be neglected, the first term on the
right-hand-side of the above equation will {\em only} incorporate
contributions from (Anderson) localization. One therefore writes
\begin{eqnarray}
  \left<\ln \left|1+t(\omega)\right|^2\right>
   &=& const. -\frac{L}{\ell(\omega)}.
\end{eqnarray}
Hence, under the assumption that leakage can be neglected the
localization length, $\ell(\omega)$, can be obtained from a straight
line fit to $\left<\ln \left|1+t(\omega)\right|^2\right>$ vs. system
size $L$.

In Fig.~\ref{Fig:SPP-lnT} we present such a plot resulting from
numerical simulations using the same surface parameters that lead to
the results shown in Fig.~\ref{Fig:SPP-R_power_2} except that now the
length of the rough portion of the surface was different. These
simulations were performed on the basis of the reduced Rayleigh
equation that $T(q,\omega)$ satisfies~\cite{MP2},
\begin{eqnarray}
   T(p,\omega) &=& V(p|k_{sp}(\omega))
         + \int^{\infty}_{-\infty} \frac{dq}{2\pi}\; V(p|q)
         G_0(q,\omega) T(q,\omega),
\end{eqnarray}
where $V(p|k_{sp}(\omega))$ is the scattering potential defined in
Ref.~\citen{MP2}, and from which $t(\omega)$ can be calculated
according to Eq.~\r{Eq:log:t}. Notice that to perform such a
calculation of $t(\omega)$ is not at all straight forward. The reason
being that the reduced Rayleigh equation, through the Green's function
$G_0(q,\omega)$, has poles at $q=\pm k_{sp}(\omega)$. Details on the
numerical method used for such calculations can be found in
Ref.~\citen{MP2} and will therefore not be given here.

As can be see from Fig.~\ref{Fig:SPP-lnT}, $\left<\ln
  \left|1+t(\omega)\right|^2\right>$ is consistent, within the
error bars, with a linear dependence on $L$. The solid line in this figure
is a $\chi^2$-fit to the numerical data. The slope of this curve is
according to Eq.~\r{Eq:Phen:loc-length} related to the Anderson
localization length, $\ell(\omega)$, as $1/\ell(\omega)$.  Numerically
the Anderson localization length for our system is found to be 
\begin{eqnarray}
  \ell(\omega) &=& (5319\pm905) \,\lambda.
\end{eqnarray}
This length should be compared to the one due to ohmic losses, which
for our set of parameters is $\ell_{\varepsilon}(\omega) =
30.1\,\lambda$.

We have thus shown theoretically that by using specially designed
surfaces, there might be hopes of observing the localization of
surface plasmon polaritons at a randomly rough metal surface.


\sectionmark{Angular Intensity Correlations}
\subsection{Angular Intensity Correlations for the Scattered Light from
  Randomly Rough Surfaces}
\sectionmark{Angular Intensity Correlations}


It has been known for quite some time that when electromagnetic waves, all
of the same frequency, are scattered from a random system, speckle patterns
might be observed~\cite{Book:Dainty75}. Such patterns are results of
interference between waves scattered from different locations in the random
medium.  From studies off volume disordered systems, such patterns are known
to contain a rather rich structure~\cite{Book:Dainty75,Corr16}. In
particular, it was predicted theoretically~\cite{Corr16} for such scattering
systems that there should exist three types of correlations --- short-range
correlations, long-range correlations, and infinitely-range correlations.
These correlations were termed the $C^{(1)}$, $C^{(2)}$, and
$C^{(3)}$-correlations, respectively, and they have all been observed
experimentally~\cite{Corr20,Corr19,Scheffold98}.

\begin{figure}[t!] 
  \vspace*{1cm}                 
  \begin{center}
    \leavevmode
    \includegraphics*[width=14cm,height=11cm]{\myfigpath/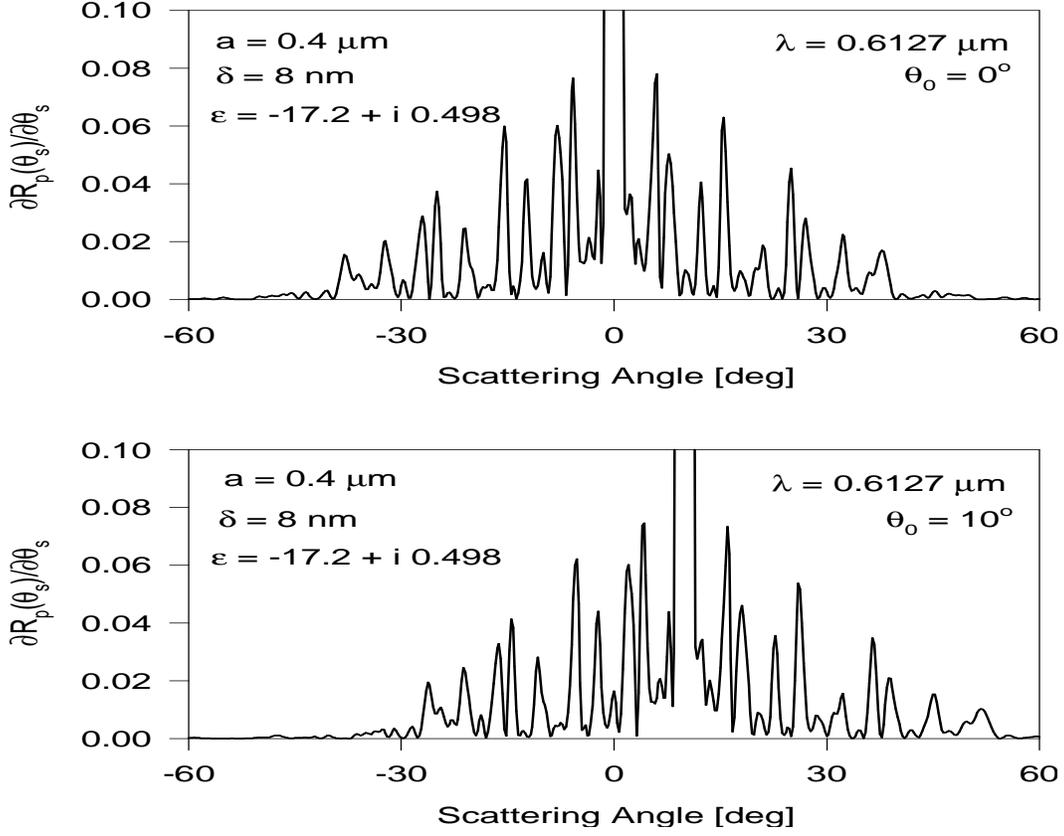}
    \caption{Speckle patterns that result from the scattering of
      light of wavelength $\lambda =612.7 {\rm nm}$ incident on a
      rough silver surface at angles (a)~$\theta_0=0^\circ$ and
      (b)~$\theta_0=10^\circ$. The Gaussian height-distributed surface
      was characterized by an {\sc rms}-height $\delta = 10 {\rm nm}$
      and a (Gaussian) correlation length $a=0.4 {\rm \mu m}$. The
      length of the surface was $L=100 {\rm \mu m}$ and the dielectric
      constant of silver at the wavelength of the incident light is
      $\eps{}=-17.2+i0.498$.}
    \label{Fig:PHEN:Speckle}
  \end{center}
\end{figure}

In this section we will discuss speckle correlations, not for light
scattered from volume disordered systems, but instead for light
scattered from randomly rough surfaces.  Examples of such speckle
patterns obtained when an electromagnetic wave is scattered from a
randomly rough surface are shown in Fig.~\ref{Fig:PHEN:Speckle}.


Let us start by considering a planar surface separating two different
materials. Since the surface is planar, the scattering is completely
understood as expressed through the celebrated Fresnel's
formulae~\cite{Book:Born,Book:Kong}. Imagine an experiment where
light is incident at an angle $\theta_0$ onto the interface.  Since
the surface is planar, all the light is scattered into the specular
direction $\theta_s=\theta_0$, and its intensity is given by Fresnel
formula. If we in a second experiment incident the light at an angle
$\theta_0'=-\theta_0$, {\it i.e.} at an angle that was the specular
direction in the first experiment, all the light will be scattered
into $\theta_s'=\theta_0'=-\theta_0$ and its intensity is again given
by Fresnel's formula.  The scattered intensities in these two
experiments are in fact equal. This is easily realized from the
Fresnel formulae~\r{Theory:Fresnel} by noting that the
$\alpha$-factors that they contain are unaffected by a change of sign
in the momentum variables.  Thus, if we know the result of the first
experiment, say, we also know the outcome of the second one.  In other
words, these two intensities are perfectly correlated.  We now
introduce the momentum variables $q$ and $k$ related in the usual way
to the scattering and incident angles respectively~(see {\em e.g.}
Eq.~\r{PHEN:Eq:mom-angle} below).  Let the notation $(q,k)$ denotes a
corresponding pair of momenta variables where $q$ is the scattered
momentum and $k$ the incident one.  For our planar surface geometry we
will thus have perfect correlation between the two scattering
processes $(q,k)$ and $(q',k')$ if $(q,k)=(-k',-q')$.  Furthermore,
since any process, of course, is correlated with itself, we in
addition will expect perfect correlation when $(q,k)=(q',k')$.

The above example is rather trivial and well-known example of
correlations in the scattered intensity from a planar surface.
However, what happens to the intensity correlations if the surface is
not planar, but instead randomly rough? This is an interesting and
non-trivial question and we will address it in this section.  In the
discussion to be presented below we will be focusing on the angular
correlations in the light scattered {\em incoherently} from a rough
surface.  Furthermore, we will mainly discuss the case where the
surface is weakly rough. In particular we will try to answer the
following question: When and under which conditions will the
intensity scattered (incoherent) into the far field for different
incident and scattering angles be related to one and other?


\subsubsection{Definition of the Angular Intensity Correlation
  Functions}

\begin{figure}[t!]
  \begin{center}
    \leavevmode
    \includegraphics[height=4.5cm,width=9cm]{\myfigpath/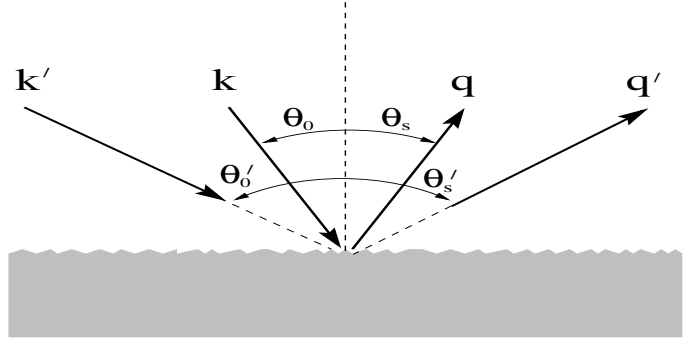}
    \caption{The scattering system considered in the study of the angular
      correlation functions. }
    \label{Fig:PHENOM:Correlation-angles}
  \end{center}
\end{figure}

Let us start by introducing the {\em unnormalized} angular correlation
function $ C(q,k|q^{\prime },k^{\prime })$, which we will define
as\footnote{We have here suppressed any explicit reference to the
  polarization (the $\nu$-index) since no confusion should result from
  doing so. All quantities in this section should be understood to be
  referring to one and the same polarization.}
\begin{eqnarray}
  \label{PHEN:Eq:Corr-def-unnorm}
  C(q,k|q^{\prime },k^{\prime })  &=&
   \left< I(q|k)I(q^{\prime }|k^{\prime})\right> 
      - \left< I(q|k)\right> \left< I(q^{\prime }|k^{\prime })\right> ,
\end{eqnarray}
where $I(q|k)$ denotes the intensity of the light scattered from the
surface, and the angle brackets denote an average taken over an
ensemble of realizations of the surface profile function $\zeta(x_1)$.
Furthermore, the (lateral) momentum variables, $q$ and $k$, are both,
in the radiative region ($|q|\leq \sqrt{\varepsilon_0}\omega/c$), understood
to be related to the scattering and incident angles $\theta_s$ and
$\theta_0$ respectively according to
\begin{eqnarray}
  \label{PHEN:Eq:mom-angle}
   k = \sqrt{\eps{0}}\, \ooc{} \sin\theta_0, \qquad 
   q = \sqrt{\eps{0}}\, \ooc{} \sin\theta_s.
\end{eqnarray}
The primed momentum variables, $q^\prime$ and $k'$, are related in a
similar way to the primed angles $\theta_s'$ and $\theta_0'$. Theses
angles, both primed and unprimed, are defined positive according to
the convention indicated in Fig.~\ref{Fig:PHENOM:Correlation-angles}.
This figure also shows our scattering system consisting of a semi-infinite
dielectric medium with a rough interface to vacuum.


Furthermore, the intensity $I(q|k)$ can be defined through the
scattering matrix $S(q|k)$ according to the formula
\begin{eqnarray}
  \label{PHEN:Eq:Intensity}
  I(q|k) &=& \frac{ \sqrt{\eps{0}} }{L_1}\left( \ooc{} \right) 
                \left|S(q|k)\right|^2,
\end{eqnarray}
where $L_1$ is the length of the $x_1$-axis covered by the random
surface.

In many cases it is convenient to work with a normalized correlation
function, $\Xi (q,k|q^{\prime },k^{\prime })$, in contrast to the
unnormalized one. The normalized angular intensity correlation function
will we define by\footnote{It should be noticed that a somewhat
  different definition for the normalized angular intensity
  correlation function is used by some authors~\protect\cite{Corr12}.
  However, the advantage of the
  definition~\protect\r{PHEN:Eq:Corr-def-norm} is that it does not
  contain any $\delta$-functions in the denominator.}
\begin{eqnarray}
  \label{PHEN:Eq:Corr-def-norm}
  \Xi (q,k|q^{\prime },k^{\prime }) &=&
     \frac{\langle I(q|k)I(q^{\prime}|k^{\prime })\rangle 
   -\langle I(q|k)\rangle \langle I(q^{\prime }|k^{\prime})\rangle 
       }{
    \langle I(q|k)\rangle \langle I(q^{\prime }|k^{\prime })\rangle }.
\end{eqnarray}

The lesson to be learned from the huge amount of research being
conducted on correlation function in the field of random (bulk)
disordered systems~\cite{Corr16,Corr17,Corr18,Corr19,Corr20} is that
there may exist correlations on many different length scales including
{\em short} to {\em infinite} range correlations.  Thus part of the
challenge we are facing will be to separate these different
contribution to $C(q,k|q^{\prime },k^{\prime })$ (or equivalently to
$\Xi(q,k|q^{\prime },k^{\prime })$) from one another.

The first step towards such a separation is to rewrite the correlation
function in terms of the S-matrix.  This is done by substituting the
expression for the intensity, Eq.~\r{PHEN:Eq:Intensity}, into the
definition of the correlation function and thus obtaining
\begin{equation}
  \label{Eq:Phen:Rewritten}
    C(q,k|q^{\prime },k^{\prime }) = 
     \frac{\varepsilon_0}{L_1^2} \ooc{2}
          \left[
             \left< 
               \left|S(q|k)\right|^2
               \left|S(q^{\prime }|k^{\prime})\right|^2
              \right> 
             -
              \left<\left|S(q|k)\right|^2\right> 
              \left<\left|S(q'|k')\right|^2\right> 
          \right]. \nn \\    
\end{equation}
Due to the stationarity of the surface profile function, the average
of the S-matrix should be diagonal in $q$ and $k$,
\begin{eqnarray}
  \label{Eq:Phen:avr-S-matrix}
  \left< S(q|k)\right> &=& 2\pi \delta(q-k)\, S(k).
\end{eqnarray}
By now taking advantage of this relation in addition to the cumulant
average~\cite{Corr21,Book:StatMech}
\begin{eqnarray}
     \{AB\} &=& \left< AB \right> - \left<A\right>\left<B\right>,
\end{eqnarray}
the correlation function~\r{Eq:Phen:Rewritten} can be written as 
\begin{subequations}
\begin{eqnarray}
  \label{Eq:Phen:C-Rewritten}
  C(q,k|q^{\prime },k^{\prime }) &=& 
  \frac{\varepsilon_0}{L_1^2} \ooc{2} \,
  \left[
      \left| \left< 
            \delta S(q|k) \delta S^*(q'|k')
      \right> \right|^2
    + 
      \left| \left< 
            \delta S(q|k) \delta S(q'|k')
      \right> \right|^2
    \nn \right. \\ && \mbox{} \left.\qquad \quad
     + \left\{
                \delta S(q|k) \delta S^{*}(q|k)
                 \delta S(q^{\prime }|k^{\prime })
                 \delta S^{*}(q^{\prime }|k^{\prime})
       \right\}
   \right]
  + s.t. 
\end{eqnarray} 
where $\delta S(q|k)$ denotes the incoherent component of the S-matrix defined as 
\begin{eqnarray}
    \delta S(q|k) &=& S(q|k) - \left< S(q|k)\right>.
\end{eqnarray} 
\end{subequations}
In Eq.~\r{Eq:Phen:C-Rewritten} the asterisks denote complex conjugate
while $s.t.$ means specular terms, {\em i.e.} terms that are
proportional to $\delta (q-k)$ and/or $\delta (q^{\prime }-k^{\prime
  })$. Such terms will not be focused on here since we will concentrate
on the incoherent part of the scattered light. With
Eq.~\r{Eq:Phen:C-Rewritten} we can now write\footnote{Notice that
  equivalent expressions can be derived for the normalized correlation
  functions based on Eq.~\protect\r{PHEN:Eq:Corr-def-norm}.}
\begin{subequations}
  \label{PHEN:Eq:Rewritten}
  \begin{eqnarray}
    C (q,k|q^{\prime },k^{\prime }) &=&
    C^{(1)} (q,k|q^{\prime },k^{\prime }) +
    C^{(10)} (q,k|q^{\prime },k^{\prime }) +
    C^{(N)} (q,k|q^{\prime },k^{\prime }), \qquad 
    \nonumber \\  
  \end{eqnarray}
  where
  \begin{eqnarray}
      \label{PHEN:Eq:C1}
     C^{(1)}(q,k|q^{\prime },k^{\prime }) &=&
     \frac{\varepsilon_0}{L_1^2} \ooc{2} \,
        \left|\left< \delta S(q|k)
                     \delta S^{*}(q^{\prime}|k^{\prime})\right>\right|^2, \\
    \label{PHEN:Eq:C10}
    C^{(10)}(q,k|q^{\prime },k^{\prime }) &=&
\frac{\varepsilon_0}{L_1^2} \ooc{2} \,
         \left| \left< \delta S(q|k)
                       \delta S(q^{\prime}|k^{\prime })
         \right> \right|^2,
  \end{eqnarray}
  and 
  \begin{eqnarray}
    \label{PHEN:Eq:CN}
        C^{(N)}(q,k|q^{\prime },k^{\prime }) &=&
        \frac{\varepsilon_0}{L_1^2} \ooc{2} \,
           \left\{
                 \delta S(q|k) \delta S^{*}(q|k)
                 \delta S(q^{\prime }|k^{\prime })
                 \delta S^{*}(q^{\prime }|k^{\prime})
           \right\}. \qquad
  \end{eqnarray}
\end{subequations}
Due to reasons which should be clear from the discussion below, the
correlation functions in Eqs.~\r{PHEN:Eq:C1} and \r{PHEN:Eq:C10} are
termed {\em short-range} correlation functions, while the one in
Eq.~\r{PHEN:Eq:CN} contains contribution from {\em long} and {\em
  infinite-range} correlations. They will now be discussed in turn.



\subsubsection{Short Range Correlations for Weakly rough Surfaces}

In this subsection the short-range correlation functions, $C^{(1)}$
and $C^{(10)}$, for weakly rough surfaces will be discussed.  These
correlation functions are to leading order in the surface profile
function a result of single scattering processes~\cite{Corr10}.
However, above leading order they will also receive contributions from
multiple scattering. The long and infinite range correlations,
$C^{(N)}$, contain at least one multiple scattering process as we will
see~\cite{Corr10}.  Therefore the ``optical paths'' involved in the
processes leading to $C^{(1)}$ and $C^{(10)}$ are typically shorter
then those giving rise to $C^{(N)}$.  This is one of the reasons why
the $C^{(1)}$ and $C^{(10)}$ correlation functions are termed
short-range correlation functions.  Another reason stems from the fact
that $C^{(1)}$ and $C^{(10)}$ are both independent of the length of
the random surface.  In the next subsection we will demonstrate
explicitly that the $C^{(N)}$-correlation function is proportional to
$1/L_1$.  Hence, in the limit of a long surface the amplitude of the
correlation function $C^{(N)}$ is neglectable compared to $C^{(1)}$
and $C^{(10)}$.

\subsubsubsection{The $C^{(1)}$ Correlations Function; 
  The Memory- and Reciprocal Memory-Effect }

At first sight, the expressions in Eqs.~\r{PHEN:Eq:Rewritten} might
not seem too useful to us. However, they are as we now will try to
explain. We will only be concerned about one-dimensional random
surfaces, $\zeta(x_1)$, that are stationary and constitutes a Gaussian
random process. Under this assumption the expression $\langle \delta
S(q|k)\delta S^{*}(q^{\prime}|k^{\prime })\rangle$, contained in
$C^{(1)}$, will be proportional to a $\delta$-function, {\it i.e.}
\begin{eqnarray}
  \label{PHEN:Eq:t1_prop}
  \langle \delta S(q|k) 
          \delta S^{*}(q^{\prime}|k^{\prime })\rangle &\propto&
  2\pi\delta(q-k-q^{\prime}+k^{\prime }).  
\end{eqnarray}
This is so due to the stationarity of the surface profile function
$\zeta(x_1)$.  To motivate this we recalling from
Sect.~\ref{Chap:Theory}, or Ref.~\citen{Maradudin93}, that to lowest
order in the surface profile function the scattering amplitude that is
proportional to the S-matrix, is proportional to $\tilde{\zeta}(q-k)$,
where $\tilde{\zeta}$ denotes the Fourier transform of the surface
profile function.  Since
$\langle\tilde{\zeta}(q)\tilde{\zeta}^*(k)\rangle=2\pi\delta(q-k)$,
Eq.~\r{PHEN:Eq:t1_prop}, to lowest order, follows immediately.

Thus, with Eq.~\r{PHEN:Eq:t1_prop} we find that the correlation
function $C^{(1)}$ can be written in the convenient form
\begin{eqnarray}
  \label{PHEN:Eq:C1_prop}
  C^{(1)}(q,k|q^{\prime },k^{\prime }) &=&
        \frac{2\pi \delta(q-k-q^{\prime}+k^{\prime })}{L_1} \;
         C^{(1)}_0(q,k|q^{\prime },q'-q+k)  . \qquad
\end{eqnarray}
Here $C^{(1)}_0$ is known as the {\em envelope function} of $C^{(1)}$
and it is {\em independent} of the length $L_1$ of the surface.
Notice that the $C^{(1)}$-correlation function can only be
non-vanishing when the argument of the $\delta$-function vanishes.
Therefore, since $2\pi\delta(0)=L_1$, the (full) $C^{(1)}$-correlation
function is also independent of the length of the surface.


\begin{figure}[t!] 
  \begin{center}
    \leavevmode
    \begin{tabular}{@{}c@{\hspace{1.0cm}}c@{}}
      \includegraphics[height=4cm,width=2.5in]{\myfigpath/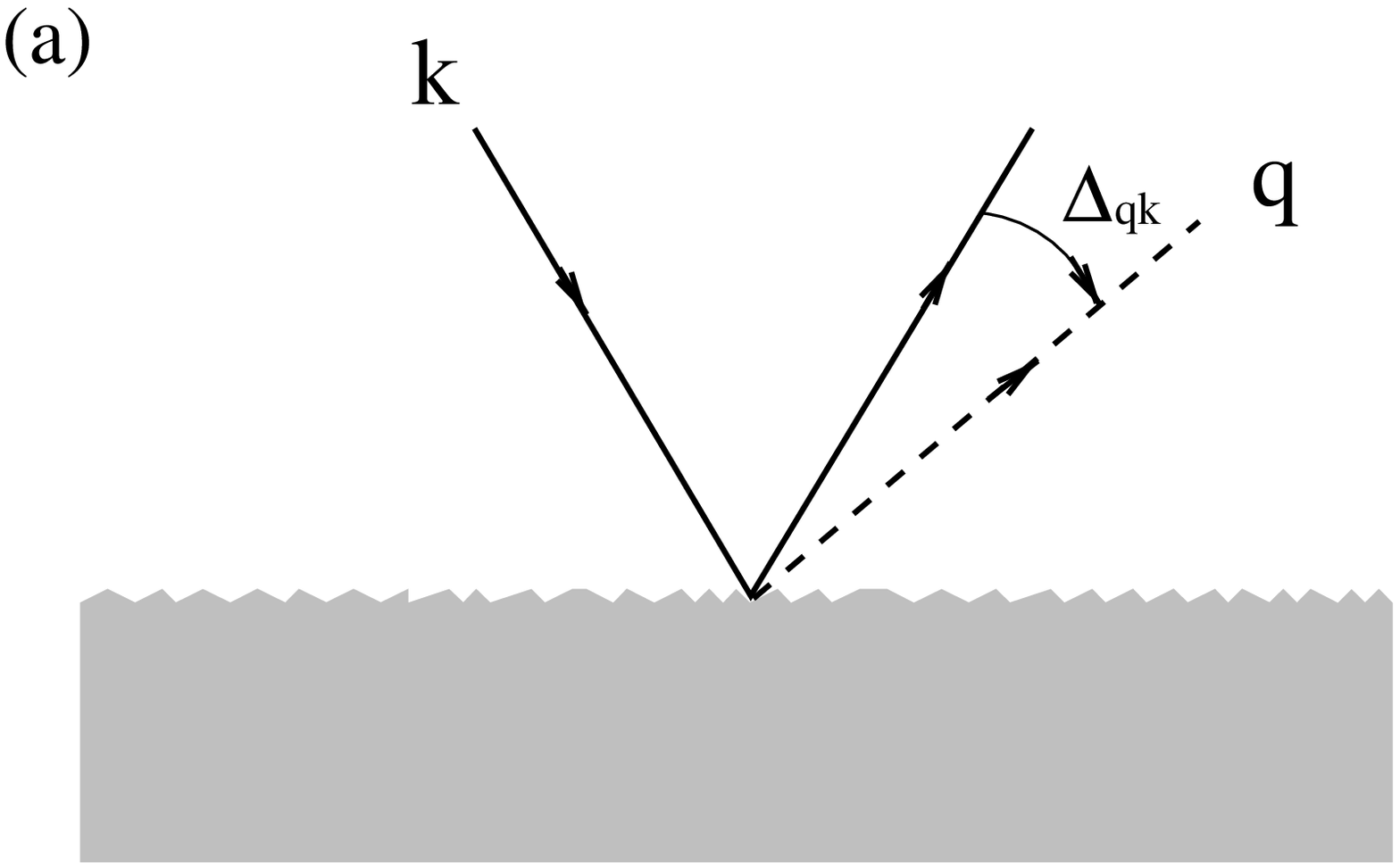} 
      & 
      \includegraphics[height=4cm,width=2.5in]{\myfigpath/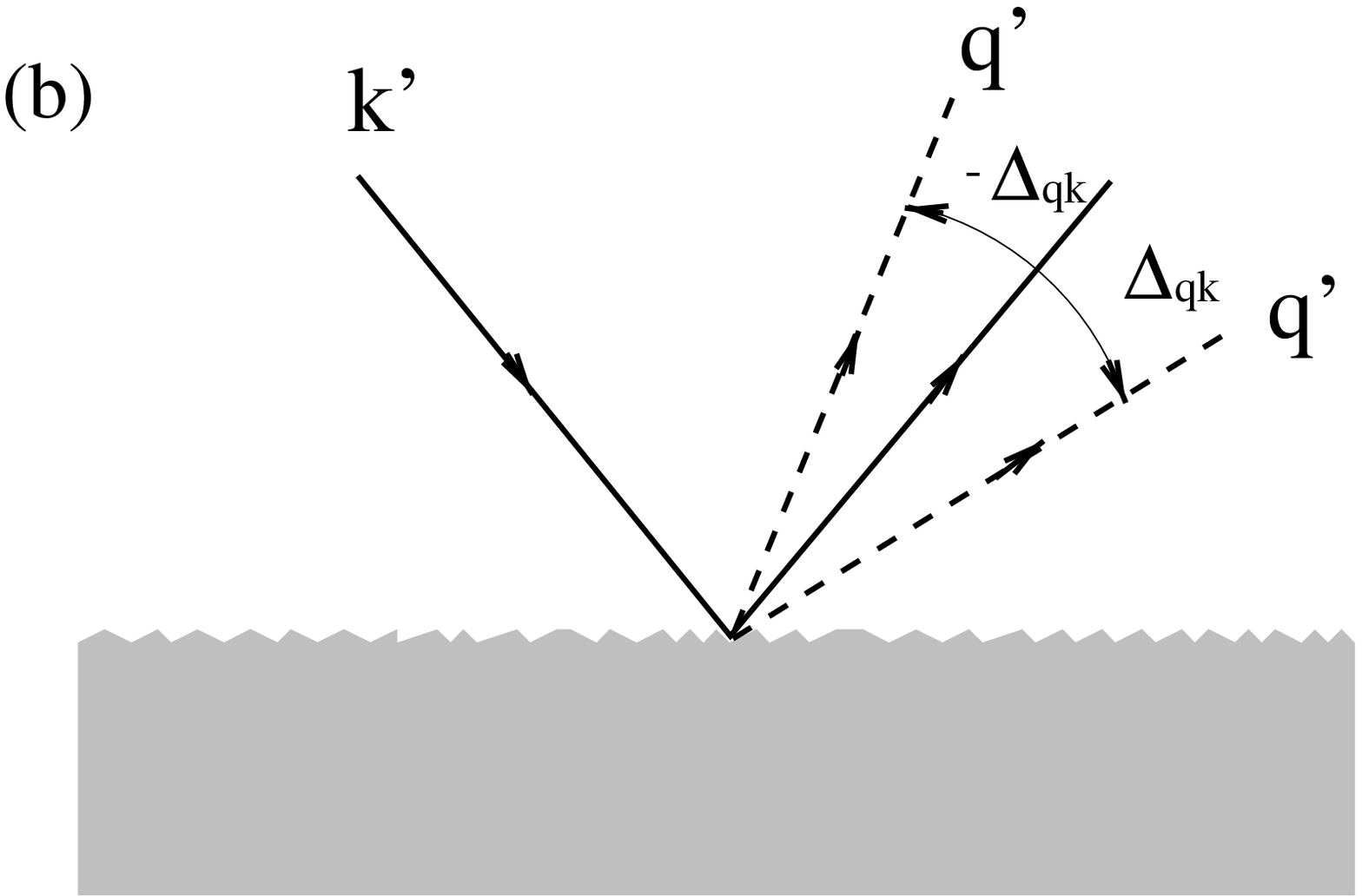} 
    \end{tabular}
    \caption{Interpretation of the correlation condition for the 
      short-range correlation functions $C^{(1)}$ and $C^{(10)}$. The
      outgoing solid lines indicate the specular direction. The
      scattering process $(q,k)$ that gives rise to the momentum
      transfer $\Delta_{qk}=q-k$ might be correlated with the process
      $(q',k')$ if $\Delta_{qk}=\Delta_{q'k'}$~($C^{(1)}$) or if
      $\Delta_{qk}=-\Delta_{q'k'}$~($C^{(10))}$). }
  \label{PHEN:Fig:PRB59_fig3}
  \end{center}
\end{figure}


To see what the $\delta$-function condition of Eq.~\r{PHEN:Eq:C1_prop}
means physically, it is convenient to introduce the momentum transfer
that can be associated with the scattering process. 
If the incident light has momentum $k$ and the scattered light is
described by the momentum variable $q$ the momentum transfer is 
\begin{eqnarray}
  \label{PHEN:Eq:mom-transfer}
 \Delta_{qk} &=& q-k.
\end{eqnarray}
Such a scattering event we recall was earlier denoted by $(q,k)$.
Thus, what Eq.~\r{PHEN:Eq:C1_prop} says is that the two scattering
processes $(q,k)$ and $(q',k')$ might have non-vanishing $C^{(1)}$
correlations if and only if the two scattering events have the same
momentum transfer, {\em i.e.}  if and only if
\begin{eqnarray}
  \label{PHEN:Eq:C1_condition}
  \Delta_{qk} &=& \Delta_{q'k'}.  
\end{eqnarray}
This condition is depicted in Fig.~\ref{PHEN:Fig:PRB59_fig3}.  From the
condition~\r{PHEN:Eq:C1_condition} it follows that if the incident
momentum is changed from say $k$ to $k'=k+\Delta k$, the entire
speckle pattern shifts in such a way that any feature initial at $q$
moves to $q'=q+\Delta q$.  In terms of the angles of incidence and
scattering, we have that if $\theta _0$ is changed to $\theta
_0^{\prime }=\theta _0+\Delta \theta _0$, any feature in the speckle
pattern originally at $\theta _s$ is shifted to $\theta _s^{\prime
  }=\theta _s+\Delta \theta _s$, where $\Delta \theta _s=\Delta
\theta_0(\cos \theta _0/\cos \theta _s)$ to first order in $\Delta
\theta_0$. This effect can indeed be seen from the speckle patters
presented in Figs.~\ref{Fig:PHEN:Speckle}.

It should in particular be noticed that
condition~\r{PHEN:Eq:C1_condition} is satisfied if ({\it i}) $k=k'$
and $q=q'$ as well as if ({\it ii}) $k=-q'$ and $q=-k'$. These choices
are the ones mentioned in the beginning of this section for the
scattering from a planar surface. It is interesting to notice that
these for a planar surface trivial correlations, also holds true for
the scattering from randomly rough surfaces, even though as should be
noticed, their physical origin is rather different. Case ({\it i}) is
kind of trivial since any scattering process should be perfectly
correlated with itself. This effect is known in the literature as the
{\em memory-effect}. Situation ({\it ii}), that doesn't seem that
obvious at first, is, in fact, a consequence of the reciprocity of the
S-matrix; $S(q|k)=S(-k|-q)$. Hence, when $k=-q'$ and $q=-k'$ there
should be perfect correlations, and the effect is known as the {\em
  reciprocal memory-effect}. If the scattering system does not possess
any damping, the system also respects time-reversal symmetry. Due to
this reason this latter effect is also known by some authors as the {\em
  time-reversed memory effect}.


\begin{figure}[t!]                                
  \begin{center}
    \leavevmode 
    \hspace*{-2cm}    
    \includegraphics[width=4in,height=3in]{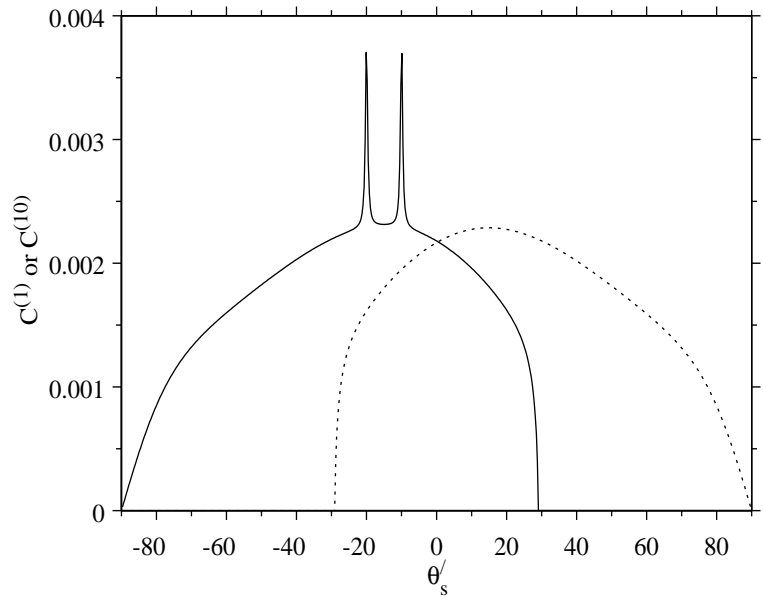}
    \caption{The envelopes of the short-range correlation
      functions $C^{(1)}(q,k|q',k')$ (solid line) and
      $C^{(10)}(q,k|q',k')$~(dashed line) as a function of the
      scattering angle $\theta_s'$ for $\theta_s=-10^\circ$ and
      $\theta_0=20^\circ$.  The angle $\theta_0'$ is determined from
      the $\delta$-function constraint.  The rough surface was a
      silver surface characterized by Gaussian height statistics of
      {\sc rms}-height $\delta=5 {\rm nm}$. The correlation function
      was also Gaussian with a correlation length of $a=100 {\rm nm}$.
      The wavelength of the incident light was $\lambda=457.9 {\rm
        nm}$.  At this wavelength the dielectric constant of silver is
      $\varepsilon(\omega)= -7.5+i0.24$.  (After
      Ref.~\protect\citen{Corr10}.) }
     \label{PHEN:C1-C10}
   \end{center}
\end{figure}


In Fig.~\ref{PHEN:C1-C10}~(solid line) we present the result of
perturbative calculations~\cite{Corr10} for the envelope of the
$C^{(1)}$ correlation function as a function of the scattering angle
$\theta_s'$ for $\theta_s=-10^\circ$ and $\theta_0=20^\circ$.  The
angle $\theta_0'$ is determined from the $\delta$-function condition
of Eq.~\r{PHEN:Eq:C1_prop}. The incident wave was $p$-polarized, and
the surface parameters are defined in the caption of this figure. Two
well-pronounced peaks at scattering angles $\theta_s=-20^\circ$ and
$\theta_s=-10^\circ$ are easily spotted in the envelope of $C^{(1)}$.
They corresponds respectively to the memory and reciprocal memory
effect.  It can in fact be shown that by instead considering the
envelope of the normalized correlation function, $\Xi^{(1)}$, one will
have perfect correlation at the maximum point of these two peaks~(see
e{\it e.g.}  Ref.~\citen{MP4}).

Before continuing, we would like to point out that the memory and
reciprocal memory effect seen in Fig.~\ref{PHEN:C1-C10} are due to
multiple scattering processes that involves surface plasmon
polaritons. Thus, for an $s$-polarized wave incident onto a weakly
rough metal surface, such peaks are not expected to be seen since in
this case the incident wave cannot excite surface plasmon polariton
at the rough surface~\cite{MP4}.  However, for scattering of an
$s$-polarized wave at a dielectric-dielectric interface the $C^{(1)}$
may show peaks~\cite{MP6} even though no surface plasmon polaritons
are involved. These peaks originate from multiple scattering
processes involving so-called lateral waves~\cite{Book:Felsen}.

\begin{figure}[t!]
  \begin{center}
    \leavevmode
    \includegraphics[height=12cm,width=14cm]{\myfigpath/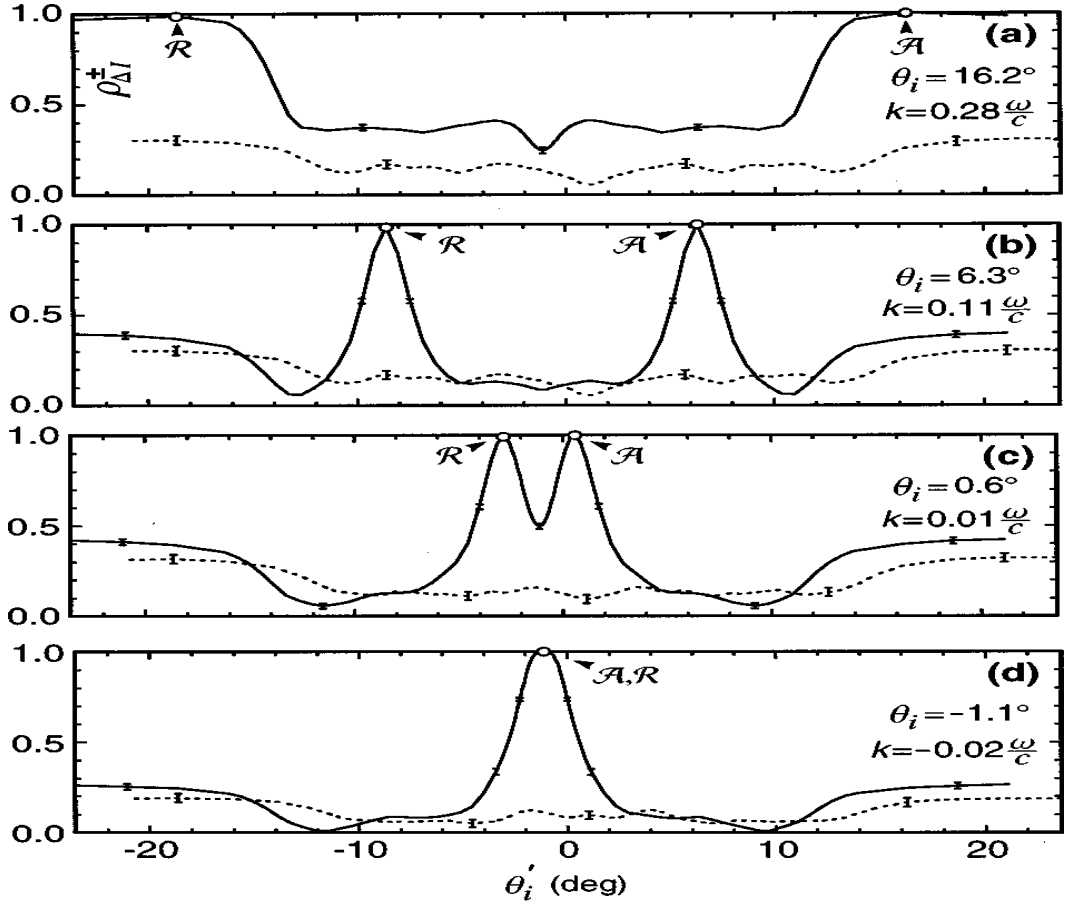}
    \caption{Experimental measurements of the normalized correlation
      functions $\rho_{\Delta I}^+(k,k',\Delta_{qk})$ (solid lines)
      and $\rho_{\Delta I}^-(k,k',\Delta_{qk})$ (dashed lines) as
      defined by West and O'Donnell~\cite{Corr12} as function of the
      angle of incidence $\theta_i'$. These correlation functions are
      these authors equivalent to our envelope functions
      $\Xi^{(1)}_0(q,k|q',k')$ and $\Xi^{(10)}_0(q,k|q',k')$ (see
      Ref.~\protect\citen{Corr12} for details). The incident light had
      wavelength $\lambda=6.12 {\rm nm}$ and the momentum transfer was
      $\Delta_{qk}=0.04(\omega/c)$.  The Gaussian height-distributed
      gold surface had {\sc rms}-height $\delta\simeq 15.5 {\rm nm}$.
      Its correlation was characterized by a West-O'Donnell
      (rectangular) power spectrum of parameters $k_-=0.83(\omega/c)$
      and $k_+ =1.30(\omega/c)$. These values satisfy $k_-<k_{sp}<k_+$
      where $k_{sp}=1.06(\omega/c)$ is the surface plasmon polariton
      wave vector, and hence an incident wave should couple strongly
      to such modes. The memory and the time-revised memory peaks are
      indicated in these figures by ${\cal A}$ and ${\cal R}$
      respectively. At these two positions we see that there are
      perfect correlations.  (After Ref.~\protect\citen{Corr12}.)}
    \label{PHEN:C1_C10-experimental-results}
  \end{center}

\end{figure}

Recently both the memory and reciprocal memory effect have been
observed experimentally by West and O'Donnell~\cite{Corr12} in the
scattering of $p$-polarized light from a weakly rough,
one-dimensional, random gold surface. We have reproduced one of their
graphs in Fig.~\ref{PHEN:C1_C10-experimental-results}.


\subsubsubsection{The $C^{(10)}$ Correlation functions} 

We now focus on the $C^{(10)}$-correlation function. This correlation
function was originally overlooked in the early studies of correlation
functions~\cite{Arsenieva1993} due to the use of the factorization
method~\cite{Corr17}.  By essentially duplicating the arguments used
in arriving at Eq.~\r{PHEN:Eq:t1_prop}, we find in an analogous way
that
\begin{eqnarray}
  \label{PHEN:Eq:t10_prop}
  \langle \delta S(q|k) \delta S(q^{\prime}|k^{\prime })\rangle &\propto&
  2\pi\delta(q-k+q^{\prime}-k^{\prime }),   
\end{eqnarray}
with the consequence that we might write
\begin{eqnarray}
  \label{PHEN:Eq:C10_prop}
  C^{(10)}(q,k|q^{\prime },k^{\prime }) &\propto&
        \frac{2\pi \delta(q-k+q^{\prime}-k^{\prime })}{L_1} \;
        C^{(10)}_0(q,k|q^{\prime },q'+q+k) . \qquad
\end{eqnarray}
Here $C^{(10)}_0(q,k|q^{\prime },q'+q+k)$ is an envelope function, and
both $C^{(10)}$ and its envelope $C^{(10)}_0$ are independent of
the length of the randomly rough surface.

The presence of the $\delta$-function on the right hand side of
Eq.~\r{PHEN:Eq:C10_prop} is in terms of the momentum transfer
equivalent to
\begin{eqnarray}
  \label{PHEN:Eq:C10_condition}
  \Delta_{qk} &=& -\Delta_{q'k'}.  
\end{eqnarray}
What this condition implies for the speckle pattern is that if we
change the angle of incidence in such a way that $k$ goes into
$k^{\prime }=k+\Delta k$, a feature originally at $q=k-\Delta q$ will be shifted
to $q^{\prime }=k^{\prime }+\Delta q$, {\it i.e.} to a point as much
to one side of the new specular direction as the original point was on
the other side of the original specular direction. For one and the
same incident beam the $C^{(10)}$ correlation function therefore
reflects the symmetry of the speckle pattern with respect to the
specular direction~(see Fig.~\ref{PHEN:Fig:PRB59_fig3}).

The dashed line in Fig.~\ref{PHEN:C1-C10} shows the angular
dependence, obtained from perturbation theory~\cite{Corr10}, for the
envelope of $C^{(10)}$. The parameters used to obtain these results
were the same used to obtain the $C^{(1)}$ correlation shown by the
solid line in the same figure. It is seen that the $C^{(10)}_0$
envelope is a smooth function of $\theta_s'$, and in particular does
not show any peaks.  Moreover, its amplitude is roughly of the same
order of magnitude as the $C^{(1)}$ correlation function. This
behavior is the same as the one found by West and
O'Donnell~\cite{Corr12} in their experimental investigation of the
$C^{(10)}_0$ envelope (Fig.~\ref{PHEN:C1_C10-experimental-results}).

It should be pointed out that the $C^{(10)}$ correlation function has
no known analogy within scattering from volume disordered system.
This new type of correlations in surface scattering was first
predicted from perturbation theory by Malyshkin {\it et al.} in
1997~\cite{Corr9,Corr10}.

\subsubsection{Long- and Infinite-Range Correlations}

We will now consider the last term of the right hand side of
Eq.~\r{PHEN:Eq:Rewritten} that gives rise to $C^{(N)}$. Due to the
stationarity of the surface 
$$
\left\{\delta S(q|k)\delta S^{*}(q|k)
       \delta S(q^{\prime }|k^{\prime})
       \delta S^{*}(q^{\prime}|k^{\prime})\right\} 
  \propto 2\pi\delta(0)=L_1.
$$
Hence, the correlation function itself, in light of
  Eq.~\r{PHEN:Eq:CN}, should behave as
\begin{eqnarray}
  \label{Eq:Phen:1-over_L}
  C^{(N)}(q,k|q^{\prime },k^{\prime }) 
        &\propto& \frac{1}{L_1}.
\end{eqnarray}
It should be noticed that the $C^{(N)}$-correlation function is not
constrained in its momentum variables through $\delta$-functions as we
saw earlier was the case for the short-range correlation functions.

Even though we will not address this point explicitly here it has
recently been shown that $C^{(N)}$ can be written as a sum of the
three following terms~\cite{Corr9, Corr10}
\begin{eqnarray}
  C^{(N)}(q,k|q^{\prime },k^{\prime })   &=&
    C^{(1.5)}(q,k|q^{\prime },k^{\prime }) +
    C^{(2)}(q,k|q^{\prime },k^{\prime }) +
    C^{(3)}(q,k|q^{\prime },k^{\prime }).  \nn 
\end{eqnarray}
Here $C^{(1.5)}$ denotes a correlation function of {\em
  intermediate-range}, $C^{(2)}$ is a correlation function of {\em
  long-range}, while $C^{(3)}$ is an {\em infinite-range} correlation
function. For explicit expressions for these three correlation
functions the interested reader is directed to Refs.~\citen{Corr9,
  Corr10} and \citen{CorrAdd}.  In scattering from bulk disordered
systems $C^{(2)}$~\cite{Corr16,Corr18,Corr19} and
$C^{(3)}$~\cite{Corr16,Corr20} have their analogies.  However, the
intermediate range correlation function, $C^{(1.5)}$, predicted
theoretically by Malyshkin {\it et al.} in 1997~\cite{Corr9,Corr10},
has no equivalent in scattering from volume disordered systems. It is
unique to scattering from randomly rough surfaces that support surface
plasmon polaritons at the frequency of the incident light. An explicit
example of such a scattering system is provided by a randomly rough
metal surface in $p$-polarization.

\begin{figure}[tbhp]
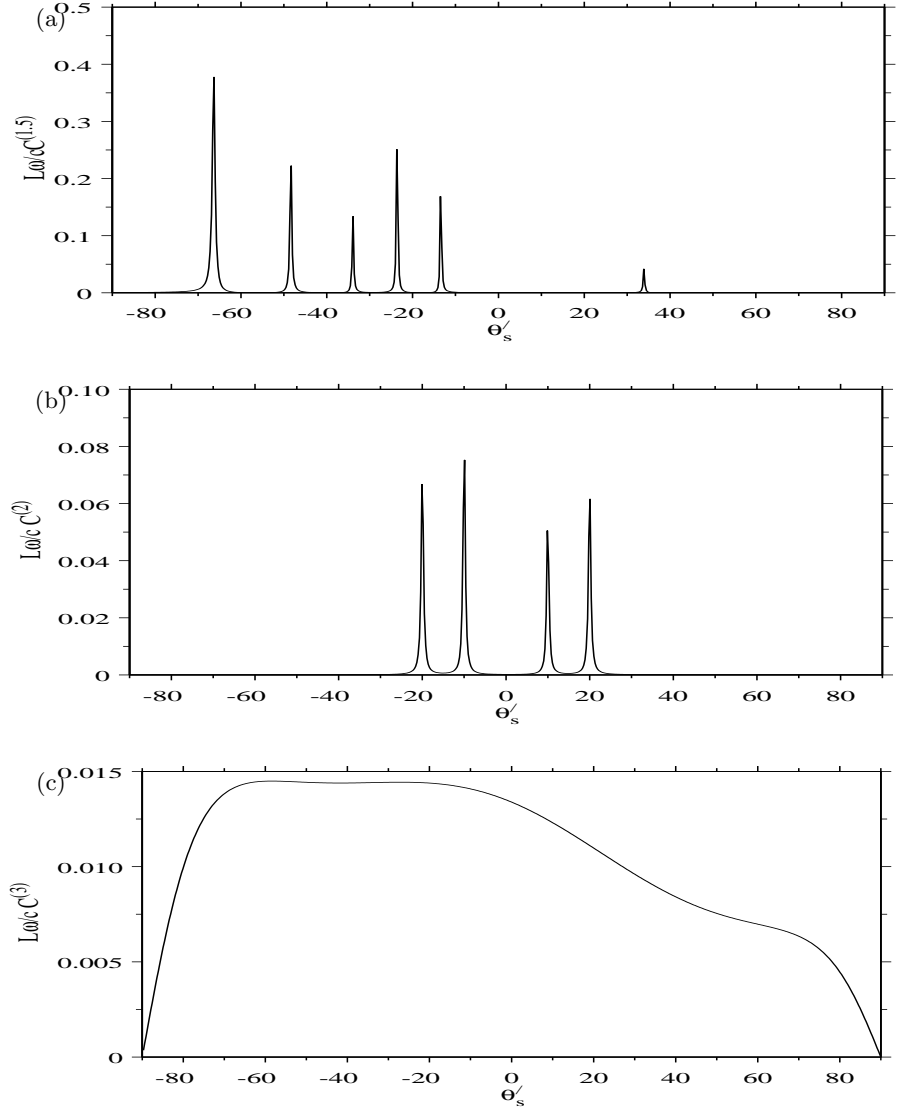

  \begin{center}
    \leavevmode
        \hspace*{-9.5cm}(a) \\ \vspace*{-0.5cm} \hspace*{-1.7cm}
        \includegraphics[height=4.4cm,width=12cm]{\myfigpath/m6.ps} \\*[0.7cm]
        \hspace*{-9.5cm}(b) \\ \vspace*{-0.5cm} \hspace*{-1.7cm}
        \includegraphics[height=4.4cm,width=12cm]{\myfigpath/m7.ps} \\*[0.7cm]
        \hspace*{-9.5cm}(c) \\ \vspace*{-0.5cm} \hspace*{-1.7cm}
        \includegraphics[height=4.4cm,width=12cm]{\myfigpath/m8.ps} \\*[0.7cm]
    \caption{Perturbative results for the  angular dependence of the
      correlation functions (a) $C^{(1.5)}(q,k|q',k')$, (b)
      $C^{(2)}(q,k|q',k')$ and (c) $C^{(3)}(q,k|q',k')$ on the
      scattering angle $\theta_s'$ for $\theta_s=-10^\circ$,
      $\theta_0=20^\circ$ and $\theta_0'=30^\circ$. The remaining
      parameters were the same used in Fig.~\protect\ref{PHEN:C1-C10}.
      (After Ref.~\protect\citen{Corr10}.) }
    \label{PHEN:Fig:CN}
  \end{center}
\end{figure}

Based on a diagrammatic perturbation theoretical study, Malyshkin {\it
  et al.}~\cite{Corr10} found that $C^{(1.5)}$ shows a rather rich
peak structure.  Peaks in $C^{(1.5)}$ are expected to occur for a
number of cases in which a linear combination of three of the momenta
$q$, $k$, $q'$ and $k'$ add up to $\pm k_{sp}$, where $k_{sp}$, is the
wave vector of the surface plasmon polariton. These condition are
summarized in Table~\ref{PHEN:Table:C15-C2}. In an expansion of
$C^{(1.5)}$ in powers of the surface profile function the leading
order is ${\cal O}(\zeta^6)$.  The intermediate-range correlation
function $C^{(1.5)}$ is therefore for a weakly rough surface a
result of correlations between a single scattering and a multiple
scattering process that involves surface plasmon polaritons. 
In Fig.~\ref{PHEN:Fig:CN}a we have plotted the intermediate-rang
correlation function $C^{(1.5)}$ for the randomly rough silver surface
that lead to the results shown earlier in Fig.\ref{PHEN:C1-C10}.  In
this graphs several peaks are easily seen.  Their positions should be
compared to the predictions that can be obtained from
Table~\ref{PHEN:Table:C15-C2}.


So far there is no experimental measurements for any of the
correlations contained in $C^{(N)}$. In fact such an experimental
confirmation represents a real challenge to the experimentalists. The
reason being that for long surface these correlations are rather small
(due to Eq.~\r{Eq:Phen:1-over_L}). In order to be able to observe
them, one probably has to use a well-focused incident beam, or a short
surface.


%
%

\begin{table}[t!]
  \begin{center}
    \leavevmode
    \begin{tabular}{lll}
      \hline 
      Correlation function &  Peak condition  \\
      \hline \hline \\
      $C^{(1.5)}$   &  $-k'+k+q'= -k_{sp}$              \\
      $C^{(1.5)}$   &  $q-q'+k' =  k_{sp}$              \\
      $C^{(1.5)}$   &  $q'+q-k' = -k_{sp}$              \\
      $C^{(1.5)}$   &  $q'-k'+q = -k_{sp}$              \\
      $C^{(1.5)}$   &  $k+k'-q' =  k_{sp}$              \\
      $C^{(1.5)}$   &  $q'-q+k  =  k_{sp}$              \\
      $C^{(2)}$     &  $q'=-k$                          \\
      $C^{(2)}$     &  $q'= q$                          \\
      $C^{(2)}$     &  $q'=-q$                          \\
      $C^{(2)}$     &  $q'= k$                          \\
      \hline
    \end{tabular}
    \caption{The peak conditions for the intermediate range
      $C^{(1.5)}$ and long range correlation function $C^{(2)}$ for a
      metallic one-dimensional surface. See text for details.} 
    \label{PHEN:Table:C15-C2}
  \end{center}
\end{table}


Malyshkin {\it et al.}~\cite{Corr10} also showed perturbatively that
the $C^{(2)}$-correlation function should have a peak structure, while
the infinite range correlation function, $C^{(3)}$, should be a smooth
function of its arguments.  This is seen from the perturbative results
plotted in Fig.~\ref{PHEN:Fig:CN}b~($C^{(2)}$) and
Fig.~\ref{PHEN:Fig:CN}c~($C^{(3)}$). The correlations described by the
$C^{(2)}$-correlations function are a result of correlation between
two multiple scattering processes. For weakly rough metal surfaces
this correlation function is dominated by double scattering processes.
Its peaks are associated with surface plasmon polaritons, as was found
to be the case also for $C^{(1.5)}$. The peak conditions for $C^{(2)}$
are that two of the four momenta involved should add/subtract to zero.
That is to say that for fixed $k$, $q$, and $k'$, peaks are expected
when $q'=\pm k'$, $q'= \pm k$ or $q'= \pm q$~(see
Table~\ref{PHEN:Table:C15-C2}). Also the infinite range correlations
are due to multiple scattering events.  What distinguish the
long-range correlation, $C^{(2)}$, from the infinite-range, $C^{(3)}$,
is that the latter involves at least one triple scattering
process\footnote{The leading contribution to $C^{(3)}$ is of order
  $\zeta^{10}$ in the surface profile function
  $\zeta(x_1)$~\cite{Corr10}.}.  For more details information about
$C^{(1.5)}$, $C^{(2)}$, and $C^{(3)}$, the reader is invited to
consult Refs.~\citen{Corr9,Corr10} and \citen{CorrAdd}.

\subsubsection{Angular Intensity Correlation Functions for Strongly Rough Surfaces}

\begin{figure}[t!]
  \begin{center}
    \leavevmode
    \begin{tabular}{@{}c@{\hspace{1.0cm}}c@{}}
      \includegraphics[width=6cm,height=7cm,angle=-90]{\myfigpath/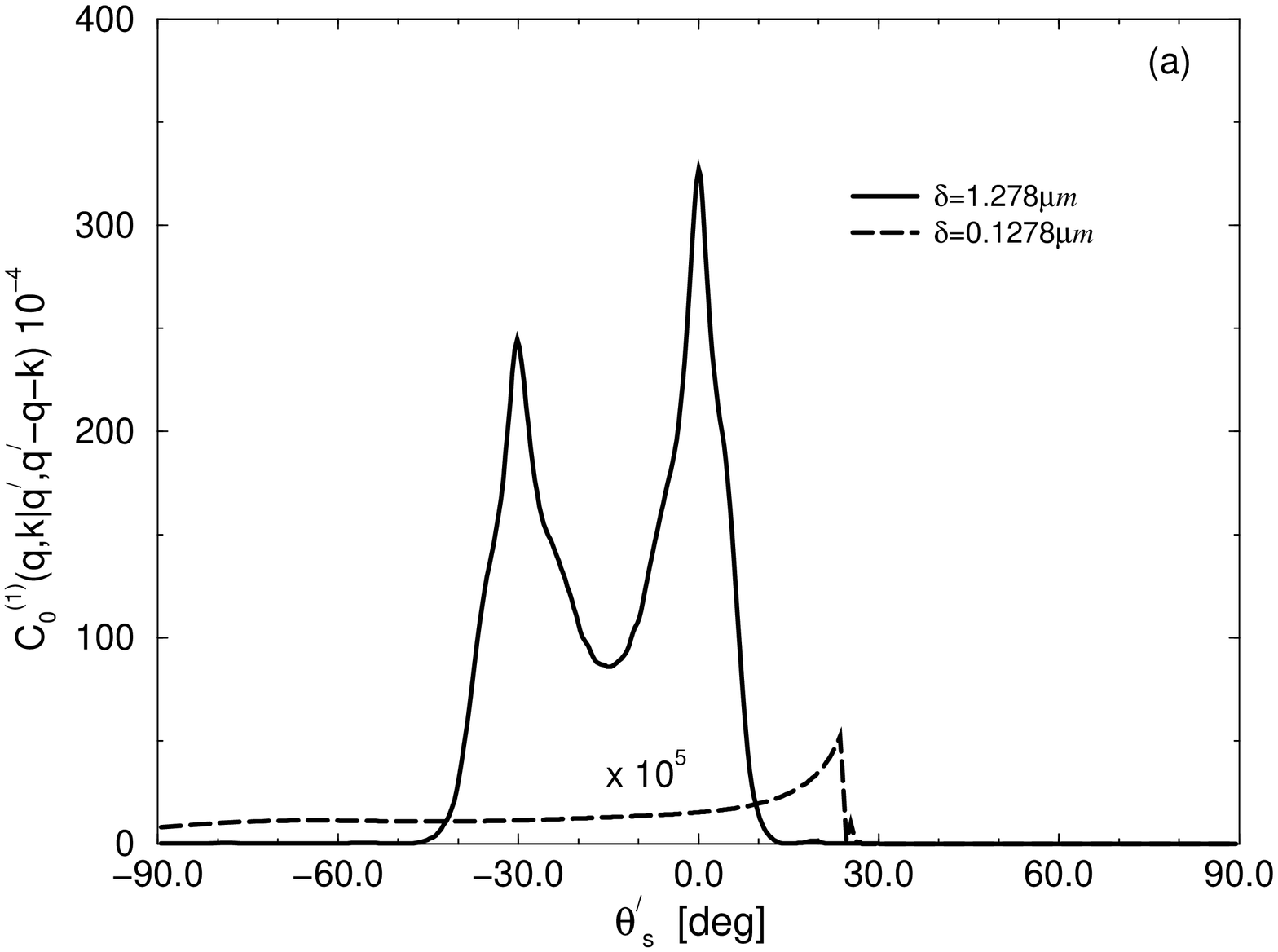}
      & 
      \includegraphics[width=6cm,height=7cm,angle=-90]{\myfigpath/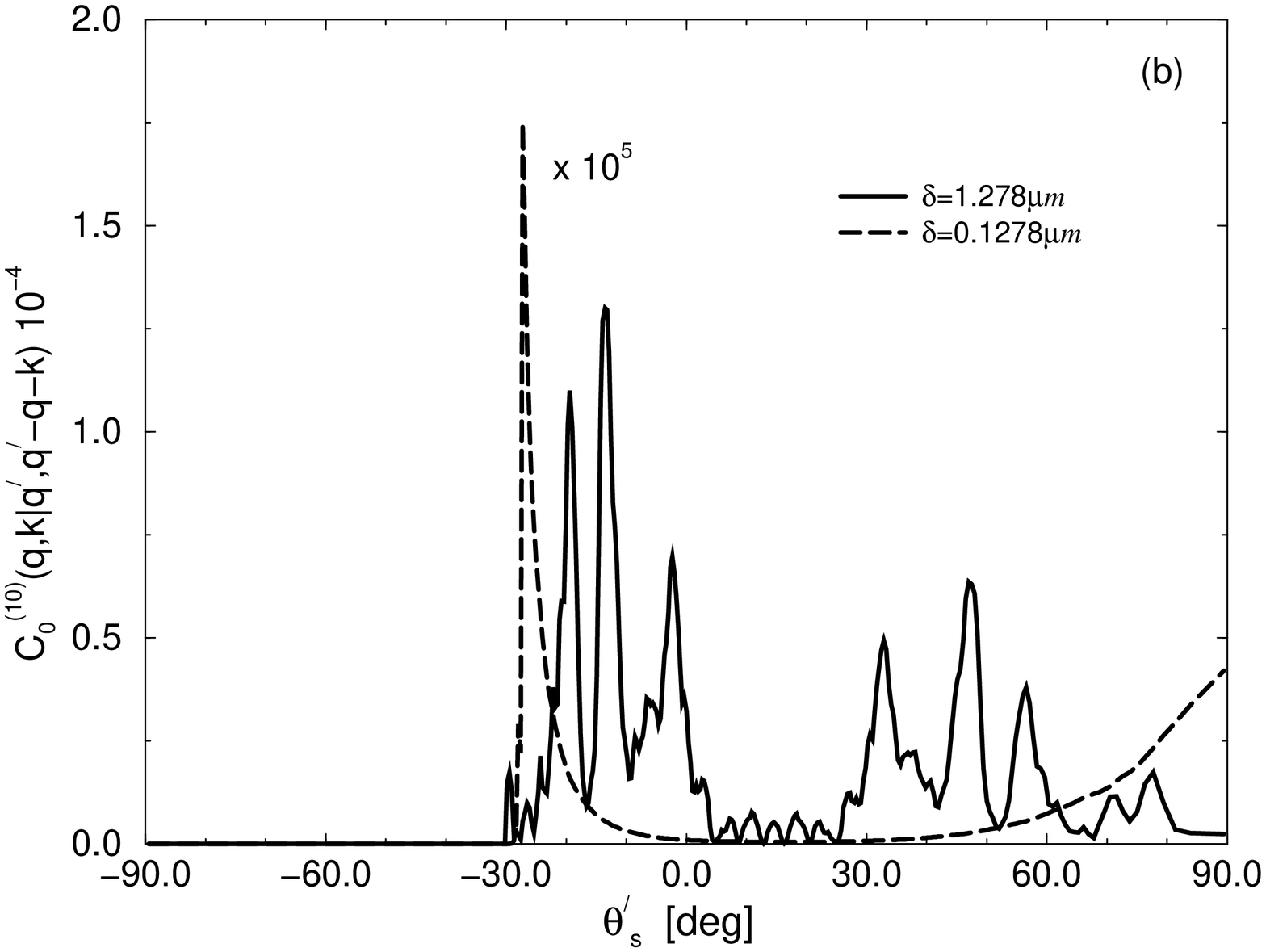} 
    \end{tabular}
    \caption{Rigorous numerical simulation results  for the (a)
      $C^{(1)}_0$ and (b) $C^{(10)}_0$ envelopes as functions of
      $\theta_s^{\prime}$ for $\theta_0=30^{\circ}$ and $\theta_s
      =0^{\circ}$. The angle $\theta'_0$ was determined from the
      $\delta$-function constraint.  The $s$-polarized incident light
      had wavelength $\lambda= {\rm 632.8  nm}$.  The randomly rough
      silver surface characterized by a (Gaussian) correlation length
      $a= 3.85{\rm \mu m}$.  The {\sc rms}-height of the Gaussian
      height-distributed surface was $\delta = 1.278 {\rm\mu
        m}$~(solid line) $\delta = 0.1278 {\rm \mu m}$~(dashed line).
      As the {\sc rms}-height is increased one observes that the
      memory and reciprocal memory peaks start appearing in the
      envelope of $C^{(1)}$.  (After Ref.~\citen{MP4}.) }
    \label{PHEN:CORR:Comp-sim}
  \end{center}
\end{figure}

Before closing this section, we would like to make a few remarks
regarding strongly rough surfaces.  Above we always assumed that the
surface was a weakly rough metal surface. We saw that many of the
interesting features of $C(q,k|q',k')$ appeared due to excitations
of surface plasmon polaritons. For strongly rough surfaces the
excitation of surface plasmon polaritons, if any, is weak, and the
dominating mechanism for multiple scattering from such surfaces is
multiple scattering of volume waves. As might have been guessed,
multiple scattering of volume waves take over for strongly rough
surfaces the role that surface plasmon polaritons had for weakly rough
surfaces. These multiple scattered volume waves give rise to the
memory and reciprocal memory effect for strongly rough surface.  This
is in fact the case for both $p$- and $s$-polarized incident light in
contrast to what is the case for weakly rough surfaces.  This is
illustrated by the rigorous computer simulation results of
Fig.~\ref{PHEN:CORR:Comp-sim}a showing the $C^{(1)}_0$ envelope for
$s$-polarized incident light~\cite{MP4}. It is seen from this figure
that as the {\sc rms}-height is increased from a value corresponding
to a weakly rough surface the memory and reciprocal memory peaks start
to emerge in the $C^{(1)}_0$ envelope due to the increased
contribution from multiple scattered volume waves.


It should also be noticed, as was realized recently~\cite{MP4}, that a
measurement of the angular intensity correlations can provide valuable
information regarding the statistical properties of the amplitude of
the scattered field. In particular, it was shown that the short-range
correlation function $C^{(10)}$ is in a sense a measure of the
non-circularity of the complex Gaussian statistics of the scattering
matrix.  If the random surface is such that only the $C^{(1)}$ and
$C^{(10)}$ correlation functions are observed, then $S(q|k)$ obeys
complex Gaussian statistics. If the random surface is such that only
$C^{(1)}$ is observed, then $S(q|k)$ obeys {\rm circular} complex
Gaussian statistics\footnote{Two complex random variables $A=A_1+iA_2$
  and $B=B_1+iB_2$ are said to be {\em circular} complex Gaussian
  if~\cite{Corr22,Corr23} $\left< A_1 B_1 \right> =\left<A_2 B_2
  \right>$ and $\left< A_1 B_2 \right> =-\left<A_2 B_1 \right>$.}.
This can indeed be seen from Fig.~\ref{PHEN:CORR:Comp-sim}b, which
shows that as the surface is made rougher, and therefore $\delta
S(q|k)$ approaches a circular complex Gaussian process, the
$C^{(10)}$-correlation vanishes as compared to $C^{(1)}$. Finally, if
the random surface is such that $C^{(1.5)}$, $C^{(2)}$ and $C^{(3)}$
are observed in addition to both $C^{(1)}$ and $C^{(10)}$, then
$\delta S(q|k)$ is not a Gaussian random process at all. However,
which kind of statistics $\delta S(q|k)$ satisfies in this case is not
clear for the moment. These results fits the findings from standard
speckle theory~\cite{Corr22, Corr23,Book:Dainty75} which assumes that
the disorder is strong and that $\delta S(q|k)$ constitutes a circular
complex Gaussian process.


\subsection{Second Harmonic Generation of Scattered Light}

%
%
%

So far in this section, we have discussed exclusively rough surface
scattering phenomena that find their explanation within linear
electromagnetic theory.  There are still many exciting
nonlinear~\cite{Book:Keller90} surface scattering effects that have to
be addressed in the future. Such nonlinear studies are still at their
early beginning.  The studies that have been conducted so far on
nonlinear surface scattering effects have mainly been related to the
angular distribution of the scattered {\em second harmonic} generated
light~\cite{Book:Kong,Book:Born}.  In particular what have been
studied are some new features in the backscattering directions of the
second harmonic light. In this section we will discuss some of these
results. The presentation given below follows closely the one given in
Ref.~\citen{Leyva-Lucero1999}.

It is well-known from solid state physics that an (infinite)
homogeneous and isotropic metal has inversion
symmetry~\cite{Book:Kitel, Book:Ascroft}. A consequence of this is
that there is no nonlinear polarization in the bulk. If, however, the
metal is semi-infinite with an interface to vacuum, say, the inversion
symmetry is broken. Thus, a nonlinear polarization, different from
zero, will exist close to the surface. As we move into the bulk of the
metal, this effect will become smaller and smaller and finally vanish.
Therefore, one might talk about a nonlinear surface layer which
through nonlinear interactions will give rise to light that is
scattered away from the rough surface at the second harmonic
frequency.

The scattering system that we will be considering is the by now
standard one depicted in Fig.~\ref{Fig:Theory:Geometry}.  This geometry
is illuminated from the vacuum side, $x_3>\zeta(x_1)$, by a
$p$-polarized planar wave of (fundamental) frequency $\omega$. Only
the $p$-polarized component of the scattered second harmonic generated
light will be considered here, even though there also will exit a weak
$s$-polarized component due to the nonlinear interaction at the
surface.  However, the $p$-polarized component represents the main
contribution to the scattered light at the second harmonic frequency
$2\omega$, and will therefore be our main concern here.  Moreover it
will be assumed that the generation of the second harmonic light does
not influence the field at the fundamental frequency in any
significant way.

\begin{figure}[t!]
  \begin{center}
    \leavevmode
    \includegraphics[height=12cm,width=14cm]{\myfigpath/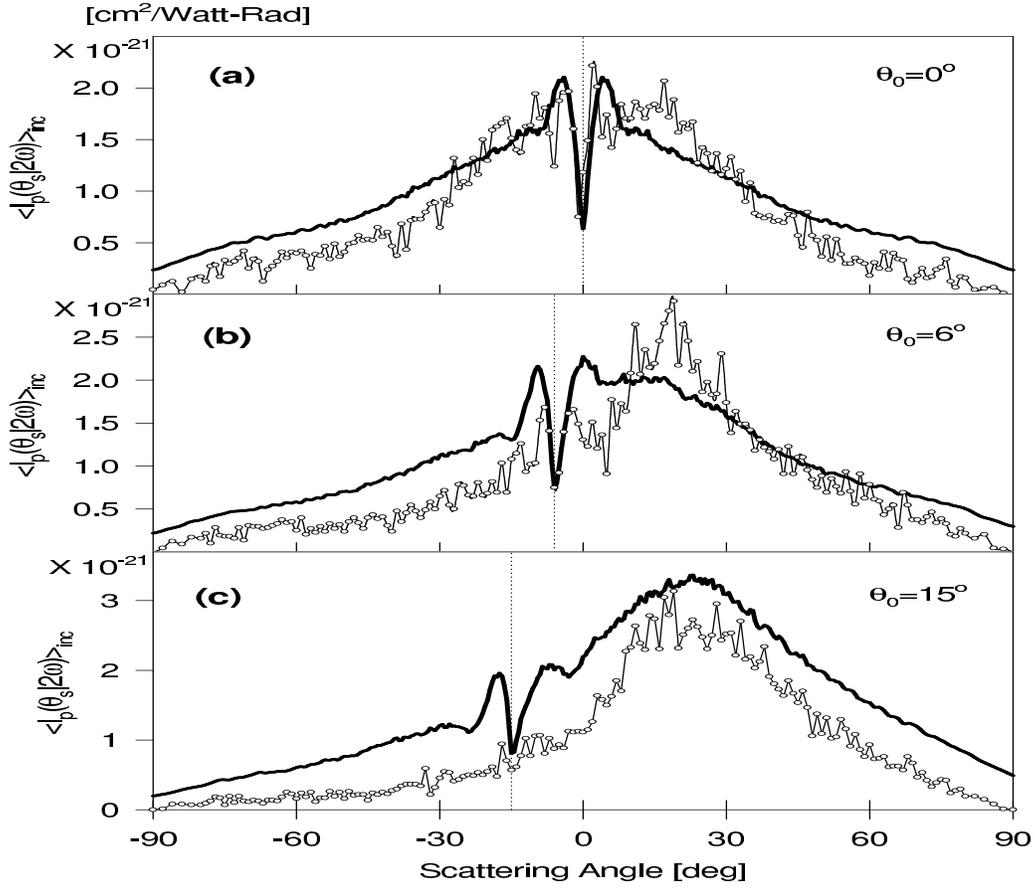}
    \vspace*{0.5cm}        
    \caption{The mean normalized  second harmonic intensity as a
      function of the scattering angle $\theta_s$ for the scattering
      of $p$-polarized light from a randomly rough silver surface. The
      surface was characterized by a Gaussian height distribution of
      {\sc rms}-height $\delta= 1.81 \rm \mu m$, as well as a Gaussian
      correlation function of correlation length $a=3.4 {\rm \mu m}$.
      The dielectric constants were at the fundamental and second
      harmonic frequency $\varepsilon(\omega)=-56.25+i0.60$ and
      $\varepsilon(2\omega)=-11.56+0.37$ respectively. The thick lines
      represent the results of numerical simulations and the open
      circles represent the experimental results of O'Donnell and
      Torre~\protect\cite{Odonnell97}. The incident plane wave had a
      wavelength $\lambda=1.064 {\rm \mu m}$.  In the numerical
      simulations the surface had length $L=40 \lambda$ and it was
      sampled with an interval $\Delta x_1=\lambda/20$.  The numerical
      results were averaging over $N_\zeta=2000$ realizations of the
      surface, and the angles of incidence were (a) $
      \theta_0=0^\circ$, (b) $\theta_0=6^\circ$, and (c)
      $\theta_0=15^\circ$. (After
      Ref.~\protect\citen{Leyva-Lucero1999}.).}
    \label{Fig:Phen:SH-experimental}
  \end{center}
\end{figure}

To motivate the study, we in Figs.~\ref{Fig:Phen:SH-experimental} show
some experimental results~(open circles) due to K.\ A.\ O'Donnell and
R.\ Torre~\cite{Odonnell97} for the so-called normalized\footnote{It
  can be shown that the total power scattered from a randomly rough
  surface at the second harmonic frequency is proportional to the
  square of the irrandiance, the incident power per unit area, on the
  surface.  One therefore defines the normalized intensity of the
  scattered second harmonic light so that it is independent of the
  incident power. The analytic expressions for this quantity can be
  found in Eq.~(34) of Ref.~\citen{Leyva-Lucero1999}.}  intensity of
the second harmonic light scattered incoherently from a strongly rough
silver surface. The surface was characterized by Gaussian height
statistics of {\sc rms}-height $\delta=1.81{\rm \mu m}$ and a Gaussian
correlation function. The transverse correlation length was $a=3.4{\rm
  \mu m}$.  The wavelength of the incident light was $\lambda=2\pi
c/\omega= 1.064 {\rm \mu m}$, while the angles of incidence considered
were $\theta_0=0^\circ$, $\theta_0=6^\circ$, and $\theta_0=15^\circ$
as indicated in Fig.~\ref{Fig:Phen:SH-experimental}.


The most noticeable feature of the experimental results~(open circles)
shown in Figs.~\ref{Fig:Phen:SH-experimental} are, without question,
the dips seen in the backscattering direction. It should be recalled
that for the linear problem one gets at this scattering angle an
enhanced backscattering {\em peak}~(result not shown) similar to the
one shown {\it e.g} in Fig.~\ref{Fig:Phen:BC-num-calc}.  So why do we
have a dip for the second harmonic light at the backscattering
direction, and not a peak?

\subsubsection{Strongly Rough Surfaces: A Numerical Simulation Approach 
  to the Second Harmonic Generated Light}
\sectionmark{Strongly Rough Surfaces}

Below we will with the help of numerical simulations try to get a
deeper understanding of what causes these dips.  The nonlinear layer
existing along the surface is of microscopic dimensions.  Since we are
working with the macroscopic Maxwell's equations it is natural to
assume that this layer is infinitely thin.  Under this assumption, the
effect of the nonlinear boundary layer is accounted for in the
boundary conditions to be satisfied by the field, and its normal
derivative, at the second harmonic frequency. These boundary
conditions have jumps at the nonlinear interface, and their degree of
discontinuity depends on the nonlinear polarization, or equivalently,
on the parameters that describes this polarization. The form of the
(nonlinear) boundary conditions at the second harmonic frequency
$2\omega$ can be shown to be~\cite{Leyva-Lucero1999}
\begin{subequations}
  \label{Eq:Phen:Nonolinear-BC}
  \begin{eqnarray}
    {\cal F}^+_\nu(x_1 |2\omega) - {\cal F}^-_\nu(x_1 |2\omega) 
          &=& {\cal A}(x_1), \\
     {\cal N}^+_\nu(x_1 |2\omega) - {\cal N}^-_\nu(x_1 |2\omega) 
          &=& {\cal B}(x_1),
  \end{eqnarray}
\end{subequations}
where the sources ${\cal F}_\nu$ and ${\cal N}_\nu$ have been defined
in Eqs.~\r{Eq:Theory:Source-func}. As before, the superscripts $\pm$
denote the sources evaluated just above~($+$) and below~($-$) the
rough surface defined by $x_3=\zeta(x_1)$. The functions ${\cal
  A}(x_1)$ and ${\cal B}(x_1)$ are related to the nonlinear
polarization ${\mathbf P}(x_1,x_3)$ through the integral of this
quantity over the nonlinear boundary layer~\cite{Leyva-Lucero1999}. To
fully specify the nonlinearity of the problem, the polarization
${\mathbf P}(x_1,x_3)$ has to be specified.  For instance for a free
electron model, that we will consider here for simplicity, it takes
on the form~\cite{Mendez95,Bloembergen68,Maystre86}
\begin{subequations}
  \label{Eq:Phen:Nonlin-pol}
\begin{eqnarray}
  {\mathbf P}(x_1,x_3)  
                       &=& \gamma{\mathbf \nabla}
                               \left({\mathbf E \cdot \mathbf E}\right)
                           +\beta{\mathbf E} 
                               \left( {\mathbf\nabla\cdot \mathbf E}\right). 
\end{eqnarray}
Here the constants $\gamma$ and $\beta$ are defined as  
\begin{eqnarray}
  \gamma &=& \frac{e^3n_0({\mathbf
                 r}_{\perp}(x_1,x_3))}{8m^2\omega^4}, \\
  \beta  &=&  \frac{e}{8\pi m\omega^2},
\end{eqnarray}
\end{subequations}
where $n_0$ is the electron number density, ${\mathbf
  r}_{\perp}(x_1,x_3)$ is a vector normal to the local surface at 
point $(x_1,x_3)$, and $e$ and $m$ are the charge and mass of the
electron respectively. The explicit expressions, in this model, for
${\cal A}(x_1)$ and ${\cal B}(x_1)$ can be found in
Ref.~\citen{Leyva-Lucero1999}.

Since the surfaces used in the experiments leading to the results
shown in Figs.~\ref{Fig:Phen:SH-experimental} are strongly rough,
perturbation theory does not apply, and one has in theoretical studies
to resort to rigorous numerical calculations of the second harmonic
scattered light. Such kind of simulations are conducted on the basis
of the rigorous simulation approach presented in
Sect.~\ref{Sect:Theory:Sect:NumSim}.  The calculations are now,
however, made out of two main steps: First, one calculates the
(linear) source functions ${\cal F}_\nu(x_1|\omega)$ and ${\cal
  N}_\nu(x_1|\omega)$; the field and it's normal derivative evaluated
on the surface at the fundamental frequency $\omega$. This is done
exactly as described in Sect.~\ref{Sect:Theory:Sect:NumSim}.  From the
knowledge of the linear sources functions at the fundamental
frequency, the right-hand-side of the boundary
conditions~\r{Eq:Phen:Nonolinear-BC} can be calculated since they
depend directly on these source functions as well as on the form of
the nonlinear polarization ${\mathbf
  P}(x_1,x_3)$~\cite{Leyva-Lucero1999}. In all numerical results to be
presented later in this section the form for the nonlinear
polarization given by Eq.~\r{Eq:Phen:Nonlin-pol} will be used. With
the functions ${\cal A}(x_1)$ and ${\cal B}(x_1)$ available, the
nonlinear sources, ${\cal F}^\pm_\nu(x_1 |2\omega)$ and ${\cal
  N}^\pm_\nu(x_1 |2\omega)$, are readily calculated from an approach
similar to the one described in detail in
Sect.~\ref{Sect:Theory:Sect:NumSim}. The only main difference is that
now the boundary conditions to be used when coupling the two integral
equations are the nonlinear boundary conditions given in
Eqs.~\r{Eq:Phen:Nonolinear-BC}.  With the source functions both for
the fundamental and second harmonic frequency available, all
interesting quantities about the scattering process, both linear and
nonlinear, are easily obtained. The full details of this approach can
be found in Ref.~\citen{Leyva-Lucero1999}.

Based on this numerical approach, we compare in
Figs.~\ref{Fig:Phen:SH-experimental} the numerical simulation
results~(solid lines) obtained by Leyva-Lucero {\it et
  al.}~\cite{Leyva-Lucero1999} to the experimental results obtained by
O'Donnell and Torre (open circles)~\cite{Odonnell97}. The dielectric
constants used in the simulations were at the fundamental frequency
$\varepsilon(\omega)=-56.25+i0.60$ and
$\varepsilon(2\omega)=-11.56+i0.37$ at the second harmonic frequency.
Indeed by comparing the experimental and theoretical results shown in
Figs.~\ref{Fig:Phen:SH-experimental}, a nice correspondence is
observed both qualitatively and quantitatively. Particular in light of
the oversimplified model used in the simulations for the nonlinear
interaction, the agreement is no less then remarkable.

From the experimental and theoretical results shown in
Figs.~\ref{Fig:Phen:SH-experimental}a, a clear dip is seen in the
incoherent component of the mean normalized second harmonic intensity
for the backscattering direction $\theta_s=0^\circ$. For the linear
scattering problem, however, there is an enhancement at the same
scattering angle. So what is the reason for the dip in the second
harmonic light?  O'Donnell and Torre~\cite{Odonnell97}, who conducted
the experiments leading to the experimental results shown in
Figs.~\ref{Fig:Phen:SH-experimental}, suggested that these dips were
due to coherent effects. In particular they suggested that the dips
originated from destructive interference between waves scattered
multiple times in the valleys of the strongly rough surface.  Since
the numerical simulation approach seems to catch the main physics of
the second harmonic generated light, it might therefore serve as a
useful tool for testing the correctness of the suggestion made by of
O'Donnell and Torre~\cite{Odonnell97}.

\begin{figure}[t!]
  \begin{center}
    \leavevmode
    \includegraphics[bb=35 80 120 216,clip=,width=10cm,height=6cm]{\myfigpath/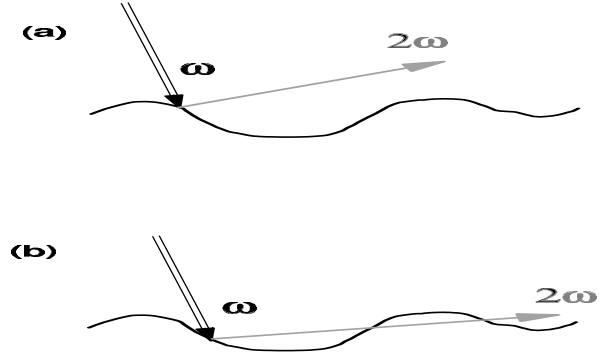}
    \caption{ Diagrams illustrating two of the single scattering processes
      that produce the second harmonic scattered light in a single
      scattering approach. The double line black arrows represent
      light of frequency $\omega$, while the thick gray arrows
      represent light of frequency $2\omega$. Notice that the process
      in Fig.~\protect\ref{Fig:Phen:SH-single-scattering-proc}b is
      unphysical.  }
    \label{Fig:Phen:SH-single-scattering-proc}
  \end{center}
\end{figure}

This can be done by applying a single scattering approximation to the
generation of the second harmonic light.  As described above, the
numerical approach leading to the theoretical results shown as solid
lines in Figs.~\ref{Fig:Phen:SH-experimental}, consists mainly of a
linear and nonlinear stage where each stage is basically solved by
some variant of the approach given in
Sect.~\ref{Sect:Theory:Sect:NumSim}.  By using a single scattering
approach, like the Kirchhoff approximation~\cite{Book:Nieto,SanGil91},
to both stages of the calculation, a single scattering approximation
for the full problem is obtained.  The single scattering processes
included in such a calculation is illustrated in
Figs.~\ref{Fig:Phen:SH-single-scattering-proc}. Notice that also
unphysical scattering processes like the one shown in
Fig.~\ref{Fig:Phen:SH-single-scattering-proc}b, are included in this
approximation.


\begin{figure}[b!]
  \begin{center}
    \leavevmode
    \includegraphics*[width=10cm,height=6cm]{\myfigpath/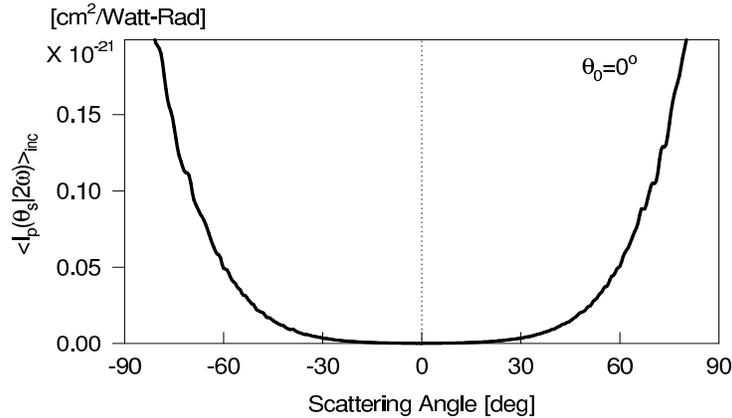}
    \caption{The mean normalized second harmonic intensity 
      $\left<I_p(\theta_s|2\omega)\right>_{inc}$ as a function of the
      scattering angle $\theta_s$ calculated in a single scattering
      approximation. The remaining parameters of the simulation were
      the same as those used in obtaining the results shown in
      Figs.~\protect\ref{Fig:Phen:SH-experimental}. The angle of
      incidence was $\theta_0 = 0^\circ$.  (After
      Ref.~\citen{Leyva-Lucero1999}.)}
    \label{Fig:Phen:SH-single-scattering}
  \end{center}
\end{figure}


In Fig.~\ref{Fig:Phen:SH-single-scattering} we present the consequence
for the angular dependence of the normalized intensity
$\left<I_p(\theta_s|2\omega)\right>$ of only including single
scattering processes in the second harmonic generation.  From this
figure it is easily seen that the intensity of the second harmonic
generated light calculated in a single scattering approximation does
{\em not} give rise to a dip (or peak) for the backscattering
direction.  In fact the overall angular dependence of
$\left<I_p(\theta_s|2\omega)\right>$ in the single scattering
approximation is quite different from the one obtained by the rigorous
approach described above.  Similar result holds for the other two
angles incidence considered in Figs.~\ref{Fig:Phen:SH-experimental}.
Hence, one may conclude that the dips present in the backscattering
direction of the incoherent component for the mean second harmonic
generated light is not due to single scattering. It therefore has to
be a multiple scattering phenomenon.

\begin{figure}[t!]
  \begin{center}
    \leavevmode
    \includegraphics[height=11cm,width=14cm]{\myfigpath/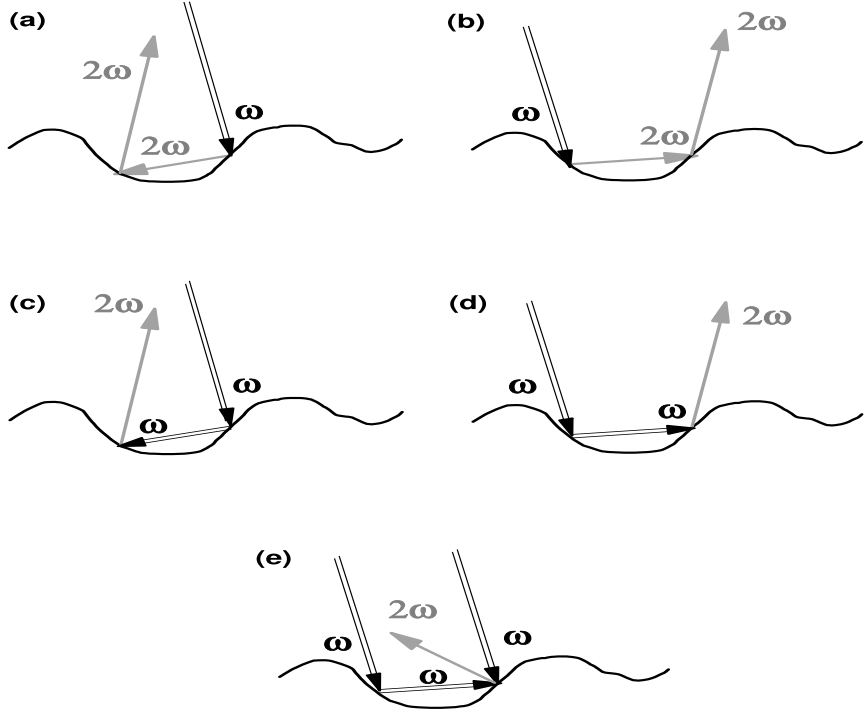}
    \caption{Diagrams illustrating some of the multiple scattering processes
      that produce the second harmonic scattered light. The double
      line black arrows represent light of frequency $\omega$, while
      the thick gray arrows represent light of frequency $2 \omega$.
      (After Ref.~\protect\citen{Leyva-Lucero1999}.).}
    \label{Fig:Phen:SH-multiple-scattering-proc}
  \end{center}
\end{figure}

To look more closely into this, the authors of
Ref.~\citen{Leyva-Lucero1999} used an iterative approach for the
linear part of the scattering problem which enabled them to calculate
the scattered fields according to the order of the scattering process.
Such a (Neumann-Liouville) iterative approach has been developed and
used earlier in the literature~\cite{Liska85,AnnPhys,Sentenac93}.  For
the nonlinear part of the calculation the rigorous simulation approach
was used and thus all higher order scattering processes were here
taken into account.  Some of the processes accounted for by this
procedure and which give rise to the second harmonic light is depicted
in Figs.~\ref{Fig:Phen:SH-multiple-scattering-proc}.


\begin{figure}[t!]
  \begin{center}
    \leavevmode
    \includegraphics[height=12cm,width=14cm]{\myfigpath/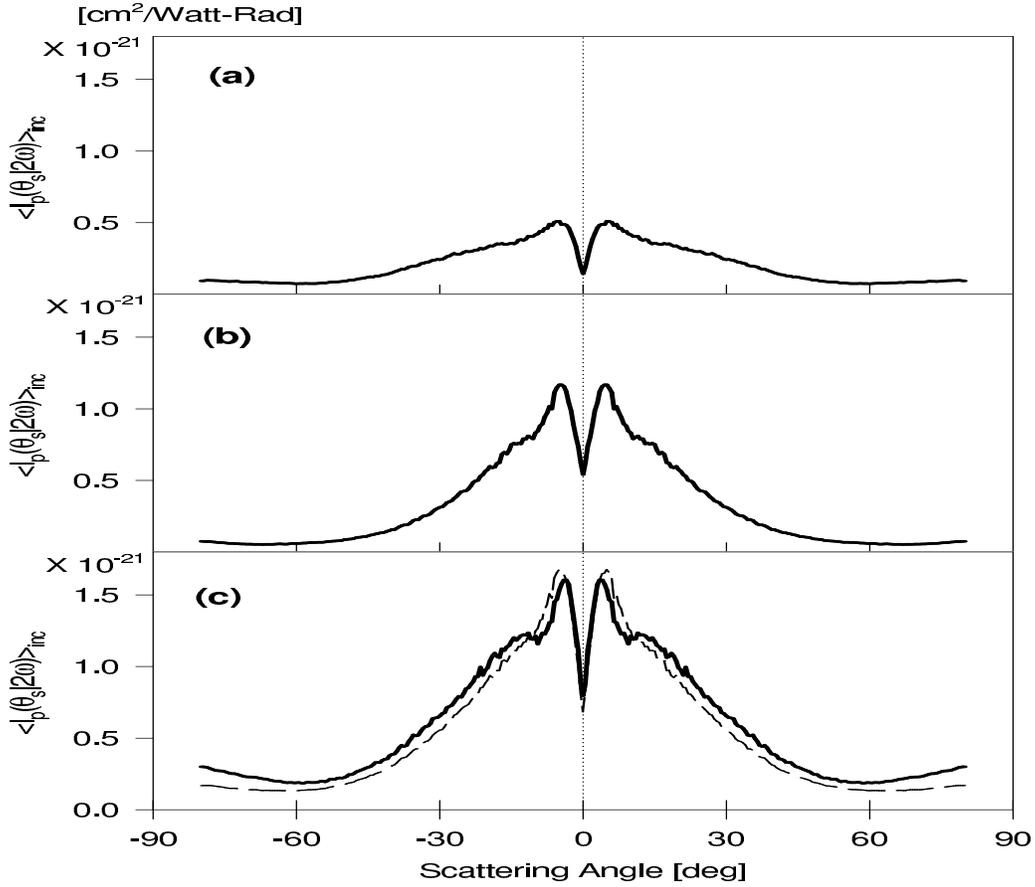}
    \vspace*{0.5cm}   
    \caption{Calculations of the mean normalized  second harmonic
      intensity as a function of the scattering angle $\theta_s$ for
      the scattering of $p$-polarized light from a random silver
      surface where the linear part of the problem was solved by
      iteration.  The incident angle of the light was
      $\theta_0=0^\circ$ and the other parameters of the simulation
      were as in Fig.~\protect\ref{Fig:Phen:SH-experimental}.  The
      curves have, (a) the single scattering contributions in the
      linear scattering and all contributions at the harmonic
      frequency, (b) pure double scattering contributions in the
      linear scattering and all contributions at the harmonic
      frequency, and (c) the single and double scattering
      contributions in the linear scattering and all contributions at
      the harmonic frequency. In (c), the curve shown with the dashed
      line represents the sum of the curves shown in (a) and (b).
      (After Ref.~\protect\citen{Leyva-Lucero1999}.)}
    \label{Fig:Phen:SH-Num}
  \end{center}
\end{figure}

We notice that the processes depicted in
Figs.~\ref{Fig:Phen:SH-multiple-scattering-proc}a and b represent
single scattering in the linear part and are thus taken properly into
account by using the standard Kirchhoff approximation~\cite{SanGil91}
(for the linear part).  However, for the paths shown in
Figs.~\ref{Fig:Phen:SH-multiple-scattering-proc}c and d, one needs to
consider a pure double scattering approximation in order to include
these processes properly.  In Figs.~\ref{Fig:Phen:SH-Num} the
simulation results for the second harmonic light
$\left<I_p(\theta_s|2\omega)\right>$ are shown for the case where a
single scattering~(Fig.~\ref{Fig:Phen:SH-Num}a) and a pure double
scattering~(Fig.~\ref{Fig:Phen:SH-Num}b) approximation is used for the
linear part of the scattering process.  In both cases dips in the
backscattering direction are observed.  In order to obtain the solid
curve of Fig.~\ref{Fig:Phen:SH-Num}c {\em both} single and double
scattering processes were taken into account for the linear part of
the calculation. This result would therefore include any interference
effect between paths like those show in
Figs.~\ref{Fig:Phen:SH-multiple-scattering-proc}a--e.  The dashed line
in Fig.~\ref{Fig:Phen:SH-Num}c is just the sum of the curves shown in
Figs.~\ref{Fig:Phen:SH-Num}a and b. It does therefore not contain any
interference effects between type~I
paths~(Figs.~\ref{Fig:Phen:SH-multiple-scattering-proc}a--b) and
type~II paths~(Figs.~\ref{Fig:Phen:SH-multiple-scattering-proc}c--d).
That the two curves shown in Fig.~\ref{Fig:Phen:SH-Num}c are so close
to each other tells us that the interference between type~I and
type~II paths are rather small (if any). Furthermore, paths of the
type illustrated in Fig.~\ref{Fig:Phen:SH-multiple-scattering-proc}e
do not seem to be important, and they do not have coherent partners.


The numerical results presented so far seem to indicate that the
behavior in the backscattering direction is affected by interference
between the paths of either type~I or type~II. In the backscattering
direction, there is no phase difference due to optical path difference
between the two type~I paths say. Similar argument hold for the
type~II paths. Hence any phase difference between the two paths has to
come from phase shifts during the reflection. In the linear
multiple-scattering processes giving rise to enhanced backscattering
the phase shifts due to reflection will be the same for the two
processes because the local Fresnel coefficients are {\em even}
functions of the angle of incidence.  Hence, the two paths in the
backscattering direction will for the full linear problem both have
the same phase and hence interferer constructively giving rise to the
celebrated enhanced backscattering pack. However, for multiple
scattering processes involving second harmonic generated light the
situation is quite different. The reason for this is that the local
nonlinear Fresnel coefficient is not an even, but an {\em odd}
function of the angle of incidence~\cite{Leyva-Lucero1999}.  Hence,
the phase difference between the two type~I paths, say, will not be
zero any more in general since the phases for these two paths will add
instead of subtract. If this phase shift is positive in
Fig.~\ref{Fig:Phen:SH-multiple-scattering-proc}a, say, then it will be
negative for the path shown in
Fig.~\ref{Fig:Phen:SH-multiple-scattering-proc}b since the local
incident angles in the two cases have different signs and the local
nonlinear Fresnel reflection coefficient is an odd function of the
incident angle. Hence in the nonlinear case the phase difference in
the backscattering direction is different from zero for the paths that
seem to interfere. From the numerical results shown in this section
they in fact seem to be close do $\pi$ out of phase resulting in
destructive interference, or a dip as compared to its background at
the backscattering direction.

\subsubsection{Weakly Rough Surfaces}

So far in this section we have presented both experimental and
numerical results for the second harmonic generated light scattered
from strongly rough surfaces.  There has also been conducted
experiments for weakly rough surfaces~\cite{Odonnell96}. The results
are quite similar to the experimental results presented in
Fig.~\ref{Fig:Phen:SH-experimental}.  In particular, also for these
weakly rough surfaces the second harmonic generated light scattered
diffusely showed a dip in the backscattering direction. However, in
theoretical studies~\cite{Mcgurn91,Leyva96,Leskova98} both dips as
well as peaks in the backscattering direction have been predicted. If
it is a peak or dip depends on the values used for the nonlinear
phenomenological constants. Even though predicted theoretically, only
dips have so far been seen in experiments.

For weakly rough surface the scattering processes giving rise to
these dips (or peaks) are believed to be different for weakly and
strongly rough surfaces. This situation resembles quite a bit the
origin of the enhanced backscattering peak for weakly and strongly
rough surfaces. Indeed, for weakly
rough surfaces the origin of the dip in the intensity of the diffusely
scattered light at frequency $2\omega$ is intimately related to the
excitation of surface plasmon polaritons at this
frequency~\cite{Leyva96,Leskova98}. Thus such dips are not to be
expected for the second harmonic light generated in $s$-polarization
from weakly rough surfaces.


\section{Directions for Future Research}

We have in the introduction to this review tried to give some glimpses
of the many multiple scattering phenomena that may take place when
electromagnetic waves are scattered by a randomly rough surface
separating two media of different dielectric properties. Even though
much is understood today when it comes to the rough surface scattering
problem, there are still, after a century of research efforts, many
questions that have not been addressed and answered properly. Below we
will therefore try to sketch out some directions for further research.

We have exclusively considered one-dimensional surfaces. Naturally
occurring surfaces are mostly two-dimensional.  Thus, the advance most
needed in the field are techniques, either numerical or analytical,
that accurately and fast can handle electromagnetic wave scattering
from two-dimensional surfaces of varying {\sc rms}-height.
Two-dimensional weakly rough surfaces can be treated by perturbation
theory~\cite{McGurn96, McGurn87}, but if the surface is not weakly
rough this approach is not adequate any more.  In principle a general
solution of the scattering problem can be formulated on the basis of a
vector version of the extinction theorem~\cite{Book:Nieto, Book:Kong}.
However, the resulting system of linear equations that needs to be
solved in order to calculate the source functions is so big, and
therefore require so much computer memory, that it for the moment is
not practical in general~\cite{Tran}. Thus, one has to come up with
new and more efficient methods for solving this kind of problems, or,
the less appealing approach, to wait for advances in computer
technology to make the extinction theorem approach tractable from a
computational point of view.

The scientific community dealing with wave scattering from disordered
systems, seems to be divided into two separate groups: ({\it i}) those
that deal with surface disordered systems and ({\it ii}) those that
concentrate on systems with volume disorder. In the future these two
``groups'' have to be unified to a much higher degree then what is the
case today in order to deal with scattering systems consisting of bulk
disorder materials bounded by a random surface. Strictly speaking
there has already been published some works for such ``dual''
disordered systems~\cite{Elson84,Elson93,Pak93,Elson97}, but still
more work, and in a more general framework, need to be done for such
problems.

An area that needs to be addressed further in the future is the {\em
  inverse scattering problem}\cite{Zierau} in contrast to the forward
scattering problem that is the one that has received the most
attention in theoretical studies of wave scattering from randomly
rough surfaces.  In the inverse scattering problem one has
information, {\it e.g.} from experiments, about the angular dependence
of the scattered light and one is interested in trying to reconstruct
the surface profile function or its statistical properties. This problem
is quite difficult and huge research efforts have been spent on it in
related fields like  remote sensing and seismic in order to try to find
its solution. So far a general solution to the problem has not been
found.

\section*{Acknowledgement}

It is a pleasure to acknowledge numerous fruitful discussion with Alex
Hansen, Ola Hunderi, Jacques Jupille, R\'emi Lazzari, Tamara A.\ 
Leskova, Alexei A.\ Maradudin, Eugenio R.  M\'endez, St\'ephane Roux,
and Damien Vandembroucq.  This research was supported in part by The
Research Council of Norway (Contract No.~32690/213), Norsk Hydro ASA,
NTNU, Total Norge ASA, Centre National de la Recherche Scientifique
(CNRS), and the Army Research Office (DAAD19-99-1-0321).

\appendix


\section{Matrix Elements}
\label{App:Matrix-element-expansion}

In this Appendix, some calculational details are presented for the
matrix elements appearing in the matrix
equations~\r{Eq:Theory:matrix-eq} used to determine the source
functions needed  in the rigorous numerical simulation approach given in
Sect.~\ref{Sect:Theory:Sect:NumSim}.

From this section, Eqs.~\r{Eq:Theory:Matrix-elements}, we recall that
these matrix elements are  defined as
\begin{subequations}
  \label{APP:Matrix-elements}
  \begin{eqnarray}
    {\cal A}^\pm_{mn} &=&
        \int^{\xi_n+\Delta\xi/2}_{\xi_n-\Delta\xi/2} dx_1'\;
                    A_\pm(\xi_m|x_1') ,  \nonumber \\
       &=& 
        \int^{\Delta\xi/2}_{-\Delta\xi/2} du\;
                    A_\pm(\xi_m|\xi_n+u),             
\qquad  \\
    {\cal B}^\pm_{mn} &=&
        \int^{\xi_n+\Delta\xi/2}_{\xi_n-\Delta\xi/2} dx_1'\;
                    B_\pm(\xi_m|x_1') ,  \nonumber \\
            &=&
        \int^{\Delta\xi/2}_{-\Delta\xi/2} du\;
                    B_\pm(\xi_m|\xi_n+u)  
       \qquad  
  \end{eqnarray}
\end{subequations}
where we in the last transition have made a change of variable
$u=x_1-\xi_n$ and where the kernels, according to
Eqs.~\r{Eq:Theory:Kernels} and \r{Eq:theory:kernel}, are given by
\begin{subequations}
\label{App:Kernels}
  \begin{eqnarray}
    A_\pm(x_1|x_1';\omega) &=& \lim_{\eta\rightarrow 0^+} 
            \left. 
              \frac{1}{4\pi}\; \gamma(x_1')
                \partial_{n'} G_\pm({\mathbf r}|{\mathbf r}'; \omega)
            \right|_{\begin{array}{l}
                           x_3=\zeta(x_1)+\eta \\
                           x_3'=\zeta(x')
                     \end{array} }
           , \qquad \\
    B_\pm(x_1|x_1';\omega) &=&  \lim_{\eta\rightarrow 0^+} 
    \left.
      \frac{1}{4\pi}\;G_\pm({\mathbf r}|{\mathbf r}'; \omega)
            \right|_{\begin{array}{l}
                           x_3=\zeta(x_1)+\eta \\
                           x_3'=\zeta(x')
                     \end{array} } ,
  \end{eqnarray}
\end{subequations}
with ${\mathbf r}=(x_1,x_3)$ and a similar expression holds for
${\mathbf r}'$, and
\begin{eqnarray}
  \xi_n = -\frac{L}{2}+\left(n-\frac{1}{2} \right) \Delta\xi,
  \qquad   n=1,2,3,\ldots, N, \qquad
\end{eqnarray}
with $\Delta\xi=L/N$.  In the above expressions $G_\pm({\mathbf
  r}|{\mathbf r}';\omega )$ denote the free-space Green's functions
for the Helmholtz equation. In 2-dimensions, as we will be considering
here, it can be written as~\cite{Book:Morse}
\begin{eqnarray}
  \label{App:Greens-func}
  G_\pm({\mathbf r} | {\mathbf r}' ; \omega) &=& 
   i \pi H_0^{(1)} \;
      \left( 
        \varepsilon_\pm\ooc{}\left|{\mathbf r}-{\mathbf r}'\right|
      \right),
\end{eqnarray}
where $H_0^{(1)}(z)$ is the Hankel function of the first kind and
zeroth-order~\cite{Book:Morse,Stegun}.
By substituting this expression for the Green's function into
Eqs.~\r{App:Kernels} for the kernels, one gets
\begin{subequations}
  \label{App:Kernel-via-Hankel}
\begin{eqnarray}
  \label{App:Kernel-via-Hankel-A}
    A_\pm (x_1|x_1';\omega) &=&  \lim_{\eta\rightarrow 0^+} 
       \left(-\frac{i}{4} \right)
        \varepsilon_\pm \frac{\omega^2}{c^2}
        \frac{ H^{(1)}_1\!\!\left( 
                            \chi_\pm(x_1|x_1') 
                      \right)
                   }{
                            \chi_\pm (x_1|x_1')    
                    } 
   \nonumber\\
   & & \qquad  \times             
           \left[ (x_1-x_1')\zeta'(x_1') 
                    - (\zeta(x_1)-\zeta(x_1')+\eta)\right],
           \qquad    \\
  \label{App:Kernel-via-Hankel-B}
    B_\pm(x_1|x_1'; \omega) &=&   \lim_{\eta\rightarrow 0^+} 
     \left(- \frac{i}{4}\right) 
            H^{(1)}_0\!\!\left( 
                           \chi_\pm(x_1|x_1') 
                       \right),
\end{eqnarray}
where we have defined 
\begin{eqnarray}
  \chi_\pm(x_1|x_1') 
          &=& \sqrt{\varepsilon_\pm(\omega)}\; \ooc{}\;
               \sqrt{(x_1-x_1')^2+(\zeta(x_1)-\zeta(x_1')+\eta)^2}. 
         \qquad   
\end{eqnarray}
\end{subequations}
Notice that since the Hankel functions are divergent for vanishing
argument~\cite{Book:Morse,Stegun}, so are the kernels $A_\pm
(x_1|x_1';\omega)$ and $B_\pm (x_1|x_1';\omega)$. However, fortunately
these singularities are integrable, so the matrix elements ${\cal
  A}^\pm_{mn}$ and ${\cal B}^\pm_{mn}$ are in fact non-singular
everywhere and in particular when $\xi_m=\xi_n$.  We will now show
this and obtain explicit expressions for these matrix elements.

We start by considering the off-diagonal elements where the kernels
are non-singular. In this case, one may approximate the integrals in
Eqs.~\r{APP:Matrix-elements} by for example the midpoint
method~\cite{NR} with the result that~($m \neq n$)
\begin{subequations}
\begin{eqnarray}
    {\cal A}^\pm_{mn} &=&  \Delta \xi \;  A_\pm (\xi_m|\xi_n; \omega), \\
    {\cal B}^\pm_{mn} &=&  \Delta \xi \;  B_\pm (\xi_m|\xi_n; \omega), 
  \end{eqnarray}
\end{subequations}
where the expressions for the kernels are understood to be taken in
the form Eqs.~\r{App:Kernel-via-Hankel}.

So now what about the diagonal elements where the kernels are
singular? In order to calculate these elements, we start by noting
that $\chi_\pm(\xi_m|\xi_m+u)$, needed in order to evaluate the matrix
element, can be written as
\begin{eqnarray}
  \label{chi-exp}
    \chi_\pm(\xi_m|\xi_m+u) \!
         &=& \!  \sqrt{\varepsilon_\pm}\; \ooc{}\;
             \sqrt{ u^2 + \left(\zeta'(\xi_m)u +
         \frac{1}{2}\zeta''(\xi_m)u^2+\ldots+\eta\right)^2},
              \nonumber \\
        &=& \sqrt{ \gamma(\xi_m) u^2 -2\eta \zeta'(\xi_m)u +\eta^2
         +\ldots }
          \nn \\
        &=& \sqrt{\varepsilon_\pm}\; \ooc{} \gamma(\xi_m) |u|+\ldots
\end{eqnarray}
where we have Taylor expanded $\zeta(\xi_m+u)$ and where we recall from
Eq.~\r{Eq:Theory:gamma-def} that $\gamma(x_1)=\sqrt{1+\zeta'^2(x_1)}$.
Furthermore, by advantage of the following (small argument) asymptotic
expansions for the Hankel functions~\cite{Stegun}
\begin{subequations}
\label{App:H-expansion}
\begin{eqnarray}
  \label{Hankel-10}
  H^{(1)}_0(z) &=& \frac{2i}{\pi}\left( \ln\frac{z}{2}+\gamma\right)+1+ 
       {\cal O}(z^2\ln z), \\
  \label{Hankel11}     
  \frac{H^{(1)}_1(z)}{z} &=& -\frac{2i}{\pi} \frac{1}{z^2} +
  \frac{i}{\pi} \left( \ln\frac{z}{2} + \gamma +\frac{1}{2} \right)
  - \frac{1}{2} + {\cal O}(z^2\ln z), \qquad 
\end{eqnarray}
\end{subequations}
where $\gamma=0.5772157\ldots$ is the Euler constant.

With these expressions, it is now rather straight forward to derive,
to obtain the matrix elements by integrating the resulting expressions
term-by-term. To demonstrate this we start with the ${\cal
  B}^\pm_{mm}$ matrix element. With Eqs.~\r{chi-exp} and \r{Hankel-10}
and passing to limit $\eta\rightarrow 0^+$ whenever no singularities
results and one gets
\begin{eqnarray}
  {\cal B}^\pm_{mm} &=& 2\int_0^{\Delta\xi/2} du \;B_\pm(\xi_m|\xi_m+u)
  \nn \\
    &\simeq&  -\frac{i}{2} \int_0^{\Delta\xi/2} du  \; H^{(1)}_0\left(
  \sqrt{\varepsilon_\pm} \ooc{} \gamma(\xi_m)u \right)
   \nn \\
    &=& -\frac{i}{2} \int_0^{\Delta\xi/2} du  
          \left[ \frac{2i}{\pi} \left\{ \ln
  \left(\sqrt{\varepsilon_\pm}\ooc{} \gamma(\xi_m)u \right\} 
             + \gamma \right) +1+ \ldots \right]
   \nn \\
    &=& - \frac{i}{2} \frac{\Delta\xi}{2}   
          \left[ \frac{2i}{\pi} \left\{ 
       \ln \left(\sqrt{\varepsilon_\pm}\ooc{}\frac{\gamma(\xi_m)\Delta\xi}{2e} \right)
             +\gamma\right\} +1  + \ldots  
   \right]
   \nn \\
     &\simeq& -\frac{i}{4} \Delta\xi \, H^{(1)}_0\left(
                   \sqrt{\varepsilon_\pm} \ooc{} \frac{\gamma(\xi_m)\Delta\xi}{2e} \right).
\end{eqnarray}
Here in the last transition we have Eq.~\r{Hankel-10} one
more. 

Furthermore, for the leading term of the diagonal elements of ${\cal
  A}$ one gets in a similar way from Eqs.~\r{Hankel11} and \r{chi-exp}
\begin{eqnarray}
    {\cal a}^\pm_{mm} &=& \int^{\Delta\xi/2}_{-\Delta\xi/2}  du \;A_\pm(\xi_m|\xi_m+u)
  \nn \\ 
 &=& 
    \frac{i}{4} \varepsilon_\pm\ooc{2}
    \lim_{\eta\rightarrow 0^+} 
   \int^{\frac{\Delta \xi}{2}}_{-\frac{\Delta \xi}{2}} du
    \left[
      -\frac{2i}{\pi} \frac{1}{\chi^2_\pm(\xi_m|\xi_m+u)}
      + \ldots
    \right]
  \nn \\ && \hspace{3.5cm}\times
    \left[
      \eta+ \frac{1}{2} \zeta''(\xi_m)u^2+\ldots
    \right] 
   \nn \\
   &=& 
    \lim_{\eta\rightarrow 0^+} \frac{1}{2\pi} 
   \int^{\frac{\Delta \xi}{2\eta}}_{-\frac{\Delta \xi}{2\eta}} du\;
      \frac{1}{\gamma^2(\xi_m)u^2-2\zeta'(\xi_m)u+1} 
    \nn \\ && \mbox{}
    +  \frac{1}{4\pi}\frac{\zeta''(\xi_m)}{\gamma^2(\xi_m)}
       \int^{\frac{\Delta \xi}{2}}_{-\frac{\Delta \xi}{2}} du
       \nn \\ &=& 
   \frac{1}{2\pi}\lim_{\eta\rightarrow 0^+} 
   \left[
       \tan^{-1}\left(-\zeta'(\xi_m)+\gamma(\xi_m)u\right)
   \right]^{\frac{\Delta \xi}{2\eta}}_{u=-\frac{\Delta \xi}{2\eta}}
     + \Delta\xi\frac{\zeta''(\xi_m)}{4\pi\gamma^2(\xi_m)}
       \nn \\ &=& 
       \frac{1}{2} + \Delta\xi\frac{\zeta''(\xi_m)}{4\pi\gamma^2(\xi_m)}
\end{eqnarray}

To sum up we have for the matrix elements
\begin{eqnarray}
  \label{App:A-matrix-elemnets}
    {\cal A}^\pm_{mn} 
       &=&
       \left\{
     \begin{array}{ll}
      \Delta \xi \, A_\pm(\xi_m|\xi_n),   & \quad m\neq n, \\
            \frac{1}{2} + \Delta \xi\;
            \frac{\zeta''(x_m)}{4\pi \gamma^2(\xi_m)},
              & \quad  m=n, \\
   \end{array}
       \right. \qquad
    \label{A-matrix} 
\end{eqnarray}
and 
\begin{eqnarray}
  \label{App:B-matrix-elemnets}
    {\cal B}^{\pm}_{mn} 
       &=&
       \left\{
     \begin{array}{ll}
      \Delta \xi \, B_\pm(\xi_m|\xi_n),   & \quad m \neq n \\
      -\frac{i}{4} \Delta\xi \, H^{(1)}_0\left(
          \sqrt{\varepsilon_\pm} \ooc{} \frac{\gamma(\xi_m)\Delta\xi}{2e} \right),
     & \quad  m=n. \\
   \end{array}
       \right. \qquad
    \label{B-matrix} 
\end{eqnarray}
In these equations $A_\pm(\xi_m|\xi_n)$ and $B_\pm(\xi_m|\xi_n)$ are
given by Eqs.~\r{App:Kernels}


\section{The $\chi$-functions used in Small Amplitude Perturbation Theory}
\label{APP:SAPT}

In this appendix some of the lengthy formulae found in small amplitude
perturbation theory, Sect.~\ref{Sect:Theory:SAPT}, are given. In
particular we here give the first few $\chi$-functions found in
Eqs.~\r{Eq:theory:SAPT:R-first-terms}. We will now in the next two
subsection explicitly give these functions for $p$ and
$s$-polarization.  All explicit reference to the frequency $\omega$
has been suppressed. We have also for completeness used
$\varepsilon_0$ for the dielectric constant of the upper medium. In
the case of vacuum this constant is  $\varepsilon_0=1$.   

\subsection{P-polarization}
The three first functions in the set $\{\chi_p^{(n)}\}$
are~\cite{Shchegrov-PhD-thesis}:
\begin{subequations}
  \begin{eqnarray}
    \chi^{(1)}_p(q|k) &=& i
           \frac{\varepsilon_{0}-\varepsilon_{}
              }{
                   \varepsilon_{}\alpha_0(q)+\varepsilon_{0}\alpha(q)
               } 
            \left[\varepsilon_{0}\alpha(q)\alpha(k)-\varepsilon_{}qk\right]
         \nn  \hspace*{1cm} \\
       &&  \hspace*{2cm} \times  
         \frac{2\alpha_0(k)}{\varepsilon_{}\alpha_0(k)+\varepsilon_{0}\alpha(k)},
  \end{eqnarray}
  \begin{eqnarray}
    \chi^{(2)}_p(q|p_1|k) &=& 
           \frac{
                \varepsilon_{0}-\varepsilon_{}
              }{
                \varepsilon_{}\alpha_0(q)+\varepsilon_{0}\alpha(q)
               }
         \left\{
              \varepsilon_{}\alpha(q)\left[\alpha^2_0(k)-qk\right]
          +   \varepsilon_{0}\alpha(k)\left[\alpha^2(q)-qk\right] 
         \right\} 
         \frac{2\alpha_0(k)}{\varepsilon_{}\alpha_0(k)+\varepsilon_{0}\alpha(q)}  
    \nn \\ && \mbox{}
        + 2 \frac{ 
                  \left(\varepsilon_{0}-\varepsilon_{}\right)^2
              }{
                   \varepsilon_{}\alpha_0(q)+\varepsilon_{0}\alpha(q)
               }
            \frac{
                  \alpha(q)\alpha_0(p_1)+qp_1
               }{
                  \varepsilon_{}\alpha_0(p_1)+\varepsilon_{0}\alpha(p_1)
                }
            \frac{
                  2\alpha_0(k)
                   \left[
                       \varepsilon_{0}\alpha(p_1)\alpha(k)-\varepsilon_{}p_1k
                   \right]    
               }{
                  \varepsilon_{}\alpha_0(k)+\varepsilon_{0}\alpha(k)
                }
  \end{eqnarray}
  \begin{eqnarray}
    \chi^{(3)}_p(q|p_1|p_2|k) &=& -i
         \frac{
                \varepsilon_{0}-\varepsilon_{}
              }{
                \varepsilon_{}\alpha_0(q)+\varepsilon_{0}\alpha(q)
               }
         \left\{
              2\varepsilon_{}\alpha^2(q)\alpha^2_0(k) 
             + \left[ \alpha^2(q) + \alpha^2_0(k) \right]
               \left[
         \varepsilon_{0}\alpha(q)\alpha(k)-\varepsilon_{}qk
         \right]
      \nn \right. \\ && \mbox{} \hspace{3cm} \left.
             -2\varepsilon_{0}qk\alpha(q)\alpha(k) 
         \right\} 
          \frac{
                  2\alpha_0(k)  
               }{
                  \varepsilon\alpha_0(k)+\varepsilon_{0}\alpha(k)
                }
      \nn \\ && \mbox{} 
        - 3i  \frac{
                \varepsilon_{0}-\varepsilon_{}
              }{
                \varepsilon_{}\alpha_0(q)+\varepsilon_{0}\alpha(q)
               }
             \left[\alpha(q)\alpha_0(p_1)+qp_1\right]
            \frac{
                \varepsilon_{0}-\varepsilon_{}
              }{
                \varepsilon_{}\alpha_0(p_1)+\varepsilon_{0}\alpha(p_1)
               } 
     \nn  \\ && \mbox{} \qquad  \times
           \left\{
               \varepsilon_{}\alpha(p_1)\left[\alpha^2_0(k)-p_1k\right]
             + \varepsilon_{0}\alpha(k)\left[\alpha^2(p_1)-p_1k\right]
           \right\}
           \frac{2\alpha_0(k)}{\varepsilon_{}\alpha_0(k)+\varepsilon_{0}\alpha(k)}
      \nn \\ && \mbox{} 
          -i \left\{
               3 \frac{
                      \varepsilon_{0}-\varepsilon_{}
                    }{
                      \varepsilon_{}\alpha_0(q)+\varepsilon_{0}\alpha(q)
                     }
                    \left[\alpha(q)\alpha_0(p_2)+qp_2\right]
                    \left[\alpha(q)-\alpha_0(p_2)\right]
      \nn \right. \\ && \mbox{}  \qquad 
             +6 \frac{
                      \varepsilon_{0}-\varepsilon_{}
                    }{
                      \varepsilon_{}\alpha_0(q)+\varepsilon_{0}\alpha(q)
                     }
                   \left[\alpha(q)\alpha_0(p_1)+qp_1\right]
      \nn  \\ && \mbox{}  \qquad \quad \times \left.
                   \frac{
                         \varepsilon_{0}-\varepsilon_{}
                        }{
                         \varepsilon_{}\alpha_0(p_1)+\varepsilon_{0}\alpha(p_1)
                        }
                      \left[\alpha(p_1)\alpha_0(p_2)+p_1p_2\right]
           \right\}
     \nn  \\ && \mbox{}  \quad \times 
           \frac{
                   \varepsilon_{0}-\varepsilon_{}
                 }{
                    \varepsilon_{}\alpha_0(p_2)+\varepsilon_{0}\alpha(p_2)
                  }
                \left[\varepsilon_0\alpha(p_2)\alpha(k)-\varepsilon p_1k\right]
                \frac{2\alpha_0(k)}{\varepsilon_{}\alpha_0(k)+\varepsilon_{0}\alpha(k)},
  \end{eqnarray}
\end{subequations}

\subsection{S-polarization}

Here follows the corresponding expressions for 
$s$-polarization~\cite{Shchegrov-PhD-thesis}:
\begin{subequations}
  \begin{eqnarray}
    \chi^{(1)}_s(q|k) &=& -i \ooc{2}
           \frac{\varepsilon_{0}-\varepsilon
              }{
                   \alpha_0(q)+\alpha(q)
               } 
         \frac{2\alpha_0(k)}{\alpha_0(k)+\alpha(k)},
  \end{eqnarray}
  \begin{eqnarray}
    \chi^{(2)}_s(q|p_1|k) &=& -\ooc{2}
           \frac{
                \varepsilon_{0}-\varepsilon_{}
              }{
                \alpha_0(q)+\alpha(q)
               }
         \left(
              \alpha(q)+\alpha(k)
             +2\ooc{2} \frac{
                              \varepsilon_{0}-\varepsilon_{}
                            }{
                              \alpha_0(p_1)+\alpha(p_1)
                             }
         \right) 
         \frac{2\alpha_0(k)}{\alpha_0(k)+\alpha(k)}  
   \end{eqnarray}
  \begin{eqnarray}
    \chi^{(3)}_s(q|p_1|p_2|k) &=& i \ooc{2}
         \frac{
                \varepsilon_{0}-\varepsilon_{}
              }{
                \alpha_0(q)+\alpha(q)
               }
         \left\{
               \alpha^2(q)+2\alpha(q)\alpha(k)+\alpha_0^2(q)
          +3  \ooc{2}
              \frac{
                      \varepsilon_{0}-\varepsilon_{}
                   }{
                      \alpha_0(p_1)+\alpha(p_1)
                    }
             \left[\alpha(p_1)+\alpha(k)\right]
      \nn \right. \\ && \mbox{} \hspace{3cm}\left.
       + 3\left[ \alpha(q)-\alpha_0(p_1)\right]
               + 2\ooc{2}
                    \frac{
                           \varepsilon_{0}-\varepsilon_{}
                         }{
                            \alpha_0(p_1)+\alpha(p_1)
                          }
                  \ooc{2}
                    \frac{
                           \varepsilon_{0}-\varepsilon_{}
                         }{
                            \alpha_0(p_2)+\alpha(p_2)
                          } 
        \right\}
         \frac{2\alpha_0(k)}{\alpha_0(k)+\alpha(k)} 
      \nn \\        
   \end{eqnarray}
\end{subequations}

With these expressions we close this appendix.

\bibliographystyle{plain}

\end{document}